\documentclass[a4paper,11pt]{article}
\pdfoutput=1
\usepackage{jheppub}
\usepackage[T1]{fontenc}
\usepackage{amsmath, amsthm, amssymb, mathtools}
\usepackage{pifont}
\usepackage{dsfont}
\usepackage{amsfonts}
\usepackage{url}
\usepackage{bbm}
\usepackage{booktabs}
\usepackage{multirow}
\usepackage{bm}
\usepackage{amssymb}
\usepackage{mathrsfs}
\usepackage{amsthm}
\usepackage{adjustbox}
\usepackage{harpoon}
\usepackage{enumitem}
\usepackage{colortbl}
\usepackage{lscape}
\usepackage{makecell}
\usepackage{pdflscape}
\usepackage{adjustbox}
\usepackage{hhline}
\usepackage{longtable}
\usepackage{diagbox}
\usepackage{rotating}
\newcommand{\ignore}[1]{}

\def\c#1#2{\ensuremath{c_{#1#2}}}

\def\nn{\nonumber}

\newcommand{\newc}{\newcommand}
\newc{\be}{\begin{equation}}
\newc{\ee}{\end{equation}}
\newc{\bea}{\begin{eqnarray}}
\newc{\eea}{\end{eqnarray}}
\newc{\simlt}{~\mbox{\smaller\(\lesssim\)}~}
\newc{\simgt}{~\mbox{\smaller\(\gtrsim\)}~}

\def\beq{\begin{equation}}
\def\eeq#1{\label{#1}\end{equation}}
\def\eeqn{\end{equation}}
\def\beqa{\begin{eqnarray}}
\def\eeqa#1{\label{#1}\end{eqnarray}}
\def\eeqan{\end{eqnarray}}

\newcommand{\CP}{$\mathcal{CP}$\,}

\allowdisplaybreaks

\title{\Large Neutrino Mass and Mixing with Modular Symmetry\note{Review article
submitted for publication in Reports on Progress in Physics}}

\author[a]{Gui-Jun Ding}
\author[b]{Stephen F. King}
\affiliation[a]{Department of Modern Physics, University of Science and Technology of China,\\
Hefei, Anhui 230026, China}
\affiliation[b]{Department of Physics and Astronomy,
University of Southampton,\\
Southampton, SO17 1BJ, U.K.}
\emailAdd{dinggj@ustc.edu.cn}
\emailAdd{king@soton.ac.uk}


\abstract{This is a review article about neutrino mass and mixing and flavour model building strategies based on modular symmetry. After a brief survey of neutrino mass and lepton mixing, and various Majorana seesaw mechanisms, we construct and parameterise the lepton mixing matrix and summarise the latest global fits, before discussing the flavour problem of the Standard Model. We then introduce some simple patterns of lepton mixing, introduce family (or flavour) symmetries, and show how they may be applied to direct, semi-direct and tri-direct CP models, where the simple patterns of lepton mixing, or corrected versions of them, may be enforced by the full family symmetry or a part of it, leading to mixing sum rules. We then turn to the main subject of this review, namely a pedagogical introduction to modular symmetry as a candidate for family symmetry, from the bottom-up point of view. After an informal introduction to modular symmetry, we introduce the modular group, and discuss its fixed points and residual symmetry, assuming supersymmetry throughout. We then introduce finite modular groups of level $N$ and modular forms with integer or rational modular weights, corresponding to simple geometric groups or their double or metaplectic covers, including the most general finite modular groups and vector-valued modular forms, with detailed results for $N=2, 3, 4, 5$. The interplay between modular symmetry and generalized CP symmetry is discussed, deriving CP transformations on matter multiplets and modular forms, highlighting the CP fixed points and their implications. In general, compactification of extra dimensions generally leads to a number of moduli, and modular invariance with factorizable and non-factorizable multiple moduli based on symplectic modular invariance and automorphic forms is reviewed. Modular strategies for understanding fermion mass hierarchies are discussed, including the weighton mechanism, small deviations from fixed points, and texture zeroes. Then examples of modular models are discussed based on single modulus $A_4$ models, a minimal $S'_4$ model of leptons (and quarks), and a multiple moduli model based on three $S_4$ groups capable of reproducing the Littlest Seesaw model. We then extend the discussion to include Grand Unified Theories (GUTs) based on modular (flipped) $SU(5)$ and $SO(10)$. Finally we briefly mention some issues related to top-down approaches based on string theory, including eclectic flavour symmetry and moduli stabilisation, before concluding. }

\begin{document}
\maketitle
\flushbottom

\section{\label{Intro}Introduction to neutrino mass and lepton mixing}

Neutrino physics holds a unique place in particle physics as the only sector which demands new physics beyond the Standard Model (BSM). We know this is the case since the origin of neutrino mass and mixing is not present in the SM and is so far unknown. Moreover, neutrino physics forms a part of the flavour puzzle, the unexplained origin of the three fermion famlies, and their puzzling pattern of hierarchical masses and mixing parameters, namely the mixing angles and CP violating phases. In this section we first give a brief survey of neutrino mass and lepton mixing, and then discuss various Majorana seesaw mechanisms. We then construct and parameterise the lepton mixing matrix and summarise the latest global fits, before discussing the flavour problem of the Standard Model.

\subsection{A brief survey of neutrino mass and mixing}

A brief history of neutrino oscillation discoveries in the period 1998-2012 indicates the rapid development
of this field, from the SM starting point of zero neutrino mass and mixing, to the BSM situation where
three neutrino mixing is required:
\begin{itemize}
\item 1998 Atmospheric $\nu_{\mu}$ disappear, large atmospheric mixing (SK)
\item  2002 Solar $\nu_{e}$ disappear, large solar mixing (SK, after Homestake and Gallium)
\item 2002 Solar $\nu_{e}$ appear as $\nu_{\mu}$ and $\nu_{\tau}$ (SNO)
\item 2004 Reactor $\overline{\nu_{e}}$ oscillations observed (KamLAND)
\item 2004 Accelerator $\nu_{\mu}$ disappear (K2K)
\item 2006 Accelerator $\nu_{\mu}$ disappearance confirmed (MINOS)
\item 2010 Accelerator $\nu_{\mu}$ appear as $\nu_{\tau}$  (OPERA)
\item 2011 Accelerator $\nu_{\mu}$ appear as $\nu_{e}$ (T2K, MINOS)
\item 2012 Reactor $\overline{\nu_{e}}$ disappear, reactor mixing measured (Daya Bay, RENO)
\end{itemize}

\begin{figure}[tb]
\centering
\includegraphics[width=0.6\textwidth]{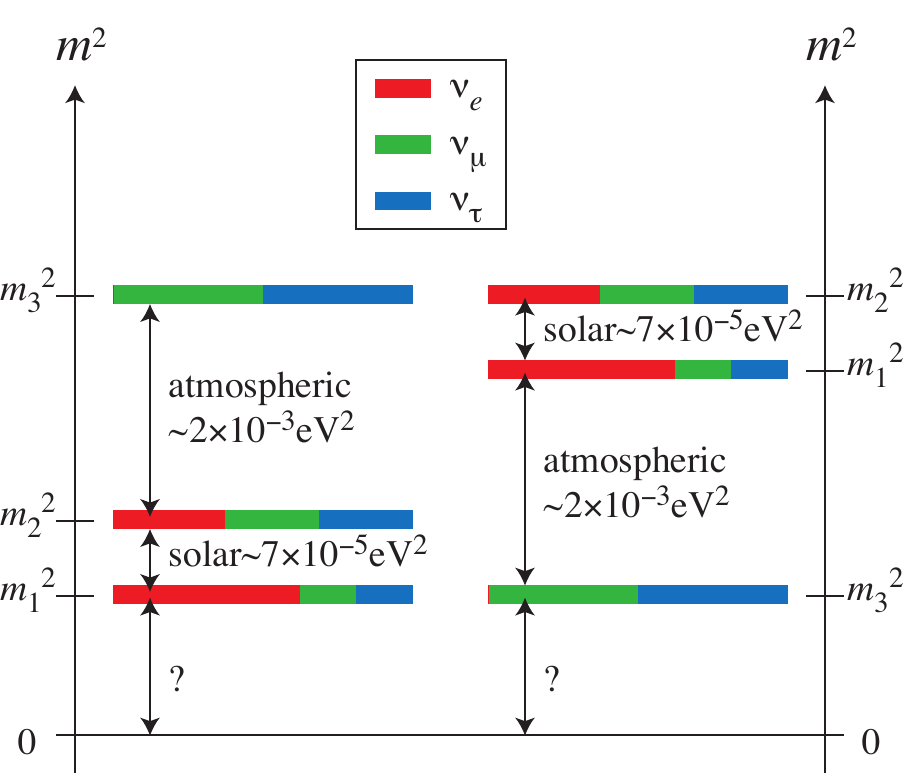}
 \caption{\label{fig:massorderings}\small{Neutrino oscillation data allows two possible orderings
 of neutrino squared masses (normal on the left and inverted on the right). In each case the colour coded
 $\nu_e , \nu_{\mu} , \nu_{\tau}$
 probability composition
 of each of the mass eigenstates is shown. Neutrino oscillation data is consistent with an approximate tri-bimaximal mixing pattern, namely roughly
 equal $\nu_{\mu} - \nu_{\tau}$ composition for one mass eigenstate and roughly
 equal $\nu_e - \nu_{\mu} - \nu_{\tau}$ for another.
 Neutrino oscillation experiments do not determine the absolute neutrino mass scale,
 only the mass squared differences.}}
\end{figure}

It is known that neutrino oscillations only depend on the two mass squared differences $\Delta m^2_{21}\equiv m_2^2-m_1^2$, which is constrained by data to be positive, and $\Delta m^2_{31}\equiv m_3^2-m_1^2$, which current data allows to take a positive (normal) or negative (inverted) value.
The neutrino oscillation data may be further parameterised by three lepton mixing angles, and one CP violating phase, whose precise definitions will be discussed later. The 1998 atmospheric neutrino oscillation data was consistent with equal bi-maximal $\nu_{\mu}- \nu_{\tau}$ mixing,
corresponding to a maximal atmospheric mixing angle $\theta_{23}$ of roughly $45^{\circ}$~\cite{Super-Kamiokande:1998kpq}. The 2002 solar neutrino oscillation data was consistent with equal tri-maximal $\nu_e-\nu_{\mu}-\nu_{\tau}$ mixing, corresponding to a solar mixing angle $\theta_{12}$ determined to be roughly $35^{\circ}$~\cite{SNO:2001kpb}. The 2012 reactor oscillation experiments were consistent with a reactor mixing angle $\theta_{13}$ of around $8.5^{\circ}$~\cite{DayaBay:2012fng,RENO:2012mkc}. Note that no single experiment to date is able to determine the CP violating oscillation phase.
Our present understanding of neutrino mass and mixing is summarised in figure~\ref{fig:massorderings}.

These milestones have led to a remarkable transformation of our knowledge of the neutrino sector, from the pre-1998 situation where neutrino masses could be zero, in accordance with the SM expectation, to our present understanding which may be characterised as follows:
\begin{itemize}
\item Neutrinos have exceptionally small masses, a million times smaller than $m_e$.
\item At least two neutrino masses are not so hierarchical, unlike charged fermion masses.
\item Neutrino masses break individual lepton numbers $L_e , L_{\mu} , L_{\tau}$
\item Neutrino masses may or may not respect total lepton number $L=L_e + L_{\mu} + L_{\tau}$,
depending on them being Dirac or Majorana in nature.
\item Neutrinos are observed to mix a lot, within a three-neutrino mixing paradigm, where the largest (atmospheric
and solar) mixing angles are much larger than any mixing in the quark sector, and the smallest lepton mixing angle
(the reactor angle)  being comparable to the largest quark mixing angle (the Cabibbo angle).
\end{itemize}

Although we have learned quite a lot about the lepton mixing angles and neutrino masses, it is worth summarising what we still don't know:
\begin{itemize}
\item Is leptonic \CP symmetry violated?
\item Does $\theta_{23}$ belong to the first octant ($<45^o$) or the second octant ($>45^o$)?
\item Are the neutrino mass squareds normal ordered (NO) or inverted ordered (IO)?
\item What is the lightest neutrino mass value?
\item Are the neutrino masses of the Dirac or Majorana type?
\end{itemize}

In the SM, there are three neutrinos $\nu_e$, $\nu_{\mu}$, $\nu_{\tau}$ which all massless and are distinguished by separate lepton numbers $L_e$, $L_{\mu}$, $L_{\tau}$. The neutrinos and antineutrinos are distinguished by total lepton number $L=\pm 1$. Clearly we must go beyond the SM to understand the origin of the tiny neutrino masses. It is important to recall that, in the SM, neutrinos are massless for three reasons:
\begin{itemize}
\item There are no right-handed (RH) sterile neutrinos $\nu_R$ in the SM;
\item In the SM there are no Higgs in $SU(2)_L$ triplet representation;
\item The SM Lagrangian is renormalisable.
\end{itemize}
In order to account for neutrino mass and mixing, at least one or more of these conditions must therefore be relaxed. For instance, if RH (sterile) neutrinos $\nu_R$ are included, then the usual Higgs mechanism of the SM
yields Dirac neutrino mass in the same way as for the electron mass $m_e$.
This would break $L_e$, $L_{\mu}$, $L_{\tau}$, but preserve $L$ and $B-L$. According to this simplest possibility, the Yukawa term would be
$Y_{\nu}\overline{L}H_u\nu_R$, in the standard notation, and $Y_{\nu}\sim 10^{-12}$. By comparison, the electron mass has a Yukawa coupling eigenvalue $Y_e$ of about $10^{-6}$.

According to the above argument, Dirac neutrino Yukawa couplings are a million times smaller than that of the electron. However there have been many attempts which yield Dirac neutrinos without relying on such tiny Yukawa couplings
\cite{Arkani-Hamed:2000oup,Murayama:2002je,Thomas:2006gr,Gu:2006dc,Gu:2007gy,Langacker:2011bi,Memenga:2013vc,Chen:2012baa,Farzan:2012sa,Ding:2013eca,Aranda:2013gga,Wang:2016lve,Kanemura:2016ixx,Fujimoto:2016gfu,Valle:2016kyz,Ma:2016mwh,Borah:2017leo,Wang:2017mcy,Yao:2017vtm,Yao:2018ekp,Bolton:2019bou,Saad:2019bqf,Earl:2019wjw}.
For example, a Dirac seesaw mechanism has been proposed as an ultraviolet completion of the dimension-five operator $\frac{1}{\Lambda}\bar{L} \nu_R H S$, where $S$ is a scalar and $\Lambda$ denotes the scale of heavy intermediate particles, and renormalisable Dirac terms are forbidden by an extra symmetry. Although there are several types of Dirac seesaw mechanisms, the minimal models all rely on a $Z_2$   symmetry~\cite{Yao:2018ekp}, under which $\nu_R$ and $S$ are odd, where the $S$ vacuum expectation value (VEV) spontaneously breaks the symmetry.
This leads to cosmological domain wall formation, where a soft $Z_2$ breaking term leads to their annihilation, resulting in gravitational waves with a peaked spectrum~\cite{King:2023cgv}, unlike those produced from spontaneous breaking of $U(1)_{B-L}$ or $U(1)_{L}$ in Majorana models which leads to cosmic strings and a flatter spectrum.

As mentioned above, neutrinos may have a Majorana mass which would break $L$ and $B-L$, leading to neutrinoless double beta decay. Indeed, having introduced RH neutrinos (also called sterile neutrinos, since they are SM singlets), something must prevent (large) Majorana mass terms $M_{R} \nu_R \nu_R$ where $M_{R}$ could take any value up to the Planck scale. A conserved symmetry such as $U(1)_{B-L}$ would forbid RH neutrino masses,
but if gauged (in order to be a robust symmetry) it would have to be broken at the TeV scale or higher, allowing Majorana masses at the $U(1)_{B-L}$ breaking scale. If the symmetry is broken spontaneously this would also lead to cosmic strings and gravitational waves.

Indeed it is possible to generate left-handed (LH) Majorana neutrino masses,
even without RH neutrinos. For instance, introducing a Higgs triplet $\Delta$ (written as a $2\times 2$ matrix), LH Majorana neutrino masses arise from the term $y_M  L^T (\Delta ) L$, where $y_M$ is a dimensionless coupling. Majorana masses occur once the lepton doublets $L$ are contracted with the neutral component of the Higgs triplet which develops a VEV.

Majorana mass can also arise from non-renormalisable dimension five operators proportional to a dimensionful coupling constant $\kappa$, first proposed by Weinberg~\cite{Weinberg:1980bf},
\begin{equation}
-\kappa L^T (HH) L= -\frac{1}{2}\left( \frac{\lambda}{\Lambda} \right) L^T (HH) L\,,
\label{dim5}
\end{equation}
where $\lambda $ is a dimensionless coupling constant, $\Lambda$ is a mass scale and $(HH)$ is an $SU(2)_L$ triplet combination of two Higgs doublets (written as a $2\times 2$ matrix). This is a non-renormalisable operator, which is the lowest dimension operator which may be added to the renormalisable SM Lagrangian. We require
$\left( \frac{\lambda}{\Lambda} \right) \sim 1/(10^{14} {\rm GeV}) $ for $m^{\nu}\sim 0.1$ eV. The elegant type I seesaw mechanism~\cite{Minkowski:1977sc,Gell-Mann:1979vob,Yanagida:1979as,Glashow:1979nm,Mohapatra:1979ia,Schechter:1980gr} identifies the mass scale $\Lambda$ with the RH neutrino Majorana mass $\Lambda=M_N$, and $\lambda $ with the product of Dirac Yukawa couplings $\lambda=Y_{\nu}^2$. Of course the situation is rather more complicated in practice since there may be three RH neutrinos and both $M_N$ and $Y_{\nu}$ may be a $3\times 3 $ matrices. In this case, after integrating out the RH neutrinos~\cite{Minkowski:1977sc,Gell-Mann:1979vob,Yanagida:1979as,Glashow:1979nm,Mohapatra:1979ia,Schechter:1980gr},
we reproduce the Weinberg operator in Eq.~\eqref{dim5}. This is called the type I seesaw mechanism and is discussed in more detail in the next subsection.

\begin{figure}[tb]
\centering
\includegraphics[width=0.56\textwidth]{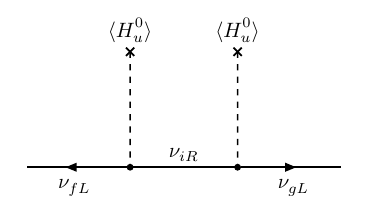}
 \caption{\label{fig:TypeIDiagrams} Diagram illustrating the type I (or more precisely type Ia) seesaw mechanism. The type Ib seesaw mechanism involves replacing one of the two Higgs doublets $H_u$ with a charge conjugated second Higgs doublet $H^c_d$. In the type III seesaw mechanism, the electroweak singlet RH neutrinos are replaced with fermion electroweak triplets.}
\end{figure}

In general there are three classes of proposals in the literature
for the new physics at the scale $\Lambda$:
\begin{itemize}
\item Several types of Majorana seesaw mechanisms~\cite{Minkowski:1977sc,Gell-Mann:1979vob,Yanagida:1979as,Glashow:1979nm,Mohapatra:1979ia,Schechter:1980gr,Magg:1980ut,Lazarides:1980nt,Mohapatra:1980yp,Wetterich:1981bx,Foot:1988aq,Hernandez-Garcia:2019uof};
also in addition low (TeV) scale seesaw mechanisms~\cite{King:2004cx}
(with the Weinberg operator resulting from the mass $M$ of a heavy particle exchanged at tree-level with $\Lambda=M$). For example the type I seesaw mechanism is illustrated diagrammatically in figure~\ref{fig:TypeIDiagrams}. To be precise, this is the type Ia seesaw mechanism since it involves two insertions of the same Higgs doublet $H_u$. The type Ib seesaw mechanism involves replacing one of the Higgs doublets
by a charge conjugated second Higgs doublet $H^c_d$ in the framework of a two Higgs doublet model~\cite{Hernandez-Garcia:2019uof}, leading to a new kind of effective Weinberg operator involving the factor $(H_uH^c_d)$.
\item $R$-parity violating supersymmetry~\cite{Drees:1997id}
($\Lambda=$TeV Majorana mass neutralinos $\chi$).
\item Loop mechanisms involving new fields with masses of order the TeV-scale~\cite{Zee:1980ai,Babu:1988ki}
(in which the Weinberg operator arises from loop diagrams involving
additional Higgs doublets/singlets);
\end{itemize}
In addition there are two classes of early\footnote{We shall discuss some recent developments later.} string-inspired explanations for neutrino mass:
\begin{itemize}
\item Extra dimensions~\cite{Arkani-Hamed:1998jmv,Dienes:1998sb,Arkani-Hamed:1998wuz,Dvali:1999cn,Barbieri:2000mg} with RH neutrinos in the bulk leading to suppressed Dirac Yukawa $Y_{\nu}$;
\item
String inspired mechanisms~\cite{Mohapatra:1986bd,Faraggi:1993zh,Haba:1993yj,Cleaver:1997nj}.
\end{itemize}

\subsection{\label{sec:seesaw}Majorana seesaw mechanisms}
There are several different kinds of seesaw mechanism in the
literature. Here we shall focus on the type I and II seesaw mechanisms, which are used later in the review, although we shall also briefly discuss other seesaw mechanisms.

\subsubsection{\label{sec:seesaw-1}Type I seesaw}
If we introduce right-handed neutrino fields then there are two sorts
of additional neutrino mass terms that are possible. There are
additional Majorana masses of the form
\begin{equation}
M_{N}\overline{\nu_R^c}\nu_R^{} \sim M_{N}\nu_R \nu_R \,,
\label{MRR}
\end{equation}
where $\nu_R$ is a right-handed neutrino field and $\nu_R^c$ is
the CP conjugate of a right-handed neutrino field, in other words
a left-handed antineutrino field. In addition there are
Dirac masses arising from the Yukawa couplings which, after the Higgs VEVs are inserted, take the form in LR convention,
\begin{equation}
M_{D}\overline{\nu_L}\nu_R\,.
\label{mLR}
\end{equation}
Such Dirac mass terms conserve lepton number, and are not forbidden
by electric charge conservation even for the charged leptons and
quarks.

Once this is done then the types of neutrino mass discussed
in Eqs.~(\ref{MRR},\ref{mLR}) are permitted, and we have the mass matrix
\begin{equation}
\left(\begin{array}{cc} \overline{\nu_L} & \overline{\nu^c_R}
\end{array} \\ \right)
\left(\begin{array}{cc}
0 & M_{D}\\
M_{D}^T & M_{N} \\
\end{array}\right)
\left(\begin{array}{c} \nu_L^c \\ \nu_R \end{array} \\ \right)\,.
\label{matrix}
\end{equation}
Since the right-handed neutrinos are electroweak singlets
the Majorana masses of the right-handed neutrinos $M_{N}$
may be orders of magnitude larger than the electroweak
scale. In the approximation that $M_{N}\gg M_{D}$
the matrix in Eq.~(\ref{matrix}) may be diagonalised to
yield the approximate light effective $3\times 3$ Majorana mass matrix,
\begin{equation}
M_{\nu}\approx -M_{D}M_{N}^{-1}M_{D}^T\,.
\label{seesaw}
\end{equation}
The effective left-handed Majorana masses $M_{\nu}$ are naturally
suppressed by the heavy scale $M_{N}$. In a one family example if we take $M_{D}\sim M_W$ (where $M_W$ is the mass of the $W$ boson) and $M_{N}\sim M_{\mathrm{GUT}}$ then we find $M_{\nu}\sim 10^{-3}$ eV which looks good for solar neutrinos. Atmospheric neutrino masses would require
a right-handed neutrino with a mass below the GUT scale.

The type I seesaw mechanism can be formally derived from the following Lagrangian
\begin{equation}
\mathcal{L}=-\overline{\nu_L}M_{D}\nu_R-\frac{1}{2}\nu_R^TM_{N}\nu_R+\text{h.c.} \,,
\end{equation}
where $\nu_L$ represents left-handed neutrino fields
(arising from electroweak doublets), $\nu_R$ represents right-handed
neutrino fields (arising from electroweak singlets), in a matrix
notation where the $M_{D}$ matrix elements
are typically of order the charged lepton masses,
while the $M_{N}$ matrix elements may be much larger than the
electroweak scale, and maybe up to the Planck scale.
The number of right-handed neutrinos is not fixed, but the number
of left-handed neutrinos is equal to three.
Below the mass scale of the right-handed neutrinos we can
integrate them out using the equations of motion
\begin{equation}
\frac{d{\cal L}}{d\nu_R}=0 \ ,
\end{equation}
which gives
\begin{equation}
\nu_R^T=-\overline{\nu_L}M_{D}M_{N}^{-1}\ ,\ \quad
\nu_R=-M_{N}^{-1}M_{D}^T\overline{\nu_L}^T \ .
\end{equation}
Substituting back into the original Lagrangian we find
\begin{equation}
\mathcal{L}=-\frac{1}{2}\overline{\nu_L}M_{\nu}\nu_L^c+\text{h.c.} \ ,
\end{equation}
with $M_{\nu}$ as in Eq.~(\ref{seesaw}).
Note that if we had assumed a RL convention for Dirac masses this would have led to a charge conjugated definition of $M_{\nu}$, where both conventions appear frequently in the literature.

\subsubsection{\label{sec:seesaw-2}Minimal seesaw extension of the Standard Model}

We now briefly discuss what the Standard Model looks like, assuming a minimal seesaw extension. In the Standard Model Dirac mass terms for charged leptons and quarks are generated from Yukawa type couplings to Higgs doublets whose vacuum expectation value gives the Dirac mass term. Neutrino masses are zero in the Standard Model because right-handed neutrinos are not present, and also because the Majorana mass terms of left-handed neutrinos require Higgs triplets in order to be generated at the renormalisable level. The simplest way to generate neutrino masses from a renormalisable theory is to introduce right-handed neutrinos,
as in the type I seesaw mechanism, which we assume here. The Lagrangian for the lepton sector of the Standard Model containing three right-handed neutrinos with heavy Majorana masses in Weyl fermion notation is\footnote{We introduce two Higgs doublets as in the supersymmetric Standard Model,
since the usual formulation of modular symmetry as discussed later requires supersymmetry. For the same reason we express the Standard Model Lagrangian
in terms of left-handed fields, replacing right-handed fields $\psi_R$
by their CP conjugates $\psi^c$, where the subscript $R$ has been dropped.
In the case of the minimal standard type Ia seesaw model with only one Higgs doublet, the other Higgs doublet in Eq.~(\ref{SM}) is obtained by charge
conjugation, i.e. $H_u\equiv H_d^c$.}
\begin{equation}
\mathcal{L}_{\mathrm{mass}}=-\left[\epsilon_{\alpha\beta}\tilde{Y}_e^{ij}H_d^\alpha L_i^\beta
  e^c_j -\epsilon_{\alpha\beta}\tilde{Y}_{\nu}^{ij}H_u^\alpha L_i^\beta \nu^c_j +
\frac{1}{2} \nu^c_i\tilde{M}_N^{ij}\nu^c_j
\right] +\text{h.c.} \ ,
\label{SM}
\end{equation}
where $\epsilon_{\alpha\beta}=-\epsilon_{\beta\alpha}$, $\epsilon_{12}=1$,
and the remaining notation is standard except that
the $3$ right-handed neutrinos $\nu_R^i$ have been replaced by their
CP conjugates $\nu^c_i$, and $\tilde{M}_N^{ij}$ is a complex symmetric
Majorana matrix.
When the two Higgs doublets get their
VEVs $\langle H_u^{\alpha =2}\rangle=v_u$, $\langle H_d^{\alpha=1}\rangle=v_d$, where the ratio of VEVs
is defined to be $\tan\beta \equiv v_u/v_d$,
we find the terms
\begin{equation}
\mathcal{L}_{\mathrm{mass}}= -v_d\tilde{Y}_e^{ij}e_ie^c_j
-v_u\tilde{Y}_{\nu}^{ij}\nu_i\nu^c_j -
\frac{1}{2}\tilde{M}_N^{ij}\nu^c_i\nu^c_j +\text{h.c.} \,.
\end{equation}
Replacing CP conjugate fields we can write in a matrix notation
\begin{equation}
\mathcal{L}_{\mathrm{mass}}=-\overline{e_L}v_d {\tilde{Y}_e}^\ast e_R
-\overline{\nu_L}v_u{\tilde{Y}_\nu}^\ast \nu_R -
\frac{1}{2}\nu^T_R M_N \nu_R +\text{h.c.} \,, \label{eq:Mast-MRR}
\end{equation}
where $M_N=\tilde{M}_N^\ast$. It is convenient to work in the diagonal charged lepton basis
\begin{equation}
{\rm diag}(m_e,m_{\mu},m_{\tau})
= V^{\dagger}_{e_L}v_d\tilde{Y}_e^\ast V_{e_R} \, ,
\end{equation}
and the diagonal right-handed neutrino basis
\begin{equation}
{\rm diag}(M_{1},M_{2},M_{3})=
V_{\nu_R}^{T}M_N V_{\nu_R} \,,
\end{equation}
where $V_{e_L},V_{e_R},V_{\nu_R}$ are unitary transformations.
This is commonly referred to as the flavour basis.
In the flavour basis the neutrino Yukawa couplings are given by
\begin{equation}
Y_{\nu}=V^{\dagger}_{e_L} \tilde{Y}_\nu^\ast V_{\nu_R}\,,
\end{equation}
and the Lagrangian in this basis is
\begin{eqnarray}
{\cal L}_{\mathrm{mass}}&=-&(\overline{e_L} \;\overline{\mu_L} \;\overline{\tau_L})
{\rm diag}{(m_e,m_{\mu},m_{\tau})} ({e}_R \,{\mu}_R\, {\tau}_R)^T
\nonumber \\
&-&(\overline{\nu_{e L}}\; \overline{\nu_{\mu L}} \; \overline{\nu_{\tau L}})v_uY_{\nu}
(\nu_{R1}\, \nu_{R2} \, \nu_{R3} )^T
\nonumber \\
&-&\frac{1}{2} (\nu_{R1} \,\nu_{R2}\, \nu_{R3} ){\rm diag}{(M_{1},M_{2},M_{3})}
(\nu_{R1} \,\nu_{R2} \,\nu_{R3} )^T
+\text{h.c.} \,.
\end{eqnarray}
After integrating out the right-handed neutrinos (the seesaw mechanism)
we find
\begin{eqnarray}
{\cal L}_{\mathrm{mass}}&=-&(\overline{e_L} \;\overline{\mu_L} \;\overline{\tau_L})
\text{diag}{(m_e,m_{\mu},m_{\tau})} ({e}_R \,{\mu}_R \,{\tau}_R)^T
\nonumber \\
&-&\frac{1}{2}(\overline{\nu_{eL}}\; \overline{\nu_{\mu L}} \;\overline{\nu_{\tau L}} )
M_{\nu} ({\nu_e}_L^c \,{\nu_\mu}_L^c \,{\nu_\tau}_L^c)^T
+\text{h.c.} \,,
\label{Lmass2}
\end{eqnarray}
where the light effective left-handed Majorana
neutrino mass matrix in the above basis is given by the
following seesaw formula which is equivalent to Eq.~(\ref{seesaw}),
\begin{equation}
M_{\nu}=-v_u^2\,Y_{\nu}\,
{\rm diag}{(M_{1}^{-1},M_{2}^{-1},M_{3}^{-1})}  \,Y_{\nu}^T \ .
\label{seesaw1}
\end{equation}

\subsubsection{(Constrained) Sequential Dominance}

Although the type I seesaw mechanism can qualitatively explain the smallness of neutrino masses through the heavy right-handed neutrinos (RHNs), if one doesn't make other assumptions, it contains too many parameters to make any particular predictions for neutrino mass and mixing. The sequential dominance (SD)~\cite{King:1998jw,King:1999cm} of right-handed neutrinos proposes that
the mass spectrum of heavy Majorana neutrinos in the basis of Eq.~\eqref{seesaw1} is strongly hierarchical, i.e. $M_1\ll M_2\ll M_3$,
where the lightest RHN with mass $M_1$ is dominantly
responsible for the heaviest physical neutrino mass $m_3$, that with mass $M_2$ is mainly responsible for
the second heaviest physical mass $m_2$, and a third largely decoupled RHN of mass $M_3$
gives a very suppressed lightest neutrino mass. It leads to an effective two right-handed neutrino (2RHN) model ~\cite{King:1999mb,Frampton:2002qc} with a natural explanation for the physical neutrino mass hierarchy, with normal ordering and the lightest neutrino being approximately massless, $m_1=0$. This is a good starting point to make predictions for the mixing angles, as follows.

After decoupling the third right-handed neutrino, the Dirac neutrino mass matrix or Yukawa matrix $Y^{\nu}$ in the flavour basis consists of a $3\times 2$ matrix with the elements of the first column $(d,e,f)$ describing the Yukawa couplings of $({\nu_e}_L\, {\nu_\mu}_L\, {\nu_\tau}_L)$ to the first RH neutrino of mass $M_1$, and the elements of the second column
$(a,b,c)$ describing the Yukawa couplings of $({\nu_e}_L, \,{\nu_\mu}_L, \,{\nu_\tau}_L)$ to the second RH neutrino of mass $M_2$. Then according to SD, we expect that the atmospheric mixing angle is given by $\tan \theta_{23}\sim |e|/|f|$, and the solar mixing angle is given by $\tan \theta_{12}\sim\sqrt{2}\,|a|/|b-c|$. Maximal atmospheric mixing suggests that $|e|\sim |f|$ and tri-maximal solar mixing suggests that $|b-c|\sim 2|a|$. Assuming $d=0$ then the reactor angle is given by $\tan \theta_{13}\lesssim m_2/m_3$ in agreement with the data.

A very predictive structure for the minimal type I seesaw model with two right-handed neutrinos and one texture zero is the so-called constrained sequential dominance (CSD) model~\cite{King:2005bj,Antusch:2011ic,King:2013iva,King:2015dvf,King:2016yvg,Ballett:2016yod,King:2018fqh,King:2013xba,King:2013hoa,Bjorkeroth:2014vha}.
Motivated by the SD results above, the CSD($n$) scheme
assumes that the two columns of the Dirac neutrino mass matrix or Yukawa matrix $Y^{\nu}$
are accurately proportional to $(d,e,f)\propto (0,1, 1)$ which yields approximately maximal atmospheric mixing and
$(a,b,c)\propto (1, n, n-2)$ or $(a,b,c)\propto (1, n-2, n)$
where the parameter $n$ is a real number which cancels
in the formula $\tan \theta_{12}\sim\sqrt{2}|a|/|b-c|$
to yield approximate tri-maximal solar mixing. The reactor angle is given by
$\tan \theta_{13}\sim (n-1)\sqrt{2}m_2/(3m_3)$.
A phenomenologically viable reactor angle requires $n\approx 3$,
and such models are called the Littlest Seesaw.
For example the CSD($3$)~\cite{King:2013iva,King:2015dvf,King:2016yvg,Ballett:2016yod,King:2018fqh}, CSD($2.5$)~\cite{Chen:2019oey}
and CSD($1+\sqrt{6}$) $\approx$ CSD($3.45$)~\cite{Ding:2019gof,Ding:2021zbg,deMedeirosVarzielas:2022fbw,deAnda:2023udh},
as suggested by modular symmetry, can give rise to phenomenologically viable predictions for lepton mixing parameters and the two neutrino
mass squared differences $\Delta m^2_{21}$ and $\Delta m^2_{31}$, in terms of three real input parameters~\cite{Costa:2023bxw}, as discussed later in the review.

\subsubsection{\label{sec:seesaw-4}Type II seesaw mechanism}

\begin{figure}[tb]
\centering
\includegraphics[width=0.46\textwidth]{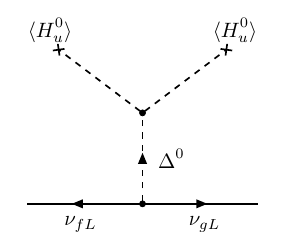}
 \caption{\label{fig:TypeIIDiagrams}\small{Diagram leading to a type II
contribution $M_L$ to the neutrino mass matrix via an induced VEV of the neutral component of a triplet Higgs $\Delta$.}}
\end{figure}

It is also possible to generate the dimension 5 operator in Eq.~(\ref{dim5}) by the exchange of heavy Higgs triplets of $SU(2)_L$, referred to as the type II seesaw mechanism~\cite{Schechter:1980gr,Magg:1980ut,Lazarides:1980nt,Mohapatra:1980yp}. In the type II seesaw the general neutrino mass matrix is given by
\begin{equation}
\left(\begin{array}{cc} \overline{\nu_L} & \overline{\nu^c_R}
\end{array} \\ \right)
\left(\begin{array}{cc}
M_L & M_{D}\\
M_{D}^T & M_{N} \\
\end{array}\right)
\left(\begin{array}{c} \nu_L^c \\ \nu_R \end{array} \\ \right)\,.
\label{matrix-II}
\end{equation}
Under the assumption that the mass eigenvalues $M_{i}$ of $M_{{N}}$ are very large compared to the components of  $M_L$ and $M_{{D}}$, the mass matrix can approximately be diagonalised yielding the light effective Majorana $3\times 3$ mass matrix
\begin{eqnarray}\label{eq:TypIIMassMatrix}
M_{\nu} \approx
M_L- M_{{D}}
\,M^{-1}_{{N}}\,M^{ T}_{{D}}\;
\label{typeII}
\end{eqnarray}
which may be compared to the type I result in Eq.~\eqref{seesaw}.
The new direct mass term $M_L$ can provide a naturally small contribution to the light neutrino masses if it stems e.g.~from a seesaw suppressed induced VEV, see figure~\ref{fig:TypeIIDiagrams}. The general case,  where both possibilities are allowed, is generally referred to as the type II seesaw mechanism, although people also talk about type II dominance or type I dominance, if the first or second term in Eq.~\eqref{typeII} dominates.

\subsubsection{Other Majorana seesaw mechanisms}

It is possible to implement the Majorana seesaw in a two-stage process
by introducing additional neutrino singlets $S$ beyond the three right-handed neutrinos $\nu^i_R$ that we have considered so far. If the singlets have Majorana masses $M_{S}$, but the right-handed neutrinos
have a zero Majorana mass $M_{N}=0$, the seesaw mechanism may proceed via mass couplings of singlets to right-handed neutrinos $M$. In the basis $(\nu_L^c,\nu_R,S)$ the mass matrix is
\begin{equation}
\label{double}
\left( \begin{array}{ccc}
0& M_{D} & 0    \\
M_{D}^T & 0 & M \\
0 & M^T & M_{S}
\end{array}
\right)\,.
\end{equation}
There are two different cases often considered:

(i) Assuming $M_{S}\gg M$
the light physical left-handed
Majorana neutrino masses are then,
\begin{equation}
M_{\nu}= -M_{D}M_{N}^{-1}M_{D}^T \ ,
\end{equation}
where
\begin{equation}
M_{N} = - M M_{S}^{-1}M^T.
\end{equation}
The mass matrix of the light physical left-handed Majorana neutrino masses is then,
\begin{equation}
\label{eq:inverse-seesaw}M_{\nu} = M_{D}{(M^{T})^{-1}} M_{S}M^{-1}M_{D}^T,
\end{equation}
which has a double suppression. This is called the double seesaw mechanism~\cite{Mohapatra:1986aw}. It is often used in GUT or string inspired neutrino mass
models to explain why $M_{N}$ is below the GUT or string scale.

(ii) Assuming $M_{S}\ll M$, the matrix has one light and two heavy
quasi-degenerate states for each generation.
In the limit that $M_{S}\rightarrow 0$ all neutrinos become massless and
lepton number symmetry is restored. Close to this limit $M_{S}\approx 0$ one may have
acceptable light neutrino masses for $M\sim~1$~TeV, allowing a testable
low energy seesaw mechanism referred to as the inverse seesaw mechanism~\cite{Mohapatra:1986bd}. If one allows the 1-3 elements of Eq.~(\ref{double}) to be filled
in by a matrix $M'$~\cite{Akhmedov:1995vm} then one obtains another version of the low
energy seesaw mechanism called the linear seesaw mechanism~\cite{Akhmedov:1995ip,Akhmedov:1995vm} with
\begin{equation}
M_{\nu} = M_{D}({M'}M^{-1})^T +   (M' M^{-1})M_{D}^T .
\end{equation}

\subsection{Constructing the Lepton Mixing Matrix}
In this subsection we discuss lepton mixing from first principles.
For definiteness we consider Majorana masses, since Dirac neutrinos
are completely analogous to
the SM description of quarks.
Consider the effective Lagrangian, in some general basis,
\begin{equation}
\mathcal{L}^{\rm mass}_{\rm lepton} =
-v_dY^e_{ij}\overline e^i_{L} e^j_{R}
-\frac{1}{2} M^{\nu}_{ij} \overline{{\nu}^i_{L}} {\nu}_{L}^{cj}
+ \mathrm{h.c.}\,,
\label{lepton}
\end{equation}
which is valid below the EW symmetry breaking scale,
where $i,j$ are flavour indices.
We do not yet specify the mechanism responsible for the above complex symmetric
Majorana neutrino matrix\footnote{For example it could arise from a seesaw mechanism as in Eq.~\eqref{Lmass2} where here we are not in the flavour basis but in some general basis. The choice of superscript or subscript $\nu$ has no significance, thus $M^{\nu}=M_{\nu}$.} $M^{\nu}$.
The mass matrices may be diagonalised by unitary matrices,
\begin{eqnarray}
U^{\dagger}_{e_L}Y^eU_{e_R}=
\left(\begin{array}{ccc}
y_e&0&0\\
0&y_{\mu}&0\\
0&0&y_{\tau}
\end{array}\right), \ \ \ \
U^{\dagger}_{{\nu}_L}M^{\nu}U_{{\nu}_L}^{*}=
\left(\begin{array}{ccc}
m_1&0&0\\
0&m_2&0\\
0&0&m_3
\end{array}\right).
\end{eqnarray}
The charged current (CC) couplings to $W^-$ in the flavour basis are given by
$-\frac{g}{\sqrt{2}}\overline{e}^i_L\gamma^{\mu}W_{\mu}^-{\nu}^i_{L}$,
which becomes in the mass basis, where $e_L,\mu_L,\tau_L$ and $\nu_i$ are now mass eigenstates,
\begin{eqnarray}
\mathcal{L}^{CC}_{\rm lepton}= -\frac{g}{\sqrt{2}}
\left(\begin{array}{ccc}
\overline{e}_L  & \overline{\mu}_L &  \overline{\tau}_L
\end{array}\right)
U_{\rm PMNS}
\gamma^{\mu}W_{\mu}^-
\left(\begin{array}{c}
{\nu}_{1L} \\
{\nu}_{2L} \\
{\nu}_{3L}
\end{array}\right)+\text{h.c.}\, .
\end{eqnarray}
The Pontecorvo-Maki-Nakagawa-Sakata (PMNS) lepton mixing matrix is identified as\footnote{Different physically equivalent conventions appear in the literature, we follow the conventions in~\cite{King:2002nf}.},
\begin{equation}
\label{enu}
U_{\rm PMNS}=U^{\dagger}_{e_L}U_{{\nu}_{L}}\,.
\end{equation}
It is possible to remove three of the lepton phases in $U_{\rm PMNS}$, using the phase invariance of $m_e, m_{\mu}, m_{\tau}$. For example, $m_e\overline e_{L}e_{R}$, is unchanged by $e_{L}\rightarrow e^{i\phi_e}e_{L}$ and $e_{R}\rightarrow e^{i\phi_e}e_{R}$. The three such phases $\phi_e, \phi_{\mu}, \phi_{\tau}$ may be chosen in various ways to yield an assortment of possible PMNS parameterisations one of which is the PDG standard choice discussed below. This does not apply to the Majorana mass terms
$-\frac{1}{2} {m_i} \overline{{\nu}_{iL}} {\nu}_{iL}^{c}$ where $m_i$ are real and positive, and thus the PMNS matrix may be parametrised as in Eq.~\eqref{eq:matrix} but with an extra Majorana phase matrix~\cite{Workman:2022ynf}:
\begin{eqnarray}
\label{eq:matrix_pmns}
U_{\rm PMNS}=\left(\begin{array}{ccc}
c_{12} c_{13} & s_{12} c_{13} & s_{13} e^{-i\delta} \\
-s_{12}c_{23}-c_{12} s_{13} s_{23} e^{i\delta} &  c_{12} c_{23}-s_{12} s_{13} s_{23} e^{i\delta} & c_{13} s_{23}  \\
s_{12} s_{23}-c_{12} s_{13} c_{23} e^{i\delta} & -c_{12}s_{23}-s_{12} s_{13} c_{23} e^{i\delta}  & c_{13} c_{23}  \end{array}\right)\left(\begin{array}{ccc}
1 & 0  & 0 \\
0 & e^{i\frac{\alpha_{21}}{2}}  &  0 \\
0 & 0  & e^{i\frac{\alpha_{31}}{2}}
\end{array}\right)\,,~~~~~
\end{eqnarray}
where $\alpha_{21}$ and $\alpha_{31}$ are irremovable Majorana phases.
The mixing angles $\theta_{13}$ and $\theta_{23}$ must lie between $0$ and ${\pi}/{2}$, while (after reordering the masses) $\theta_{12}$ lies between $0$ and ${\pi}/{4}$. The phases all lie between $0$ and $2 \pi$, however we shall equivalently express $\delta$ in the range $-\pi$ to $\pi$. There is no current constraint on the Majorana phases, $\alpha_{21}$ and $\alpha_{31}$,
nor is there likely to be in the forseeable future. The first step will be to experimentally show that neutrinos are Majorana particles, which will most likely require neutrinoless double beta decay to be discovered. Then, only after precision studies of neutrinoless double beta decay rates, will there be any hope of determining the Majorana phases $\alpha_{21}$ and $\alpha_{31}$~\cite{King:2013psa}.

\subsection{Neutrino mass and lepton mixing - the global fits}
\begin{figure}[htb]
\centering
\includegraphics[width=0.7\textwidth]{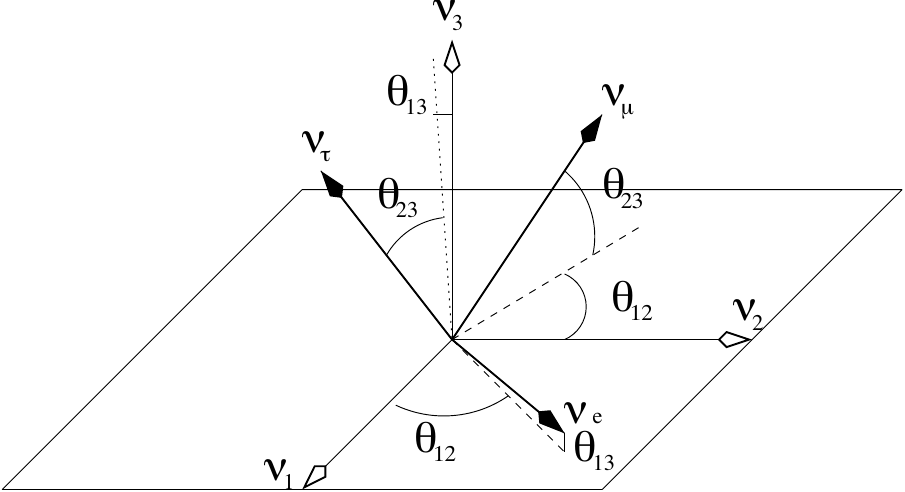}
\caption{Neutrino mixing angles may be represented as Euler angles relating the states in the charged lepton mass basis $(\nu_e, \nu_{\mu}, \nu_{\tau})$ to the mass eigenstate basis states $(\nu_1, \nu_2, \nu_3)$. \label{angles}}
\end{figure}

Neutrino physics has made remarkable progress since the discovery of neutrino mass and mixing in 1998. The reactor angle is accurately measured by Daya Bay: $\theta_{13}\approx 8.2^{\circ}-8.9^{\circ}$~\cite{DayaBay:2018yms}.
The other lepton mixing angles are determined from global fits
 to be in the three sigma ranges: $\theta_{12}\approx 31^{\circ}-36^{\circ} $ and $\theta_{23}\approx 37^{\circ}-52^{\circ}$,
but the CP-violating (CPV) phase is allowed in the three sigma
range $\delta \approx 108^{\circ}-404^{\circ}$ (corresponding to a sizeable three sigma excluded region
$\delta \approx 45^{\circ}-107^{\circ}$).
The best global fit values with one sigma errors are given in table~\ref{tab:nufit}~\cite{Esteban:2020cvm} where
the meaning of the angles is given in figure~\ref{angles}.

\begin{table}\centering
\begin{footnotesize}
\begin{tabular}{c|l|cc|cc}\hline\hline
\multirow{11}{*}{\begin{sideways}\hspace*{-7em}without SK atmospheric data\end{sideways}} & & \multicolumn{2}{c|}{Normal Ordering (best fit)} & \multicolumn{2}{c}{Inverted Ordering ($\Delta\chi^2=2.7$)} \\ \cline{3-6}
&& bfp $\pm 1\sigma$ & $3\sigma$ range
& bfp $\pm 1\sigma$ & $3\sigma$ range \\ \cline{2-6}
\rule{0pt}{4mm}\ignorespaces & $\sin^2\theta_{12}$ & $0.303_{-0.011}^{+0.012}$ & $0.270 \rightarrow 0.341$ & $0.303_{-0.011}^{+0.012}$ & $0.270 \rightarrow 0.341$ \\[1mm]
& $\theta_{12}/^\circ$ & $33.41_{-0.72}^{+0.75}$ & $31.31 \rightarrow 35.74$ & $33.41_{-0.72}^{+0.75}$ & $31.31 \rightarrow 35.74$ \\[3mm]
& $\sin^2\theta_{23}$ & $0.572_{-0.023}^{+0.018}$ & $0.406\rightarrow0.620$ & $0.578_{-0.021}^{+0.016}$ & $0.412\rightarrow0.623$ \\[1mm]
& $\theta_{23}/^\circ$ & $49.1_{-1.3}^{+1.0}$ & $39.6 \rightarrow51.9$ & $49.5_{-1.2}^{+0.9}$ & $39.9\rightarrow 52.1$  \\[3mm]
& $\sin^2\theta_{13}$ & $0.02203_{-0.00059}^{+0.00056}$ & $0.02029\rightarrow 0.02391$ & $0.02219_{-0.00057}^{+0.00060}$ & $0.02047 \rightarrow 0.02396$  \\[1mm]
& $\theta_{13}/^\circ$ & $8.54_{-0.12}^{+0.11}$ & $8.19\rightarrow8.89$ & $8.57_{-0.11}^{+0.12}$ & $8.23 \rightarrow 8.90$  \\[3mm]
& $\delta_{\mathrm{CP}}/^{\circ}$ & $197_{-25}^{+42}$ & $108 \rightarrow404$ & $286_{-32}^{+27}$ & $192\rightarrow360$  \\[3mm]
& $\dfrac{\Delta m^2_{21}}{10^{-5}~\text{eV}^2}$ & $7.41_{-0.20}^{+0.21}$ & $6.82 \rightarrow 8.03$ & $7.41_{-0.20}^{+0.21}$ & $6.82 \rightarrow 8.03$  \\[3mm]
& $\dfrac{\Delta m^2_{3\ell}}{10^{-3}~\text{eV}^2}$
& $+2.511_{-0.027}^{+0.028}$ & $+2.428\rightarrow +2.597$ & $-2.498_{-0.025}^{+0.032}$ & $-2.581\rightarrow-2.408$  \\[2mm] \hline\hline
\multirow{11}*{\begin{sideways}\hspace*{-7em}with SK atmospheric data\end{sideways}} & & \multicolumn{2}{c|}{Normal Ordering (best fit)}
& \multicolumn{2}{c}{Inverted Ordering ($\Delta\chi^2=7.1$)} \\ \cline{3-6}
&& bfp $\pm 1\sigma$ & $3\sigma$ range & bfp $\pm 1\sigma$ & $3\sigma$ range  \\ \cline{2-6}
\rule{0pt}{4mm}\ignorespaces & $\sin^2\theta_{12}$ &   $0.303_{-0.012}^{+0.012}$ & $0.270 \rightarrow 0.341$  & $0.303_{-0.011}^{+0.012}$ & $0.270\rightarrow 0.341$ \\[1mm]
& $\theta_{12}/^\circ$ & $33.41_{-0.72}^{+0.75}$ & $31.31\rightarrow35.74$  & $33.41_{-0.72}^{+0.75}$ & $31.31\rightarrow 35.74$  \\[3mm]
& $\sin^2\theta_{23}$ & $0.451_{-0.016}^{+0.019}$ & $0.408 \rightarrow0.603$ & $0.569_{-0.021}^{+0.016}$ & $0.412\rightarrow0.613$ \\[1mm]
& $\theta_{23}/^\circ$ & $42.2_{-0.9}^{+1.1}$ & $39.7\rightarrow 51.0$ & $49.0_{-1.2}^{+1.0}$ & $39.9\rightarrow 51.5$  \\[3mm]
& $\sin^2\theta_{13}$ & $0.02225_{-0.00059}^{+0.00056}$ & $0.02052\rightarrow 0.02398$ & $0.02223_{-0.00058}^{+0.00058}$ & $0.02048 \rightarrow 0.02416$ \\[1mm]
& $\theta_{13}/^\circ$  & $8.58_{-0.11}^{+0.11}$ & $8.23\rightarrow 8.91$ & $8.57_{-0.11}^{+0.11}$ & $8.23\rightarrow 8.94$  \\[3mm]
& $\delta_{\mathrm{CP}}/^{\circ}$ & $232_{-26}^{+36}$ & $144 \rightarrow 350$  & $276_{-29}^{+22}$ & $194\rightarrow 344$  \\[3mm]
& $\dfrac{\Delta m^2_{21}}{10^{-5}~\text{eV}^2}$   & $7.41_{-0.20}^{+0.21}$ & $6.82 \rightarrow 8.03$   & $7.41_{-0.20}^{+0.21}$ & $6.82\rightarrow 8.03$  \\[3mm]
& $\dfrac{\Delta m^2_{3\ell}}{10^{-3}~\text{eV}^2}$  & $+2.507_{-0.027}^{+0.026}$ & $+2.427\rightarrow +2.590$   & $-2.486_{-0.028}^{+0.025}$ & $-2.570 \rightarrow-2.406$ \\[2mm] \hline\hline
\end{tabular}
\end{footnotesize}
\caption{\label{tab:nufit}The NuFIT 5.2 results~\cite{Esteban:2020cvm}. }
\end{table}

The measurement of the reactor angle had a major impact on models of
neutrino mass and mixing as reviewed in~\cite{King:2013eh,King:2014nza,King:2015aea,Feruglio:2019ybq,Xing:2020ijf}
(for earlier reviews see e.g.~\cite{King:2003jb,Altarelli:2010gt,Ishimori:2010au}). Despite the enormous progress in the measurement of neutrino mass and mixing, further higher precision measurements are
required in order to test the hypothesis that the lepton sector is controlled by symmetry, as discussed in~\cite{Costa:2023bxw}.
Fortunately, a large program of future neutrino experiments is in progress and higher accuracy is expected over the next few years.

\subsection{The flavour puzzle of the Standard Model}

In PDG convention, the CKM matrix for quarks has a similar form to the PMNS matrix but without the Majorana phases:
\begin{eqnarray}
 \label{eq:matrix}
\!\!\!\!\!\!\!\!\!\!\!\!\!\!\!\!\!
U_{\rm CKM} = \left(\begin{array}{ccc}
    c^q_{12} c^q_{13}
    & s^q_{12} c^q_{13}
    & s^q_{13} e^{-i\delta^q}
    \\
    - s^q_{12} c^q_{23} - c^q_{12} s^q_{13} s^q_{23} e^{i\delta^q}
    & \hphantom{+} c^q_{12} c^q_{23} - s^q_{12} s^q_{13} s^q_{23}
    e^{i\delta^q}
    & c^q_{13} s^q_{23} \hspace*{5.5mm}
    \\
    \hphantom{+} s^q_{12} s^q_{23} - c^q_{12} s^q_{13} c^q_{23} e^{i\delta^q}
    & - c^q_{12} s^q_{23} - s^q_{12} s^q_{13} c^q_{23} e^{i\delta^q}
    & c^q_{13} c^q_{23} \hspace*{5.5mm}
    \end{array}\right)
   \end{eqnarray}
where $s^q_{13}=\sin \theta^q_{13}$, etc. are quark mixing angles angles which are very different from
the lepton mixing angles in Eq.~\eqref{eq:matrix_pmns}.

It is interesting to compare quark mixing, which is small,
\begin{equation}
s^q_{12}= \lambda , \ \  s^q_{23}\sim \lambda^2, \ \  s^q_{13}\sim \lambda^3
\end{equation}
where the Wolfenstein parameter is $\lambda = 0.226\pm 0.001$,
to lepton mixing, which is large\footnote{As in section~\ref{Intro} lepton parameters are denoted without a superscript $l$.},
\begin{equation}
s_{13}\sim \lambda /\sqrt{2} , \ \  s_{23}\sim 1/\sqrt{2}, \ \  s_{12}\sim 1/\sqrt{3}.
\end{equation}
The smallest lepton mixing angle $\theta_{13}$ (the reactor angle), is of order the largest quark mixing angle $\theta^q_{12}=\theta_C=13.0^\circ$ (the Cabibbo angle, where $\sin \theta_C= \lambda$). There have been attempts to relate quark and lepton mixing angles such as postulating a reactor angle $\theta_{13}=\theta_C/\sqrt{2}$~\cite{Minakata:2004xt}, and the CP violating lepton phase $\delta \sim -\pi/2$ (c.f. the well measured CP violating quark phase $\delta^q \sim (\pi/2)/\sqrt{2}$).

\section{Neutrino Mass and Mixing Models without Modular Symmetry}

\subsection{Simple patterns of lepton mixing}

Various simple ansatzes for the PMNS matrix were proposed, the most simple one involving a zero reactor angle and bimaximal atmospheric mixing, $s_{13}=0$ and $s_{23}=c_{23}=1/\sqrt{2}$, leading to a PMNS matrix of the form,
\begin{eqnarray}
U_0 =\left( \begin{array}{ccc}
c_{12} ~& s_{12} ~& 0\\
 -\frac{s_{12}}{\sqrt{2}}  ~&  \frac{c_{12}}{\sqrt{2}} ~& \frac{1}{\sqrt{2}}\\
\frac{s_{12}}{\sqrt{2}}  ~&  -\frac{c_{12}}{\sqrt{2}} ~& \frac{1}{\sqrt{2}}
\end{array}
\right)\,,
\label{GR}
\end{eqnarray}
where the zero subscript reminds us that this form has $\theta_{13}=0$ (and $\theta_{23}=45^\circ$).

For tri-bimaximal (TB) mixing~\cite{Harrison:2002er,Xing:2002sw,He:2003rm}, one assumes
$s_{12}=1/\sqrt{3}$, $c_{12}=\sqrt{2/3}$ ($\theta_{12}=35.26^\circ$) in Eq.~(\ref{GR}),
\begin{eqnarray}
U_{\mathrm{TB}} =
\left(\begin{array}{ccc}
\sqrt{\frac{2}{3}} ~& \frac{1}{\sqrt{3}} ~& 0 \\
-\frac{1}{\sqrt{6}} ~& \frac{1}{\sqrt{3}} ~& \frac{1}{\sqrt{2}} \\
\frac{1}{\sqrt{6}}  ~& -\frac{1}{\sqrt{3}} ~& \frac{1}{\sqrt{2}}
\end{array}
\right).
\label{TB}
\end{eqnarray}

For bimaximal (BM) mixing (see e.g.~\cite{Barger:1998ta,Davidson:1998bi,Altarelli:2009gn,Meloni:2011fx} and references therein),
one has $s_{12}=c_{12}=1/\sqrt{2}$ ($\theta_{12}=45^\circ$) into Eq.~(\ref{GR}),
\begin{eqnarray}
U_{\mathrm{BM}} =
\left( \begin{array}{ccc}
\frac{1}{\sqrt{2}} ~& \frac{1}{\sqrt{2}} ~& 0\\
-\frac{1}{2}  ~& \frac{1}{2} ~& \frac{1}{\sqrt{2}} \\
\frac{1}{2}  ~& -\frac{1}{2} ~& \frac{1}{\sqrt{2}}
\end{array}
\right).
\label{BM}
\end{eqnarray}

For golden ratio (GRa) mixing~\cite{Datta:2003qg,Kajiyama:2007gx,Everett:2008et,Feruglio:2011qq,Ding:2011cm}, the solar angle is given by $\tan \theta_{12}=1/\phi$, where $\phi = (1+\sqrt{5})/2$ is the golden ratio which implies $\theta_{12}=31.7^\circ$. There are two alternative versions where $\cos \theta_{12} =\phi/2$ and $\theta_{12}=36^\circ$~\cite{Rodejohann:2008ir} which we refer to as GRb mixing, and GRc where $\cos \theta_{12} =\phi/\sqrt{3}$ and $\theta_{12} \approx 20.9^{\circ}$.

Finally another pattern studied in the literature with $\theta_{13}=0$ (and $\theta_{23}=45^\circ$) is the hexagonal mixing (HEX) where $\theta_{12} = \pi/6$~\cite{Albright:2010ap,Kim:2010zub}.

As we discuss in the next subsection, these simple patterns may be enforced by discrete non-Abelian family symmetry.
Although these simple patterns are excluded by current data, mainly because of the non-zero reactor angle, it is possible
that some relic of these patterns may survive, either due to charged lepton mixing corrections, or due to the first or second column of these matrices surviving, where these situations correspond to a controlled symmetry breaking as discussed in the next subsection.

\subsection{Symmetry of the lepton mass matrices}
The starting point for family symmetry models is to consider the symmetry of the mass matrices.
In a basis where the charged lepton mass matrix $M_e$ is diagonal,
the symmetry is,
\begin{equation}
T^{\dagger}(M_e^{\dagger}M_e)T= M_e^{\dagger}M_e
\end{equation}
where $T={\rm diag}(1, \omega^2 , \omega)$ and $\omega = e^{i2\pi /n}$.
For example for $n=3$ clearly $T$ generates a cyclic group $Z^T_3$.

In the diagonal charged lepton mass basis, assuming $U_{e_L}=I$,
\begin{equation}
\label{diag_e}
U_{\rm PMNS}^{\dagger} m^{\nu} U_{\rm PMNS}^*= {\rm diag}(m_1,m_2,m_3)
\end{equation}
and the neutrino mass matrix in this basis may be expressed as
\begin{equation}
\label{eq:diag_e}
m^{\nu} = U_{\rm PMNS}\,{\rm diag}(m_1,m_2,m_3)U_{\rm PMNS}^T= m_1 G_1+m_2 G_2+m_3 G_3
\end{equation}
where $G_i=G_i^T=\Phi_i \Phi_i^T$ and $\Phi_i$ are the three columns of $U_{\rm PMNS}\equiv (\Phi_1, \Phi_2, \Phi_3)$ with $\Phi_i^{\dagger} \Phi_j=\delta_{ij}$.

The Klein symmetry $Z^S_2\times Z^U_2$ of the light Majorana neutrino mass matrix defined in Eq.~\eqref{lepton} is given by the four element group $(I,S,U,SU)$~\cite{King:2009ap},
\begin{equation}
m^{\nu}= S^{\dagger}m^{\nu} S^{*}, \ \ \ \  m^{\nu}= U^{\dagger}m^{\nu} U^{*}, \ \ \ \ m^{\nu}= (SU)^{\dagger}m^{\nu} (SU)^{*}
\end{equation}
where
\begin{eqnarray}
S= U_{\rm PMNS}\ {\rm diag}(-1,+1,-1)\ U_{\rm PMNS}^{\dagger}=-G'_1+G'_2-G'_3 \label{Sd}\\
U= U_{\rm PMNS}\ {\rm diag}(-1,-1,+1)\ U_{\rm PMNS}^{\dagger}= -G'_1-G'_2+G'_3 \label{Ud}\\
SU= U_{\rm PMNS}\ {\rm diag}(+1,-1,-1)\ U_{\rm PMNS}^{\dagger}= G'_1-G'_2-G'_3 \label{SUd}
\end{eqnarray}
with $G'_i=G'^{\dagger}_i=\Phi_i \Phi_i^{\dagger}$ and $G'_iG'_j=\delta_{ij}G'_i$. Note that we have $G_i=G'_i$ in the limit that the lepton mixing matrix $U_{\rm PMNS}$ is real. One can check that $S$ and $U$ generate a Klein four group and they satisfy the following identities:
\begin{equation}
S^2=U^2=\mathds{1},\qquad SU=US\,.
\end{equation}
If the generators $S,U,T$ are identified with the generators of $S_4$, then the Klein symmetry enforces TB mixing.
Note also that the $S_4$ subgroups $Z^S_2$ and $Z^{SU}_2$ enforce TM$_2$ and TM$_1$ mixing, respectively, where the preserved column of the TB matrix in each case is given by the eigenvector associated with the $+1$ eigenvalue which preserves the symmetry.

\subsection{Direct Models}

\begin{figure}[htb]
\centering
\includegraphics[width=0.50\textwidth]{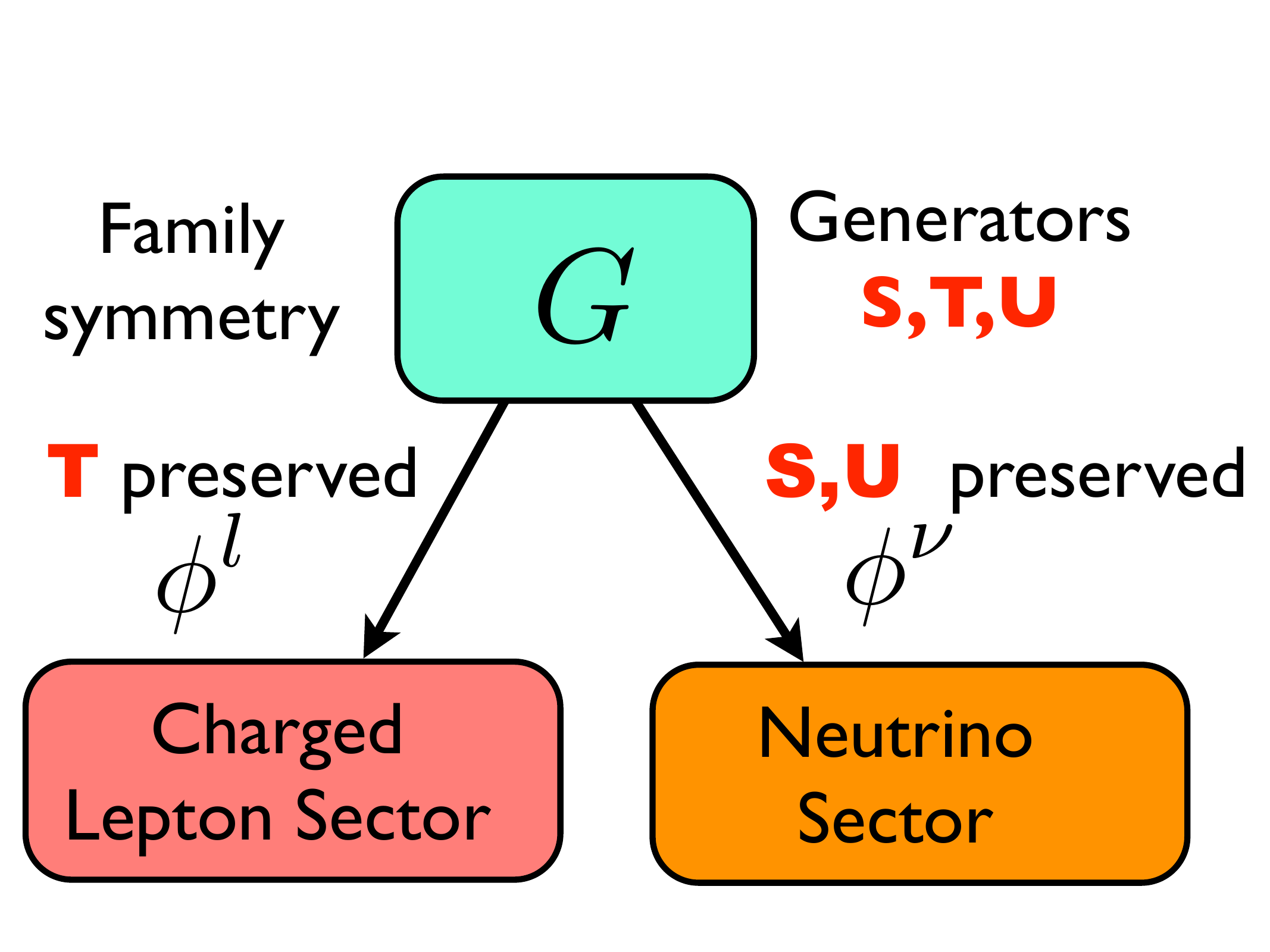}
\caption{The diagram illustrates the so called direct approach to models of lepton mixing. For example, for the flavor symmetry group $G=S_4$, this structure leads to tri-bimaximal mixing. To avoid the bad prediction that
$\theta_{13}=0$, one or more of the generators $S,T,U$ must be broken, as discussed in the main text.} \label{discrete1}
\end{figure}

\begin{figure}[htb]
\centering
\includegraphics[width=0.90\textwidth]{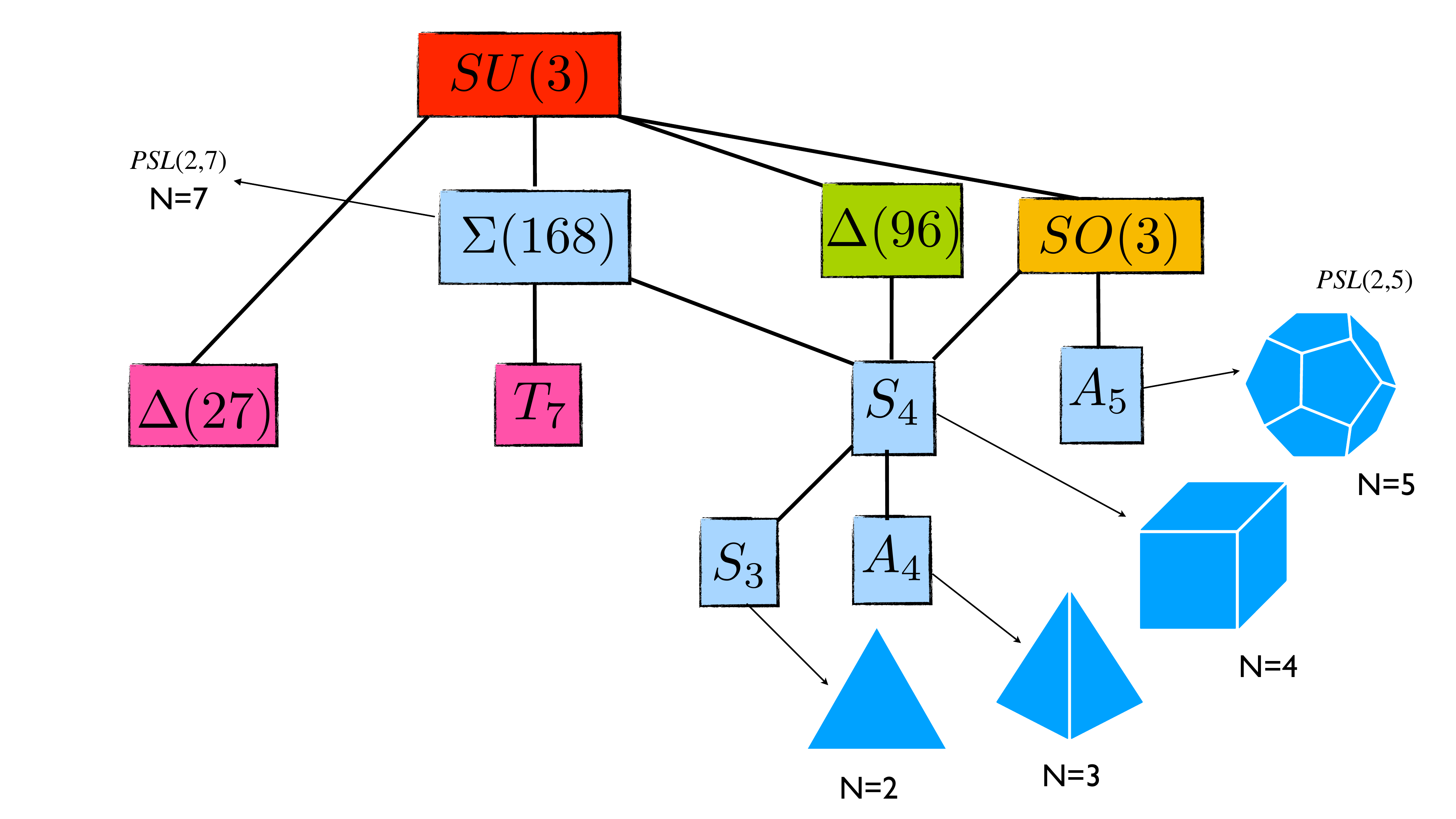}
\caption{The diagram shows some possible choices of the flavor symmetry group $G$. As discussed later in this review, the blue groups can emerge at level $N=2,3,4,5,7$ of the finite modular symmetry group $PSL(2,N)$, which is the projective special linear group of $2\times 2$ matrices with integer elements mod $N$, with unit determinant, where positive and negative matrices are identified. The coloured shapes for the real groups are the regular geometric shapes which enjoy the symmetries indicated. The simple group $\Sigma(168)$ has complex representations, so does not have any simple geometrical interpretation.} \label{discrete2}
\end{figure}

The idea of ``direct models''~\cite{King:2013eh}, illustrated in figure~\ref{discrete1}, is that the three generators $S,T,U$ introduced above are embedded into a discrete family symmetry $G$ which is broken by new Higgs fields called ``flavons'' of two types: $\phi^l$ whose VEVs preserve $T$ and $\phi^{\nu}$ whose VEVs preserve $S,U$. These flavons are segregated such that $\phi^l$ only appears in the charged lepton sector and $\phi^{\nu}$ only appears in the neutrino sector, thereby enforcing the symmetries of the mass matrices. Note that the full Klein symmetry $Z^S_2\times Z^U_2$ of the neutrino mass matrix is enforced by symmetry in the direct approach.

There are many choices of the group $G$, with some examples given in figure~\ref{discrete2} (right panel),  with each choice leading to different lepton mixing being predicted. For example, consider the group $S_4$ whose irreducible triplet representations are\footnote{There are precise group theory rules for establishing the irreducible representations of any group, but here we shall only state the results for
$S_4$ in the $T$-diagonal basis, see \cite{Ishimori:2010au} for
proofs, other examples and bases (e.g. dropping the $U$ generator leads to the $A_4$ subgroup).}:
\begin{eqnarray}
 \label{eq:matrixU}
S=\frac{1}{3}\begin{pmatrix}
-1 ~& 2~&2 \\
2 ~&-1~&2 \\
2 ~&2 ~& -1
\end{pmatrix},\qquad
T= \begin{pmatrix}
1~&0~&0\\
0&~\omega^2 ~&0 \\
0~&0~&\omega
\end{pmatrix},\qquad
U= \mp \begin{pmatrix}
1 ~& 0 ~& 0 \\
0 ~& 0 ~& 1 \\
0 ~& 1 ~& 0
\end{pmatrix}\,,
\end{eqnarray}
where $\omega = e^{i2\pi /3}$. Assuming these $S_4$ matrices,
the $Z^T_3$ symmetry of the charged lepton mass matrix and the Klein symmetry $Z^S_2\times Z^U_2$ of the neutrino mass matrix leads to the prediction of TB mixing (indeed one can check that $S$ and $U$ are diagonalised by $U_{TB}$ as in Eqs.~(\ref{Sd},\ref{Ud})).

\subsection{Semi-direct and tri-direct CP models}

In the ``semi-direct'' approach~\cite{King:2013eh}, in order
to obtain a non-zero reactor angle, one of the generators $T$ or $U$ of the residual symmetry is assumed to be broken. For example, consider the following two interesting possibilities:
\begin{enumerate}
\item {The $Z_3^T$ symmetry of the charged lepton
 mass matrix is broken, but the full Klein symmetry $Z_2^S\times Z_2^U$
 in the neutrino sector is respected. This corresponds to having
 charged lepton corrections, with solar sum rules as discussed recently~\cite{Costa:2023bxw}.}
\item{The $Z_2^U$ symmetry of the neutrino mass matrix is broken, but the
$Z_3^T$ symmetry of the charged lepton
mass matrix is unbroken. In addition either $Z_2^S$ or $Z_2^{SU}$ (with $SU$ being the product of $S$ and $U$) is preserved.
This leads to either $\rm{TM}_1$ mixing (if $Z_2^{SU}$ is preserved~\cite{Luhn:2013lkn});
or $\rm{TM}_2$ mixing (if $Z_2^S$ is preserved~\cite{King:2011zj}).
Then we have the atmospheric sum rules as discussed recently~\cite{Costa:2023bxw}.}
\end{enumerate}

The ``semi-direct approach'' may be extended to include a generalised CP symmetry $X$ such that $(M^{\nu})^*= X^TM^{\nu} X$, with a separate flavour and CP symmetry in the neutrino and charged lepton sectors~\cite{Feruglio:2012cw,Holthausen:2012dk} (see also~\cite{Ding:2013hpa,Feruglio:2013hia,Ding:2013bpa,Li:2013jya,King:2014rwa,Hagedorn:2014wha,Ding:2014ora,Ding:2015rwa,Li:2015jxa,DiIura:2015kfa,Ballett:2015wia,Li:2016ppt,Li:2016nap,Yao:2016zev}). Such models typically tend to predict maximal CP violation $\delta =\pm \pi /2$  (the first example of such generalised CP symmetry is mu-tau reflection symmetry discussed in the following subsection). The combination of flavor symmetry and generalized CP symmetry allows to predict both lepton mixing angles and CP violation phases~\cite{Chen:2014wxa,Chen:2015nha,Everett:2015oka}, it provides a plenty of possible residual symmetries which constrain the lepton mixing matrix to depended on few free parameters~\cite{Lu:2016jit,Lu:2018oxc,Rong:2016cpk}. It is notable that both quark and lepton mixing could be explained if the flavor and CP symmetry are broken down to $Z_2\times CP$ subgroups in the up type quark sector, down type quark sector, charged lepton sector and neutrino sectors~\cite{Li:2017abz,Lu:2019gqp}. The resulting quark and lepton mixing matrices only depend on two rotation angles separately, and the minimal flavor symmetry group is the dihedral group $D_{14}$~\cite{Lu:2019gqp}.

In the ``tri-direct'' CP approach~\cite{Ding:2018fyz,Ding:2018tuj,Chen:2019oey}, a separate flavour and CP symmetry is assumed for each
right-handed neutrino sector (in the framework of two right-handed neutrino models~\cite{King:1999mb})
in addition to the charged lepton sector.

While early family symmetry models focussed on continuous non-Abelian gauge theories such as
$SO(3)$~\cite{King:2005bj} or $SU(3)$~\cite{King:2001uz}, non-Abelian discrete symmetries~\cite{Ishimori:2010au}
have been widely used to account for the large lepton mixing angles, as discussed above.
However, the proliferation of flavons is rather undesirable, and the theoretical origin of such
non-Abelian discrete symmetries remains as open questions. Recently modular symmetry has been proposed to address both
of these questions, and this forms the subject of the remainder of this review.

\section{Introduction to modular symmetry}
In this section, we first give an informal introduction to modular symmetry. We then introduce the modular group, and discuss its fixed points and residual symmetry. Finally we write down the form of the supersymmetric action,
including the K\"ahler potential and the conditions that the superpotential must satisfy.

\subsection{What is modular symmetry?}

As we have already discussed, the Standard Model (SM), despite its many successes, does not account for the origin of neutrino mass nor the
quark and lepton family replication, and gives no insight into the fermion masses and mixing parameters.
We have already introduced the idea of a family symmetry which may be a finite discrete or continuous, gauged or global, Abelian or non-Abelian, and we have noted that
large lepton mixing has motivated studies of non-Abelian finite discrete groups such as $A_4,S_4,A_5$ (for reviews see e.g.~\cite{King:2013eh,King:2017guk}). However, as we have seen, such family symmetries must eventually be
spontaneously broken by new Higgs fields called flavons, and it turns out that the vacuum alignment of such flavon fields plays
a crucial role determining the physical predictions of such models.

In the remainder of this review we shall turn to another interesting class of symmetries which arise from the modular group $SL(2,\mathbb{Z})$, which is the group of familiar $2\times 2 $
matrices with positive or negative integer elements, but with unit determinant.
Geometrically, this is the symmetry of a torus, which has a flat geometry in two dimensions when it is cut open.
The symmetry $SL(2,\mathbb{Z})$ corresponds to the discrete coordinate transformations which leave the torus invariant,
which allows for different choices of two dimensional lattice vectors $\tau$ describing the same torus.
In fact it turns out that an overall minus sign in the matrices makes no difference to the transformation and when such matrices are identified the
group is called the projective special linear group $PSL(2,\mathbb{Z})$, often denoted as $\overline{\Gamma}$.

The two dimensional space of the torus may be identified as
the real and imaginary directions of the complex plane, where the lattice vectors $\tau$
correspond to complex vectors in the upper half of the Argand plane.
This is sufficient to describe all the
symmetries of the torus, and this leads to the the idea of holomorphicity, where complex conjugation is forbidden, as in supersymmetry. The principal congruence subgroup of level $N$ corresponds to the subset of matrices $\Gamma (N)$ which belong to $SL(2,\mathbb{Z})$ and which are equal to the unit matrix mod $N$, where these matrices also act on the complex variable
$\tau$ in the upper half of the complex plane. If the positive and negative unit matrices are identified then the resulting group is called
$\overline{\Gamma} (N)$.

At first sight the modular symmetry groups do not look like a promising starting point for a family symmetry.
Firstly they are all infinite groups, since there are an infinite number of $2\times 2 $ matrices with integer elements and unit determinant. Secondly, it might seem that the symmetry of a torus has nothing to do with particle physics.
However in the framework of superstring theory and extra dimensions, the second point seems more reasonable
since orbifold compactifications of six extra dimensions are often performed as three factorisable products of
two extra dimensions compactified on three tori~\cite{Ferrara:1989bc,Ferrara:1989qb}.
Each torus may be described by a single lattice vector (where by convention
the other lattice vector has unit length and lies along the real axis) which is
identified as a modulus field $\tau$, whose vacuum expectation value (VEV) is determined by a
supergravity potential which fixes the geometry of the torus \cite{Ishiguro:2021ccl,Cremades:2004wa,Ishiguro:2020tmo}. As regards the first point, it is possible to obtain a finite discrete group from the infinite modular group as discussed below.

For a given choice of level $N>2$, the infinite modular symmetry may be rendered finite by removing the infinite matrices which are
trivially related to the unit matrix, leaving only a small finite number of interesting matrices which form a closed finite group.
This is achieved by considering the
quotient group $\Gamma_N=PSL(2,\mathbb{Z})/\overline{\Gamma} (N)$ which is finite and may be identified with the groups
$\Gamma_N=A_4,S_4,A_5$ for levels $N=3,4,5$, and so on for higher levels~\cite{Feruglio:2017spp}.
These finite modular groups, may then be applied to neutrino and flavour models in the usual way, except that now
such theories offer the promise that the only flavon present is the single modulus field $\tau$, whose VEV fixes the value of Yukawa couplings which form representations of $\Gamma_N$ and
are modular forms, leading to very predictive theories~\cite{Feruglio:2017spp}.

Following the above observations~\cite{Feruglio:2017spp}, there has been considerable activity in applying modular symmetry to flavour models, and also in extending the framework to more general settings. There have already appeared some very good early reviews~\cite{Feruglio:2019ybq} in the context of neutrino physics. The purpose of the present review is to provide a dedicated
and comprehensive review of the literature relating to the bottom-up approach to modular symmetry flavour models, starting from the basic ideas, through the most recent more sophisticated approaches, before discussing some applications to neutrino and flavour models.

\subsection{The modular group}

\begin{figure}
\centering
\includegraphics[width=0.8\textwidth]{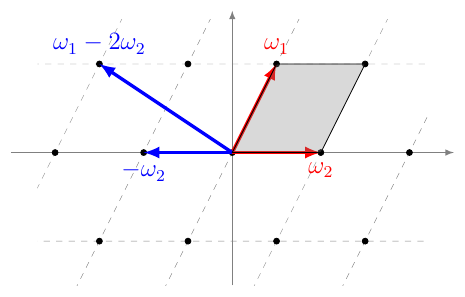}
\caption{\label{fig:Lattice} Two equivalent lattices generated by the base vectors $\left\{\omega_1, \omega_2\right\}$ and $\left\{-\omega_2, \omega_1-2\omega_2\right\}$, the corresponding modular transformation is $\tau\rightarrow-1/(\tau-2)$. }
\end{figure}

Following the informal introduction to the modular group of the previous subsection, we now provide a more rigorous exposition, setting out the mathematical formalism in some detail. As mentioned above, the modular group is ubiquitous in string theory. It is the invariance group of a lattice $\Lambda=\left\{m_1 \omega_1+m_2 \omega_2|~m_{1,2}\in \mathbb{Z}\right\}$ in the complex plane $\mathbb{C}$, where $\omega_1$ and $\omega_2$ are the basis vectors of the lattice with $\tau\equiv\omega_1/\omega_2$ and we may assume $\text{Im}(\tau)>0$ by swapping $\omega_1$ and $\omega_2$ if necessary.  As shown in figure~\ref{fig:Lattice}, the two lattices $\Lambda=\left\{m_1 \omega_1+m_2 \omega_2|~m_{1,2}\in \mathbb{Z}\right\}$ and $\Lambda'=\left\{m_1 \omega'_1+m_2 \omega'_2|~m_{1,2}\in \mathbb{Z}\right\}$ are identical if and only if
\begin{equation}
\begin{pmatrix}
\omega'_1 \\ \omega'_2
\end{pmatrix}=\begin{pmatrix}
a ~&~ b \\ c ~&~ d
\end{pmatrix}\begin{pmatrix}
\omega_1 \\ \omega_2
\end{pmatrix}\,,
\end{equation}
which implies\footnote{This modular transformation is well-defined, as it fulfills $\texttt{Im}(\gamma(\tau))=\frac{\texttt{Im}(\tau)}{|c\tau+d|^2}>0$ and $(\gamma\gamma')(\tau)=\gamma(\gamma'(\tau))$.}
\begin{equation}
\label{eq:linear-fractional-trans}\tau\mapsto\gamma\tau=\gamma(\tau)=\frac{a\tau+b}{c\tau+d},~~~\texttt{Im}(\tau) >0\,.
\end{equation}
where $a$, $b$, $c$, $d$ are integers and they fulfill $ad-bc=1$. A complex torus is a quotient $\mathbb{C}/\Lambda$ of the complex plane $\mathbb{C}$ by a lattice $\Lambda$, it is obtained by gluing both opposite pairs of edges of the fundamental parallelogram depicted in gray in figure~\ref{fig:Lattice}. Obviously each linear fractional transformation of Eq.~\eqref{eq:linear-fractional-trans} is associated with a $2\times2$ matrix $\gamma=\begin{pmatrix}a  &  b \\
c  &  d \end{pmatrix}$ with integer coefficients and determinant 1. All the linear fractional transformations form the full modular group $\Gamma$ which is isomorphic to $SL(2,\mathbb{Z})$, i.e.
\begin{equation}
SL(2,\mathbb{Z})=\left\{\begin{pmatrix}
a  &  b \\
c  &  d
\end{pmatrix}\bigg|a,b,c,d\in\mathbb{Z}, ad-bc=1
\right\}\,.
\end{equation}
Notice that $\gamma$ and $-\gamma$ act in the same way on the modulus $\tau$, the faithful action group is the projective special linear group $\overline{\Gamma}\equiv PSL(2,\mathbb{Z})\cong SL(2,\mathbb{Z})/\{\mathds{1}_2, -\mathds{1}_2\}$, where $\mathds{1}_2$ stands for the two-dimensional identity matrix. Note that the modular group is defined to be $\overline{\Gamma}$ in some literature. The modular group is an infinite discrete group and it can be generated by two elements $S$ and $T$~\cite{Bruinier2008The,diamond2005first}
\begin{equation}
\label{eq:S-T-Mod-Gen} S=\begin{pmatrix}
0 ~&~ 1 \\
-1 ~&~ 0
\end{pmatrix},  ~~~~  T=\begin{pmatrix}
1 ~&~ 1 \\
0 ~&~ 1
\end{pmatrix}\,.
\end{equation}
Note that $S$ and $T$ are often referred to as modular inversion and  translation respectively,
\begin{equation}
S:~ \tau\mapsto-\frac{1}{\tau}\,,~~~~~T:~\tau\mapsto \tau+1\,.
\end{equation}
It is straightforward to check that the two generators satisfy the following relations
\begin{equation}
\label{eq:ST-full-ModG} S^4=(ST)^3=\mathds{1}_2,~~~~~S^2T=TS^2
\end{equation}
and also $(TS)^3=\mathds{1}_2$ which is equivalent to $(ST)^3=\mathds{1}_2$. The corresponding relations in $\overline{\Gamma}$ are $S^2=(ST)^3=\mathds{1}_2$, since $\mathds{1}_2$ and $-\mathds{1}_2$ are indistinguishable in $\overline{\Gamma}$. Moreover, one can find that the inverse of a modular transformation is
\begin{equation}
\gamma^{-1}=\begin{pmatrix}
d  ~&~ -b \\
-c ~&~  a
\end{pmatrix},~~~~\gamma=\begin{pmatrix}
a  ~&~ b \\
c ~&~  d
\end{pmatrix}\,.
\end{equation}

The modular group $\Gamma$ has a series of infinite normal subgroups
$\Gamma(N)$ ($N=1,2,3,\ldots$) defined by,
\begin{equation}
\label{eq:Gamma-N-PCSG}\Gamma(N)=\left\{\left(\begin{array}{cc}a&b\\c&d\end{array}\right)\in SL(2,\mathbb{Z}),~~ \left(\begin{array}{cc}a&b\\c&d\end{array}\right)=\begin{pmatrix}
1  ~&~  0 \\
0  ~&~ 1
\end{pmatrix}~(\texttt{mod}~N)
\right\}\,,
\end{equation}
which is the principal congruence subgroup of level $N$.  Note that for $N=1$ the principal congruence subgroup is just equal to the full modular group, $\Gamma=\Gamma(1)$. Also the element $T^N$ belongs to $\Gamma(N)$, i.e. $T^N\in\Gamma(N)$.

The groups $\overline{\Gamma}(N)$ of linear fractional transformations are slightly different from the groups $\Gamma(N)$. We have $\overline{\Gamma}(N)=\Gamma(N)/\{\mathds{1}_2, -\mathds{1}_2\}$ for $N=1, 2$,  while $\overline{\Gamma}(N)=\Gamma(N)$ for $N>2$ because in this case $-\mathds{1}_2$ doesn't belong to $\Gamma(N)$.

The quotient groups
\begin{equation}
\label{eq:Gamma_N-def}
\Gamma_N\equiv
PSL(2,\mathbb{Z})
/\overline{\Gamma}(N) \equiv
\overline{\Gamma}/\overline{\Gamma}(N)
\end{equation}
are the so-called inhomogeneous finite modular groups and are isomorphic to $PSL(2, N)$.
The group $\Gamma_N$ can be generated by two element $S$ and $T$ satisfying
\begin{equation}
\label{eq:GammaN-rules}S^2=(ST)^3=T^N=1\, ,
\end{equation}
where here $1$ denotes the identity element.
Notice that additional relations are necessary in order to render the group $\Gamma_N$ finite for $N\geq6$~\cite{deAdelhartToorop:2011re}. We see that $\Gamma_1$ is a trivial group comprising only the identity element, $\Gamma_2$ is isomorphic to $S_3$. For small values of $N$, the groups $\Gamma_N$ are isomorphic to permutation groups: $\Gamma_3\cong A_4$, $\Gamma_4\cong S_4$ and $\Gamma_5\cong A_5$~\cite{deAdelhartToorop:2011re}. The finite modular groups $\Gamma_N$ as flavor symmetry have been widely studied to explain neutrino mixing.

The quotient groups
\begin{equation}
\label{eq:Gammap_N-def}\Gamma'_N\equiv SL(2,\mathbb{Z})/\Gamma(N)
\end{equation}
are the so-called homogeneous finite modular groups. They can be regarded as the group of two-by-two matrices with entries that are integers modulo $N$ and determinant equal to one modulo $N$, and they are isomorphic to $SL(2, N)$.

We see $\Gamma_2\cong\Gamma'_2$, and $\Gamma_N$ for $N>2$ is isomorphic to the quotient of $\Gamma'_N$ over its center $\{\mathds{1}_2, -\mathds{1}_2\}$, i.e., $\Gamma_{N}\cong\Gamma'_N/\{\mathds{1}_2, -\mathds{1}_2\}$ in matrix form. Hence $\Gamma'_N$ has double the number of group elements as $\Gamma_N$ with $|\Gamma'_N|=2|\Gamma_N|$.  The group $\Gamma'_N$ can be obtained from $\Gamma_N$ by including another generator $R$ which is related to $-\mathds{1}_2\in SL(2,\mathbb{Z})$ and commutes with all elements of the $SL(2,\mathbb{Z})$ group, such that the generators $S$, $T$ and $R$ of $\Gamma'_N$ obey the following relations\footnote{The multiplication rules of $\Gamma'_N$ can also be written as $S^4=(ST)^3=T^N=1, S^2T=TS^2$.}~\cite{Liu:2019khw},
\begin{equation}
\label{eq:GammaNp-rules}S^{2}=R,~~~(S T)^{3}=T^{N}=R^{2}=1,~~~RT=TR\,.
\end{equation}
Note that additional relations are needed for $N\geq6$. We summarize the finite modular groups $\Gamma_N$, $\Gamma'_N$ and their orders in table~\ref{tab:finite-modular-groups}.

\begin{table}[ht!]
\centering
\renewcommand{\arraystretch}{1.25}
\begin{tabular}{cccccc}
\hline\hline
$N$&  $\Gamma_N$ & $|\Gamma_N|$ & $\Gamma'_N$  & $|\Gamma'_N|$  \\
\midrule
2& $S_3$ & 6 & $S_3$  & 6 \\
3&  $A_4$   &  12 & $T'$  & 24  \\
4&  $S_4$ &  24  & $S'_4$ & 48  \\
5&  $A_5$ & 60 & $A'_5$ & 120  \\
6&  $S_3\times A_4$ & 72 & $S_3\times T'$ & 144  \\
7&  $PSL(2, Z_7)\cong \Sigma(168)$ & 168 & $SL(2, Z_7)$ & 336 \\
\hline\hline
\end{tabular}
\caption{The finite modular groups $\Gamma_N$ and $\Gamma'_N$ and their orders up to $N=7$. \label{tab:finite-modular-groups}}
\end{table}

\begin{figure}
\centering
\includegraphics[width=0.9\textwidth]{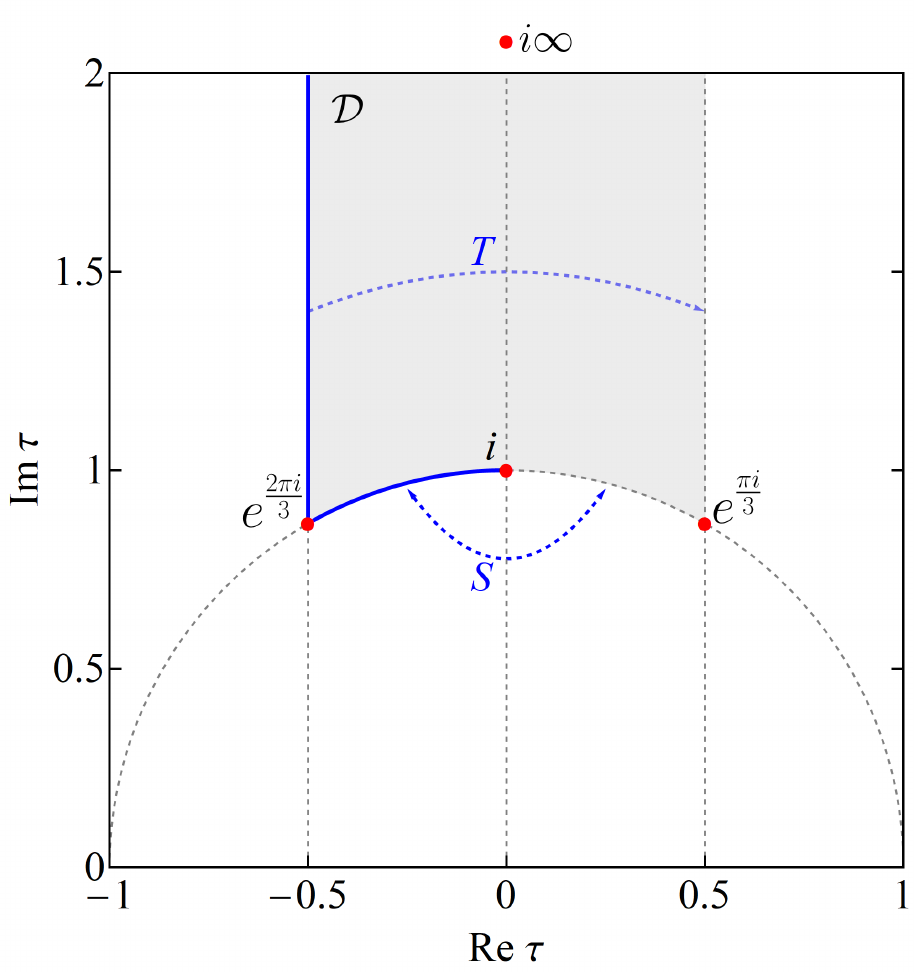}
\caption{\label{fig:fundamental-domain}The fundamental domain $\mathcal{D}$ of the modular group $\Gamma$. }
\end{figure}

As shown in figure~\ref{fig:fundamental-domain}, the $\Gamma$ orbit of every modulus $\tau$ has a representative in the standard fundamental domain $\mathcal{D}$~\footnote{More precisely, each orbit has a unique representative in the standard fundamental domain
\begin{equation}
\mathcal{D}=\Big\{\tau\Big||\tau|>1, -\frac{1}{2}\leq\texttt{Re}(\tau)<\frac{1}{2}\Big\}\cup \Big\{\tau\Big||\tau|=1,\texttt{Re}(\tau)\leq0 \Big\}\,,
\end{equation}}\,.
\begin{equation}
\mathcal{D}=\left\{\tau|\texttt{Im}(\tau)>0, |\texttt{Re}(\tau)|\leq\frac{1}{2}, |\tau|\geq1 \right\}\,,
\end{equation}
which is bounded by the vertical lines $\texttt{Re}(\tau)=-\frac{1}{2}$, $\texttt{Re}(\tau)=\frac{1}{2}$ and the circle $|\tau|=1$ in the upper half plane $\mathcal{H}$. Every point in the upper half plane is equivalent to a point of $\mathcal{D}$ under the action of $SL(2, \mathbb{Z})$, and no two points inside $\mathcal{D}$ differ by a linear fraction transformation. The transformation $T$ pairs the two vertical lines $\texttt{Re}(\tau)=\pm\frac{1}{2}$, and the transformation $S$ maps the arc of $|\tau|=1$ from $i$ to $e^{\pi i/3}$ into the arc from $i$ to $e^{2\pi i/3}$. Notice that the fundamental domain is not unique, the transformed region $\gamma\mathcal{D}$ by any element $\gamma$ of $\Gamma$ can also be taken as the fundamental domain.

\subsection{\label{subsec:fixed-points-SL2Z}Fixed points of modulus and residual modular symmetry}

The modular symmetry is spontaneously broken by the vacuum expectation value (VEV) of the modulus $\tau$. There is no allowed value of $\tau$ which preserves the whole modular group (for example $\tau=0$ is not allowed since it is not in the fundamental domain). However, some value $\tau_0$ of the modulus is invariant under the action of certain $SL(2, \mathbb{Z})$ transformation $\gamma_0$, i.e.
\begin{equation}
\label{eq:gamm0-tau0}\gamma_0\tau_0=\tau_0\,.
\end{equation}
Thus the modular group $\Gamma$ is partially broken and the residual symmetry generated by $\gamma_0$ is preserved. We call $\tau_0$ is the fixed point of $\gamma_0$ and the subgroup generated by $\gamma_0$ as the stabilizer of $\tau_0$ with
\begin{equation} \texttt{Stab}(\tau_0)\equiv\left\{\gamma_0\in\Gamma|\gamma_0\tau_0=\tau_0\right\}.
\end{equation}	
Using the identity $\Im(\gamma\tau)=\dfrac{\Im\tau}{|c\tau+d|^2}$, from Eq.~\eqref{eq:gamm0-tau0} we find $|c_0\tau_0+d_0|=1$ and consequently $c_0\tau_0+d_0$ must be a phase. Moreover the explicit expression of Eq.~\eqref{eq:gamm0-tau0} for the fixed point $\tau_0$ is
\begin{equation}
\dfrac{a_0\tau_0+b_0}{c_0\tau_0+d_0}=\tau_0\,.
\end{equation}
Thus $\tau_0$ is determined to be
\begin{equation}
\tau_0=\left\{\begin{array}{cc}
\dfrac{b_0}{d_0-a_0}, ~&~ c_0=0\\[0.1in]
\dfrac{a_0-d_0\pm\sqrt{(a_0+d_0)^{2}-4}}{2c_0},   ~&~ c_0\neq0
\end{array}
\right.\,.
\end{equation}
In the fundamental region $\mathcal{D}$, the fixed points $\tau_0$ and the corresponding stabilizers $\gamma_0$ are summarized in table~\ref{tab:stabilizer-modular}~\cite{Ding:2019gof}. It is remarkable that the subgroup $Z^{S^2}_2$ is unbroken for any value of $\tau$. The fixed points $\tau_S=i$, $\tau_{ST}=e^{2\pi i/3}=-\frac{1}{2}+i\frac{\sqrt{3}}{2}$ and $\tau_T=i\infty$ are additionally invariant under the action of $S$, $ST$ and $T$. Notice the fixed point of $TS$ is $\tau_{TS}=e^{\pi i/3}=\frac{1}{2}+i\frac{\sqrt{3}}{2}$ which is related to $\tau_{ST}$ by $T$ transformation~\cite{Ding:2019gof}. Moreover, we find that all the infinite fixed points $\tau_f$ in the upper half complex plane and the corresponding modular transformation $\gamma_f$ satisfying $\gamma_f\tau_f=\tau_f$, are given by~\cite{Ding:2019gof,deMedeirosVarzielas:2020kji},
\begin{equation}
\label{eq:fp-general}\tau_f=\gamma'\tau_0,~~~~~\gamma_f=\gamma'\gamma_0\gamma'^{-1},~~~~\gamma'\in\Gamma\,,
\end{equation}
where $\gamma'$ is an arbitrary modular symmetry element. Hence all fixed points are related to $\tau_S$, $\tau_{ST}$ and  $\tau_T$ by modular transformations. In the case of single modulus, they are equivalent to the fixed points in the fundamental domain. However, in the case of multiple moduli, not all the fixed points of moduli can be moved to the fundamental domains in the presence of flavons, as discussed in sections~\ref{sec:multiple-moduli} and~\ref{subsec:modular-littlest-seesaw}. From Eq.~\eqref{eq:fp-general} we see that only the modular symmetry transformation conjugate to $\gamma_0$ can have fixed point. In other words, $\gamma_f$ and $\gamma_0$ must belong to the same conjugacy class. It is straightforward to see that the stabilizer $\texttt{Stab}(\tau_f)=\gamma'\texttt{Stab}(\tau_0)\gamma'^{-1}$ is isomorphic $\texttt{Stab}(\tau_0)$, and the isomorphism is given by a conjugation with $\gamma'$. The alignment of the modular form at the symmetric points are fixed so that the lepton mass matrices and mixing parameters are strongly constrained, the phenomenological implications of the residual symmetry fixed points would be discussed later.

\begin{table}[t!]
\centering
\begin{tabular}{|c|c|c|c|c|c|}
\hline\hline
$\tau_0$ & $\gamma_0$ & $\texttt{Stab}(\tau_0)$  \\
\hline
$i$ & $S$ &  $Z^S_4=\left\{1, S, S^2, S^3\right\}$     \\ \hline

$e^{2\pi i/3}$ & $ST, S^2$ & $Z^{ST}_3\times Z^{S^2}_2=\left\{1, ST, (ST)^2, S^2, S^3T, S^2(ST)^2 \right\}$     \\ \hline

$i\infty$ & $T, S^2$ & $Z^{T}\times Z^{S^2}_2=\left\{1, S^2, T, S^2T, T^2, S^2T^2, \ldots\right\}$     \\ \hline

others &  $S^2$  & $Z^{S^2}_2=\left\{1, S^2\right\}$     \\ \hline\hline
\end{tabular}
\caption{\label{tab:stabilizer-modular}The fixed points $\tau_0$ in the fundamental domain $\mathcal{D}$ and the corresponding stabilizers $\texttt{Stab}(\tau_0)$ which are abelian subgroups of $\Gamma$~\cite{Ding:2019gof}. Notice that the cyclic $Z^{g}_m$ has the presentation rule $Z^{g}_{m}=\left\{g| g^m=1\right\}$. The stabilizer $Z^{ST}_3\times Z^{S^2}_2$ of $\tau_0=e^{2\pi i/3}$ is isomorphic to the cyclic group $Z^{S^3T}_6$, and the $Z^{T}$ denotes the infinite cyclic group generated by the translation $T$.}
\end{table}

\subsection{\label{subsec:modular-invariant-theory}Modular invariant supersymmetric theories}

We work in the framework of the modular invariant supersymmetric theory~\cite{Ferrara:1989bc,Ferrara:1989qb,Feruglio:2017spp}. In the context of $N=1$ global supersymmetry, the most general form of the action is
\begin{equation}
\mathcal{S}=\int d^4xd^2\theta d^2\bar{\theta} \, \mathcal{K}(\Phi_I,\bar{\Phi}_I,\tau,\bar{\tau}) + \left[\int d^4x d^2\theta \mathcal{W}(\Phi_I,\tau) + \text{h.c.}\right]\,,
\end{equation}
where $\mathcal{K}(\Phi_I, \bar{\Phi}_I,\tau,\bar{\tau})$ is the K\"ahler potential, it is a real gauge invariant function of the chiral superfields $\Phi_I$, the modulus $\tau$ and their hermitian conjugates $\bar{\Phi}$, $\bar{\tau}$. $\mathcal{W}(\Phi,\tau)$ stands for the superpotential, and it is a holomorphic gauge invariant function of the chiral superfields $\Phi_I$ and $\tau$. The action $\mathcal{S}$ should be modular invariant and respect the SM (or GUT) gauge symmetry. The transformation properties of $\Phi_I$ are specified by its modular weight $-k_I$ and the representation $\mathbf{r}_I$ under $\Gamma'_N$,
\begin{equation}
\tau\rightarrow\gamma\tau=\frac{a\tau+b}{c\tau+d},~~~~~\Phi_I\rightarrow (c\tau+d)^{-k_I}\rho_{\mathbf{r}_I}(\gamma)\Phi_I\,.
\end{equation}
The K\"ahler potential to be the minimal form~\cite{Feruglio:2017spp},
\begin{equation}
\label{eq:Kahler-min}\mathcal{K}_{\rm min}(\Phi_I,\bar{\Phi}_I,\tau,\bar{\tau}) = -h\Lambda^2\log(-i\tau+i\bar{\tau})+\sum_I(-i\tau+i\bar{\tau})^{-k_I}|\Phi_I|^2\,,
\end{equation}
where $h$ is a positive constant. Given the modular transformation $(\tau-\bar{\tau})\rightarrow(\tau-\bar{\tau})/|c\tau+d|^2$ shown in the footnote 6,  one can see that the minimal K\"ahler potential $\mathcal{K}_{\rm min}$ is invariant under modular symmetry up to K\"ahler transformation, i.e.
\begin{equation}
\mathcal{K}_{\rm min}\rightarrow \mathcal{K}_{\rm min}+h\Lambda^2\log(c\tau+d)+h\Lambda^2\log(c\bar{\tau}+d)\,.
\end{equation}
The last two terms give null contribution after integration over Grassmannian coordinates $\theta$ and $\bar{\theta}$. Once the modulus $\tau$ gets a vacuum expectation, this K\"ahler potential gives the kinetic terms for the scalar components of the supermultiplet $\Phi_I$ and the modulus field $\tau$. Notice the K\"ahler potential is loosely constrained by the modular symmetry, there are additional terms consistent with modular symmetry~\cite{Chen:2019ewa}. However, the K\"ahler potential $\mathcal{K}$ is subject to strong constraint in some top-down models motivated by string theory~\cite{Nilles:2020nnc,Nilles:2020kgo,Ohki:2020bpo}, and the above minimal K\"ahler potential as the leading order contribution could possibly be achieved. The superpotential $\mathcal{W}$ can be expanded into power series of supermultiplets $\Phi_I$
\begin{equation}
\mathcal{W}(\Phi_I,\tau)=\sum_n Y_{I_1...I_n}(\tau)\Phi_{I_1}...\Phi_{I_n}\,.
\end{equation}
Modular invariance requires the function $Y_{I_1...I_n}(\tau)$ should be a modular form (defined in the next section) of weight $k_Y$ of level $N$ and in the representation $\mathbf{r}_Y$ of $\Gamma'_N$:
\begin{equation}
Y(\tau)\to Y(\gamma\tau)=(c\tau+d)^{k_Y}\rho_{\mathbf{r}_Y}(\gamma)Y(\tau)\,,
\end{equation}
where $k_Y$ and $\mathbf{r}_Y$ should satisfy the conditions
\begin{equation}
k_Y=k_{1}+...+k_{n},~~~~\rho_{\mathbf{r}_Y}\otimes \rho_{\mathbf{r}_{I_1}}\otimes\ldots\otimes\rho_{\mathbf{r}_{I_n}} \ni \mathbf{1}\,.
\end{equation}
Clearly the concept of modular forms is a crucial element which enables the modular transformations of matter fields to be to compensated so that modular invariance is preserved. Accordingly, the next section is devoted to modular forms.

\section{\label{sec:MF-MG}Modular forms and finite modular groups}

In this section we discuss the concept of modular forms in some detail, especially in relation to the finite modular groups. We first give a formal definion of modular forms. We then discuss the simplest even weight modular forms, before progressing to integer, half integer and general fractional weight modular forms, before discussing general vector valued modular forms associated with general finite modular groups which are defined by the quotient procedure using more general normal subgroups than the principal congruence subgroup.

\subsection{Modular forms}
A modular form $f(\tau)$ of weight $k$ and level $N$ may be defined as a holomorphic function of the complex variable $\tau$, where under $\Gamma(N)$ it transforms in the following way
\begin{equation}
\label{eq:def-MF}f\left(h\tau\right)=(c \tau+d)^{k} f(\tau) \quad\text{for}\quad \forall~~ h=\begin{pmatrix}
a  &  b \\
c  &  d
\end{pmatrix}\in \Gamma(N)\,,
\end{equation}
where $k\ge 0$ is a non-negative integer.
As mentioned previously for $N=1$ we have $\Gamma = \Gamma (1)$, so that the above transformation also applies to the full modular group in this case. It is notable that each
modular form of the full modular group $\Gamma$ can be written as a polynomial of $E_4(\tau)$ and $E_6(\tau)$ which are Eisenstein series of weight 4 and 6 respectively~\cite{Bruinier2008The}.

The square of the modular symmetry generator $S$ can be represented by $S^2=-\mathds{1}_2\in\Gamma(N)$ for levels $N=1, 2$ (see the comment above Eq.~\eqref{eq:Gamma_N-def}). In this case for $N=1, 2$,  Eq.~\eqref{eq:def-MF} with $h=S^2$ leads to
\begin{equation}
\label{eq:vanishing_cond}f(\tau)=(-1)^{k}f(\tau)\,.
\end{equation}
We conclude that there are no non-vanishing modular forms of odd weight at both level 1 and level 2.

However, the groups $\Gamma(N)$ for $N>2$ have non-vanishing modular forms with odd weight because $S^2\notin \Gamma(N>2)$ and the condition in Eq.~\eqref{eq:vanishing_cond} is not necessary to be fulfilled. The modular forms of weight $k$ and level $N$ form a linear space $\mathcal{M}_{k}(\Gamma(N))$, and its dimension is~\cite{Bruinier2008The,schultz2015notes},
\begin{subequations}
\begin{eqnarray}
\label{eq:dime_N2}&&\texttt{dim}\mathcal{M}_{2k}(\Gamma(2))=k+1 ,\quad  N=2,\,k\geq1\,,\\
\label{eq:dimNLa}&&\texttt{dim}\mathcal{M}_{k}(\Gamma(N))=\dfrac{(k-1)N+6}{24}N^2 \prod_{p|N}(1-\dfrac{1}{p^2}), \quad~N>2,\,k\geq 2 \,,
\end{eqnarray}
\end{subequations}
Notice that there is no general dimension formula for weight one modular form, but Eq.~\eqref{eq:dimNLa} is still applicable to the case of $N<6$. The linear space $\mathcal{M}_{k}(\Gamma(N))$ of the modular form has been constructed explicitly in~\cite{schultz2015notes}. Let us denote the linearly independent modular forms of integral weight $k$ and level $N$ as $f_{i}(\tau)$ with $i=1, 2, \ldots, \texttt{dim}\mathcal{M}_{k}(\Gamma(N))$, and we define $F_{i\gamma}(\tau)\equiv J^{-k}(\gamma,\,\tau) f_i(\gamma\tau)$ for any element $\gamma\in\Gamma$, and $J(\gamma, \tau)$ is the so-called automorphy factor defined as
\begin{equation}
\label{eq:J-automorphy-fac}J(\gamma, \tau)=c\tau+d\,.
\end{equation}
It is straightforward to check that $J(h,\tau)$ satisfies the following properties
\begin{subequations}
\begin{eqnarray}
\label{eq:J-prop-1}&&J(\gamma_1\gamma_2,\,\tau)=J(\gamma_1,\,\gamma_2\tau)J(\gamma_2,\,\tau),\quad \gamma_1,\gamma_2 \in \Gamma\,,\\
\label{eq:J-prop-2}&&J(\gamma^{-1},\gamma\tau)=J^{-1}(\gamma, \tau)\,.
\end{eqnarray}
\end{subequations}
Under a generic modular transformation $h\in\Gamma(N)$ and using Eqs.~(\ref{eq:J-prop-1}, \ref{eq:J-prop-2}), one find\footnote{
\begin{eqnarray*}
\nonumber F_{i\gamma}(h\tau)&=& J^{-k}(\gamma,\,h\tau)f_i(\gamma h\tau)=J^{-k}(\gamma,\,h\tau)f_i(\gamma h \gamma^{-1} \gamma\tau)=J^{-k}(\gamma,\,h\tau)J^k(\gamma h \gamma^{-1},\,\gamma \tau) f_i(\gamma \tau)\\
&=&J^k(h\gamma^{-1},\,\gamma\tau)f_i(\gamma\tau)=J^k(h,\,\tau)J^{-k}(\gamma,\,\tau)f_i(\gamma\tau)=J^k(h,\,\tau)F_{i\gamma}(\tau)\,.
\end{eqnarray*}}
\begin{equation}
F_{i\gamma}(h\tau)=J^k(h,\,\tau)F_{i\gamma}(\tau)\,.
\end{equation}
This implies that the holomorphic functions $F_{i\gamma}(\tau)$ are weight $k$ modular forms of level $N$. Hence $F_{i\gamma}(\tau)$ can be expressed as linear combinations of $f_i(\tau)$, i.e.
\begin{equation}
F_{i\gamma}(\tau)=\rho_{ij}(\gamma)f_{j}(\tau)\,,
\end{equation}
which leads to
\begin{equation}
\label{eq:MF-rho}f_i(\gamma\tau)=J^k(\gamma,\,\tau)\rho_{ij}(\gamma)f_j(\tau)\,.
\end{equation}
The linear combination matrix $\rho(\gamma)$ depends on the modular transformation $\gamma$, and from Eq.~\eqref{eq:J-prop-2} we know  $\rho(\gamma)$ is a representation of the modular group and it satisfies\footnote{Using Eq.~\eqref{eq:MF-rho}, we can obtain
\begin{equation*}
f(\gamma_1\gamma_2\tau)=J_k(\gamma_1\gamma_2, \tau)\rho(\gamma_1\gamma_2)f(\tau)\,,
\end{equation*}
and
\begin{eqnarray*}
f(\gamma_1\gamma_2\tau)=J_k(\gamma_1,\,\gamma_2\tau)\rho(\gamma_1)f(\gamma_2\tau)=J_k(\gamma_1,\,\gamma_2\tau)J_k(\gamma_2, \tau)\rho(\gamma_1)\rho(\gamma_2)f(\tau)=J_k(\gamma_1\gamma_2, \tau)\rho(\gamma_1)\rho(\gamma_2)f(\tau)\,.
\end{eqnarray*}
Comparing the above two formulas, we arrive at the following result,
\begin{equation*}
\rho(\gamma_1\gamma_2)=\rho(\gamma_1)\rho(\gamma_2)\,.
\end{equation*}},
\begin{equation}
\label{eq:homomorphism}\rho(\gamma_1\gamma_2)=\rho(\gamma_1)\rho(\gamma_2)\,.
\end{equation}
Moreover, comparing Eq.~\eqref{eq:MF-rho} with the definition of modular form in Eq.~\eqref{eq:def-MF}, we obtain that $\rho(h)$ should be unit matrix of dimension $\texttt{dim}\mathcal{M}_{k}(\Gamma(N))$, i.e.,
\begin{equation}
\label{eq:id_h}\rho(h)=1,~~~\forall h\in\Gamma(N)\,,
\end{equation}
Hence the representation matrix $\rho(\gamma)$ differs from identity matrix only if the element $\gamma$ is in the quotient group $\Gamma'_N=\Gamma/\Gamma(N)$. Given that $S^4=(ST)^3=1$, $RT=TR$ and $T^N\in\Gamma(N)$ with $R=S^2$, we have
\begin{equation}
\label{eq:rho-GammaNp}\rho^4(S)=\rho^3(ST)=\rho^N(T)=1,\qquad \rho(R)\rho(T)=\rho(T)\rho(R)\,.
\end{equation}
The representation $\rho$ satisfies the same rules of Eq.~\eqref{eq:GammaNp-rules} as $\Gamma'_N$. Therefore $\rho$ essentially is a linear representation of the homogeneous finite modular group $\Gamma'_N$. Since each representation of the finite group can be decomposed into a direct sum of irreducible unitary representations. As a result, by properly choosing basis, $\rho$ can be written into a block diagonal form,
\begin{equation}
\label{rho-reduce}\rho \sim \rho_{\mathbf{r_1}} \oplus \rho_{\mathbf{r_2}} \oplus \dots \,,~~~\text{with}~~~ \sum_i\texttt{dim}\,\rho_{\mathbf{r}_i} = \texttt{dim}\mathcal{M}_{k}(\Gamma(N))\,,
\end{equation}
where $\rho_{\mathbf{r}_i}$ denotes an irreducible unitary representation of $\Gamma'_N$. Consequently it is always possible to choose a basis $Y_{\mathbf{r}}(\tau)\equiv\left(f_1(\tau),\,f_2(\tau),\,\dots\right)^T$ in the modular form space $\mathcal{M}_{k}(\Gamma(N))$ so that
$Y_{\mathbf{r}}(\tau)$ transforms under the full modular group $\Gamma$ as~\cite{Feruglio:2017spp,Liu:2019khw}
\begin{equation}
\label{eq:MF-decomp}Y_\mathbf{r}(\gamma\tau)= (c\tau + d)^k\rho_{\mathbf{r}}(\gamma)Y_\mathbf{r}(\tau),~~~\gamma \in \Gamma\,,
\end{equation}
where $\rho_{\mathbf{r}}$ is an irreducible representation of $\Gamma'_N$ and $\gamma$ is a representative element of $\Gamma'_N$. In practice, it is sufficient to focus on the modular generators $S$ and $T$ such that we have,
\begin{equation}
\label{eq:decom-ST}Y_\mathbf{r}(S\tau)= (-\tau)^k\rho_{\mathbf{r}}(S)Y_\mathbf{r}(\tau)\,,~~~~ Y_\mathbf{r}(T\tau)= \rho_{\mathbf{r}}(T)Y_\mathbf{r}(\tau)\,.
\end{equation}
Applying Eq.~\eqref{eq:MF-decomp} for $\gamma=R$, we find that the non-vanishing modular form $Y_\mathbf{r}(\tau)$ requires
\begin{equation}
\label{eq:R-rep}\rho_{\mathbf{r}}(R)=(-1)^k\,,
\end{equation}
which is $+1$ and $-1$ for even $k$ and odd $k$ respectively.

At the fixed point $\tau_f$,  Eq.~\eqref{eq:MF-decomp} implies that the modular multiplet $Y_\mathbf{r}(\tau_f)$ is an eigenvector of $\rho_{\mathbf{r}}(\gamma_f)$ with the eigenvalue $J^{-k}(\gamma_f, \tau_f)$, where $J(\gamma, \tau)=c\tau+d$ is the automorphy factor. It is remarkable that $J(\gamma_f, \tau_f)$ is a phase with unit absolute value\footnote{From the identity $\texttt{Im}(\gamma_f(\tau_f))=\texttt{Im}(\tau_f)/|c_f\tau_f+d_f|^2=\texttt{Im}(\tau_f)$ we know  $|c_f\tau_f+d_f|=1$ and consequently $c_f\tau_f+d_f$ must be a phase with unit absolute value.} and particularly the identity $J(\gamma_f, \tau_f)=J(\gamma_0, \tau_0)$ is fulfilled with
\begin{equation}
J(S, \tau_S)=-i,~~~J(ST, \tau_{ST})=e^{4\pi i/3},~~~J(T, \tau_T)=1\,.
\end{equation}

\begin{table}[h]
\centering
\renewcommand{\arraystretch}{1.25}
\resizebox{1.0\textwidth}{!}{
\begin{tabular}{|c|c|c|c|c|c|c|}\hline \hline
\multirow{2}{*}{$N$}&\multirow{2}{*}{$\dim\mathcal{M}_{k}(\Gamma(N)) $}
&\multirow{2}{*}{$\Gamma_N$} & \multicolumn{3}{c|}{modular multiplets} \\ \cline{4-6}
& & & $k=2$ & $k=4$ & $k=6$ \\ \hline
2& $k/2 + 1$  &$S_3$  & $Y^{(2)}_{\mathbf{2}}$ & $Y^{(4)}_{\mathbf{1}}, Y^{(4)}_{\mathbf{2}}$  & $Y^{(6)}_{\mathbf{1}}, Y^{(6)}_{\mathbf{1'}}, Y^{(6)}_{\mathbf{2}}$  \\ \hline
3 & $k+1$ & $A_4$  & $Y^{(2)}_{\mathbf{3}}$ & $Y^{(4)}_{\mathbf{1}}, Y^{(4)}_{\mathbf{1'}}, Y^{(4)}_{\mathbf{3}}$ & $Y^{(6)}_{\mathbf{1}}, Y^{(6)}_{\mathbf{3}I}, Y^{(6)}_{\mathbf{3}II}$\\ \hline

\multirow{2}{*}{4} & \multirow{2}{*}{$2k+1$} & \multirow{2}{*}{$S_4$} & \multirow{2}{*}{$Y^{(2)}_{\mathbf{2}}$, $Y^{(2)}_{\mathbf{3}}$} & \multirow{2}{*}{$Y^{(4)}_{\mathbf{1}}$, $Y^{(4)}_{\mathbf{2}}$, $Y^{(4)}_{\mathbf{3}}$, $Y^{(4)}_{\mathbf{3'}}$} & $Y^{(6)}_{\mathbf{1}}$, $Y^{(6)}_{\mathbf{1'}}$, $Y^{(6)}_{\mathbf{2}}$,   \\

&  &  &  &  &  $Y^{(6)}_{\mathbf{3}I}$, $Y^{(6)}_{\mathbf{3}II}, Y^{(6)}_{\mathbf{3'}}$ \\ \hline

\multirow{3}{*}{5} & \multirow{3}{*}{$5k+1$} & \multirow{3}{*}{$A_5$} & \multirow{3}{*}{$Y^{(2)}_{\mathbf{3}},Y^{(2)}_{\mathbf{3}'}, Y^{(2)}_{\mathbf{5}}$} &  \multirow{2}{*}[-0.04in]{$Y^{(4)}_{\mathbf{1}}$, $Y^{(4)}_{\mathbf{3}}$, $Y^{(4)}_{\mathbf{3'}}$,  } & $Y^{(6)}_{\mathbf{1}}$, $Y^{(6)}_{\mathbf{3}I}$, $Y^{(6)}_{\mathbf{3}II}$, \\

&  & & & & $Y^{(6)}_{\mathbf{3'}I}$, $Y^{(6)}_{\mathbf{3'}II}$, $Y^{(6)}_{\mathbf{4}I}$, \\

&  & & &  \multirow{1}{*}[0.12in]{$Y^{(4)}_{\mathbf{4}}$, $Y^{(4)}_{\mathbf{5}I}$, $Y^{(4)}_{\mathbf{5}II}$} & $Y^{(6)}_{\mathbf{4}II}$, $Y^{(6)}_{\mathbf{5}I}$, $Y^{(6)}_{\mathbf{5}II}$\\  \hline

\multirow{3}{*}{6} & \multirow{3}{*}{$6k$} &  \multirow{3}{*}{$S_3\times A_4$}  & \multirow{3}{*}{$Y^{(2)}_{\bm{1^1_2}}$, $Y^{(2)}_{\bm{2_0}}$, $Y^{(2)}_{\bm{3^0}}$, $Y^{(2)}_{\bm{6}}$}  & \multirow{2}{*}[-0.06in]{ $Y^{(4)}_{\bm{1^{0}_{0}}}$, $Y^{(4)}_{\bm{1^{0}_{1}}} $, $Y^{(4)}_{\bm{2_{0}}} $, $Y^{(4)}_{\bm{2_{2}}} $,}  & $Y^{(6)}_{\bm{1^{0}_{0}}} $, $Y^{(6)}_{\bm{1^{1}_{0}}} $, $Y^{(6)}_{\bm{1^{1}_{2}}} $, $Y^{(6)}_{\bm{2_{0}}}$,\\

&  & & &  \multirow{1}{*}[-0.16in]{$Y^{(4)}_{\bm{3^{0}}}$, $Y^{(4)}_{\bm{3^{1}}}$, $Y^{(4)}_{\bm{6I}}$, $Y^{(4)}_{\bm{6ii}}$} &  $Y^{(6)}_{\bm{2_{1}}}$,  $Y^{(6)}_{\bm{2_{2}}}$, $Y^{(6)}_{\bm{3^{0}i}}$, $Y^{(6)}_{\bm{3^{0}ii}}$, \\

&  & & &  &  $Y^{(6)}_{\bm{3^{1}}}$, $Y^{(6)}_{\bm{6i}}$, $Y^{(6)}_{\bm{6ii}}$, $Y^{(6)}_{\bm{6iii}}$\\   \hline

\multirow{4}{*}{7} &\multirow{4}{*}{$14k-2$} & \multirow{4}{*}{$\Sigma(168)$} &  \multirow{4}{*}{$Y^{(2)}_{\mathbf{3}}$, $Y^{(2)}_{\mathbf{7}}$, $Y^{(2)}_{\mathbf{8}a}$, $Y^{(2)}_{\mathbf{8}b}$}  &   \multirow{2}{*}[-0.04in]{$Y^{(4)}_{\mathbf{1}a}$, $Y^{(4)}_{\mathbf{3}a}$, $Y^{(4)}_{\mathbf{6}a}$,} & $Y^{(6)}_{\mathbf{1}}$, $Y^{(6)}_{\mathbf{3}a}$, $Y^{(6)}_{\mathbf{3}b}$, $Y^{(6)}_{\mathbf{\bar{3}}}$,\\

&  & & & \multirow{1}{*}[-0.14in]{$Y^{(4)}_{\mathbf{6}b}$, $Y^{(4)}_{\mathbf{7}a}$, $Y^{(4)}_{\mathbf{7}b}$,}  &   $Y^{(6)}_{\mathbf{6}a}$, $Y^{(6)}_{\mathbf{6}b}$, $Y^{(6)}_{\mathbf{7}a}$,  $Y^{(6)}_{\mathbf{7}b}$,\\

&  & & & \multirow{1}{*}[-0.12in]{$Y^{(4)}_{\mathbf{8}a}$, $Y^{(4)}_{\mathbf{8}b}$, $Y^{(4)}_{\mathbf{8}c}$} &   $Y^{(6)}_{\mathbf{7}c}$, $Y^{(6)}_{\mathbf{7}d}$, $Y^{(6)}_{\mathbf{8}a}$, $Y^{(6)}_{\mathbf{8}b}$, \\

&  & & & & $Y^{(6)}_{\mathbf{8}c} $, $Y^{(6)}_{\mathbf{8}d}$
\\ \hline \hline
\end{tabular}}
\caption{Even weight modular forms of level $N$ up to weight 6 and the decomposition under the inhomogeneous finite modular group $\Gamma_N$. The conventions for the irreducible representations follow~\cite{Kobayashi:2018vbk} for $\Gamma_2$, \cite{Yao:2020qyy} for $\Gamma_3$, ~\cite{Qu:2021jdy} for $\Gamma_4$, ~\cite{Ding:2019xna} for $\Gamma_5$, ~\cite{Li:2021buv} for $\Gamma_6$ and ~\cite{Ding:2020msi} for $\Gamma_7$. \label{tab:even-MF-ModGroup} }
\end{table}

\subsection{Even weight modular forms and inhomogeneous finite modular group}

If the modular weight $k$ is an even positive integer, Eq.~\eqref{eq:R-rep} gives us
\begin{equation}
\rho_{\mathbf{r}}(R)=(-1)^k=1\,.
\end{equation}
Thus the homogeneous finite modular group $\Gamma'_N$ reduces to inhomogeneous finite modular group $\Gamma_N$, and the even weight modular forms can be arranged into multiplets of $\Gamma_N$ up to the automorphy factor $(c\tau+d)^{k}$. Remarkably for $N\leq 5$, the inhomogeneous finite modular groups are isomorphic to permutation groups: $\Gamma_2\cong S_3$, $\Gamma_3\cong A_4$, $\Gamma_4\cong S_4$ and $\Gamma_5\cong A_5$, as shown in table~\ref{tab:finite-modular-groups}.

There are no modular forms of negative weights, and the weight zero modular form is a constant. Modular forms of weight $2k$ and level $N$
form a linear space ${\cal M}_{2k}(\Gamma(N))$ of finite dimension, and the dimensions $\dim\mathcal{M}_{2k}(\Gamma(N))$ for lower levels $N$ are shown in table~\ref{tab:even-MF-ModGroup}. The product of two modular forms of level $N$ and weights $2k$, $2k'$ is a modular form of level $N$ and weight $2(k+k')$ and the set ${\cal M}(\Gamma(N))$ of all modular forms of level $N$ is a ring
\begin{equation}
{\cal M}(\Gamma(N))=\bigoplus_{k=0}^{\infty}{\cal M}_{2k}(\Gamma(N))~~~,
\end{equation}
generated by few elements. For instance ${\cal M}(\Gamma)$ is generated by two modular forms $E_4(\tau)$ and $E_6(\tau)$ of weight 4 and 6 respectively,
so that each modular form in ${\cal M}_{2k}(\Gamma)$ can be written as a polynomial $\sum_{ij} c_{ij}~ E_4(\tau)^{n_i} E_6(\tau)^{n_j}$, with powers satisfying $2k=4 n_i+6 n_j$. Hence one only needs to know the lowest weight two modular form to construct the even weight modular forms of level $N$.

\subsubsection{Constructing even weight modular forms of level 3}
As an example of the general procedure, we now explicitly construct the even weight modular forms of level 3.
The Dedekind eta-function $\eta(\tau)$ is frequently used to construct modular forms, it was introduced by Dedekind in 1877 and it is defined as follow~\cite{diamond2005first,Bruinier2008The,lang2012introduction},
\begin{equation}
\label{eq:eta-func}\eta(\tau)=q^{1/24}\prod_{n=1}^\infty \left(1-q^n \right),\qquad q\equiv e^{i 2 \pi\tau}\,.
\end{equation}
The $\eta(\tau)$ function can also be written into the following infinite series,
\begin{equation}
\eta(\tau)=q^{1/24}\sum^{+\infty}_{n=-\infty} (-1)^nq^{n(3n-1)/2}\,.
\end{equation}
Under the $S$ and $T$ transformations, $\eta(\tau)$ behaves as~\cite{diamond2005first,Bruinier2008The,lang2012introduction}
\begin{equation}
\label{eq:eta-S-T}\eta(\tau+1)=e^{i \pi/12}\eta(\tau),\qquad \eta(-1/\tau)=\sqrt{-i \tau}~\eta(\tau)\,.
\end{equation}
Hence $\eta^{24}(\tau)$ is a modular form of weight 12.

In this section, we follow the method of Ref.~\cite{Feruglio:2017spp} to explicitly construct a basis of weight 2 for $\Gamma(3)$. From table~\ref{tab:even-MF-ModGroup} we know the dimension $\dim\mathcal{M}_{2}(\Gamma(3))=3$, in the following we find out three linearly independent modular forms of weight 2 and level 3. We start by observing that if $f(\tau)$ transforms as\footnote{Notice that here $f(\tau)$ is not a modular form of weight $k$ because of the presence of the constant factor $e^{i\alpha}$.}
\begin{equation}
\label{eq:f-phase}f(\tau)\to e^{i\alpha}~(c\tau+d)^k~f(\tau)\,,
\end{equation}
then
\begin{equation}
\label{eq:Ferruccio1}\frac{d}{d\tau}\log f(\tau)\to (c\tau+d)^2 \frac{d}{d\tau}\log f(\tau)+ kc\,(c\tau+d)\,.
\end{equation}
The deviation of this formula is given in Appendix~\ref{app:derivation-Ferruccio}. Furthermore,  we can linearly combine several $f_i(\tau)$ with weights $k_i$ as follow
\begin{equation}
\label{eq:Ferruccio}\frac{d}{d\tau}\sum_i\alpha_i\log f_i(\tau)\to (c\tau+d)^2 \frac{d}{d\tau}\sum_i\alpha_i\log f_i(\tau)+ \left(\sum_i\alpha_i k_i\right) c(c\tau+d)~~~,
\end{equation}
with
\begin{equation}
\sum_i\alpha_i k_i=0\,.
\end{equation}
Then the inhomogeneous term can be removed and consequently weight 2 modular forms can be obtained. The seed functions for the modular forms of weight 2 and level 3 are given by~\cite{Feruglio:2017spp}
\begin{eqnarray}
\nonumber&&f_1(\tau)=\eta\left(\frac{\tau}{3}\right),~~~~f_2(\tau)=\eta\left(\frac{\tau+1}{3}\right)\\
&&f_3(\tau)=\eta\left(\frac{\tau+2}{3}\right),~~~~f_4(\tau)=\eta(3\tau)\,.
\end{eqnarray}
They are closed under the modular group, and each of them is mapped to itself under the action of $\Gamma(3)$. Moreover, under $T$ they transform as
\begin{eqnarray}
\hskip-0.1in f_1(\tau)\stackrel{T}{\longmapsto} f_2(\tau)\,,~~f_2(\tau)\stackrel{T}{\longmapsto} f_3(\tau)\,,~~f_3(\tau)\stackrel{T}{\longmapsto} e^{i\pi/12}f_1(\tau),~~f_4(\tau)\stackrel{T}{\longmapsto} e^{i\pi/4}f_4(\tau)\,.
\end{eqnarray}
From Eq.~\eqref{eq:eta-S-T}, we known that they transform under $S$ in the following way,
\begin{eqnarray}
\nonumber&&f_1(\tau)\stackrel{S}{\longmapsto}\sqrt{3}~\sqrt{-i \tau}\,f_4(\tau)\,,~~~
f_2(\tau)\stackrel{S}{\longmapsto} e^{-i\pi/12}~\sqrt{-i\tau}~f_3(\tau)\,,\\
&&f_3(\tau) \stackrel{S}{\longmapsto} e^{i\pi/12}~\sqrt{-i\tau}~f_2(\tau)\,,~~~
f_4(\tau)\stackrel{S}{\longmapsto}\sqrt{1/3}~\sqrt{-i \tau}~f_1(\tau)
\end{eqnarray}
Hence the most general form of weight 2 and level 3 modular forms can be written as
\begin{equation}
\mathcal{F}(\alpha_1,\alpha_2,\alpha_3,\alpha_4|\tau)=\frac{d}{d\tau}\sum_i\alpha_i\log f_i(\tau)
\end{equation}
with $\sum^4_{i=1}\alpha_i=0$ to eliminate the inhomogeneous term. Under the action of modular generators $S$ and $T$, we have
\begin{eqnarray}
\nonumber&&\mathcal{F}(\alpha_1,\alpha_2,\alpha_3,\alpha_4|\tau)\stackrel{S}{\longmapsto} \tau^2~ \mathcal{F}(\alpha_4,\alpha_3,\alpha_2,\alpha_1|\tau)\\
&&\mathcal{F}(\alpha_1,\alpha_2,\alpha_3,\alpha_4|\tau)\stackrel{T}{\longmapsto} \mathcal{F}(\alpha_3,\alpha_1,\alpha_2,\alpha_4|\tau)\,.
\end{eqnarray}
As shown in Eq.~\eqref{eq:MF-decomp}, it is always possible to choose a basis in the linear space of modular form such that the independent basis vectors form irreducible multiplets of the finite modular group up to the automorphy factor. For the present case, there are three independent modular forms $Y_i(\tau)$ which can be arranged into a $A_4$ triplet $Y^{(2)}_{\mathbf{3}}(\tau)=(Y_1, Y_2, Y_3)^{T}$ satisfying
\begin{equation}
\label{eq:weight2-MF-level3}Y^{(2)}_{\mathbf{3}}(-1/\tau)=\tau^2~\rho_{\mathbf{3}}(S) Y^{(2)}_{\mathbf{3}}(\tau)\,,~~~~Y^{(2)}_{\mathbf{3}}(\tau+1)=\rho_{\mathbf{3}}(T) Y^{(2)}_{\mathbf{3}}(\tau)\,,
\end{equation}
where $\rho_{\mathbf{3}}(S)$ and $\rho_{\mathbf{3}}(T)$ are the unitary representation matrices of $S$ and $T$ in the triplet representation of $A_4$ given in Appendix~\ref{app:group-MF-N3},
\begin{equation}
\label{eq:rho-S-T-A4}\rho_{\mathbf{3}}(S)=\frac{1}{3}
\begin{pmatrix}
-1~&~2~&~2\\
2~&~-1~&~2\\
2~&~2~&~-1
\end{pmatrix}\,,~~~
\rho_{\mathbf{3}}(T)=
\begin{pmatrix}
1~&~0~&~0\\
0~&~\omega~&~0\\
0~&~0~&~\omega^2
\end{pmatrix}\,,~~~~\omega=e^{2\pi i/3}\,.
\end{equation}
Solving Eq.~\eqref{eq:weight2-MF-level3}, we can obtain the explicit form of modular forms $Y_i(\tau)$ as follows
\begin{eqnarray}
\nonumber Y_1(\tau)&=&\frac{3i}{2\pi}~\mathcal{F}(1,1,1,-3\vert\tau)=\frac{i}{2 \pi}
\left[\frac{\eta'\left(\frac{\tau }{3}\right)}{\eta \left(\frac{\tau}{3}\right)} +\frac{\eta'\left(\frac{\tau+1}{3}\right)}{\eta \left(\frac{\tau+1}{3}\right)} +\frac{\eta'\left(\frac{\tau +2}{3}\right)}{\eta\left(\frac{\tau +2}{3}\right)}-\frac{27 \eta'(3 \tau )}{\eta (3 \tau)}\right]\,,\\
\nonumber Y_2(\tau)&=&-\frac{3i}{\pi}~\mathcal{F}(1,\omega^2,\omega,0\vert\tau)=\frac{-i}{\pi}
\left[\frac{\eta'\left(\frac{\tau }{3}\right)}{\eta \left(\frac{\tau}{3}\right)}+\omega^2~\frac{\eta'\left(\frac{\tau+1}{3}\right)}{\eta \left(\frac{\tau+1}{3}\right)}+\omega~\frac{\eta'\left(\frac{\tau +2}{3}\right)}{\eta\left(\frac{\tau +2}{3}\right)}\right]\,,\\
\label{eq:MF-level-3-w2}Y_3(\tau)&=&-\frac{3i}{\pi}~\mathcal{F}(1,\omega,\omega^2,0\vert\tau)=\frac{-i}{\pi}
\left[\frac{\eta'\left(\frac{\tau }{3}\right)}{\eta \left(\frac{\tau}{3}\right)}+\omega~\frac{\eta'\left(\frac{\tau+1}{3}\right)}{\eta \left(\frac{\tau+1}{3}\right)} +\omega^2~\frac{\eta'\left(\frac{\tau +2}{3}\right)}{\eta\left(\frac{\tau +2}{3}\right)}\right]\,,
\end{eqnarray}
up to an overall constant which can be chosen freely. In numerical calculation, one usually needs $q$-expansion of $Y_i$,
\begin{eqnarray}
\nonumber&&Y_1(\tau)=1 + 12q + 36q^2 + 12q^3 + 84q^4 + 72q^5+36q^6+96q^7+180q^8+12q^9+216q^{10}+\dots \,, \\
\nonumber&&Y_2(\tau)=-6q^{1/3}\left(1 + 7q + 8q^2 + 18q^3 + 14q^4+31q^5+20q^6+36q^7+31q^8+56q^9+32q^{10} +\dots\right)\,, \\
\label{eq:Y1-Y2-Y3-Feruglio}&&Y_3(\tau)=-18q^{2/3}\left(1 + 2q + 5q^2 + 4q^3 + 8q^4 +6q^5+14q^6+8q^7+14q^8+10q^9+21q^{10}+\dots\right)~~\,.
\end{eqnarray}
From the above $q$-expansion we see that the modular forms $Y_i(\tau)$ satisfy the constraint:
\begin{equation}
\label{eq:constraint-level3MF}Y_2^2+2 Y_1 Y_3=0\,.
\end{equation}
The whole ring of even weight modular forms of level 3 can be generated from the products of the weight 2 modular forms $Y_{1,2,3}(\tau)$. At weight 4, the tensor product of $Y^{(2)}_{\mathbf{3}}$ gives rise to three independent modular multiplets,
\begin{eqnarray}
\nonumber Y^{(4)}_{\mathbf{1}}&=&(Y^{(2)}_{\mathbf{3}}Y^{(2)}_{\mathbf{3}})_{\mathbf{1}}=Y_1^2+2 Y_2 Y_3\,, \\
\nonumber Y^{(4)}_{\mathbf{1}'}&=&(Y^{(2)}_{\mathbf{3}}Y^{(2)}_{\mathbf{3}})_{\mathbf{1}'}=Y_3^2+2 Y_1 Y_2\,,\\
\label{eq:MF-w4l3}Y^{(4)}_{\mathbf{3}}&=&\frac{1}{2}(Y^{(2)}_{\mathbf{3}}Y^{(2)}_{\mathbf{3}})_{\mathbf{3}_S}=
\begin{pmatrix}
Y_1^2-Y_2 Y_3\\
Y_3^2-Y_1 Y_2\\
Y_2^2-Y_1 Y_3
\end{pmatrix}\,.
\end{eqnarray}
There are seven linearly independent modular forms of level 3 and weight 6 and they decompose as $\mathbf{1}\oplus\mathbf{3}\oplus\mathbf{3}$ under $A_4$,
\begin{eqnarray}
Y^{(6)}_{\mathbf{1}}&=&(Y^{(2)}_{\mathbf{3}}Y^{(4)}_{\mathbf{3}})_{\mathbf{1}}=Y_1^3+Y_2^3+Y_3^3-3 Y_1 Y_2 Y_3\,,\nonumber\\
Y^{(6)}_{\mathbf{3}I}&=&Y^{(2)}_{\mathbf{3}}Y^{(4)}_{\mathbf{1}}=(Y_1^2+2Y_2Y_3)\begin{pmatrix}
Y_1\\
Y_2\\
Y_3
\end{pmatrix}\,,\nonumber\\
\label{eq:MF-w6}Y^{(6)}_{\mathbf{3}II}&=&Y^{(2)}_{\mathbf{3}}Y^{(4)}_{\mathbf{1}'}=
(Y_3^2+2 Y_1Y_2)\begin{pmatrix}
Y_3\\
Y_1\\
Y_2
\end{pmatrix}\,.
\end{eqnarray}
The weight 8 modular forms can be arranged into three singlets $\mathbf{1}$, $\mathbf{1}'$, $\mathbf{1}''$ and two triplets $\mathbf{3}$ of $A_4$,
\begin{eqnarray}
\nonumber Y^{(8)}_{\mathbf{1}}&=&(Y^{(2)}_{\mathbf{3}}Y^{(6)}_{\mathbf{3}I})_{\mathbf{1}}=(Y_1^2+2 Y_2 Y_3)^2\,,\\
\nonumber Y^{(8)}_{\mathbf{1'}}&=&(Y^{(2)}_{\mathbf{3}}Y^{(6)}_{\mathbf{3}I})_{\mathbf{1'}}=(Y_1^2+2 Y_2 Y_3)(Y_3^2+2 Y_1 Y_2)\,, \\
\nonumber Y^{(8)}_{\mathbf{1''}}&=&(Y^{(2)}_{\mathbf{3}}Y^{(6)}_{\mathbf{3}II})_{\mathbf{1''}}=(Y_3^2+2 Y_1 Y_2)^2\,,\\
\nonumber Y^{(8)}_{\mathbf{3}I}&=&Y^{(2)}_{\mathbf{3}}Y^{(6)}_{\mathbf{1}}=(Y_1^3+Y_2^3+Y_3^3-3 Y_1 Y_2 Y_3)\begin{pmatrix}
Y_1 \\
Y_2\\
Y_3
\end{pmatrix}\,,\\
\label{eq:MF-w8}Y^{(8)}_{\mathbf{3}II}&=&(Y^{(2)}_{\mathbf{3}}Y^{(6)}_{\mathbf{3}II})_{\mathbf{3}_A}=(Y_3^2+2 Y_1Y_2)\begin{pmatrix}
Y^2_2-Y_1Y_3\\
Y^2_1-Y_2Y_3\\
Y^2_3-Y_1Y_2
\end{pmatrix}\,.
\end{eqnarray}
The weight 10 modular forms of level 3 decompose as $\mathbf{1}\oplus\mathbf{1}'\oplus\mathbf{3}\oplus\mathbf{3}\oplus\mathbf{3}$ under $A_4$, and they are
\begin{eqnarray}
\nonumber Y^{(10)}_{\mathbf{1}}&=&(Y^{(2)}_{\mathbf{3}}Y^{(8)}_{\mathbf{3}I})_{\mathbf{1}}=(Y_1^2+2 Y_2 Y_3)(Y_1^3+Y_2^3+Y_3^3-3 Y_1 Y_2 Y_3)\,,\\
\nonumber Y^{(10)}_{\mathbf{1'}}&=&(Y^{(2)}_{\mathbf{3}}Y^{(8)}_{\mathbf{3}I})_{\mathbf{1'}}=(Y_3^2+2 Y_1 Y_2)(Y_1^3+Y_2^3+Y_3^3-3 Y_1 Y_2 Y_3)\,, \\
\nonumber Y^{(10)}_{\mathbf{3}I}&=&Y^{(2)}_{\mathbf{3}}Y^{(8)}_{\mathbf{1}}=(Y_1^2+2 Y_2 Y_3)^2\begin{pmatrix}
Y_1 \\
Y_2\\
Y_3
\end{pmatrix}\,,\\
\nonumber Y^{(10)}_{\mathbf{3}II}&=&Y^{(2)}_{\mathbf{3}}Y^{(8)}_{\mathbf{1}'}=(Y_1^2+2 Y_2 Y_3)(Y_3^2+2 Y_1 Y_2)\begin{pmatrix}
Y_3 \\
Y_1\\
Y_2
\end{pmatrix}\,,\\
\label{eq:MF-w10}Y^{(10)}_{\mathbf{3}III}&=&Y^{(2)}_{\mathbf{3}}Y^{(8)}_{\mathbf{1}''}=(Y_3^2+2 Y_1 Y_2)^2\begin{pmatrix}
Y_2 \\
Y_3\\
Y_1
\end{pmatrix}\,.
\end{eqnarray}
Higher weight modular forms can be constructed analogously, and one should consider the nonlinear constraints such as Eq.~\eqref{eq:constraint-level3MF} to pick out the linearly independent modular forms from all possible tensor products. It is straightforward although tedious.

In the same fashion, the weight two modular forms of level 2 and level 4 can be constructed. There are two independent modular forms of weight 2 at level 2~\cite{Kobayashi:2018vbk},
\begin{eqnarray}
\frac{i}{4\pi}\left[\frac{\eta'\left(\frac{\tau}{2}\right)}{\eta\left(\frac{\tau}{2}\right)}  +\frac{\eta'\left(\frac{\tau+1}{2}\right)}{\eta\left(\frac{\tau+1}{2}\right)}
- \frac{8\eta'(2\tau)}{\eta(2\tau)} \right]\,,~~~~~~~\frac{\sqrt{3}i}{4\pi}\left[ \frac{\eta'\left(\frac{\tau}{2}\right)}{\eta\left(\frac{\tau}{2}\right)}  -\frac{\eta'\left(\frac{\tau+1}{2}\right)}{\eta\left(\frac{\tau+1}{2}\right)}   \right]\,,
\end{eqnarray}
which form a $S_3$ doublet. The basis of the weight 2 modular forms at level 4 can be chosen as~\cite{Penedo:2018nmg,Novichkov:2018ovf}
\begin{eqnarray}
\nonumber && -\frac{i}{8\pi}\left[\frac{\eta'\left(\frac{\tau}{4}\right)}{\eta\left(\frac{\tau}{4}\right)}+\frac{\eta'\left(\frac{\tau+1}{4}\right)}{\eta\left(\frac{\tau+1}{4}\right)}
+\frac{\eta'\left(\frac{\tau+2}{4}\right)}{\eta\left(\frac{\tau+2}{4}\right)}+\frac{\eta'\left(\frac{\tau+3}{4}\right)}{\eta\left(\frac{\tau+3}{4}\right)}-\frac{8\eta'\left(\tau+\frac{1}{2}\right)}{\eta\left(\tau+\frac{1}{2}\right)}
-\frac{32\eta'(4\tau)}{\eta(4\tau)} \right]\,,\\
\nonumber &&\frac{\sqrt{3}\,i}{8\pi}\left[\frac{\eta'\left(\frac{\tau}{4}\right)}{\eta\left(\frac{\tau}{4}\right)}-\frac{\eta'\left(\frac{\tau+1}{4}\right)}{\eta\left(\frac{\tau+1}{4}\right)}
+\frac{\eta'\left(\frac{\tau+2}{4}\right)}{\eta\left(\frac{\tau+2}{4}\right)}-\frac{\eta'\left(\frac{\tau+3}{4}\right)}{\eta\left(\frac{\tau+3}{4}\right)} \right]\,,\\
\nonumber && \frac{i}{\pi}\left[\frac{\eta'\left(\tau+\frac{1}{2}\right)}{\eta\left(\tau+\frac{1}{2}\right)}
-\frac{4\eta'(4\tau)}{\eta(4\tau)} \right]\,,\\
\nonumber &&\frac{-\sqrt{2}\,i}{8\pi}\left[\frac{\eta'\left(\frac{\tau}{4}\right)}{\eta\left(\frac{\tau}{4}\right)}-i\frac{\eta'\left(\frac{\tau+1}{4}\right)}{\eta\left(\frac{\tau+1}{4}\right)}
-\frac{\eta'\left(\frac{\tau+2}{4}\right)}{\eta\left(\frac{\tau+2}{4}\right)}+i\frac{\eta'\left(\frac{\tau+3}{4}\right)}{\eta\left(\frac{\tau+3}{4}\right)} \right]\,,\\
&&\frac{-\sqrt{2}\,i}{8\pi}\left[\frac{\eta'\left(\frac{\tau}{4}\right)}{\eta\left(\frac{\tau}{4}\right)}+i\frac{\eta'\left(\frac{\tau+1}{4}\right)}{\eta\left(\frac{\tau+1}{4}\right)}
-\frac{\eta'\left(\frac{\tau+2}{4}\right)}{\eta\left(\frac{\tau+2}{4}\right)}-i\frac{\eta'\left(\frac{\tau+3}{4}\right)}{\eta\left(\frac{\tau+3}{4}\right)} \right]\,,
\end{eqnarray}
which can be organized into a doublet and a triplet of $S_4$~\cite{Penedo:2018nmg,Novichkov:2018ovf}. For $N\geq 5$, the linear combinations of the logarithmic derivatives of $\eta$ function is not enough generate the whole modular form space, and consequently other seed functions suh as Klein forms and Jacobi constants are necessary. See Refs.~\cite{Novichkov:2018nkm,Ding:2019xna} and~\cite{Ding:2020msi} for the construction of modular forms of level 5 and level 7 respectively. We summarize the even weight modular forms and their decompositions under $\Gamma_N$ in table~\ref{tab:even-MF-ModGroup} for $N\leq7$.

\subsection{Integer weight modular forms and homogeneous finite modular groups}

\begin{table}[h]
\centering
\renewcommand{\arraystretch}{1.25}
\resizebox{1.0\textwidth}{!}{
\begin{tabular}{|c|c|c|c|c|c|c|}\hline \hline
\multirow{2}{*}{$N$}&\multirow{2}{*}{$\dim\mathcal{M}_{k}(\Gamma(N)) $}
&\multirow{2}{*}{$\Gamma'_N$} & \multicolumn{3}{c|}{modular multiplets} \\ \cline{4-6}
& & & $k=1$ & $k=2$ & $k=3$ \\ \hline

3 &$k+1$& $T'$ & $Y^{(1)}_{\mathbf{2}}$ & $Y^{(2)}_{\mathbf{3}}$ & $Y^{(3)}_{\mathbf{2}},Y^{(3)}_{\mathbf{2}''}$ \\ \hline

4&$2k+1$&$S'_4 $ & $Y^{(1)}_{\mathbf{\hat{3}}'}$ & $Y^{(2)}_{\mathbf{2}},Y^{(2)}_{\mathbf{3}}$ & $Y^{(3)}_{\mathbf{\hat{1}'}},Y^{(3)}_{\mathbf{\hat{3}}},Y^{(3)}_{\mathbf{\hat{3}'}}$ \\ \hline

5&$5k+1$&$A'_5$ &$Y^{(1)}_{\mathbf{6}}$ & $Y^{(2)}_{\mathbf{3}},Y^{(2)}_{\mathbf{3}'},Y^{(2)}_{\mathbf{5}}$ & $Y^{(3)}_{\mathbf{4}'},Y^{(3)}_{\mathbf{6}I},Y^{(3)}_{\mathbf{6}II}$ \\ \hline

\multirow{2}{*}{6} &\multirow{2}{*}{$6k$} & \multirow{2}{*}{$S_3\times T'$} & \multirow{2}{*}{$Y_{\bm{2^0_2}}^{(1)}$,  $Y^{(1)}_{\bm{4_1}}$}  & \multirow{2}{*}{$Y^{(2)}_{\bm{1^1_2}}$, $Y^{(2)}_{\bm{2_0}}$, $Y^{(2)}_{\bm{3^0}}$, $Y^{(2)}_{\bm{6}}$} &  $Y^{(3)}_{\bm{2^0_1}}$, $Y^{(3)}_{\bm{2^0_2}}$, $Y^{(3)}_{\bm{2^1_1}}$,\\

&  &  & &  & $Y^{(3)}_{\bm{4_0}}$, $Y^{(3)}_{\bm{4_1}}$, $Y^{(3)}_{\bm{4_2}}$  \\ \hline \hline

\end{tabular}}
\caption{Integer weight modular forms of level $N$ and the decomposition under the homogeneous finite modular group $\Gamma'_N$. Notice that odd weight modular form doesn't exist for $\Gamma(2)$. The conventions for the irreducible representations follow~\cite{Liu:2019khw} for $\Gamma'_3$, ~\cite{Liu:2020akv} for $\Gamma'_4$, ~\cite{Yao:2020zml} for $\Gamma'_5$ and ~\cite{Li:2021buv} for $\Gamma'_6$.\label{tab:integer-MF-ModGroup}}
\end{table}

As shown in the beginning of section~\ref{sec:MF-MG}, the principal congruence subgroups $\Gamma(N)$ for $N>2$ have odd weight modular forms because of $S^2\notin \Gamma(N>2)$, and the general integer weight modular forms of level $N$ can be arranged into irreducible multiplets of the homogeneous finite modular group $\Gamma'_N$ which is the double cover of $\Gamma_N$. The groups $\Gamma'_N$ for lower levels $N$ are summarized in in table~\ref{tab:finite-modular-groups}, they could be realized in orbifold compactifications on $T^{2}\times T^2$ with magnetic fluxes~\cite{Kikuchi:2020nxn}. Notice that the odd weight and even weight modular forms are in the representations $\rho_{\mathbf{r}}(R)=-1$ and $\rho_{\mathbf{r}}(R)=1$ respectively. All the modular forms can be obtained from the products of the lowest weight 1 modular forms and they can be constructed from the Dedekind $\eta$ function and Klein forms for small levels $N\leq6$. In the following, we present the linearly independent basis for the linear space $\mathcal{M}_{1}(\Gamma(N))$ of weight 1 modular forms~\cite{schultz2015notes,Li:2021buv},
\begin{eqnarray}
\nonumber \mathcal{M}_{1}(\Gamma(3))&=&\left\{\frac{\eta^{3}(3\tau)}{\eta(\tau)},~ \frac{\eta^{3}(\tau/3)}{\eta(\tau)}\right\}\,, \\
\nonumber \mathcal{M}_{1}(\Gamma(4))&=&\left\{\frac{\eta^{4}(4\tau)}{\eta^{2}(2\tau)},~ \frac{\eta^{10}(2\tau)}{\eta^{4}(4\tau)\eta^{4}(\tau)},~ \frac{\eta^{4}(2\tau)}{\eta^{2}(2\tau)}\right\}\,, \\
\nonumber \mathcal{M}_{1}(\Gamma(5))&=&\Big\{\frac{\eta^{15}(5\tau)}{\eta^3(\tau)}\mathfrak{k}^{5}_{\frac{2}{5}, 0}(5\tau),~\frac{\eta^{15}(5\tau)}{\eta^3(\tau)}\mathfrak{k}_{\frac{1}{5}, 0}(5\tau)\mathfrak{k}^{4}_{\frac{2}{5}, 0}(5\tau),~\frac{\eta^{15}(5\tau)}{\eta^3(\tau)}\mathfrak{k}^{2}_{\frac{1}{5}, 0}(5\tau)\mathfrak{k}^{3}_{\frac{2}{5}, 0}(5\tau),\\
\nonumber&&~ \frac{\eta^{15}(5\tau)}{\eta^3(\tau)}\mathfrak{k}^{3}_{\frac{1}{5}, 0}(5\tau)\mathfrak{k}^{2}_{\frac{2}{5}, 0}(5\tau),~ \frac{\eta^{15}(5\tau)}{\eta^3(\tau)}\mathfrak{k}^{4}_{\frac{1}{5}, 0}(5\tau)\mathfrak{k}_{\frac{2}{5}, 0}(5\tau),~\frac{\eta^{15}(5\tau)}{\eta^3(\tau)}\mathfrak{k}^{5}_{\frac{1}{5}, 0}(5\tau)\,.\Big\}\,, \\
\mathcal{M}_{1}(\Gamma(6))&=&\left\{\frac{\eta^3(3\tau)}{\eta(\tau)},~ \frac{\eta^3(\tau/3)}{\eta(\tau)},~ \frac{\eta^3(6\tau)}{\eta(2\tau)},~
\frac{\eta^3(\tau/6)}{\eta(\tau/2)},~ \frac{\eta^3(2\tau/3)}{\eta(2\tau)},~ \frac{\eta^3(3\tau/2)}{\eta(\tau/2)}\right\}\,,
\end{eqnarray}
where the Klein form $\mathfrak{k}_{(r_1,r_2)}(\tau)$ is a holomorphic function which has no zeros and poles on the upper half complex plane. The Klein form can be written into an infinite product expansion~\cite{K_lang1981,lang1987elliptic,lang2012introduction,eum2011modularity}:
\begin{equation}
\label{KleinForm}
\mathfrak{k}_{(r_1,r_2)}(\tau)=q^{(r_1-1)/2}_z(1-q_z)\prod_{n=1}^\infty(1-q^nq_z)(1-q^nq_z^{-1})(1-q^n)^{-2}\,,
\end{equation}
with $q_z=e^{2\pi iz}$ with $z=r_1\tau+r_2$. The integer weight modular forms up to weight 3 and the decomposition under $\Gamma'_N$ are listed in table~\ref{tab:integer-MF-ModGroup}.

\subsection{Half-integer weight modular form and metaplectic cover}

Let us first consider the problem for half-integer weight modular forms for the purposes of illustration. Then later we provide the solution for half-integer weight modular forms using the metaplectic cover.

Similar to integer weight modular form, one may attempt to define the half-integer weight modular form as a holomorphic function of $\tau$ and satisfying
\begin{equation}
\label{eq:wrong-def-Half-MF}Y(\gamma\tau)=(c\tau+d)^{k/2}Y(\tau)=J^{k/2}(\gamma,\tau)Y(\tau),~~~~ \gamma=\begin{pmatrix}
a & b \\ c & d
\end{pmatrix} \in SL(2,\mathbb{Z})
\end{equation}
for a positive integer $k$. We see that the square root of the $c\tau+d$ appears in the above transformation formula. It is known that the function of the square root has two branches $\sqrt{z}=\left\{z^{1/2}, -z^{1/2}\right\}$ for a complex number $z$. It is crucial to deal with the two branches for the square root in a systematic way.
The most common choice and the one we will always use is to choose the principal branch of the square root, i.e.,
\begin{equation}
z^{1/2}\equiv|z|^{1/2}\exp\left(i\frac{\texttt{Arg}(z)}{2}\right),~~~-\pi<\texttt{Arg}(z)\leq\pi\,.
\end{equation}
Then $z^{1/2}$ for real $z< 0$ is a pure positive imaginary number and $(-1)^{1/2}=i$. Hence $(z_1 z_2)^{1/2}$ is equal to $z_1^{1/2} z_2^{1/2}$ only up to a sign $\pm 1$, i.e.,
\begin{eqnarray}
(z_1z_2)^{1/2}= \left\{
\begin{array}{cc}
z_1^{1/2} z_2^{1/2}, ~&~ -\pi < \texttt{Arg}(z_1) + \texttt{Arg}(z_2) \leq \pi\\
-z_1^{1/2} z_2^{1/2},~&~ \text{otherwise}
\end{array}
\right.
\end{eqnarray}
For an (even or odd) integer $k$, $z^{k/2}$ always refer to $(z^{1/2})^k$. Note that this is not always equal to $(z^k)^{1/2}$ for $k$ odd. Because of the multivalues of the square root function, $J^{k/2}(\gamma, \tau)=(c\tau+d)^{k/2}$ doesn't satisfy the cocycle relation~\cite{stromberg2013weil,cohen2017modular}
\begin{equation}
J^{k/2}(\gamma_1,\gamma_2 \tau) J^{k/2}(\gamma_2, \tau)=\zeta^{k}_{1/2}(\gamma_1,\gamma_2)J^{k/2}(\gamma_1 \gamma_2,\tau)\neq J^{k/2}(\gamma_1 \gamma_2,\tau), \quad \gamma_{1,2} \in \Gamma\,,
\end{equation}
where the two-cocycle $\zeta_{1/2}(\gamma_1,\gamma_2)$ can only take values $+1$ and $-1$, and its explicit expression is given in~\cite{Liu:2020msy}. Notice that $\zeta^{k}_{1/2}(\gamma_1,\gamma_2)$ is always equal to 1 for any values of $\gamma_1$ and $\gamma_2$ if $k$ is even. As a consequence, and the definition in Eq.~\eqref{eq:wrong-def-Half-MF} is inconsistent, i.e.
\begin{equation}
Y(\gamma_1(\gamma_2(\tau)))\neq Y(\gamma_1\gamma_2(\tau))\,.
\end{equation}
Hence $(c\tau+d)^{k/2}$ is not the automorphy factor anymore. It turns out that the half-integral weight modular forms can only be defined on the principal congruence subgroup $\Gamma(4N)$, and a multiplier is generally needed. The weight $k/2$ weight modular form $Y(\tau)$ with $k$ positive integer is a holomorphic function of $\tau$ and satisfies the following condition~\cite{shimura1973modular}:
\begin{equation}
\label{eq:MF-Gamma4N}Y(h\tau)=v^k(h)(c\tau+d)^{k/2}Y(\tau)= v^k(h)J^{k/2}(h,\tau)Y(\tau),\qquad h=\begin{pmatrix}
a & b \\ c & d
\end{pmatrix} \in \Gamma(4N)\,,
\end{equation}
where $v(h)=(\frac{c}{d})$ is the Kronecker symbol, it is $1$ or $-1$ here and its more detailed properties can be found in~\cite{Liu:2020msy}. The Kronecker symbol $v(h)$ fulfills the following identity~\cite{stromberg2013weil,cohen2017modular}:
\begin{equation}
v(h_1 h_2)=\zeta_{1/2}(h_1,h_2)v(h_1)v(h_2),~~~h_{1,2} \in \Gamma(4N)\,.
\end{equation}
Thus one can check that $v^k(h)(c\tau+d)^{k/2}$ is the correct automorphy factor and it satisfies the cocycle relation, and the ambiguity caused by half-integral weight is eliminated by the factor $v(h)$.

Now we provide the solution to the problem of modular forms for half-integer modular weights using metaplectic cover.
In order to consistently define the half-integer modular forms, one should consider both branches of the square root function and consider the metaplectic (twofold) cover group of $SL(2, \mathbb{Z})$~\cite{shimura1973modular}, and it usually denoted as $\text{Mp}_2(\mathbb{Z})$ in the literature. For notation simplicity, we shall denote $\text{Mp}_2(\mathbb{Z})$ as $\widetilde{\Gamma}$ in the following. It is well-known that can be realized as the group of pairs $(\gamma, \epsilon)$, where $\epsilon=\pm1$ stands for the branch of the square root function. Consequently the group element of $\widetilde{\Gamma}$ can be written in the form~\cite{shimura1973modular,Duncan:2018wbw,Kikuchi:2020frp}:
\begin{equation}
\label{eq:Gamma-tilde}\widetilde{\Gamma} \equiv \left\{ [\gamma, \epsilon] \bigl| \gamma \in \Gamma, \ \epsilon \in \{ \pm 1 \} \right\}\,.
\end{equation}
Obviously each element $\gamma\in\Gamma$ corresponds to two elements $\widetilde{\gamma}=\left[\gamma, \pm \right]$ of the metaplectic group $\widetilde{\Gamma}$. The multiplication law of arbitrary two elements, $[\gamma_1, \epsilon_1], [\gamma_2, \epsilon_2] \in \widetilde{\Gamma}$, is defined by\footnote{Alternative but equivalent definition of $\widetilde{\Gamma}$ is given by~\cite{shimura1973modular}
\begin{eqnarray*}
\widetilde{\Gamma} = \Big\{ \widetilde{\gamma}=(\gamma, \epsilon J^{1/2}(\gamma,\tau)) ~\Big|~ \gamma=\begin{pmatrix}
a  &  b  \\
c   &  d
\end{pmatrix} \in \Gamma,~~\epsilon=\pm 1 \Big\}\,.
\end{eqnarray*}
The multiplication rule accordingly is
\begin{eqnarray*}
\nonumber(\gamma_1,\epsilon_1J^{1/2}(\gamma_1,\tau))(\gamma_2, \epsilon_2J^{1/2}(\gamma_2,\tau))&=&(\gamma_1\gamma_2, \epsilon_1\epsilon_2 J^{1/2}(\gamma_1, \gamma_2\tau) J^{1/2}(\gamma_2,\tau))\\
&=&(\gamma_1\gamma_2, \epsilon_1\epsilon_2\zeta_{1/2}(\gamma_1,\gamma_2)J^{1/2}(\gamma_1\gamma_2,\tau))\,,
\end{eqnarray*}}
\begin{equation}
\label{eq:MP2z-multirule}[\gamma_1, \epsilon_1] [\gamma_2, \epsilon_2] = [\gamma_1\gamma_2, \zeta_{1/2}(\gamma_1,\gamma_2)\epsilon_1\epsilon_2]\,.
\end{equation}
Using the generators $S$ and $T$ of $SL(2, \mathbb{Z})$, the metaplectic group $\widetilde{\Gamma}$ can also be generated by two generators $\widetilde{S}$ and $\widetilde{T}$~\cite{bruinier2010weil,stromberg2013weil}:
\begin{equation}
\widetilde{S}= \left[\begin{pmatrix}
0 ~& 1\\ -1 ~& 0 \end{pmatrix} , -1\right], ~\quad~ \widetilde{T}=\left[\begin{pmatrix}
1 ~& 1\\ 0 ~& 1 \end{pmatrix} , +1\right]\,.
\end{equation}
One can check that the generators $\widetilde{S}$ and $\widetilde{T}$ obey the following relations~\cite{Liu:2020msy}
\begin{equation}
\label{eq:STtilde-relations}\widetilde{S}^8=(\widetilde{S}\widetilde{T})^3=1\,,
\end{equation}
Consequently $\widetilde{S}$ is of order 8, and particularly we have
\begin{equation}
\widetilde{R} \equiv  \widetilde{S}^2 = \left[\begin{pmatrix}
 -1 & 0\\ 0 & -1 \end{pmatrix}, -1\right]\,,~~~\widetilde{S}\widetilde{R}=\widetilde{R}\widetilde{S},~~~\widetilde{T}\widetilde{R}=\widetilde{R}\widetilde{T}\,.
\end{equation}
Consequently the center of $\widetilde{\Gamma}$ is the $Z_4$ subgroup generated by $\widetilde{R}$. Moreover, we have the identities
\begin{equation}
\widetilde{R}^2=\left[\begin{pmatrix}
1 & 0\\ 0 & 1 \end{pmatrix}, -1\right]\,,~~~\widetilde{R}^2[\gamma, \epsilon] =[\gamma, -\epsilon]
\end{equation}
Therefore the modular group $SL(2, \mathbb{Z})$ is isomorphic to the quotient of $\text{Mp}_{2}(\mathbb{Z})$ over the $Z_2$ subgroup $Z^{\widetilde{R}^2}_2=\{1, \widetilde{R}^2\}$,
\begin{equation}
\text{Mp}_{2}(\mathbb{Z})/Z^{\widetilde{R}^2}_2\cong\text{SL}(2,\mathbb{Z})\,.
\end{equation}
It is known that metaplectic congruence subgroup of level $4N$ is~\cite{bruinier2010weil,stromberg2013weil}:
\begin{equation}
\label{eq:meta-congruence}\widetilde{\Gamma}(4N) = \Big\{ \widetilde{h}=\left[h, v(h)\right] ~|~ h \in \Gamma(4N) \Big\}\,,
\end{equation}
$\widetilde{\Gamma}(4N)$ is an infinite normal subgroup of $\widetilde{\Gamma}$ and it is isomorphic to the principal congruence subgroup $\Gamma(4N)$.
Likewise the finite metaplectic group is the quotient group $\widetilde{\Gamma}_{4N} \equiv \widetilde{\Gamma}/\widetilde{\Gamma}(4N)$. We can check $\widetilde{T}^{4N}$ is an element of $\widetilde{\Gamma}(4N)$:
\begin{equation}
\widetilde{T}^{4N}=\left[\begin{pmatrix}
1 ~&~ 4N \\ 0 ~&~ 1 \end{pmatrix} ,~ +1 \right]\in \widetilde{\Gamma}(4N)\,.
\end{equation}
Hence the generator $\widetilde{T}$ of the finite metaplectic group $\widetilde{\Gamma}_{4N}$ additionally fulfills
\begin{equation}
\label{eq:T-Gamma-sub-4N}\widetilde{T}^{4N}=1\,.
\end{equation}
For the smallest value $N=1$, the finite metaplectic group $\widetilde{\Gamma}_{4}$ denoted as $\widetilde{S}_4$ is a group of order 96 with group ID $[96, 67]$ in the computer algebra program \texttt{GAP}~\cite{GAP}. See~\cite{Liu:2020msy} for more details about group theory of $\widetilde{S}_4$. For larger $N$, the relations in Eqs.~(\ref{eq:STtilde-relations}, \ref{eq:T-Gamma-sub-4N}) are not sufficient and addition relations are needed to render the group $\widetilde{\Gamma}_{4N}$ finite. For example, in the case of $N=2$,
the multiplication rules of $\widetilde{\Gamma}_{8}$ for the generators $\widetilde{S}$ and $\widetilde{T}$ are~\cite{Liu:2020msy}
\begin{equation}
\widetilde{S}^2 =\widetilde{R},~~~ (\widetilde{S}\widetilde{T})^3=\widetilde{R}^4=\widetilde{T}^8=\widetilde{R}^2\widetilde{S}\widetilde{T}^6\widetilde{S}\widetilde{T}^4\widetilde{S}\widetilde{T}^2\widetilde{S}\widetilde{T}^4=1,
~~~\widetilde{T}\widetilde{R}=\widetilde{R}\widetilde{T}\,.
\end{equation}
Thus $\widetilde{\Gamma}_{8}$ is a group of order 768 and its group \texttt{ID} is [768,\, 1085324] in \texttt{GAP}.

The weight $k/2$ modular forms of $\widetilde{\Gamma}(4N)$
span a linear space $\mathcal{M}_{k/2}(\widetilde{\Gamma}(4N))$ of finite dimension $k+1$, where $k$ is a non-negative integer. In the same manner as shown in the beginning of section~\ref{sec:MF-MG}, it is always possible to choose the basis of $\mathcal{M}_{k/2}(\widetilde{\Gamma}(4N))$ so that the modular forms of weight $k/2$ can be arranged into multiplets of the finite metaplectic group $\widetilde{\Gamma}_{4N}$~\cite{Liu:2020msy}, i.e.
\begin{equation}
\label{eq:MF-RW-decomp}Y_\mathbf{r}(\widetilde{\gamma}\tau)=(\epsilon\sqrt{c\tau +d})^k\rho_{\mathbf{r}}(\widetilde{\gamma})Y_\mathbf{r}(\tau),~~~\widetilde{\gamma}=\left[\gamma, \epsilon\right]\in\widetilde{\Gamma}\,,
\end{equation}
where $\widetilde{\gamma}\tau=\gamma\tau$, and $\rho_{\mathbf{r}}$ is an irreducible representation of $\widetilde{\Gamma}_{4N}$. Applying Eq.~\eqref{eq:MF-RW-decomp} for $\widetilde{\gamma}=\widetilde{R}$, we obtain
\begin{equation}
Y_{\mathbf{r}}(\widetilde{R}\tau)=Y_{\mathbf{r}}(\tau)= (-i)^k \rho_{\mathbf{r}}(\widetilde{R})Y_{\mathbf{r}}(\tau)\,,
\end{equation}
which implies
\begin{equation}
\rho_{\mathbf{r}}(\widetilde{R})=i^{k},~~~~\begin{cases}
k ~\texttt{odd}\,~: ~\rho^4(\widetilde{R})=1\,, \\
k ~\texttt{even}: ~\rho^2(\widetilde{R})=1\,.
\end{cases}
\end{equation}
From the dimension formula, we know that there are only two linearly independent weight $1/2$ modular forms of level 4 and they can be expressed in terms of the Jacobi theta functions~\cite{Novichkov:2020eep,Liu:2020msy}:
\begin{equation}
Y_1(\tau)=\theta_3(0|2\tau),~~~Y_2(\tau)=-\theta_2(0|2\tau),
\end{equation}
with
\begin{eqnarray}
\nonumber\theta_2(0|2\tau)&=&\sum_{m \in \mathbb{Z}} e^{2\pi i \tau  (m+\frac{1}{2})^2 } =2q^{1/4}(1+q^2+q^6+q^{12}+\dots)\,, \\
\label{eq:theta-function-def} \theta_3(0|2\tau)&=&\sum_{m \in \mathbb{Z}} e^{2\pi i \tau  m^2}=1+2q+2q^4+2q^9+2q^{16}+\dots \,,
\end{eqnarray}
It turns out that $Y_1(\tau)$ and $Y_2(\tau)$ can be arranged into a doublet $\mathbf{\hat{2}}$ of the finite metaplectic group $\widetilde{\Gamma}_4 \equiv \widetilde{S_4}$~\cite{Liu:2020msy},
\begin{equation}
Y^{(\frac{1}{2})}_{\mathbf{\hat{2}}}(\tau)\equiv
\begin{pmatrix}
Y_1(\tau) \\
Y_2(\tau)
\end{pmatrix}\,,
\end{equation}
which transforms in the two-dimensional irreducible representation $\mathbf{\hat{2}}$ of $\widetilde{S_4}$,
\begin{eqnarray}
\nonumber&&Y^{(\frac{1}{2})}_{\mathbf{\hat{2}}}(\tau)\stackrel{S}{\longmapsto} Y^{(\frac{1}{2})}_{\mathbf{\hat{2}}}(-1/\tau)=-\sqrt{-\tau}\,\rho_{\mathbf{\hat{2}}}(\widetilde{S})Y^{(\frac{1}{2})}_{\mathbf{\hat{2}}}(\tau),\,,\\
&&Y^{(\frac{1}{2})}_{\mathbf{\hat{2}}}(\tau)\stackrel{T}{\longmapsto}
Y^{(\frac{1}{2})}_{\mathbf{\hat{2}}}(\tau+1)=\rho_{\mathbf{\hat{2}}}(\widetilde{T})Y^{(\frac{1}{2})}_{\mathbf{\hat{2}}}(\tau)\,,
\end{eqnarray}
where the unitary representation matrices $\rho_{\mathbf{\hat{2}}}(\widetilde{S})$ and $\rho_{\mathbf{\hat{2}}}(\widetilde{T})$ are given in Ref.~\cite{Liu:2020msy}. All half-integer and integer weight modular forms of level 4 can be generated from the tensor products of $Y^{(\frac{1}{2})}_{\mathbf{\hat{2}}}(\tau)$. For example, we find the weight 1 modular forms make up a triplet $\mathbf{\hat{3}'}$ of $\widetilde{S}_4$,
\begin{equation}
Y_{\mathbf{\hat{3}'}}^{(1)}=\frac{1}{\sqrt{2}}\left(Y_{\mathbf{\hat{2}}}^{(\frac{1}{2})}Y_{\mathbf{\hat{2}}}^{(\frac{1}{2})}\right)_{\mathbf{\hat{3}'}}=\begin{pmatrix}
\sqrt{2}\;Y_1Y_2\\
Y_1^2\\
-Y_2^2\\
\end{pmatrix}\,.
\end{equation}
It is straightforward to check that $Y^{(1)}_{\mathbf{\hat{3}'}}$ is the same as the original weight one modular forms given in~\cite{Novichkov:2020eep} up to a permutation, the discrepancy arises from the different convention for the representation matrices of the generators $S$ and $T$. In a similar fashion, we can obtain higher weights modular forms and decompose them into different irreducible multiplets of $\widetilde{S}_4$, see Ref.~\cite{Liu:2020msy} for the details. It is remarkable that $Y_1(\tau)$ and $Y_{2}(\tau)$ are algebraically independent, and each modular form of integral weight $k/2$ and level 4 can be written as a polynomial of degree $k+1$ in $Y_1(\tau)$ and $Y_{2}(\tau)$:
\begin{equation}
\sum_{i=0}^{k} c_i Y^i_{1}(\tau)Y^{k-i}_2(\tau)\,,
\end{equation}
where $c_i$ are free coefficients. As a result, the construction of higher weight modular forms from $Y_{1}(\tau)$ and $Y_2(\tau)$ naturally bypasses the need to search for non-linear constraints relating redundant higher weight multiplets.

\subsection{Multiplier system and explicit form of the rational weight modular forms}

Now we turn to the solution to the problem of modular forms for general rational modular weights.
The previous procedure may be followed analogously for the case of modular forms of rational weight $r$ for certain congruence subgroups. Similar to Eq.~\eqref{eq:MF-Gamma4N}, a multiplier system $v(\gamma)$ is needed such that $v(\gamma)(c\tau+d)^{r}$ is the correct automorphy factor satisfying the cocycle relation, and the ambiguity of multi-valued branches caused by the rational power is properly eliminated. For the principal congruence subgroup $\Gamma(N)$ of level odd integer $N\geq5$, a unified construction of multiplier systems denote by $v_N(\gamma)$ has been given in~\cite{ibukiyama2000modular}, and it is found that the corresponding modular forms are of weight $(N-3)/(2N)$. Specifically, $v_N$ is given by the following formula
\begin{equation}
v_N(\gamma)=
\begin{cases}
1~~~~~~~~~~~~~~~~~~~~~~~~~~~~~~~~~~~~~~~~~~~~~~~~~~~~~~~~~~~~~~~~~~~~~\qquad\text{if}~~ c=0 \,,\\[0.1in]
\exp\left(-2\pi i\dfrac{3\,\text{sign}(c)(N^2-1)}{8N}\right)\exp\left(2\pi i\dfrac{N^2-1}{8N}\Phi(\gamma)\right) \qquad\text{if}~~ c \neq 0 \,,
\end{cases}
\end{equation}
where $\gamma=\begin{pmatrix}
a&b\\c& d \end{pmatrix} \in \Gamma(N)$, and $\Phi(\gamma)$ is given by
\begin{equation}
\Phi(\gamma)=
\begin{cases}
\dfrac{b}{d} ~~~~~~~~~~~~~~~~~~~\qquad\qquad\qquad\text{if}~~ c=0 \,,\\[0.2in] \dfrac{a+d}{c}-12\, \text{sign}(c)\, s(d,|c|) \qquad\text{if}~~ c \neq 0 \,,
\end{cases}
\end{equation}
where $s(d,|c|)$ is the Dedekind sum with
\begin{equation}
s(h,k)=\sum^k_{\mu=1} \left(\left(\frac{h\mu}{k}\right)\right)\left(\left(\frac{\mu}{k}\right)\right)\,,
\end{equation}
for integers $h, k(k\neq 0)$.
Here $((x))$ is the sawtooth function defined by
\begin{equation}
((x)) =\begin{cases}  x- [x]-\frac{1}{2} ~\quad~ \text{if}~ x \notin \mathbb{Z}\,, \\
0  ~~~~~~~~~~~~~\quad~~ \text{if}~ x \in \mathbb{Z}\,,
\end{cases}
\end{equation}
with $[x]$ the floor function. Note that the multiplier system $v_N(\gamma)$ is an $N$-th root of unity, consequently $v_N(\gamma)^N=1$ for
all $\gamma \in \Gamma(N)$. In short,  $v_N(\gamma)(c\tau+d)^{(N-3)/2N}$ is the automorphy factor for the modular form of weight $(N-3)/(2N)$ at level odd integer $N\geq5$.

Moreover, for any odd integer $5 \leq N \leq 13$, the ring of the modular forms of rational weight $r=(N-3)/(2N)$ for the principal congruence subgroup $\Gamma(N)$ can be constructed from the holomorphic functions $f^{(N)}_n(\tau)$~\cite{ibukiyama2000modular},
\begin{equation}
\label{eq:rationalMF}
f^{(N)}_n(\tau)=\theta_{(\frac{n}{2N},\frac{1}{2})}(N\tau)/\eta(\tau)^{\frac{3}{N}}\,,
\end{equation}
where $n$ are odd integers with $1 \leq n \leq N-2$, and the theta constant with characteristic $(m',m'')$ is defined by
\begin{equation}
\label{eq:theta_constans}
\theta_{(m',m'')}(\tau)=\sum_{m\in \mathbb{Z}} e^{2\pi i \tau (\frac{1}{2}(m+m')^2\tau+(m+m')m'')}\,,
\end{equation}
and Dedekind eta function is
\begin{equation}
\eta(\tau)=e^{\pi i \tau/12} \prod_{n=1}^{\infty} (1-e^{2\pi i n\tau})\,.
\end{equation}
Consequently there are $(N-1)/2$ linearly independent modular forms $f^{(N)}_1, f^{(N)}_3, \dots f^{(N)}_{N-2}$ of rational weight $r=(N-3)/(2N)$, and the graded rings of modular forms $\mathcal{M}(\Gamma(N)) = \bigoplus_{m\geq 1} \mathcal{M}_{m\frac{N-3}{2N}}(\Gamma(N))$ can be generated by the tensor products of these lowest weight modular forms.
The dimension formula of $ \mathcal{M}_{m\frac{N-3}{2N}}(\Gamma(N))$ for any odd integer $N \geq 5$ and any integer $m > \frac{4(N-6)}{N-3}$ is given by~\cite{ibukiyama2000modular,ibukiyama2020graded}
\begin{equation}
\dim \mathcal{M}_{m\frac{(N-3)}{2N}}(\Gamma(N)) =\frac{N^2\left[m(N-3)-2(N-6)\right]}{48}\prod_{p | N} (1-\frac{1}{p^2})\,,
\end{equation}
where the product is over the prime divisors $p$ of $N$. As shown in Eq.~\eqref{eq:MF-RW-decomp}, we expect that rational weight modular forms can be organized into different irreducible multiplets of the finite metaplectic congruence subgroup. We summarize the dimension formula, modular forms of rational weight $r=(N-3)/(2N)$ and the corresponding finite metaplectic group in table~\ref{tab:MFspace-and-finite-group}. The finite metaplectic groups can arise from compactifications on tori with magnetic background fluxes~\cite{Kikuchi:2020frp,Almumin:2021fbk}. The theory of modular forms with real weight and even complex weight has been developed as well~\cite{bruggeman2018holomorphic}, and then the modular group $SL(2, \mathbb{Z})$ should be extended to the universal covering groups. Some concrete examples are given in~\cite{aoki2017jacobi,manin2018modular}.

\begin{table}[t!]
\centering
\resizebox{1.0\textwidth}{!}{
\begin{tabular}{|c|c|c|c|c|c|c|}
\hline\hline
$N$ & weight $r$ & $\texttt{dim}\mathcal{M}_{r}(\Gamma(N)) $  & $\mathcal{M}_{r}(\Gamma(N))|_{k=1}$ &  $\widetilde{\Gamma}_{N}$ & $|\widetilde{\Gamma}_{N}|$ & $\texttt{GAP~ID}$   \rule[-2ex]{0pt}{5ex} \\ \hline
4 & $k/2$& $k+1$ & $\Big\{\theta_3(0|2\tau),~ \theta_2(0|2\tau) \Big\}$ & $\widetilde{S}_4$ & 96 & [96,67]  \rule[-2ex]{0pt}{5ex} \\ \hline
5 & $k/5$ &$k+1$ & $\Big\{f^{(5)}_1(\tau), ~f^{(5)}_3(\tau) \Big\}$ &  $Z_5\times \Gamma'_5$   &  600 & [600,54]  \rule[-2ex]{0pt}{5ex} \\ \hline
7 & $2k/7$ & $\begin{cases} 4k-2 ~\,(\text{for}~ k \geq 2) \\ 3~\, (\text{for}~ k=1) \end{cases}$  & $\Big\{f^{(7)}_1(\tau), ~f^{(7)}_3(\tau), ~f^{(7)}_5(\tau) \Big\}$ & $Z_7\times \Gamma_7$ &  1176 & [1176,212]   \rule[-2ex]{0pt}{5ex} \\ \hline
9 & $k/3$ & $\begin{cases} 9k-9 ~\,(\text{for}~ k \geq 3) \\ 10~\, (\text{for}~ k=2) \\ 4 ~~~ (\text{for}~ k=1) \end{cases}$ & $\Big\{f^{(9)}_1(\tau), ~f^{(9)}_3(\tau), ~f^{(9)}_5(\tau),~f^{(9)}_7(\tau) \Big\}$ &  $\widetilde{\Gamma}_9$ & 1944 & [1944,2976]  \rule[-2ex]{0pt}{5ex}\\ \hline \hline
\end{tabular}}
\caption{The dimension formula of the rational weight modualr form linear  $\texttt{dim}\mathcal{M}_{r}(\Gamma(N))$ for $N=4,5,7,9$, the linear space $\mathcal{M}_{r}(\Gamma(N))|_{k=1}$ of the lowest fractional weight modular forms, and the finite metaplectic group $\widetilde{\Gamma}_{N}$. Notice that $\theta_{2,3}(0|2\tau)$ and $f^{(N)}_n(\tau)$ are defined in Eq.~\eqref{eq:theta-function-def} and Eq.~\eqref{eq:rationalMF} respectively.
\label{tab:MFspace-and-finite-group} }
\end{table}

\subsection{Most general finite modular group and vector-valued modular form (VVMF)}

Let $\rho$ denote a $d$-dimensional representation of $SL(2,\mathbb{Z})$.
Mathematically this means that each element $\gamma$ of $SL(2,\mathbb{Z})$ is mapped onto its image consisting of one $d\times d$ matrix $\rho (\gamma)$.
A vector-valued modular form $Y(\tau)=(Y_1(\tau),\dots,Y_d(\tau))^T$ of weight $k$ in representation $\rho$ is a holomorphic vector-valued function in the upper
half-plane $\mathcal{H}$, and transforms as~\cite{knopp2004vector,bantay2007vector,marks2010structure,gannon2014theory,franc2016hypergeometric,franc2018structure}
\begin{equation}
\label{eq:defVVMF}
Y(\gamma\tau)=(c\tau+d)^k\rho(\gamma)Y(\tau),~~~\gamma\in\Gamma\,,
\end{equation}
where $k$ is called the modular weight and $\rho(\gamma)$ is the irreducible representation matrix of $SL(2,\mathbb{Z})$. Moreover,  $Y(\tau)$ is generally required to satisfy the so-called moderate growth condition, i.e., there exists an integer $n$ such that for each components $Y_j(\tau)$ we have $|Y_j(\tau)|<\texttt{Im}(\tau)^n$ for any fixed $\tau$ with $\texttt{Im}(\tau)\gg 0$. In other word, $Y_j(\tau)$ is bounded as $\texttt{Im}(\tau)\rightarrow+\infty$.

In the present work, we will focus on the case that $k$ is a non-negative integer and $\rho$ has finite image. Finite image means that, although there are infinitely many elements $\gamma$, there are only a finite number of representation matrices $\rho (\gamma)$.
Note that $\rho$ can be taken to be unitary since the image of $\rho$ is finite, and it is also completely reducible by Maschke’s theorem~\cite{rao2006linear}, i.e., it can be decomposed into a direct sum of irreducible representations. Therefore, we can further restrict $\rho$ to the unitary irreducible representations without loss of generality. Also the representation matrix $\rho(\gamma)$ should have finite order for any $\gamma\in SL(2,\mathbb{Z})$ so that one can always work in the basis where $\rho(T)$ is diagonal, and the eigenvalues of $\rho(T)$ are roots of unity. Finally, the finiteness of the image of $\rho$ implies that the kernel $\ker(\rho)$ is a normal subgroup of $SL(2,\mathbb{Z})$ with finite index.

In the original paradigm of modular flavor symmetry~\cite{Feruglio:2017spp}, the modular form of level $N$ satisfying the condition in Eq.~\eqref{eq:def-MF} is assumed, $\ker(\rho)$ is chosen to be the principal subgroup $\Gamma(N)$, and the representation matrices $\rho(\gamma)$ form finite modular group $\Gamma_N$ or $\Gamma'_N$. In the framework of VVMF, $\Gamma(N)$ and the corresponding modular forms are not required, and we start from the more general VVMF obeying Eq.~\eqref{eq:defVVMF}.

The theory of the VVMF has been developed by Mason et al~\cite{marks2009classification,marks2010structure} in recent years, and the explicit form of VVMF can be obtained by solving the so-called modular linear differential equation~\cite{Liu:2021gwa,marks2009classification,marks2010structure}. In the following, we report the results for VVMF of low dimensions~\cite{Liu:2021gwa}.

\begin{itemize}
\item{one-dimensional irreducible VVMFs}

It is known that the modular generators $S$ and $T$ the relations $S^4=(ST)^3=1$. Consequently the $SL(2,\mathbb{Z})$ group has 12 one-dimensional irreducible representations denoted as $\mathbf{1}_p$ with
\begin{equation}
\label{eq:SL2Zsinglets}
\mathbf{1}_p:~~~~~	\rho_{\mathbf{1}_p}(S)=i^p\,,\qquad \rho_{\mathbf{1}_p}(T)=e^{\frac{i\pi}{6} p}\,,
\end{equation}
where $p=0,1,\dots,11$. The lowest weight of the VVMFs in the representation $\mathbf{1}_p$ is equal to $p$ and accordingly the minimal weight VVMF $Y$ is found to take the following form
\begin{equation}
\label{eq:1stMLDEsolution}
Y = \eta^{2p}\,.
\end{equation}
We would like to mention that $\eta$ is the famous Dedekind-eta function defined in Eq.~\eqref{eq:eta-func}. One can check that $\eta^{2p}(\tau)$ satisfies the following transformation form
\begin{align}
\label{eq:eta2p_tra}
\nonumber&\eta^{2p}(S\tau)=(-\tau)^{p} i^{p}\eta^{2p}(\tau) = (-\tau)^{p} \rho_{\mathbf{1}_{p}}(S)\eta^{2p}(\tau),\\
&\eta^{2p}(T\tau)=e^{\frac{i\pi}{6}p}\eta^{2p}(\tau)=\rho_{\mathbf{1}_{p}}(T)\eta^{2p}(\tau)\,.
\end{align}
Hence $\eta^{2p}(\tau)$ is indeed the weight $p$ one-dimensional VVMFs  in the singlet irrep $\mathbf{1}_{p}$. The free module theorem implies that all one-dimensional VVMFs can be expressed as~\cite{marks2010structure}:
\begin{equation}
\mathcal{M}(\mathbf{1}_p)=\mathbb{C}[E_4,E_6]\eta^{2p}\,,\qquad p=0,1,\dots,11\,.
\end{equation}
Here $\mathbb{C}[E_4,E_6]$ denotes the polynomial of the Eisenstein series $E_4(\tau)$ and $E_6(\tau)$~\cite{cohen2017modular}:
\begin{eqnarray}
\nonumber\hskip-0.4in  &&E_4(\tau)=1+240\sum_{n=1}^{\infty}\sigma_3(n)q^n= 1+240 q+2160 q^2+6720 q^3+17520 q^4+\ldots\,,\\
\hskip-0.4in &&E_6(\tau)=1-504\sum_{n=1}^{\infty}\sigma_5(n)q^n=1-504 q-16632 q^2-122976 q^3-532728 q^4+\ldots\,,
\end{eqnarray}
where $\sigma_k(n) = \sum_{d|n} d^k$ is the sum of the $k$th power of the divisors of $n$ and $q=e^{2\pi i\tau}$. Notice that $E_4(\tau)$ and $E_6(\tau)$ are scalar-valued modular forms of $SL(2, \mathbb{Z})$ with modular weight 4 and 6 respectively.

\item{two-dimensional irreducible VVMFs}

All the 2-d irreducible unitary representations of $SL(2,\mathbb{Z})$ with finite image have been fully obtained in Ref.~\cite{mason20082}. It is known that there are 54 inequivalent 2-d irreps and up to a possible similar transformation the representation matrices are given by
\begin{eqnarray}
\label{eq:rho-S-T-2D}\nonumber &\rho(T)=\begin{pmatrix}
 e^{2\pi i r_1} & 0 \\
 0 & e^{2\pi i r_2}
\end{pmatrix},  \qquad 0\leq r_1,r_2 < 1\,, \\
&\rho(S)=\frac{(\lambda_1\lambda_2)^2}{\lambda_2-\lambda_1}\begin{pmatrix}
1 &\sqrt{-(\lambda_1\lambda_2)^5(\lambda_1-\lambda_2)^2-1} \\
	\sqrt{-(\lambda_1\lambda_2)^5(\lambda_1-\lambda_2)^2-1} & -1
\end{pmatrix}\,,
\end{eqnarray}
with $\lambda_{1,2}=e^{2\pi i r_{1,2}}$. We will denote the 2-d irreducible representation (irrep) as $\mathbf{2}_{(r_1,r_2)}$, and the allowed values of the unordered pair $(r_1,r_2)$ are given in~\cite{Liu:2021gwa}. We can read out the exponent matrix $L$ as $L=\texttt{diag}(r_1,r_2)$ and the minimal weight $k_0=6\text{Tr}(L)-1=6(r_1+r_2)-1$. The corresponding minimal weight VVMF in the doublet representation $\mathbf{2}_{(r_1,r_2)}$ is found to be~\cite{Liu:2021gwa}:
\begin{equation}
\label{eq:2dVVMF}
Y(\tau)=
\begin{pmatrix}
~~\eta^{12(r_1+r_2)-2}K^{\frac{6(r_1-r_2)+1}{12}}~ {}_2F_1(\frac{6(r_1-r_2)+1}{12},\frac{6(r_1-r_2)+5}{12};r_1-r_2+1;K) \\
C \eta^{12(r_1+r_2)-2}K^{\frac{6(r_2-r_1)+1}{12}}~ {}_2F_1(\frac{6(r_2-r_1)+1}{12},\frac{6(r_2-r_1)+5}{12};r_2-r_1+1;K) \\
\end{pmatrix}
\end{equation}
up to an overall irrelevant constant.  The coefficient $C$ depends on the explicit form of the representation matrix $\rho(S)$, and its value can be fixed to satisfy the condition Eq.~\eqref{eq:defVVMF} of VVMF for $\gamma=S$. The function $K(\tau)$ is defined as $K(\tau)=1728/j(\tau)$, where $j(\tau)$ is the modular $j$-invariant~\cite{cohen2017modular} with the $q$-expansion:
\begin{equation}
j(\tau)=q^{-1}+744+196884q+21493760q^2+\dots\,.
\end{equation}
${}_nF_{n-1}$ is the generalized hypergeometric series and it is defined by the formula
\begin{equation}
\label{eq:HyperSeries}
{}_nF_{n-1}(a_1,\dots,a_n;b_1,\dots, b_{n-1}; z)= \sum_{m\geq 0}^{\infty} \dfrac{\prod_{j=1}^{n}(a_j)_m}{\prod_{k=1}^{n-1}(b_k)_m} \dfrac{z^m}{m!}\,,
\end{equation}
and $(a)_m$ is the Pochhammer symbol defined by
\begin{equation}
(a)_m = \begin{cases}
1 \,, ~~~~~~~~~~~~~~~~~~~~~~~~~~~\quad\qquad m=0 \,, \\
a(a+1)\dots(a+m-1) \,,\ \qquad m\geq 1\,.
\end{cases}
\end{equation}
Hence we know the module of two-dimensional VVMFs is given by
\begin{equation}
\mathcal{M}(\mathbf{2}_{(r_1,r_2)})=\mathbb{C}[E_4,E_6]Y\oplus \mathbb{C}[E_4,E_6]D_{k_0}Y\,,
\end{equation}
where $D_{k_0}$ is the modular differential operators acting on the VVMFs of weight $k_{0}$~\cite{Bruinier2008The},
\begin{equation}
D_{k_0}\equiv \theta - \frac{k_0E_2}{12}\,,\quad \theta\equiv q\frac{d}{dq}=\frac{1}{2\pi i} \frac{d}{d\tau},\quad k_0\in\mathbb{Z}\,,
\end{equation}
with $E_2(\tau)$ being the well-known quasi-modular Eisenstein series of weight 2~\cite{cohen2017modular}
\begin{equation}
\label{eq:E2}
E_2(\tau)=1-24\sum_{n=1}^{\infty}\sigma_1(n)q^n=1-24q-72q^2-96q^3-168q^4-144q^5+\ldots \,.
\end{equation}
For example, the group $\Gamma_2\cong S_3$ has doublet irreducible representation with~\cite{Kobayashi:2018vbk}
\begin{equation}
\rho(S)=\frac{1}{2}\begin{pmatrix}
-1 ~&~ -\sqrt{3}\\
-\sqrt{3} ~&~1
\end{pmatrix},~~~~\rho(T)=\begin{pmatrix}
1  ~&~0 \\
0  ~&~-1
\end{pmatrix}\,,
\end{equation}
which implies $r_1=0$ and $r_2=1/2$ and the weight $k_0=6(r_1+r_2)-1=2$. From the general expression of two-dimensional VVMF in Eq.~\eqref{eq:2dVVMF} and considering the $S$ transformation, we can fix the parameter $C=\frac{1}{3}$ and
\begin{equation}
Y_{\mathbf{2}_{(0, 1/2)}}(\tau)=
2\sqrt{3}\begin{pmatrix}
~~\eta^{4}K^{-1/6}~ {}_2F_1(-\frac{1}{6},\frac{1}{6};\frac{1}{2};K) \\
\frac{1}{3}\eta^{4}K^{1/3}~ {}_2F_1(\frac{1}{3},\frac{2}{3};\frac{3}{2};K) \\
\end{pmatrix}\,,
\end{equation}
where an overall constant $2\sqrt{3}$ is multiplied. The $q$-expansion of $Y_{\mathbf{2}_{(0, 1/2)}}(\tau)$ is found to be
\begin{equation}
Y_{\mathbf{2}_{(0, 1/2)}}(\tau)=\begin{pmatrix}
1+24 q+24 q^2+96 q^3+24 q^4+144 q^5+\ldots\\
8\sqrt{3}\,q^{1/2}\left(1+4q+6 q^2+8 q^3+13 q^4+12 q^5+\ldots\right)
\end{pmatrix}\,.
\end{equation}
It agrees, up to an overall factor, with the $q$-expansion derived in~\cite{Kobayashi:2018vbk}, where the
$Y_{(0, 1/2)}(\tau)$ is the weight 2 modular form of level 2 and it is expressed in terms of the eta-function and its derivative.

\item{three-dimensional irreducible VVMFs}

Similar to the 2-d case, the 3-d irreps of $SL(2,\mathbb{Z})$ are determined by the three rational numbers $r_1,r_2$ and $r_3$~\cite{tuba2001representations}. In the $T$ diagonal basis, the 3-d representation matrix $\rho(T)$ can be parameterized as
\begin{equation}
\label{eq:DiagT}
\rho(T)=\begin{pmatrix}
e^{2\pi i r_1} ~& 0 ~& 0 \\
0 ~& e^{2\pi i r_2} ~&  0 \\
0 ~& 0 ~& e^{2\pi i r_3}
\end{pmatrix}\,,  \qquad 0\leq r_1,r_2,r_3 < 1\,,
\end{equation}
where the parameters $r_{1,2,3}$ have to fulfill the following constraint:
\begin{equation}
r_1+r_2+r_3 \in \frac{1}{12}\mathbb{Z} \,.
\end{equation}
This 3-d irrep is denoted as $\mathbf{3}_{(r_1,r_2, r_3)}$. The exponential matrix is $L=\texttt{diag}(r_1,r_2,r_3)$ and the minimal weight $k_0=4\text{Tr}(L)-2=4(r_1+r_2+r_3)-2$. The minimal weight VVMF $Y$ in the triplet representation $\mathbf{3}_{(r_1,r_2,r_3)}$ is determined to be~\cite{Liu:2021gwa}
\begin{equation}
\label{eq:3dVVMF}
Y(\tau)=
\begin{pmatrix}
~~\eta^{8(r_1+r_2+r_3)-4}K^{\frac{a_1+1}{6}}~ {}_3F_2(\frac{a_1+1}{6},\frac{a_1+3}{6},\frac{a_1+5}{6};r_1-r_2+1,r_1-r_3+1;K) \\
C_1\eta^{8(r_1+r_2+r_3)-4}K^{\frac{a_2+1}{6}}~ {}_3F_2(\frac{a_2+1}{6},\frac{a_2+3}{6},\frac{a_2+5}{6};r_2-r_3+1,r_2-r_1+1;K) \\
C_2\eta^{8(r_1+r_2+r_3)-4}K^{\frac{a_3+1}{6}}~ {}_3F_2(\frac{a_3+1}{6},\frac{a_3+3}{6},\frac{a_3+5}{6};r_3-r_1+1,r_3-r_2+1;K)
\end{pmatrix}\,,
\end{equation}
with $a_1=4r_1-2r_2-2r_3,a_2=4r_2-2r_1-2r_3$ and $a_3=4r_3-2r_1-2r_2$. The constants $C_{1,2}$ are fixed by the representation matrix of $\rho(S)$ in the $T$-diagonal basis. Hence the 3-d VVMFs module is of the form
\begin{equation}
\mathcal{M}(\mathbf{3}_{(r_1,r_2,r_3)})=\mathbb{C}[E_4,E_6]Y\oplus \mathbb{C}[E_4,E_6]D_{k_0}Y\oplus \mathbb{C}[E_4,E_6]D^2_{k_0}Y\,.
\end{equation}
It is known that there are three linearly independent weight 2 modular forms of level 3 and they can be arranged into a $A_4$ triplet with the following representation matrices for the modular generators,
\begin{eqnarray}
\rho(S)=\frac{1}{3}
\begin{pmatrix}
-1~&~2 ~&~ 2\\
2~&~-1~&~2 \\
2~&~2~&~-1
\end{pmatrix},~~~~\rho(T)=\begin{pmatrix}
1~&~0~&~0\\
0~&~\omega~&~ 0\\
0~&~0~&~\omega^2
\end{pmatrix},~~~\omega=e^{2\pi i/3}\,,
\end{eqnarray}
which is given by Eq.~\eqref{eq:irr-Tp}. This implies $r_1=0$, $r_2=\frac{1}{3}$ and $r_3=\frac{2}{3}$ and the modular weight $k_0=4(r_1+r_2+r_3)-2=2$. Taking into account the modular transformation $S$, we find $C_1=-\frac{1}{2}$, $C_2=-\frac{1}{8}$, and the VVMF in the triplet representation of $A_4$ is given by
\begin{equation}
Y_{\mathbf{3}_{(0, 1/3, 2/3)}}(\tau)=2\sqrt{3}
\begin{pmatrix}
~~\eta^{4}K^{-1/6}~ {}_3F_2(-\frac{1}{6},\frac{1}{6},\frac{1}{2}; \frac{2}{3}, \frac{1}{3}; K) \\
-\frac{1}{2}\eta^{4}K^{1/6}~ {}_3F_2(\frac{1}{6},\frac{1}{2},\frac{5}{6}; \frac{2}{3}, \frac{4}{3}; K) \\
-\frac{1}{8}\eta^{4}K^{1/2}~ {}_3F_2(\frac{1}{2}, \frac{5}{6}, \frac{7}{6}; \frac{5}{3}, \frac{4}{3}; K)
\end{pmatrix}\,,
\end{equation}
where the overall factor $2\sqrt{3}$ is included. We find that the $q$-expansion of $Y_{\mathbf{3}_{(0, 1/3, 2/3)}}(\tau)$ reads as
\begin{eqnarray}
Y_{\mathbf{3}_{(0, 1/3, 2/3)}}(\tau)=\begin{pmatrix}
1+12 q+36 q^2+12 q^3+84 q^4+72 q^5+\ldots\\
-6q^{1/3}\left(1+7 q+8 q^2+18 q^3+14 q^4+31 q^5+\ldots\right)\\
-18q^{2/3}\left(1+2 q+5 q^2+4 q^3+8 q^4+6 q^5+\ldots\right)
\end{pmatrix}\,,
\end{eqnarray}
which exactly coincides with the $q$-expansion derived in~\cite{Feruglio:2017spp}, where $Y_{\mathbf{3}_{(0, 1/3, 2/3)}}(\tau)$ is expressed in terms of the eta-function and its derivative. We see that the theory of VVMF provides a convenient way to obtain the explicit expression of modular forms which is the crucial element of modular invariant model.

\end{itemize}

Finally we summarise some more general finite modular groups.
Since the irrep $\rho$ of
$\Gamma = SL(2,\mathbb{Z})$ has finite image, the image $\text{Im}(\rho)$ forms a discrete finite group and it can play the role of discrete flavor symmetry. Due to a
fundamental theorem of homomorphism we can define
$\text{Im}(\rho)$ as the quotient group
$\text{Im}(\rho)\cong \Gamma/\ker(\rho)$
where $\Gamma = SL(2,\mathbb{Z})$ and
$\ker(\rho)$ is a normal subgroup determined by the representation $\rho$.
The $\ker(\rho)$ consists of the inverse image of the identity elements.

In the paradigm of modular flavor symmetry~\cite{Feruglio:2017spp}, $\ker(\rho)$ was only restricted to principle congruence subgroups $\Gamma(N)$. The image $\text{Im}(\rho)$ is the homogeneous or inhomogeneous finite modular group $\Gamma'_N=\Gamma/\Gamma(N)$ or $\Gamma_N=\Gamma/\pm\Gamma(N)$  and it is taken as the flavor symmetry to address the flavor structure of quarks and leptons. From the view of VVMF, the kernel $\ker(\rho)$ can be a general normal subgroup of $SL(2,\mathbb{Z})$ instead of $\Gamma(N)$, and accordingly the finite modular groups are not necessarily $\Gamma_N$ or $\Gamma'_N$. Thus the framework of modular invariance can be extended significantly by considering VVMFs, and we have more possible choices for the finite modular groups to construct modular invariant models.

In table~\ref{tab:NorSubgroupSL2Z} we list the normal subgroups of $SL(2,\mathbb{Z})$ with index $\leq 72$ as well as the corresponding finite modular group, where we omit the few cases for which the finite modular group is abelian. Besides the known principal congruence subgroups $\Gamma(N)$ as well as the inhomogeneous and homogeneous finite modular groups, we see other normal subgroups of $SL(2,\mathbb{Z})$. Additional relators are elements of $\ker(\rho)$, when they are added to the $SL(2,\mathbb{Z})$ presentation relations $S^4=(ST)^3=1$,  $S^2T=T S^2$, the finite modular groups $\text{Im}(\rho)\cong \Gamma/\ker(\rho)$ are produced. Moreover, if the element $S^2\in \ker(\rho)$, we have $\rho(S^2)=\mathds{1}_d$. Then the corresponding finite modular group has only even representations, and the VVMFs in these irreps must be of even weights.

It is known that the number of VVMFs in a given irrep $\rho_Y$ generally increases with the modular weight $k_Y$. However, the module $\mathcal{M}(\rho_Y)$ is finitely generated by $d=\dim\rho_Y$ independent bases denoted by $\{Y_1,\dots, Y_d\}$, thus $Y_{I_1...I_n}(\tau)$ can be written uniformly in the form of $Y^{(k_Y)}_{\rho_Y}=\alpha_1 Y_1+\dots +\alpha_d Y_d$ with $\alpha_i\in\mathbb{C}[E_4,E_6]$, note that $\alpha_i$ are polynomials of $E_4$ and $E_6$. Some $\alpha_i$ for vanishing for small value of modular weight $k_Y$, nevertheless there are at most $d$ independent alignments $Y_1$, $Y_2$,\ldots, $Y_d$. This implies that the superpotential is strongly constrained by the modular symmetry, and it can only take a finite number of possible forms for a given $\mathcal{G}_f$, although there are infinite possible weight and representation assignments for the matter fields.

\begin{table}[t!]
\centering
\begin{tabular}{|c|c|c||c|c|c|c|c|c|c|c|c|c|c|c|c|}\hline\hline
\multicolumn{3}{|c||}{Normal subgroups $\ker(\rho)$} & \multicolumn{2}{c|}{Finite modular groups $\Gamma/\ker(\rho)\cong\text{Im}(\rho)$} \\ \hline
Index& Label & Additional relations & Group structure  & \texttt{GAP} Id \\
\hline		
6 & $\Gamma(2)$ & $T^2$ & $S_3$ & $[6,1]$  \\ \hline	
\multirow{2}{*}{12} & $-$ & $S^2 T^2$ &  $ Z_3\rtimes Z_4\cong 2D_{3}$ & $[12,1]$  \\ \cline{2-5}
& $\pm\Gamma(3)$ & $S^2, T^3$ & $A_4$ & $[12,3]$  \\ \hline
18 & $-$  & $ST^{-2}ST^2$ & $S_3\times Z_3$ & $[18,3]$  \\ \hline
\multirow{4}{*}{24} & $\Gamma(3)$ & $T^3$ & \multirow{2}{*}{$T'$} & \multirow{2}{*}{$[24,3]$}  \\ \cline{2-3}
& $-$ & $S^2T^3$ &  &   \\  \cline{2-5}
& $\pm\Gamma(4)$ & $S^2,T^4$ & $S_4$ & $[24,12]$  \\ \cline{2-5}
& $-$ &  $S^2,(ST^{-1}ST)^2$ & $A_4\times Z_2$ & $[24,13]$  \\ \hline
36	& $-$ &  $S^3T^{-2}ST^2$ & $(Z_3\rtimes Z_4) \times Z_3$ & $[36,6]$  \\\hline		
\multirow{2}{*}{42} & $-$ & $T^6,(ST^{-1}S)^2TST^{-1}ST^2$ & \multirow{2}{*}{$Z_7 \rtimes Z_6$} & \multirow{2}{*}{$[42,1]$}  \\ \cline{2-3}
& $-$ & $T^6,ST^{-1}ST(ST^{-1}S)^2T^2$ &  &   \\ \hline
\multirow{6}{*}{48} & $-$ & $S^2T^4$ & $2O$ & $[48,28]$  \\ \cline{2-5}
& $-$ & $T^8,ST^4ST^{-4}$ & $GL(2,3)$ & $[48,29]$  \\ \cline{2-5}
& $\Gamma(4)$ & $T^4$ & $A_4\rtimes Z_4\cong S'_4$ & $[48,30]$  \\ \cline{2-5}
& $-$ & $(ST^{-1}ST)^2$ & $A_4\times Z_4$ & $[48,31]$  \\ \cline{2-5} 	
& $-$ & $S^2(ST^{-1}ST)^2$ & $T'\times Z_2$ & $[48,32]$  \\ \cline{2-5}
& $-$ & $T^{12},ST^3ST^{-3}$ & $((Z_4\times Z_2)\rtimes Z_2) \rtimes Z_3$ & $[48,33]$  \\ \hline
54 & $-$ & $T^{6},(ST^{-1}ST)^3$ & $(Z_3\times Z_3)\rtimes Z_6 $ & $[54,5]$  \\ \hline
60 & $\pm\Gamma(5)$ & $S^2, T^{5}$ & $A_5$ & $[60,5]$  \\ \hline
\multirow{2}{*}{72} & $-$ & $T^{12},ST^4ST^{-4}$ & $S_4\times Z_3$ & $[72,42]$  \\ \cline{2-5}
& $\pm\Gamma(6)$ & $S^2,T^{6},(ST^{-1}STST^{-1}S)^2T^2$ & $A_4\times S_3$ & $[72,44]$  \\ \hline\hline
\end{tabular}
\caption{\label{tab:NorSubgroupSL2Z} The normal subgroups of $SL(2,\mathbb{Z})$ with index less than $78$ and the corresponding finite modular groups. It is defined $\pm\Gamma(N)=\left\{\pm\gamma, \gamma\in\Gamma(N)\right\}$. Note that $2D_{3}$ is the binary dihedral group of order 12 and $2O$ is the binary octahedral group which is the Schur cover of permutation group $S_4$ of type ``$-$''. Additional relators are some elements of $\ker(\rho)$, the quotient group $\Gamma/\ker(\rho)$ is produced when the relations ``Additional relators =1'' together with $S^4=(ST)^3=1$ and $S^2 T=T S^2$ are imposed.}
\end{table}

\section{\label{ch:05-gCP}CP symmetry and modular invariance}

The interplay between the generalized CP symmetry (gCP) and modular symmetry is nontrivial. It is found that the CP transformation can be consistently defined in modular invariant supersymmetric theories with a single modulus~\cite{Baur:2019kwi,Baur:2019iai,Novichkov:2019sqv}. In particular, the modular symmetry can constrain the CP transformation laws of the modulus $\tau$, of chiral matter multiplets and of modular forms~\cite{Novichkov:2019sqv}, and several modular invariant models
where CP is spontaneously broken have been constructed~\cite{Novichkov:2019sqv,Kobayashi:2019uyt,Novichkov:2020eep,Liu:2020akv,Yao:2020zml,Yao:2020qyy,Okada:2020brs}.
Analogously for multidimensional moduli space, consistent CP definitions have been examined in the context of symplectic modular invariant theories, where the relevant flavour group is the Siegel modular group~\cite{Ishiguro:2020nuf,Baur:2020yjl,Ding:2021iqp}. The CP-conserving vacua in Calabi-Yau compactifications are discussed in Ref.~\cite{Ishiguro:2020nuf}.
In this section we formulate the conditions for generalised CP-symmetry consistent with modular invariance, and derive the modular transformations of matter fields and modular forms, before discussing CP fixed points and their implications for flavor theories.

\subsection{\label{subsec:gCP-modular-symmetry}Generalized CP symmetry compatible with modular invariance}

In a theory invariant under $\Gamma=SL(2,\mathbb{Z})$, consistent CP transformations correspond to outer automorphism $u(\gamma)$ of $\Gamma$~\cite{Holthausen:2012dk,Chen:2014tpa}:
\begin{equation}
\label{eq:gCP-con}
\mathcal{CP}~\gamma~\mathcal{CP}^{-1}=u(\gamma)~~~.
\end{equation}
Each outer automorphism $u(\gamma)$ of $\Gamma$ can be described by~\cite{Reiner}:
\begin{equation}
\label{auto}
u(\gamma)=\chi(\gamma)~U~ \gamma ~U^{-1}\,,~~~~\text{or}~~~~ u(\gamma)=\chi(\gamma)\gamma\,,
\end{equation}
with
\begin{equation}
\label{eq:CP-U}U=\begin{pmatrix}
1  ~&~ 0\\
0 ~&~-1
\end{pmatrix}
\end{equation}
and the map $\chi(\gamma)$, called character of the modular group, is a homomorphism of $\Gamma$ into $\{+1, -1\}$. From the relations $S^4=(ST)^3=1$ satisfied by the modular generators $S$ and $T$, we see the modular group has a trivial character with
\begin{equation}
\label{eq:chi1-2nd}\chi_1(S)=\chi_1(T)=1
\end{equation}
and another nontrivial character with
\begin{equation}
\label{eq:chi2-2nd}\chi_2(S)=\chi_2(T)=-1\,.
\end{equation}
Hence the modular group $SL(2, \mathbb{Z})$ has two independent outer automorphisms $u_{1,2}=\chi_{1,2}(\gamma)~ U~ \gamma ~U^{-1}$, satisfying
the following relations:
\begin{equation}
u^2_1=u^2_2=(u_1u_2)^2=\mathds{1}\,.
\end{equation}
Hence the outer automorphism group is isomorphic to a Klein group $K_4\cong Z_2\times Z_2=\{u_1,u_2,u_3=u_1 u_2,u_4=\mathds{1}\}$\footnote{We denote by $\mathds{1}$ the identity element of the outer automorphism.}. We can derive the action of the outer automorphisms on the modular generators $S$ and $T$ as follows~\cite{Ding:2021iqp}:
\begin{eqnarray}
\nonumber&&~u_1(S)=S^{-1}\,,\qquad u_1(T)=T^{-1}\,, \\
\nonumber&&~u_2(S)=-S^{-1}\,,\qquad u_2(T)=-T^{-1}\,, \\
\nonumber&&~u_3(S)=-S\,,\qquad u_3(T)=-T\,, \\
\label{eq:u1-u2-u3-u4}&&~u_4(S)=S\,,\qquad u_4(T)=T\,.
\end{eqnarray}
We assume that the CP transformation of modulus $\tau$ is linear and, for convenience, we write:
\begin{equation}
\tau\xrightarrow{{\cal CP}}({\cal CP})\tau\equiv\tau_{{\cal CP}}=p_1~\tau^*+p_2\,,
\end{equation}
where $p_1$ and $p_2$ are parameters to be determined. The inverse CP transformation reads:
\begin{eqnarray}
\tau^{*}\xrightarrow{{\cal CP}^{-1}}\tau^{*}_{{\cal CP}^{-1}}=\frac{\tau-p_2}{p_1}\,.
\end{eqnarray}
Enforcing the ``consistency condition chain'' of Eq.~\eqref{eq:gCP-con} on the modular generators $S$ and $T$ for the automorphisms
$u_a(\gamma)(a=1,...,4)$, we can obtain:
\begin{eqnarray}
\nonumber&&\frac{p_2\tau-p^2_1-p^2_2}{\tau-p_2}=u_a(S)\tau\,,\\
\label{eq:CP-g-CPInv}&&\tau+p_1=u_a(T)\tau\,.
\end{eqnarray}
The elements $u_1$ and $u_2$ have the same action on $\tau$ and similarly  for $u_{3,4}$. The above equation can be straightforwardly solved to obtain
\begin{equation}
p_1=\left\{\begin{array}{cc}
-1,~&~ \text{for $u_{1,2}$}\\
+1,~&~ \text{for $u_{3,4}$}
\end{array}
\right.\,,\qquad~~p_2=0\,.
\end{equation}
Thus we have $\tau_{{\cal CP}}=-\tau^{*}$ for the automorphisms $u_{1,2}$ and $\tau_{{\cal CP}}=\tau^{*}$ for $u_{3,4}$. Since ${\tt Im}(\tau_{{\cal CP}})>0$, the admissible CP transformation of the modulus $\tau$ is~\cite{Baur:2019kwi,Baur:2019iai,Novichkov:2019sqv,Acharya:1995ag,Dent:2001cc,Giedt:2002ns}
\begin{equation}
\label{eq:linearCP}
\tau\xrightarrow{{\cal CP}}\tau_{{\cal CP}}=-\tau^*~\,.
\end{equation}
This represents the correct transformation law of the modulus for both $u_{1,2}$ outer automorphisms and we should instead discard $u_{3,4}$. Notice that the same gCP transformation of $\tau$ can be derived from the consistency condition, as shown in the Appendix~\ref{app:tau-gCP-transform}. From Eq.~\eqref{eq:linearCP}, we see $\tau^*_{\mathcal{CP}^{-1}}=-\tau$ and the action of ${\cal CP}$ on the modulus is involutive with ${\cal CP}^2\tau=\tau$.

Now let us check explicitly that the gCP transformation of Eq.~\eqref{eq:linearCP} is indeed compatible with Eq.~\eqref{eq:gCP-con}. Applying the consistency transformation chain $\mathcal{CP}\rightarrow\gamma\rightarrow\mathcal{CP}^{-1}$ for the modulus $\tau$, we have
\begin{equation}
\label{eq:tau-CP-gamma-CPInv}\tau
\stackrel{\mathcal{CP}}{\longrightarrow}-\tau^\ast
\stackrel{\gamma}{\longrightarrow}  -\frac{a\tau^\ast+b}{c\tau^\ast+d}
\xrightarrow{\mathcal{CP}^{-1}}
= \frac{a\tau-b}{-c\tau+d}=\gamma'\tau\,.
\end{equation}
Since $\gamma$ and $-\gamma$ give the same action on modulus $\tau$, we see that the generalized CP transformation indeed corresponds to the automorphism $u_1$ or $u_2$ of the modular group with
\begin{equation}
\label{eq:gCP-automorf-u}u_{1,2}: \gamma\rightarrow \gamma' =u_{1,2}(\gamma)=\chi_{1,2}(\gamma)~U~ \gamma ~U^{-1}=\chi_{1,2}(\gamma)\begin{pmatrix}
a  ~&  -b\\
-c ~& d
\end{pmatrix},~~~\gamma=\begin{pmatrix}
a &  b \\
c  & d
\end{pmatrix}\in\Gamma\,,
\end{equation}
where $\chi_1(\gamma)$ is the trivial character given in Eq.~\eqref{eq:chi1-2nd} and $\chi_2(\gamma)$ is the nontrivial character in Eq.~\eqref{eq:chi2-2nd}.

If we perform a modular transformation $\gamma$ and then the CP transformation of Eq.~\eqref{eq:linearCP}, we get
\begin{equation}
\label{eq:CP-generd}
\tau\xrightarrow{{\gamma}}\frac{a\tau+b}{c\tau+d}
\xrightarrow{{\cal CP}}(\gamma\circ{\cal CP})\tau=\frac{-a\tau^{*}+b}{-c\tau^{*}+d}\,,
\end{equation}
which is another allowed CP transformation. Since the theory is invariant under $\Gamma$, this choice should not be view as independent from \eqref{eq:linearCP}, which we take as representative element
in the class \eqref{eq:linearCP} with $a=d=1$, $b=c=0$. We can rewrite Eq.~\eqref{eq:CP-generd} as
\begin{equation}
\tau\xrightarrow{{\gamma\circ{\cal CP}}}\frac{a'\tau^{*}+b'}{c'\tau^{*}+d'}~
\end{equation}
with
\begin{equation}
a'=-a,~~~b'=b,~~~c'=-c,~~~d'=d\,.
\end{equation}
One can check
\begin{equation}
a'd'-b'c'=-ad+bc=-(ad-bc)=-1\,.
\end{equation}
Note that the gCP transformation on $\tau$ always corresponds to minus determinant with $a'd'-b'c'=-1$.

By combining CP and the modular transformations we get the extended modular group $\Gamma^*=GL(2,\mathbb{Z})$ and the full symmetry transformation of the complex modulus is~\cite{Novichkov:2019sqv}
\begin{equation}
\label{eq:modular-gCP}  \begin{pmatrix} a & b \\ c & d \end{pmatrix}
  \in \Gamma^*:\quad
  \begin{cases}
    \tau \to \cfrac{a\tau + b}{c\tau + d} &\quad \text{if} \quad ad - bc = 1\,, \\
    \tau \to \cfrac{a\tau^{*} + b}{c\tau^{*} + d}
&\quad \text{if} \quad ad - bc = -1\,.
  \end{cases}
\end{equation}
where $\gamma=\begin{pmatrix}
a &  b \\
c  & d
\end{pmatrix}$.
The first line of \eqref{eq:modular-gCP} refers to a modular transformation, while the second line refers to a CP transformation. For the case of the CP transformation in \eqref{eq:linearCP} the matrix $\gamma$ is represented by $U$ in Eq.~\eqref{eq:CP-U}. Notice that the action of $\gamma$ and $-\gamma$ on $\tau$ is the same and the full symmetry group acting on moduli is isomorphic to $PGL(2,\mathbb{Z})\equiv GL(2,\mathbb{Z})/\{\pm \mathds{1}_{2}\}$. Here $PGL(2,\mathbb{Z})$ is the group of integral $2 \times 2$ matrices with determinant $\pm 1$, with the matrices $M$ and $-M$ being identified.

\subsection{\label{subsec:gcp-transfm-matter}CP transformations of matter chiral multiplets $\varphi$}

For a generic chiral supermultiplets $\varphi$, its transfromation properties are characterized by $-k_{\varphi}$ and $\rho_{\mathbf{r}}$, where $-k_{\varphi}$ is the modular weight of $\varphi$ and $\rho_{\mathbf{r}}$ is an irreducible representation of the finite modular group $\Gamma_{N}$ or $\Gamma'_{N}$. Under a modular transformation $\gamma$, the multiplet $\varphi(x)$ transforms according to
\begin{equation}
\label{eq:varphi-mod-trans}\varphi(x)\stackrel{\gamma}{\longrightarrow} (c\tau+d)^{-k_{\varphi}}\rho_{\mathbf{r}}(\gamma)\varphi(x)\,,~~~\gamma=\begin{pmatrix}
a ~& b \\
c ~& d
\end{pmatrix}\in\Gamma\,.
\end{equation}
The action of generalized CP symmetry on the multiplet $\varphi$ is
\begin{equation}
\label{eq:varphi-gCP-trans}\varphi(x)\stackrel{\mathcal{CP}}{\longrightarrow} X_{\mathbf{r}}\overline{\varphi}(x_{\mathcal{P}})\,,
\end{equation}
where $x_{\mathcal{P}}=(t, -\vec{x})$, a bar denotes the hermitian conjugate superfield, and $X_{\mathbf{r}}$ is a unitary matrix that acts on flavor space. If we perform a CP transformation, followed by a modular transformation and subsequently an inverse CP transformation, we end up with
\begin{equation}
\varphi(x)\xrightarrow{{\cal CP}} X_{\mathbf{r}}\overline{\varphi}(x_{\mathcal{P}}) \xrightarrow{\gamma}  X_{\mathbf{r}} (c\tau^*+d)^{-k_{\varphi}}\rho^*_\mathbf{r}(\gamma)\overline{\varphi}(x_{\mathcal{P}}) \xrightarrow{{\cal CP}^{-1}}  (c\tau^*_{\mathcal{CP}^{-1}}+d)^{-k_{\varphi}} X_{\mathbf{r}} \rho^*_\mathbf{r}(\gamma) X_{\mathbf{r}}^{-1} \varphi(x) \,,
\end{equation}
where $\tau_{\mathcal{CP}^{-1}}$ is the result of applying the inverse CP transformation $\mathcal{CP}^{-1}$ to the modulus $\tau$. Since the theory is assumed to be invariant under both the modular and CP transformations, the resulting transformation should be amount to another modular transformation $\gamma'$, i.e.,
\begin{equation}
\label{eq:consistency-chain-matter} (c\tau^*_{\mathcal{CP}^{-1}}+d)^{-k_{\varphi}} X_{\mathbf{r}} \rho^*_\mathbf{r}(\gamma) X_{\mathbf{r}}^{-1} \varphi(x)=(c_{\gamma'}\tau+d_{\gamma'})^{-k_{\varphi}} \rho_\mathbf{r}(\gamma')\varphi(x)
\end{equation}
which implies
\begin{equation}
\label{eq:consistency-cond1}
X_{\mathbf{r}} \rho^*_\mathbf{r}(\gamma) X_{\mathbf{r}}^{-1} =\left(\frac{c_{\gamma'}\tau+d_{\gamma'}}{c\tau^*_{\mathcal{CP}^{-1}}+d}\right)^{-k_{\varphi}} \rho_\mathbf{r}(\gamma')\,,
\end{equation}
where we have denoted
\begin{equation}
\gamma'=\begin{pmatrix}
a_{\gamma'}  &  b_{\gamma'} \\
c_{\gamma'}  &  d_{\gamma'}
\end{pmatrix}\in\Gamma\,.
\end{equation}
From the transformation of the modulus $\tau$ in Eq.~\eqref{eq:tau-CP-gamma-CPInv}, we know the element $\gamma'=u_{1,2}(\gamma)$ given in Eq.~\eqref{eq:gCP-automorf-u}, thus we can read out
\begin{equation}
c_{\gamma'}=-\chi_{1,2}(\gamma)c,~~~d_{\gamma'}=\chi_{1,2}(\gamma)d\,.
\end{equation}
Considering $\tau^*_{\mathcal{CP}^{-1}}=-\tau$, the consistency condition of Eq.~\eqref{eq:consistency-cond1} for the modular symmetry and generalized CP symmetry becomes
\begin{equation}
\label{eq:consistency-cond-chi}X_{\mathbf{r}}\rho^{*}_{\mathbf{r}}(\gamma)X^{-1}_{\mathbf{r}}=\chi_{1,2}^{-k_{\varphi}}(\gamma)\rho_{\mathbf{r}}(u_{1,2}(\gamma))\,.
\end{equation}
Since the modular group has two possible characters $\chi_{1,2}(\gamma)$ given in Eqs.~(\ref{eq:chi1-2nd},\ref{eq:chi2-2nd}), two kinds of generalized CP transformations could be consistently defined in the context of $SL(2, \mathbb{Z})$. The coressponding consistency conditions are given by~\cite{Novichkov:2019sqv,Ding:2021iqp}
\begin{equation}
\label{eq:Xr-1st}X_{\mathbf{r}}\rho^{*}_{\mathbf{r}}(S)X^{-1}_{\mathbf{r}}=\rho_{\mathbf{r}}(S^{-1}),~~~X_{\mathbf{r}}\rho^{*}_{\mathbf{r}}(T)X^{-1}_{\mathbf{r}}=\rho_{\mathbf{r}}(T^{-1})\,.
\end{equation}
for the gCP transformation corresponding to the automorphism $u_1$ with trivial character $\chi_1$ in Eq.~\eqref{eq:u1-u2-u3-u4}. A second gCP transfromation corresponding to $u_2$ with nontrivial character $\chi_2$ could be defined\footnote{The consistency condition for the second gCP transformation reads as:
\begin{equation*}
\label{eq:Xr-2nd}X_{\mathbf{r}}\rho^{*}_{\mathbf{r}}(S)X^{-1}_{\mathbf{r}}=\sigma\rho_{\mathbf{r}}(S^{-1}),~~~X_{\mathbf{r}}\rho^{*}_{\mathbf{r}}(S)X^{-1}_{\mathbf{r}}=\sigma\rho_{\mathbf{r}}(T^{-1}),~~~\sigma=(-1)^{-k_{\varphi}}\rho_{\mathbf{r}}(S^2)\,.
\end{equation*}
Obviously the second gCP transformation $X_{\mathbf{r}}$ depends on $\rho_{\mathbf{r}}(S^2)$ and the modualr weight $k_{\varphi}$, and it can be defined if and only if the level $N$ is even, the dimension of representation $\rho_{\mathbf{r}}$ is even, and the traces of $\rho_{\mathbf{r}}(T)$ and $\rho_{\mathbf{r}}(S)$ are vanishing~\cite{Novichkov:2020eep}.}, nevertheless it is difficult to build phenomenologically viable models of fermion masses and mixing exploiting this second gCP transformation~\cite{Novichkov:2020eep}. Hence we shall be concerned with the gCP transformation defermined by Eq.~\eqref{eq:Xr-1st} in this work.

In the basis where both $\rho_{\mathbf{r}}(S)$ and $\rho_{\mathbf{r}}(T)$ are symemtric matrices, one has $\rho_{\mathbf{r}}(S^{-1})=\rho^{\dagger}_{\mathbf{r}}(S)=\rho^{*}_{\mathbf{r}}(S)$ and $\rho_{\mathbf{r}}(T^{-1})=\rho^{\dagger}_{\mathbf{r}}(T)=\rho^{*}_{\mathbf{r}}(T)$ so that one can determine $X_{\mathbf{r}}=\mathds{1}_{\mathbf{r}}$ up to an irrelevant phase, i.e., the gCP transformation has the canonical form. If all the Clebsch-Gordon coeffcients of the finite modular group are real in the above symemtric basis, the gCP symemtry would enforce all coupling constants in the Lagrangian to be real. Thus the real part of modulus $\tau$ would be the unique source of CP violation, and gCP symemtry can enhance the predictive power of modular invariant models. For the finite modular groups $\Gamma_N$ and $\Gamma'_N$ with level $N\leq6$, it is known that the symmetric basis can always be achieved.

\subsection{\label{subsec:gCP-MF}CP transformations of modular form $Y_{\mathbf{r}}(\tau)$}

Considering a multiplet $Y_{\mathbf{r}}(\tau)$ of modular forms of level $N$ and weight $k_Y$, it transforms in certain irreducible representation $\rho_{\mathbf{r}}$ of the finite modular group $\Gamma_N$ or $\Gamma'_N$, i.e.
\begin{equation}
\label{eq:MF-trans-GammaN}Y_\mathbf{r}(\gamma\tau)= (c\tau + d)^{k_Y}\rho_{\mathbf{r}}(\gamma)Y_\mathbf{r}(\tau),~~~\gamma \in \Gamma\,.
\end{equation}
In general, there can be several linearly independent such multiplets, particularly for higher weight corresponding large $k_Y$. We start by examining the case where there is only one. Under gCP it transforms as:
\begin{equation}
Y_{\mathbf{r}}(\tau)\xrightarrow{\cal CP}Y_{\mathbf{r}}(-\tau^*)\,.
\end{equation}
We can prove that the modular multiplet $Y_{\mathbf{r}}(\tau)$ has the same gCP transformation as that of matter fields in Eq.~\eqref{eq:varphi-gCP-trans}, i.e.,
\begin{equation}
Y_{\mathbf{r}}(-\tau^*)=X_\mathbf{r} Y_{\mathbf{r}}^*(\tau)\,.
\end{equation}
In order to show this, we define
\begin{equation}
\widetilde{Y}_{\mathbf{r}}(\tau)=X_\mathbf{r}^{-1*}Y_{\mathbf{r}}^*(-\tau^*)\,,
\end{equation}
where $X_{\mathbf{r}}$ is the solution of the consistency condition Eq.~\eqref{eq:Xr-1st} for the concerned automorphism $u_1$. The modular transfromation of $\widetilde{Y}_{\mathbf{r}}(\tau)$ under a generic element $\gamma\in \Gamma$ is
\begin{eqnarray}
\nonumber\widetilde{Y}_{\mathbf{r}}(\gamma(\tau))&=&X_\mathbf{r}^{-1*}Y_{\mathbf{r}}^*(-(\gamma(\tau))^{*})\nn\\
\nonumber&=&X_{\mathbf{r}}^{-1*}Y_{\mathbf{r}}^*(u_1(\gamma)(-\tau^*))\nn\\
\nonumber&=&X_{\mathbf{r}}^{-1*}(c\tau+d)^{k_Y}\rho^*_{\mathbf{r}}(u_1(\gamma))~Y_{\mathbf{r}}^*(-\tau^*)\nn\\
\label{eq:Y-tilde-mod-trans} &=&(c\tau+d)^{k_Y} ~\rho_{\mathbf{r}}(\gamma)~\widetilde{Y}_{\mathbf{r}}(\tau)\,,
\end{eqnarray}
where the identities $-(\gamma(\tau))^*= u(\gamma)(-\tau^*)$ and $X_{\mathbf{r}}\rho^{*}_{\mathbf{r}}(\gamma)X^{-1}_{\mathbf{r}}=\rho_{\mathbf{r}}(u_{1}(\gamma))$ have been used\footnote{From Eq.~\eqref{eq:tau-CP-gamma-CPInv}, we know $\mathcal{CP}\circ\gamma\circ\mathcal{CP}^{-1}=u_1(\gamma)$ which gives $\mathcal{CP}\circ\gamma=u_1(\gamma)\circ\mathcal{CP}$. Consequently we have  $\mathcal{CP}\circ\gamma(\tau)=u_1(\gamma)\circ\mathcal{CP}(\tau)$ which leads to $-(\gamma(\tau))^{*}=u_1(\gamma)(-\tau^*)$. Moreover,
From Eq.~\eqref{eq:consistency-cond-chi}, we
know
$X_{\mathbf{r}}\rho^{*}_{\mathbf{r}}(\gamma)X^{-1}_{\mathbf{r}}=\rho_{\mathbf{r}}(u_{1}(\gamma))$ which gives $\rho^{*}_{\mathbf{r}}(\gamma)X^{-1}_{\mathbf{r}}=X^{-1}_{\mathbf{r}}\rho_{\mathbf{r}}(u_{1}(\gamma))$ and $\rho_{\mathbf{r}}(\gamma)X^{-1*}_{\mathbf{r}}=X^{-1*}_{\mathbf{r}}\rho^{*}_{\mathbf{r}}(u_{1}(\gamma))$.}.
Since $\widetilde{Y}_{\mathbf{r}}(\tau)$ and $Y_{\mathbf{r}}(\tau)$ transform in the same way and, by assumption, there is only one linearly independent such modular form, we conclude that they are proportional, $\widetilde{Y}_{\mathbf{r}}(\tau)\propto Y_{\mathbf{r}}(\tau)$
which gives:
\begin{equation}
Y_{\mathbf{r}}(-\tau^*)=\lambda~ X_\mathbf{r} Y^*_{\mathbf{r}}(\tau)\,.
\end{equation}
By performing two gCP transformation in succsssion, we obtain
\begin{equation}
Y_{\mathbf{r}}(\tau)=|\lambda|^2~ X_\mathbf{r}X_\mathbf{r}^* Y_{\mathbf{r}}(\tau)=|\lambda|^2~ Y_{\mathbf{r}}(\tau)\,,
\end{equation}
where we have chosen the gCP transfroamtion $X_{\mathbf{r}}$ to be a constant unitary symmetric matrix fulfilling $X_\mathbf{r}X_\mathbf{r}^*=1$ which corresponds to $\mathcal{CP}^2=1$\footnote{The action of ${\cal CP}$ in the modulus, see Eq.~\eqref{eq:linearCP}, is involutive, that is ${\cal CP}^2\tau=\tau$. The action is $\mathcal{CP}^2$ on the matter field is $\varphi(x)\stackrel{\mathcal{CP}^2}{\longrightarrow} X_{\mathbf{r}}X^{*}_{\mathbf{r}}\varphi(x)$. From the consistency condtion in Eq.~\eqref{eq:varphi-gCP-trans}, we obtain:
\begin{equation*}
(c\tau+d)^{-k_\varphi} X_{\mathbf{r}} X_{\mathbf{r}}^*~\rho_{\mathbf{r}}(\gamma)~ X^{-1*}_{\mathbf{r}}X_{\mathbf{r}}^{-1}\varphi=
(c\tau+d)^{-k_\varphi} \rho_{\mathbf{r}}(\gamma)\varphi~~~.
\end{equation*}
By the Schur's Lemma, the product $X_{\mathbf{r}} X_{\mathbf{r}}^*$ is proportional to the identity. It acts without conjugating the matter fields and represents the element $\gamma_{{\cal CP}^2}$ or, more precisely, the element of the finite group $\Gamma_{N}$ or $\Gamma'_N$ that corresponds to $\gamma_{{\cal CP}^2}$. As a consequence we have:
\begin{equation*}
X_{\mathbf{r}} X_{\mathbf{r}}^*=\mathds{1}_{\mathbf{r}}~~ \text{or}~~X_{\mathbf{r}} X_{\mathbf{r}}^*= \rho_{\mathbf{r}}(S^2)=\pm \mathds{1}_{\mathbf{r}}\,.
\label{eq:noninv}
\end{equation*}
Therefore we must have $X_{\mathbf{r}} X_{\mathbf{r}}^*=\mathds{1}_{\mathbf{r}}$ for the inhomegenous finite modular group $\Gamma_{N}$.}. The non-vanishing constant $\lambda$ can be absorbed by an appropriate choice of phase of the whole multiplet $Y_{\mathbf{r}}(\tau)$ so that we have
\begin{equation}
\label{eq:Y-tau-star}Y_{\mathbf{r}}(\tau)\xrightarrow{\cal CP}Y_{\mathbf{r}}(-\tau^*)=X_\mathbf{r} Y_{\mathbf{r}}^*(\tau)\,.
\end{equation}
Hence the modular multiplets have the same gCP transformation law as that of matter fields.

If there are $s$ linearly independent multiplets $Y^{a}_{\mathbf{r}}(\tau)$ $(a=1,\ldots,s)$ transforming as in Eq.~\eqref{eq:MF-trans-GammaN}, Eq.~\eqref{eq:Y-tilde-mod-trans} holds individually for all $\widetilde{Y}^{a}_{\mathbf{r}}(\tau)=X_\mathbf{r}^{-1*}Y^{a*}_{\mathbf{r}}(-\tau^*)$ and we
have:
\begin{equation}
Y^{a}_{\mathbf{r}}(-\tau^*)=\lambda^a_{~b}~ X_\mathbf{r} Y^{b*}_{\mathbf{r}}(\tau)~~~,
\label{tYa}
\end{equation}
where $\lambda^a_{~c} \lambda^{*c}_{~~b}X_\mathbf{r} X^*_\mathbf{r}=\delta^a_{~b}\mathds{1}_\mathbf{r}$. When $X_\mathbf{r}$ is involutive with $X_\mathbf{r} X^*_\mathbf{r}=\mathds{1}_{\mathbf{r}}$ we have $\lambda^a_{~c} \lambda^{*c}_{~~b}=\delta^a_{~b}$, it is always possible to factorize the matrix $\lambda$ into $\lambda=\eta^{-1}\eta^*$ and we obtain\footnote{If $(\mathds{1}+\lambda)$ is invertible, we take $\eta=(\mathds{1}+\lambda)^{-1}$, then $\lambda\eta^{*-1}=\eta^{-1}$.
If $(\mathds{1}+\lambda)$ is not invertible, we can always find a complex number $u$ with $|u|=1$ such that $-u^2$ is not an eigenvalue of $\lambda$. Hence, $\lambda+u^2\mathds{1}$ is invertible. In this case, we take $\eta=(u^{-1}\lambda+ u\mathds{1})^{-1}$, then $\lambda\eta^{*-1}=\eta^{-1}$. This construction was given by Prof. Marc van Leeuwen~\cite{que:matrix_decom}.}:
\begin{equation}
\eta^a_{~b}Y^{b}_{\mathbf{r}}(-\tau^*)= X_\mathbf{r} [\eta^a_{~b}Y^b_{\mathbf{r}}]^*(\tau)~~~.
\end{equation}
We see that, by performing the change of basis $Y^a_{\mathbf{r}}(\tau)\to \eta^a_{~b}Y^b_{\mathbf{r}}(\tau)$, Eq.~\eqref{eq:Y-tilde-mod-trans} holds independently for each multiplet $Y^a(\tau)$. In short, when the action of gCP on the field space is involutive with $X_\mathbf{r} X^*_\mathbf{r}=\mathds{1}_{r}$, one can properly choose the basis of the modular forms such that matter fields and modular forms transform in a similar way under gCP, i.e.,
\begin{equation}
\varphi(x)\stackrel{\mathcal{CP}}{\longrightarrow} X_{\mathbf{r}}\overline{\varphi}(x_{\mathcal{P}})\,,~~~Y^a(\tau)\stackrel{\mathcal{CP}}{\longrightarrow} Y^a_{\mathbf{r}}(-\tau^*)=X_\mathbf{r} Y^{b*}_{\mathbf{r}}(\tau)\,.
\end{equation}
In this case, it is convenient to move to the basis where $X_{\mathbf{r}}=\mathds{1}_{\mathbf{r}}$\footnote{The CP transfromation matrix $X_{\mathbf{r}}$ would be a unitary and symmetric matrix if  $X_\mathbf{r} X^*_\mathbf{r}=\mathds{1}_{\mathbf{r}}$, and the corresponding Takagi factorization is $X_{\mathbf{r}}=\Omega_{\mathbf{r}}\Omega^{T}_{\mathbf{r}}$ where $\Omega_{\mathbf{r}}$ is a unitary matrix. If we perform a change of basis with a unitary matrix $\Omega_{\mathbf{r}}$ in the field space: $\varphi'=\Omega^{\dagger}_{\mathbf{r}}\varphi$, it is easy to see that the gCP transformation of $\varphi'$ is $\varphi'(x)\stackrel{\mathcal{CP}}{\longrightarrow} \overline{\varphi'}(x_{\mathcal{P}})$.}. Then the consistency condition of Eq.~\eqref{eq:Xr-1st} implies that the representation matrices $\rho_{\mathbf{r}}(S)$ and $\rho_{\mathbf{r}}(T)$ of the modular generators are symmetric. As we explained early, if both $\rho_{\mathbf{r}}(S)$ and $\rho_{\mathbf{r}}(T)$ are symmetric and unitary matrices, we would have $X_{\mathbf{r}}=\mathds{1}_{\mathbf{r}}$ which is exactly the canonical CP transformation. If the Clebsh-Gordan coefficients are all real in the adopted basis, the requirement of gCP invairance would enforces all free parameters in the superpotential to be real.

\subsection{CP fixed points }

In a modular invariant theory with gCP symmetry, the vacuum expectation value of the modulus $\tau$ would be the unique source breaking modular and gCP symmetry. However, there exist certain values of $\tau$ which conserve CP. If $\tau$ is left invariant by CP transformation of Eq.~\eqref{eq:linearCP}, i.e.
\begin{equation}
\tau\xrightarrow{{\cal CP}}\tau_{{\cal CP}}=-\tau^*=\tau\,.
\end{equation}
We see $\tau$ lies on the imaginary axis, $\Re\tau=0$. The most general gCP transfromation of $\tau$ is $\gamma\circ{\cal CP}$ given by Eq.~\eqref{eq:CP-generd}, it is the composition of the a generic modular transformation $\gamma$ and the CP transformation of Eq.~\eqref{eq:linearCP}. Thus the most general value of $\tau$ preserving the residual gCP symmetry $\gamma\circ{\cal CP}$ satisfies the following condition
\begin{equation}
\label{eq:gamma-CP-fix}(\gamma\circ{\cal CP})\tau=\tau,~~~\gamma=\begin{pmatrix}
a~&~b \\
c~&~d
\end{pmatrix}\in\Gamma
\,,
\end{equation}
which gives
\begin{equation}
\frac{-a\tau^{*}+b}{-c\tau^{*}+d}=\tau\,,
\end{equation}
or equivalently
\begin{eqnarray}
\nonumber c\left(\Re\tau\right)^2-\left(a+d\right)\Re\tau+b+c\left(\Im\tau\right)^2&=&0\,,\\
\label{eq:residual-gCP}\left(a-d\right)\Im\tau&=&0\,.
\end{eqnarray}
Thus we have
\begin{eqnarray}
\label{eq:a-d-CP}a=d\,,
\end{eqnarray}
and
\begin{eqnarray}
\label{eq:re-tau-CP}\Re\tau=\left\{\begin{array}{cc}
\dfrac{a}{c}\pm\frac{1}{c}\sqrt{1-c^2\left(\Im\tau\right)^2}\,,  ~&~ c\neq0\\
\dfrac{b}{2a}\,,     ~&~  c=0
\end{array}
\right.
\end{eqnarray}
Without loss of generality, we firstly consider that $\tau$ lies in the fundamental domain. Solving the constraints in Eqs.~(\ref{eq:a-d-CP}, \ref{eq:re-tau-CP}) together with the determinant condition $ad-bc=1$, we find the CP conserved values of $\tau$ are the imaginary axis and the boundary of the fundamental domain $\mathcal{D}$, the corresponding residual CP symmetry is summarized in table~\ref{tab:residual-CP-modulus}. At the intersection points, the residual symmetry is enhanced. For instance, the modulus $\tau=e^{2 i\pi/3}$ is invariant under both CP transformations $T^{-1}\circ\mathcal{CP}$ and $S\circ\mathcal{CP}$ so that it preserves the modular symmetry $ST$ which is generated by the two residual CP transfromations. Thus the full residual symmetry group of $\tau=e^{2 i\pi/3}$ is $\left(Z^{ST}_3\rtimes Z^{S\circ \mathcal{CP}}_2\right)\times Z^{R}_2$ defined as
\begin{eqnarray}
\nonumber \left(Z^{ST}_3\rtimes Z^{S\circ \mathcal{CP}}_2\right)\times Z^{R}_2&=&\Big\{ST, R, S\circ\mathcal{CP}| R^2=(S\circ\mathcal{CP})^2=(ST)^3=1, (ST)R=R(ST), \\ &&\hskip-0.2in (S\circ\mathcal{CP})R=R(S\circ\mathcal{CP}), (S\circ\mathcal{CP})ST(S\circ\mathcal{CP})^{-1}=(ST)^{-1}\Big\}\,.
\end{eqnarray}
Similarly the symmetric point $\tau=i$ is invariant under both modular transfromation $S$ and CP transformation $\mathcal{CP}$, and consequently the residual symmetry group is  $Z^{S}_4\rtimes Z^{\mathcal{CP}}_2=\left\{S,\mathcal{CP}| S^4=1, \mathcal{CP}^2=1, \mathcal{CP}\,S\,\mathcal{CP}^{-1}=S^{-1}\right\}$. Moreover, the symmetric point $\tau=i\infty$ is invariant under $T$, $R$ and $\mathcal{CP}$, and therefore it preserves the residual symmetry group
\begin{small}
\begin{equation}
\left(Z^{T}\rtimes\mathcal{CP}\right)\times Z^{R}_2=\left\{T, R, \mathcal{CP}| R^2=\mathcal{CP}^2=1, \mathcal{CP}\,T\,\mathcal{CP}^{-1}=T^{-1}, \mathcal{CP}\,R=R\,\mathcal{CP}, TR=RT\right\}\,.
\end{equation}
\end{small}

\begin{figure}
\centering
\includegraphics[width=0.75\textwidth]{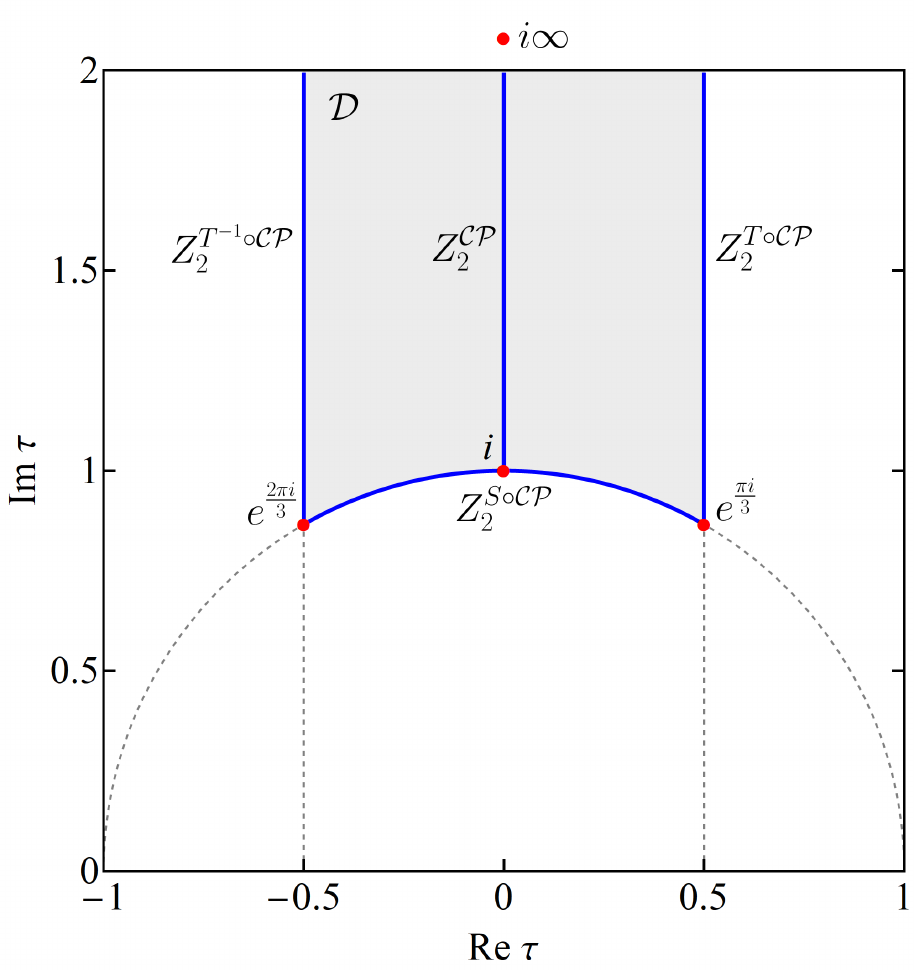}
\caption{\label{fig:gCP-fixed-points}The CP conserving lines in the fundamental domain $\mathcal{D}$ are displayed in blue and the corresponding residual CP symmetries are labelled. The intersection points $\tau=e^{\pi i/3},\, i, e^{2\pi i/3}$ labelled in red enjoy two different residual CP symmetries so that a residual modular symmetry can be generated. For instance, the self-dual point $\tau=i$ is invariant under both CP transfromations $Z^{\mathcal{CP}}_2$ and $Z^{S\circ\mathcal{CP}}_2$, consequently it is also invariant under the modular generator $S$.    }
\end{figure}

We plot the CP conserving points and the corresponding residual CP symmetries in figure~\ref{fig:gCP-fixed-points}. Furthermore we find that all the CP conserving points in the upper half complex plane are related to those in the fundamental domain by some modular transformation, and the corresponding generalized CP stabilizer given by
\begin{equation}
\tau' = \gamma_1 \tau^{\mathcal{CP}}_0 \,,~~~~ \gamma' \circ\mathcal{CP} =\gamma_1\gamma^{\mathcal{CP}}_0\circ\mathcal{CP}\,\gamma^{-1}_1\,,
\end{equation}
where $\gamma_1$ is an arbitrary modular transformation, $\tau^{\mathcal{CP}}_0$ and $\gamma^{\mathcal{CP}}_0$ take the values summarized in table~\ref{tab:residual-CP-modulus}.

\begin{table}[t!]
\centering
\begin{tabular}{|c|c|c|}
\hline\hline
$\gamma^{\mathcal{CP}}_{0}$ & $\tau^{\mathcal{CP}}_0$   & \texttt{Residual subgroup} \\
\hline
$\pm1$ & $\left\{\tau|\Re\tau=0,\Im\tau\geq1\right\}$  &  $Z^{\mathcal{CP}}_2\times Z^{R}_2$   \\ \hline

$\pm T$ & $\left\{\tau|\Re\tau=1/2,\Im\tau\geq\sqrt{3}/2\right\}$  &  $Z^{T\circ\mathcal{CP}}_2\times Z^{R}_2$    \\ \hline

$\pm T^{-1}$ & $\left\{\tau|\Re\tau=-1/2,\Im\tau\geq\sqrt{3}/2\right\}$ &   $Z^{T^{-1}\circ\mathcal{CP}}_2\times Z^{R}_2$    \\ \hline

$\pm S$ & $\left\{\tau|(\Re\tau)^2+(\Im\tau)^2=1,|\Re\tau|\leq1/2\right\}$  &    $Z^{S\circ\mathcal{CP}}_2\times Z^{R}_2$  \\ \hline

$\pm1, \pm S$& $i$ & $Z^{S}_4\rtimes Z^{\mathcal{CP}}_2$    \\ \hline

$\pm T, \pm S$& $ e^{i\pi/3}$  &  $\left(Z^{TS}_3\rtimes Z^{S\circ \mathcal{CP}}_2\right)\times Z^{R}_2$   \\ \hline

$\pm T^{-1}, \pm S$& $e^{2i\pi/3}$ &  $\left(Z^{ST}_3\rtimes Z^{S\circ \mathcal{CP}}_2\right)\times Z^{R}_2$     \\ \hline\hline
\end{tabular}
\caption{\label{tab:residual-CP-modulus}The CP conserving values of modulus $\tau$ in the fundamental domain $\mathcal{D}$ and the corresponding residual symmetry group. The symbol $\gamma^{CP}_{0}$ is defined by the condition $(\gamma^{\mathcal{CP}}_{0}\circ{\cal CP})\tau^{\mathcal{CP}}_0=\tau^{\mathcal{CP}}_0$ which is a special case of Eq.~\eqref{eq:gamma-CP-fix} with $\tau^{\mathcal{CP}}_0$ in the fundamental domain. We would like to mention that the definition of $\gamma\circ \mathcal{CP}$ is given in Eq.~\eqref{eq:CP-generd}. Notice that the modular generators $S$ and $T$ are given in Eq.~\eqref{eq:S-T-Mod-Gen} and $R=S^2$.}
\end{table}

\subsection{Implications for flavor mixing at CP fixed points }

Considering a point $\tau^{\mathcal{CP}}_0$ of the fundamental domain where CP is conserved, then there is an element $\gamma^{\mathcal{CP}}_0$ of $\Gamma$ such that $(\gamma^{\mathcal{CP}}_{0}\circ{\cal CP})\tau^{\mathcal{CP}}_0=\tau^{\mathcal{CP}}_0$ which gives  $\gamma^{\mathcal{CP}}_{0}(-\tau^{\mathcal{CP}*}_0)=\tau^{\mathcal{CP}}_{0}$. Taking the imaginary part on both sides of this equation and using the identities $\Im(-\tau^{*})=\Im\tau$ and $\Im(\gamma\tau)=\dfrac{\Im\tau}{|c\tau+d|^2}$, we obtain the automorphy factor
\begin{equation}
|J(\gamma^{\mathcal{CP}}_0, -\tau^{\mathcal{CP}*}_0)|=1
\end{equation}
which implies that $J(\gamma^{\mathcal{CP}}_0, -\tau^{\mathcal{CP}*}_0)$ is a pure phase. Note that the automorphy factor $J(\gamma, \tau)=c\tau+d$ is defined in Eq.~\eqref{eq:J-automorphy-fac}. We assume that the lepton sector is described by the superpotential:
\begin{equation}
\mathcal{W}=-E^{c}_i\mathcal{Y}^e_{ij}(\tau)L_jH_d-\frac{1}{2\Lambda}L_i\mathcal{Y}^{\nu}_{ij}(\tau)L_jH_uH_u\,,
\end{equation}
where the neutrino masses are described by the Weinberg operator. Under the action of modular transfromation $\gamma$, the matter multiplets $\varphi$ $(\varphi=H_{u,d},E^c,L)$ transform as:
\begin{equation}
\varphi\xrightarrow{\gamma} J^{-k_\varphi}(\gamma,\tau)\rho_\varphi(\gamma)\varphi\,,~~~\varphi\xrightarrow{\mathcal{CP}} X_\varphi\overline{\varphi}(x_{\mathcal{P}})\,.
\end{equation}
The weights $k_{E^c,L}$ carry a flavour index and $J^{k_{E^c,L}}$ are diagonal matrices in flavour space. The invariance of the $\mathcal{W}$ under the modular transformation $\gamma$ implies:
\begin{align}
\mathcal{Y}^e(\gamma\tau)&=J^{k_{H_d}}\rho_{H_d}^\dagger(\gamma)~ J^{k_{E^c}}\rho_{E^c}^*(\gamma)~\mathcal{Y}^e(\tau)~ \rho_L^\dagger(\gamma)J^{k_L}~~~\,, \nn\\
\mathcal{Y}^\nu(\gamma\tau)&=J^{2k_{H_u}}\rho_{H_u}^{2\dagger}(\gamma)~ J^{k_{L}}\rho_{L}^*(\gamma)~\mathcal{Y}^\nu(\tau)~ \rho_L^\dagger(\gamma)J^{k_L}~~~\,,
\label{w1}
\end{align}
On the other hand, the invariance of $\mathcal{W}$ under the CP transformation $\mathcal{CP}$ requires:
\begin{align}
\mathcal{Y}^e(-\tau^*)&= X_{H_d}^\dagger~X_{E^c}^*~\mathcal{Y}^{e*}(\tau)~ X_L^\dagger\,,\nn\\
\mathcal{Y}^\nu(-\tau^*)&=X_{H_u}^{2\dagger}~X_{L}^*~\mathcal{Y}^{\nu*}(\tau)~ X_L^\dagger~~~.
\label{w2}
\end{align}
From eqs.~\eqref{w1} and \eqref{w2}, at the point $\tau^{\mathcal{CP}}_0$ enjoying residual CP symmetry we obtain:
\begin{align}
\mathcal{Y}^{e*}(\tau^{\mathcal{CP}}_{0})&=\Omega_{H_d}~ \Omega_{E^c}^T~\mathcal{Y}^e(\tau^{\mathcal{CP}}_{0})~ \Omega_L\,,\nn\\
\mathcal{Y}^{\nu*}(\tau^{\mathcal{CP}}_{0})&=\Omega_{H_u}^{2}~ \Omega_{L}^T~\mathcal{Y}^\nu(\tau^{\mathcal{CP}}_{0})~ \Omega_L\,,
\end{align}
where we have defined the unitary matrices:
\begin{equation}
\Omega_\varphi=J^{-k_{\varphi}}(\gamma^{\mathcal{CP}}_0, -\tau^{\mathcal{CP}*}_0)\rho_\varphi(\gamma^{\mathcal{CP}}_0)X_\varphi\,,
\qquad\qquad (\varphi=H_{u,d},E^c,L)\,.
\end{equation}
Notice that $\Omega_\varphi$ is the CP transfromation matrix of the matter multiplet $\varphi$ for the residual gCP symmetry $\gamma^{\mathcal{CP}}_{0}\circ{\cal CP}$. The charged lepton and neutrino mass matrices are given by $M_e=\mathcal{Y}^{e}v_d$ and $M_{\nu}=\mathcal{Y}^{\nu}v^2_u/\Lambda$. Thus at the point $\tau$ enjoying residual CP invariance we have
\begin{eqnarray}
\nonumber&&\Omega_L^\dagger~M^{\dagger}_e(\tau^{\mathcal{CP}}_0)M_e(\tau^{\mathcal{CP}}_0)~\Omega_L=\left[M^{\dagger}_e(\tau^{\mathcal{CP}}_0)M_e(\tau^{\mathcal{CP}}_0)\right]^{*}\,,\\
\label{eq:res-constr-mass-matrices}&&\Omega_L^\dagger~M^{\dagger}_\nu(\tau^{\mathcal{CP}}_0)M_\nu(\tau^{\mathcal{CP}}_0)~\Omega_L=\left[M^{\dagger}_\nu(\tau^{\mathcal{CP}}_0)M_{\nu}(\tau^{\mathcal{CP}}_0)\right]^{*}\,.
\end{eqnarray}
We see that the hermitian combination of the neutrino and charged lepton mass matrices $M^{\dagger}_{\nu}(\tau^{\mathcal{CP}}_0)M_{\nu}(\tau^{\mathcal{CP}}_0)$ and $M^{\dagger}_{e}(\tau^{\mathcal{CP}}_0)M_{e}(\tau^{\mathcal{CP}}_0)$ are invariant under a common transformation of the left-handed charged leptons and the left-handed neutrinos, which represents a combination of CP and modular transformations. The unitary matrix $\Omega_L$ should be symmetric otherwise the neutrino and the charged lepton mass spectrum would be constrained to be partially degenerate~\cite{Feruglio:2012cw}. Furthermore, one can always go to the basis in which $\Omega_L=1$ by performing basis transfromation. Then the conditions eq.~\eqref{eq:res-constr-mass-matrices} fulfilled at the CP fixed point $\tau^{\mathcal{CP}}_0$ imply that both Dirac and Majorana CP phases are trivial~\cite{Feruglio:2012cw}. Therefore values of moduli deviating from residual CP symmetry fixed points (the fundamental domain boundary or the imaginary axis) are required to accommodate the observed non-degenerate lepton masses and a non-trivial Dirac CP phase.

\section{\label{sec:multiple-moduli}Modular invariance with multiple moduli}

More fundamental theory at high energy generally involve extra dimension. For instance, it is known that superstring theory can only be consistently defined in 10 dimensional spacetime, the extra 6 dimensions have to be compactified in the form of a Calabi–Yau manifold. As a consequence, the low energy effective theories at our 4D universe have multiple moduli which depend on the geometry of extra 6 dimensions. In this section, we shall study the scenario of several moduli $\tau_{1}, \tau_{2}, \ldots, \tau_{M}$ from the bottom-up approach, starting with factorizable moduli, before discussing non-factorizable generalisations based on symplectic modular invariance, Siegel modular groups and automorphic forms.

\subsection{Factorizable moduli}

Firstly we consider the direct product of several modular transformation groups $\Gamma^{1}\times \Gamma^{2}\times\ldots\times \Gamma^{M}$, where each $\Gamma^{j}$ for $j=1,..., M$ is the $SL(2, \mathbb{Z})$ group. The modulus field for each modular symmetry $\Gamma^{j}$ is denoted as $\tau_j$. As shown in Eq.~\eqref{eq:linear-fractional-trans}, any modular transformation $\gamma_j$ in $\Gamma^{j}$ acts on the modulus $\tau_j$ as follows~\cite{deMedeirosVarzielas:2019cyj}
\begin{eqnarray}
\label{eq:Mod-trans-j}&&\gamma_j: \tau_j \to \gamma_j \tau_j = \frac{a_j \tau_j + b_j}{c_j \tau_j + d_j},~~~\gamma_j\in\Gamma^{j} \,,
\end{eqnarray}
while other modulus $\tau_i$ is invariant under $\gamma_j$, i.e., $\gamma_j \tau_i=\tau_i$ for $i\neq j$. Analogous to Eq.~\eqref{eq:Gamma-N-PCSG},  each modular group $\Gamma^{j}$ has the principal congruence subgroup $\Gamma^{j}(N_j)$, and a series of finite modular groups $\Gamma_{N_j}^{'j}=\Gamma^{j}/\Gamma^{j}(N_j)$ and $\Gamma_{N_j}^{j}=\Gamma^{j}/\pm\Gamma^{j}(N_j)$ can be defined. Notice that $N_i$ is unnecessary to be identical to $N_{j}$ for $i\neq j$. Thus the finite modular group is extended to the direct product of several finite modular groups $\Gamma_{N_1}^1\times \Gamma_{N_2}^2 \times \cdots \times \Gamma_{N_M}^M$, and any $\Gamma_{N_j}^j$ could be $\Gamma_{N_j}^{'j}$.

The modular transformation of a generic chiral superfield $\Phi_I$ is characterized by the modular weight $k_{I,j}$ and the unitary representation $\rho_{\mathbf{r}_I,j}$ of $\Gamma_{N_j}^j$ under each modular group $\Gamma^{j}$, i.e.,
\begin{eqnarray}
\label{eq:Mod-trans-phi-duotau}\Phi_I\rightarrow \prod_{j=1,...,M} (c_j\tau_j + d_j)^{-k_{I,j}} \bigotimes_{j=1,...,M} \rho_{\mathbf{r}_I,j}(\gamma_j) \Phi_I\,,
\end{eqnarray}
where $\bigotimes$ denotes the outer product of the representation matrices for $\rho_{\mathbf{r}_I,1}$, $\rho_{\mathbf{r}_I,2}$, ..., $\rho_{\mathbf{r}_I,M}$. In the framework of the $\mathcal{N}=1$ global supersymmetry, the most general action invariant under the extended modular group can be written as
\begin{equation}
\mathcal{S}=\int d^4xd^2\theta d^2\bar{\theta} \; \mathcal{K}(\Phi_I,\bar{\Phi}_I; \tau_1,...,\tau_M,\overline{\tau}_1,...,\overline{\tau}_M) + \left[\int d^4x d^2\theta\; \mathcal{W}(\Phi_I,\tau_1,..., \tau_M) + \text{h.c.}\right]\,.
\end{equation}
Invariance of the action $\mathcal{S}$ under Eqs.~(\ref{eq:Mod-trans-j}, \ref{eq:Mod-trans-phi-duotau}) requires the invariance of the K\"ahler potential up to a K\"ahler transformation. Similar to the case of single modulus, the requirement of invariance of the K\"ahler potential can be easily satisfied and many terms are allowed. For illustration, the minimal K\"ahler potential is taken as usual.
\begin{eqnarray}
\hspace{-5mm}
\mathcal{K}(\Phi_I, \overline{\Phi}_I;\tau_1,...,\tau_M,\overline{\tau}_1,...,\overline{\tau}_M) &=& - \sum^{M}_{j=1} h_j \log(-i\tau_j+i\overline{\tau}_j) \nonumber\\
&+& \sum_{I} \, \frac{\overline{\Phi}_I \Phi_I}{\displaystyle \prod_{j=1,...,M} (-i \tau_j + i \overline{\tau}_j)^{k_{I,j}}}
\,,
\end{eqnarray}
where all $h_j$ are positive constants. Since each modular symmetry is independent from each other, one modulus field getting a VEV leaves the rest of the K\"ahler potential still satisfying the other modular symmetries. If all modulus fields acquire VEVs, the above K\"ahler potential leads to kinetic terms for the scalar components of the supermultiplets $\phi_I$ and the modulus fields $\tau_j$ as follow
\begin{eqnarray}
\sum^{M}_{j=1} \frac{h_j}{\langle -i\tau_j + i \overline{\tau}_j \rangle^2} \partial_\mu \overline{\tau}_j \partial^\mu \tau_j +
\sum_I \frac{\partial_\mu \overline{\Phi}_I \partial^\mu \Phi_I}{ \displaystyle \prod_{j=1,...,M}\langle -i\tau_j + i \overline{\tau}_j \rangle^{k_{I,j}} }  \,.
\end{eqnarray}
The superpotential $\mathcal{W}(\Phi_I;\tau_1,..., \tau_M)$ is in general a holomorphic function of the moduli $\tau_1,\ldots,\tau_M$ and superfields $\phi_I$. Modular invariance imposes strong constraint on the superpotential $\mathcal{W}$ which can be expanded in power series of $\Phi_I$ as
\begin{eqnarray}
\mathcal{W}(\Phi_I;\tau_1,..., \tau_M) = \sum_n  \sum_{\{I_1, \cdots, I_n\}}
\left(Y_{I_{1}...I_{n}} \Phi_{I_1} \cdots \Phi_{I_n} \right)_{\mathbf{1}} \,.
\end{eqnarray}
For each term to be modular invariant, the modular forms $Y_{I_{1}...I_{n}}$ should transform as
\begin{eqnarray} \label{eq:form_transformation2}
\hspace{-5mm}
&&Y_{I_{1}...I_{n}}(\tau_1,..., \tau_M) \to Y_{I_{1}...I_{n}}(\gamma_1 \tau_1, ..., \gamma_M \tau_M) \nonumber\\
 &&\hspace{2cm}= \prod^{M}_{j=1} (c_j\tau_j + d_j)^{k_{Y,j}}
 \bigotimes^{M}_{j=1} \rho_{\mathbf{r}_Y, j}(\gamma_j) Y_{I_{1}...I_{n}}(\tau_1,..., \tau_M) \,,
\end{eqnarray}
satisfying the conditions
\begin{eqnarray}
\nonumber&& k_{Y,j} = k_{I_1,j}+ \cdots k_{I_n,j}\;,\\
&&\rho_{\mathbf{r}_Y, j}\otimes\rho_{\mathbf{r}_{I_1}, j}\otimes\ldots\otimes\rho_{\mathbf{r}_{I_n}, j}\ni \mathbf{1}\,,
\end{eqnarray}
for $j=1,...,M$.

\subsection{\label{subsec:symplectic-mod-theory}Symplectic modular invariance}

Siegel modular forms can be thought of as modular forms in more than one
variable, they generalize the usual single variable modular forms on $SL(2, \mathbb{Z})$ in that group $SL(2, \mathbb{Z})$ is replaced by the Siegel modular group $\Gamma_g=Sp(2g,\mathbb{Z})$ and the upper half plane is replaced by the Siegel upper half plane $\mathcal{H}_g$, where the integer $g\geq 1$ is called the degree or genus. The Siegel modular group $\Gamma=Sp(2g,\mathbb{Z})$ arises as the duality group in string Calabi-Yau compactifications~\cite{Cecotti:1988qn,Cecotti:1988ad,Cecotti:1989kn,Dixon:1989fj,Candelas:1990pi,Strominger:1990pd,Ferrara:1991uz,Font:1992uk,Ishiguro:2020nuf,Baur:2020yjl,Nilles:2021glx,Ishiguro:2021ccl}. Siegel modular forms are relevant in the context of string one-loop corrections ~\cite{Mayr:1995rx,Stieberger:1998yi}.

The Siegel modular group $\Gamma=Sp(2g,\mathbb{Z})$ is the group of $2g\times 2g$ matrices with integer entries and satisfy the symplectic condition
\begin{equation}
\gamma^t~J~\gamma=J,~~~J=
\begin{pmatrix}
0~&~\mathds{1}_g\\
-\mathds{1}_g~&~0
\end{pmatrix},~~~\gamma=
\begin{pmatrix}
A&B\\
C&D
\end{pmatrix}\in\Gamma_{g}\,,
\end{equation}
where the superscript $t$ denotes the transpose, and all the four blocks $A$, $B$, $C$, $D$ are $g$ dimensional matrices with integer elements. Thus blocks $A$, $B$, $C$ and $D$ satisfy the conditions,
\begin{equation}
A^t C=C^t A\,,\quad B^t D=D^t B\,,\quad A^t D-C^t B=\mathds{1}_g\,,
\end{equation}
or equivalently
\begin{equation}
AB^t=BA^t\,,\quad CD^t=DC^t\,,\quad AD^t-BC^t=\mathds{1}_g\,.
\end{equation}
For the lowest genus $g=1$, the above symplectic condition is exactly the condition of unit determinant of $\gamma$. Consequently the Siegel modular group $\Gamma_1=Sp(2,\mathbb{Z})$ is exactly identical with the usual modular group $SL(2,\mathbb{Z})$. For any element $\gamma\in\Gamma_g$, both $\gamma^{t}$ and $\gamma^{-1}$ are also elements of $\Gamma_g$ with
\begin{equation}
\gamma^{-1}= \begin{pmatrix} D^t ~&~ -B^t \\
-C^t ~&~ A^t \end{pmatrix}\,.
\end{equation}
The single modulus $\tau$ in the upper half plane is generalized to the Siegel upper half plane $\mathcal{H}_g$ which is represented by $g\times g$ symmetric complex matrices with positive definite imaginary part\footnote{The modulus $\tau$ can be parameterized as $\tau=X+i Y$ where both $X$ and $Y$ are real symmetric matrices with $Y\neq0$, and all the eigenvalues of $Y$ are positive. This is exactly the condition of $\texttt{Im} (\tau) > 0$ in Eq.~\eqref{eq:tau-siegel}.},
\begin{equation}
\label{eq:tau-siegel}\mathcal{H}_g=\left\{\tau \in GL(g,\mathbb{C})~\Big|~  \tau^t = \tau ,\quad \texttt{Im} (\tau) > 0 \right\}\,.
\end{equation}
The action of $\gamma\in\Gamma_g$ on the Siegel upper half plane $\mathcal{H}_g$ is defined as follow,
\begin{equation}
\tau\to \gamma \tau=(A \tau+B)(C\tau +D)^{-1}\,.
\end{equation}
The Siegel modular group $\Gamma_g$ can be generated by the generators $S$ and $T_i$ with
\begin{equation}
\label{eq:generators-Siegel}
S=\begin{pmatrix}
0~&~\mathds{1}_g\\
-\mathds{1}_g~&~0
\end{pmatrix}\,,\quad T_i=\begin{pmatrix}
\mathds{1}_g ~&~B_i\\
0~&~\mathds{1}_g
\end{pmatrix}\,,
\end{equation}
where $\{B_i\}$ is a basis for the $g\times g$ integer symmetric matrices.  Notice that $S$ coincides with the invariant symplectic form $J$ satisfying $S^2=-\mathds{1}_{2g}$. For the case of $g=2$, without loss of generality we could choose
\begin{equation}
B_1 =\begin{pmatrix}
1 ~& 0 \\
0 ~& 0 \\
\end{pmatrix},~~~ B_2 =\begin{pmatrix}
0 ~& 0 \\
0 ~& 1 \\
\end{pmatrix}, ~~~ B_3 =\begin{pmatrix}
0 ~& 1 \\
1 ~& 0 \\
\end{pmatrix}\,.
\end{equation}
Under the action of generators $S$ and $T_i$, the modulus $\tau$ transforms as:
\begin{equation}
\tau\xrightarrow{S}-\tau^{-1},~~~\tau\xrightarrow{T_i}\tau+B_i\,.
\end{equation}
In a similarly way as the single modulus case, we can define the principal congruence subgroups $\Gamma_g(n)$ of level $n$ as
\begin{equation}
\label{eq:Gamma-g-n}
\Gamma_g(n)=\left\{\gamma \in \Gamma_g ~\Big|~  \gamma \equiv \mathds{1}_{2g} \,~(\texttt{mod}~n)\right\}\,,
\end{equation}
where $n$ is a generic positive integer, and $\Gamma_g(1)=\Gamma_g$.
The group $\Gamma_g(n)$ is a normal subgroup of $\Gamma_g$, and the quotient group $\Gamma_{g, n}\equiv\Gamma_g/ \Gamma_g(n)$, which is known as finite Siegel modular group, has finite order~\cite{Koecher1954Zur,thesis_Fiorentino}:
\begin{equation}
\label{eq:subgroup_index}
|\Gamma_{g,n}|= n^{g(2g+1)}\prod_{p|n}\prod_{1\leq k\leq g} (1 - \dfrac{1}{p^{2k}})\,,
\end{equation}
where the product is over the prime divisors $p$ of $n$.
For the simplest case of $g=1$, we have
\begin{eqnarray}
|\Gamma_{1,n}|=n^3\prod_{p|n}(1 - \dfrac{1}{p^{2}})\,.
\end{eqnarray}
This is consistent with the dimension formula of $SL(2, \mathbb{Z}_n)$~\cite{Gunning1962,Schoeneberg}.

\begin{table}[hptb]
\centering
\begin{tabular}{|c|c|}
\hline\hline
 \texttt{Modular group} $SL(2,\mathds{Z})\equiv\Gamma$ &  \texttt{Siegel modular group} $Sp(2g,\mathds{Z})\equiv\Gamma_g$  \\
\hline
$\gamma=\begin{pmatrix}
a  & b \\
c& d
\end{pmatrix}$ & $\gamma=\begin{pmatrix}
A  & B \\
C& D
\end{pmatrix}$  \\
$a,b,c,d\in\mathds{Z}$  & $A, B, C, D\in GL(g,\mathds{Z})$ \\

$ad-bc=1$  &  $\gamma^{t}J\gamma=J$, $J=\begin{pmatrix}
0  & \mathds{1}_g\\
-\mathds{1}_g  & 0
\end{pmatrix}$ \\ \hline

$H_1=\left\{\tau\in\mathds{C}|\text{Im}(\tau)>0\right\}$ & $H_g=\left\{\tau\in GL(g, \mathds{C})|\tau^{t}=\tau, \text{Im}(\tau)>0\right\}$ \\ \hline

$\tau\rightarrow\gamma\tau=\frac{a\tau+b}{c\tau+d}$  & $\tau\rightarrow\gamma\tau=\left(A\tau+B\right)\left(C\tau+D\right)^{-1}$ \\ \hline

$\Gamma(N)=\left\{\gamma\in SL(2, \mathds{Z})|\gamma=\mathds{1}_{2} \mod{N}\right\}$  &  $\Gamma_g(N)=\left\{\gamma\in Sp(2g, \mathds{Z})|\gamma=\mathds{1}_{2g} \mod{N}\right\}$ \\ \hline

$\Gamma'_N=SL(2, \mathds{Z})/\Gamma(N)$  &  $\Gamma_{g, N}=Sp(2g, \mathds{Z})/\Gamma_g(N)$ \\\hline

$Y(\gamma\tau)=(c\tau+d)^{k}Y(\tau),~~\gamma\in\Gamma(N)$  & $Y(\gamma\tau)=\left[\text{det}(C\tau+D)\right]^{k}Y(\tau),~~\gamma\in\Gamma_g(N)$ \\ \hline \hline
\end{tabular}
\caption{\label{tab:Siegel-modular-group}The comparison between the modular group $SL(2,\mathds{Z})$ and the Siegel modular group $Sp(2g,\mathds{Z})$, where $g$ is a positive integer and it corresponds to the genus. Notice that $Sp(2, \mathds{Z})$ is isomorphic to $SL(2, \mathds{Z})$.  }
\end{table}

\subsubsection{Siegel modular forms and gCP in symplectic modular invariant theory }

The Siegel modular forms $Y(\tau)$ of integer weight $k$ and level $n$ at genus $g$ are holomorphic functions of the moduli $\tau$, and they fulfill the following transformation properties under the principal congruence subgroups $\Gamma_g(n)$:
\begin{equation}
\label{eq:SiegelForms}
Y(\gamma \tau)=[\det(C\tau+D)]^k Y(\tau) \,, \quad\quad \gamma=\begin{pmatrix}
A & B \\
C & D
\end{pmatrix} \in \Gamma_g(n) \,.
\end{equation}
For lower levels $n=1,2$, by taking $\gamma=-\mathds{1}_{2g}$ in Eq.~\eqref{eq:SiegelForms}, we can see that the Siegel modular forms at genus $g$ of weight $k$ vanish if $kg$ is odd. The Siegel modular forms of given weight $k$, level $n$ and genus $g$ expand a linear space $\mathcal{M}_k(\Gamma_g(n))$ of finite dimensional~\cite{Bruinier2008The} and there are no non-vanishing
forms of negative weight~\cite{Bruinier2008The}. The product of two Siegel modular forms of weight $k$, $k'$ at level $n$ and genus $g$ is a Siegel modular form of weight $k+k'$. The full set of Siegel modular forms with respect to $\Gamma_g(n)$ form a positive graded ring $\mathcal{M}(\Gamma_g(n))= \bigoplus_{k \geq 0} \mathcal{M}_k(\Gamma_g(n))$.

The Siegel modular forms of weight $k$, level $n$ and genus $g$ are invariant up to the automorphy factor $[\det(C\tau+D)]^k$ under $\Gamma_g(n)$ but they transform under the quotient group $\Gamma_{g, n}\equiv\Gamma_g/ \Gamma_g(n)$.
Similar to the case $g=1$~\cite{Feruglio:2017spp,Liu:2019khw}, it is always possible to choose a basis $\{Y_i(\tau)\}$ in the space $\mathcal{M}_k(\Gamma_g(n))$ such that the action of $\Gamma_g$ on the elements of the basis is described by a unitary representation $\rho_{\mathbf{r}}$ of the finite Siegel modular group $\Gamma_{g, n}=\Gamma_g/\Gamma_g(n)$~\cite{Ding:2020zxw}:
\begin{equation}
\label{eq:SMF-decomposition}
Y_i(\gamma \tau) = [\det(C\tau+D)]^k \rho_\mathbf{r}(\gamma)_{ij} Y_j(\tau), ~~~\quad \gamma=\begin{pmatrix}
A ~& B \\
C ~& D
\end{pmatrix} \in \Gamma_g\,.
\end{equation}
At variance with Eq.~\eqref{eq:SiegelForms}, where only transformations of $\Gamma_g(n)$ were considered, in the previous equation the full Siegel modular group $\Gamma_g$ is acting. The comparison between the modular group $SL(2,\mathds{Z})\equiv\Gamma$ and the Siegel modular group $Sp(2g,\mathds{Z})\equiv\Gamma_g$ is summarized in table~\ref{tab:Siegel-modular-group}. One see that $\Gamma_g$ is a natural generation of $\Gamma$ at high genus $g$, and the matrix elements $a$, $b$, $c$, $d$ are replaced by $A$, $B$, $C$, $D$ which are $g\times g$ integer matrices, and the automorphy factor $c\tau+d$ is replaced by $\det(C\tau+D)$.

Then the modular invariant theory with single modulus sketched in section~\ref{subsec:modular-invariant-theory} can be straightforwardly generalized to the case with multiple moduli based on the Siegel modular symmetry $\Gamma_g$. Analogously each terms of the superpotential should be invariant under the finite Siegel modular group $\Gamma_{g, n}$ and its modular weight should be zero. Moreover, similar to the case of single modulus discussed in section~\ref{subsec:gCP-modular-symmetry}, the consistent CP transformation of the matrix $\tau$ is determined to be~\cite{Ding:2021iqp}
\begin{equation}
\label{eq:gCP-siegel}\tau\xrightarrow{{\cal CP}}\tau_{{\cal CP}}=-\tau^*\,,
\end{equation}
up to a Siegel modular transformation. Combining the above CP and Siegel modular transformations, we get the extended Siegel modular group $\Gamma_g^*=GSp(2g,\mathbb{Z})$ and the full symmetry transformation of the complex moduli is
\begin{align}
\nonumber&\tau\xrightarrow{\gamma} (A\tau+B)(C\tau+D)^{-1}\,,\\
&\tau\xrightarrow{g{\cal CP}} (-A\tau^{*}+B)(-C\tau^{*}+D)^{-1}\,,
\end{align}
where $\gamma=\begin{pmatrix}
A &  B \\
C  & D
\end{pmatrix}$ is the Siegel modular transformation satisfying $\gamma^{t}J\gamma=J$. The CP symmetry acts on a generic matter chiral supermultiplet $\varphi$ as follows,
\begin{equation}
\label{eq:gCP-varphi-Siegel}\varphi(x)\stackrel{\mathcal{CP}}{\longrightarrow} X_{\mathbf{r}}\overline{\varphi}(x_{\mathcal{P}})\,,
\end{equation}
where $X_{\mathbf{r}}$ is a unitary matrix in the flavor space. The consistency between the CP symmetry and modular symmetry requires $X_{\mathbf{r}}$ should fulfill the following conditions~\cite{Ding:2021iqp},
\begin{equation}
X_{\mathbf{r}}~\rho_{\mathbf{r}}^*(S)~X_{\mathbf{r}}^{-1}=\rho_{\mathbf{r}}(S^{-1})\,,~~~
X_{\mathbf{r}}~\rho_{\mathbf{r}}^*(T_i)~X_{\mathbf{r}}^{-1}=\rho_{\mathbf{r}}(T_i^{-1})\,.
\end{equation}
In the same fashion as section~\ref{subsec:gCP-MF}, one can show that CP transformation of the modular form multiplets is identical with that of single modulus case, i.e.,
\begin{equation}
Y^a_{\mathbf{r}}(\tau)\stackrel{\mathcal{CP}}{\longrightarrow} Y^a_{\mathbf{r}}(-\tau^*)=\lambda^a_{~b}X_\mathbf{r} Y^{b*}_{\mathbf{r}}(\tau)\,,
\end{equation}
with $\lambda^a_{~c} \lambda^{*c}_{~b}X_\mathbf{r} X^*_\mathbf{r}=\delta^a_{~b}\mathds{1}_\mathbf{r}$, where $a, b$ label linearly independent modular form multiplets of the same type. In the basis where the matrices $\rho_{\mathbf{r}}(S)$ and $\rho_{\mathbf{r}}(T_i)$ are symmetric and unitary, the gCP transformation would coincide with the canonical CP $X_{\mathbf{r}}=\mathds{1}$ and one can always work with modular forms $Y^a_{\mathbf{r}}(\tau)$ with $\lambda^a_{~b}=\delta^a_{~b}$.

\subsubsection{Invariant loci in moduli space and reduced finite Siegel modular group }

From Eq.~\eqref{eq:subgroup_index}, we see that the order of the finite Siegel modular group $\Gamma_{g, n}$ increases quickly with the
genus $g$ and level $n$. At the genus $g=2$, we have~\cite{Ding:2020zxw}
\begin{equation}
\Gamma_{2,2}\cong S_6,~~~\Gamma_{2, 3}\cong Sp(4, F_3)\,,
\end{equation}
and
\begin{equation}
|\Gamma_{2,2}|=720,~~~|\Gamma_{2,3}|=51840\,.
\end{equation}
Note that $S_6$ is the permutation group of six objects, and $Sp(4, F_3)$ is isomorphic to the double covering of Burkhardt group. However, $S_6$ has no three dimensional irreducible representations to which the three generations of left-handed lepton fields are usually assigned\footnote{The $S_6$ group has two one-dimensional, four five-dimensional, two nine-dimensional, two ten-dimensional and one sixteen-dimensional irreducible representations.}.

Small finite Siegel modular groups can be obtained by restricting the theory to a subregion $\Omega$ of the moduli space $\mathcal{H}_g$~\cite{Ding:2020zxw}. We define a region $\Omega$ whose points $\tau$ are individually left invariant under the action of a subgroup $H$ of $\Gamma_g$, i.e.
\begin{equation}
\label{eq:fp-Siegel} h~\tau=\tau\,,~~~h\in H,~~\tau\in\Omega.
\end{equation}
Because $\gamma\tau=-\gamma\tau$, we also consider the projective group
$\bar{H}=H/\{\pm \mathds{1}_{2g}\}$. Both $H$ and $\bar{H}$ are called stabilizers. The group $N(H)$ that, as a whole, leaves the region $\Omega$ invariant includes the elements $\gamma$ of $\Gamma_g$ such that:
\begin{equation}
\gamma \tau=\tau'\,,\qquad \tau,\tau'\in\Omega\,,
\label{eq:NH-def}
\end{equation}
which requires
\begin{equation}
\label{eq:normalizer}\gamma^{-1}H\gamma=H\,.
\end{equation}
Therefore $N(H)$ is the normalizer of $H$. Analogously we can define the principal congruence subgroup of $N(H)$, denoted as $N(H, n)$:
\begin{equation}
N(H, n)=\left\{\hat{\gamma}\in N(H)~\Big|~\hat{\gamma}=\mathds{1}_{2g} \,~(\texttt{mod}~n) \right\}~~~.
\end{equation}
Obviously $N(H, n)$ is a subgroup of $\Gamma_{g}(n)$, and it is also a normal subgroup of $N(H)$. The finite modular subgroup $N_n(H)$ corresponding to the modular subgroup $N(H)$ is the quotient group
\begin{equation}
N_n(H)= N(H)/N(H, n)\,,
\end{equation}
which is a subgroup of finite Siegel modular group $\Gamma_{g,n}$. In short, we can consistently truncate the moduli space to the subspace $\Omega$, and substitute
$\Gamma_g$ and $\Gamma_{g,n}$ with $N(H)$ and $N_n(H)$, respectively. As a result, in our supersymmetric action we can restrict the moduli $\tau$ to the region $\Omega$, which supersedes the full moduli space ${\cal H}_g$,
and replace the group $\Gamma_g$ with $N(H)$. An element $\gamma$ of $N(H)$ induces the transformation laws
\begin{equation}
\left\{
\begin{array}{l}
\tau\to \gamma \tau=(A \tau+B)(C\tau +D)^{-1}
\\[0.2 cm]
\varphi_{I}\to [\det(C\tau+D)]^{k_I} \rho_{\mathbf{r}_{I}}(\gamma) \varphi_{I}~~~.
\end{array}
\right.\qquad\quad \gamma=\begin{pmatrix}
A & B \\
C & D
\end{pmatrix}\in N(H)\,,
\label{eq:siegel-trans-constr}
\end{equation}
where $\rho_{\mathbf{r}_{I}}(\gamma)$ is a unitary representation of a finite group $N_n(H)$.

The genus 2 Siegel modular invariant theories is the simplest non trivial generalization of modular invariant supersymmetric theories studied in~\cite{Ferrara:1989bc,Ferrara:1989qb,Feruglio:2017spp}. The moduli space ${\cal H}_2$ has complex dimension 3 and describes 3 moduli:
\begin{equation}
\tau=\left(
\begin{array}{cc}
\tau_1&\tau_3\\
\tau_3&\tau_2
\end{array}
\right)\,,\qquad	\det(\texttt{Im}(\tau))>0 \,,\qquad \text{Tr}(\texttt{Im}(\tau))>0\,.
\end{equation}

The invariant loci in Siegel upper half plane $\mathcal{H}_2$ have been classified by Gottschling~\cite{gottschling1961fixpunkte,gottschling1961fixpunktuntergruppen,gottschling1967uniformisierbarkeit}. It is established that  $\mathcal{H}_2$ has two inequivalent invariant loci of complex dimension two,
\begin{equation}
\label{eq:dim2-loci}\begin{pmatrix} \tau_1 & 0 \\ 0 & \tau_2 \end{pmatrix}\,,~~~\begin{pmatrix} \tau_1 & \tau_3 \\ \tau_3 & \tau_1 \end{pmatrix}\,,
\end{equation}
five inequivalent invariant loci of complex dimension one,
\begin{eqnarray}
\begin{pmatrix} i & 0 \\ 0 & \tau_2 \end{pmatrix}\,,~~~\begin{pmatrix} \omega & 0 \\ 0 & \tau_2 \end{pmatrix}\,,~~~
\begin{pmatrix} \tau_1 & 0 \\ 0 & \tau_1 \end{pmatrix}\,,~~~
\begin{pmatrix} \tau_1 & 1/2 \\ 1/2 & \tau_1 \end{pmatrix}\,,~~~
\begin{pmatrix} \tau_1 & \tau_1 /2 \\ \tau_1 /2 & \tau_1 \end{pmatrix}\,,
\end{eqnarray}
and six independent isolated fixed points,
\begin{eqnarray}
\begin{pmatrix} \eta & \frac{1}{2}(\eta -1) \\ \frac{1}{2}(\eta -1) & \eta \end{pmatrix} \,,~~~
\begin{pmatrix} i & 0 \\ 0 & i \end{pmatrix}\,,~~~
\begin{pmatrix} \omega & 0 \\ 0 & \omega \end{pmatrix}\,,~~~
\dfrac{i\sqrt{3}}{3}\begin{pmatrix} 2 & 1 \\ 1 & 2 \end{pmatrix}\,,~~~
\begin{pmatrix} \omega & 0 \\ 0 & i \end{pmatrix}\,,
\end{eqnarray}
with $\zeta= e^{2\pi i /5},~\eta=\frac{1}{3}(1+i2\sqrt{2}),~\omega= e^{2\pi i/3}$. The corresponding stabilizers $\bar{H}$, the normalizers $N(H)$ and the finite modular group $N_2(H)$ for each locus are given in Ref.~\cite{Ding:2020zxw}. It is remarkable that the restricted finite modular group $N_2(H)$ has small order and it is promising as a flavor symmetry. For instance, $N_2(H)$ is isomorphic to $(S_3\times S_3)\rtimes Z_2$ and $S_4 \times Z_2$ respectively for the two-dimensional loci in Eq.~\eqref{eq:dim2-loci}. Fermion mass models based on symplectic modular symmetry are given in Refs.~\cite{Ding:2020zxw,Ding:2021iqp,Kikuchi:2023dow}.

\subsection{Multiple moduli and automorphic forms}

The modular flavor symmetry with single modulus can be naturally generalized to more general supersymmetric modular invariant theories involving multiple moduli, where a discrete subgroup $\Gamma$ of a non-compact Lie group $G$ plays the role of flavour symmetry and the symmetry breaking sector
spans an Hermitian Symmetric Space which is a coset space of the type $\mathcal{M}=G/K$, here $K$ is a maximal compact subgroup of $G$~\cite{Ding:2020zxw}. Then the modular forms which are building blocks of Yukawa couplings, would be replaced by the more general automorphic forms~\cite{Ding:2020zxw}.

It is notable that the well-known moduli space $\mathcal{H}=\left\{\tau\in\mathbb{C}|\texttt{Im}(\tau)>0\right\}$ can be equivalently defined as the quotient space $G/K$~\cite{Feruglio:2017spp}, where $G=SL(2,\mathbb{R})$ and $K=SO(2)$ is a maximal compact subgroup of $G$. To be more specific, a generic element $\textsf{g}$ of $SL(2,\mathbb{R})$ can be decomposed into the product of two matrices:
\begin{equation}
SL(2,\mathbb{R})\ni \textsf{g}=
\begin{pmatrix}
\sqrt{y}~& x/\sqrt{y}\\
0~&1/\sqrt{y}
\end{pmatrix}
~\textsf{k}\,,~~y>0~,\qquad\textsf{k}=\left(\begin{matrix}
\cos\theta  ~& -\sin\theta\\
\sin\theta ~ &\cos\theta
\end{matrix}\right)\,,
\end{equation}
where $\textsf{k}$ belongs to $K=SO(2)$, while the first factor in the above decomposition of $\textsf{g}$ is an element of the coset $SL(2,\mathbb{R})/SO(2)$. It is obvious that $\tau_0=i$ is invariant under the action of $K=SO(2)$,
\begin{equation}
\textsf{k}\cdot i=\begin{pmatrix}
\cos\theta ~& -\sin\theta\\
\sin\theta ~&\cos\theta
\end{pmatrix}
\cdot i=\frac{\cos\theta\;i-\sin\theta}{\sin\theta\;i+\cos\theta}=i\,.
\end{equation}
Hence any modulus $\tau=x+i y$ $(y>0)$ can be reached from the fixed complex number $\tau_0=i$ by a $SL(2,\mathbb{R})$ transformation $\textsf{g}$, i.e.,
\begin{equation}
\textsf{g}~\tau_0=\left(
\begin{array}{cc}
\sqrt{y} ~& x/\sqrt{y}\\
0~&1/\sqrt{y}
\end{array}
\right)\cdot i=x+i~y=\tau\,.
\label{gi}
\end{equation}
This explicitly shows the one-to-one correspondence between the elements of ${\cal H}$ and those of $SL(2,\mathbb{R})/SO(2)$.

In a similar fashion, the generic element $\tau$ of the multi-dimensional moduli space $\mathcal{M}$ can be represented by the action of $g\in G$ on an element $\tau_0$ left invariant by $K$:
\begin{equation}
\tau=g~ \tau_0,~~~g\in G\,,
\end{equation}
with
\begin{equation}
h~\tau_0=\tau_0\,,~~\text{for any}~~h\in K\,.
\end{equation}
In finite discrete subgroup $\Gamma$ of $G$ is chosen as the candidate flavor group, and its action on the moduli $\tau$ is given by
\begin{equation}
\tau\stackrel{\gamma}{\rightarrow}\gamma\tau=\left(\gamma g\right)\tau_0,~~~\gamma\in\Gamma\,.
\end{equation}
Given a normal subgroup $\Gamma_{normal}$ of $\Gamma$ with finite
index, we define the finite group $\Gamma_{finite}=\Gamma/\Gamma_{normal}$.
Being finite, the finite group $\Gamma_{finite}$ admits unitary representations $\rho(\gamma)$. A general transformation of the supermultiplets $\varphi^{(I)}$ of each sector is characterized by the  weight $k_I$ and the representation $\rho^{(I)}$ of $\Gamma_{finite}$:
\begin{equation}
\tau\stackrel{\gamma}{\rightarrow}\gamma\tau,~~~\varphi^{(I)}\xrightarrow{\gamma} j(\gamma,\tau)^{k_I}~ \rho^{(I)}(\gamma)\varphi^{(I)}~,~~~~\gamma\in\Gamma~\,,
\end{equation}
where $j(\gamma, \tau)$ is  the automorphy factor and it satisfies the so-called cocycle condition
\begin{equation}
j(\gamma_1\gamma_2,\tau)=j(\gamma_1,\gamma_2\tau)j(\gamma_2,\tau)\,.
\end{equation}
In short, the modular invariance of single modulus can be naturally generalized to the case with multiple moduli~\cite{Feruglio:2022ivc,Feruglio:2022cgv}:

\begin{eqnarray}
\begin{array}{clc}
SL(2, \mathbb{R})  ~&~ \rightarrow  ~&~  \text{Lie group}~~G  \\
SO(2) ~&~ \rightarrow  ~&~ \text{maximal compact subgroup}~~K\subset G \\
\mathcal{H}=SL(2, \mathbb{R}) /SO(2)  ~&~ \rightarrow  ~&~ \mathcal{M}=G/K \\
SL(2, \mathbb{Z}) ~&~ \rightarrow  ~&~ \text{infinite discrete subgroup}~~\Gamma\subset G \\
\Gamma(N) ~&~ \rightarrow  ~&~ \text{normal subgroup~~} \Gamma_{normal}\subset \Gamma \\
\Gamma'_N  ~&~ \rightarrow  ~&~ \Gamma_{finite}=\Gamma/\Gamma_{normal}  \\
c\tau+d ~&~ \rightarrow  ~&~ \text{automorphy factor}~~ j(\gamma, \tau)
\end{array}\,.
\end{eqnarray}

In the rigid $\mathcal{N}=1$ supersymmetry, a candidate minimal K\"ahler potential is given by:
\begin{equation}
\label{eq:kmin-automorphic} \mathcal{K}_{\tt min}=-h\log Z(\tau,\bar\tau)+\sum_I Z(\tau,\bar\tau)^{k_I} |\varphi^{(I)}|^2\,,~~Z(\tau,\bar\tau)\equiv [j^\dagger(\textsf{g},\tau_0)j(\textsf{g},\tau_0)]^{-1}\,,
\end{equation}
where $h$ is a real constant whose sign is chosen to guarantee local positivity of the metric for the moduli $\tau$. By construction, the above potential is invariant under $\Gamma$ up to a K\"ahler transformation for a general choice of $G$, $K$, $\Gamma$, $\Gamma_{normal}$ and $j(g,\tau)$. Nevertheless this is not the most general K\"ahler potential invariant under $\Gamma$. There are other terms which are compatible with invariance under $\Gamma$ in a pure bottom-up approach. Generally additional assumptions or inputs from a top-down approach are needed in order to reduce the arbitrariness of the predictions~\cite{Nilles:2020nnc,Nilles:2020kgo,Nilles:2020tdp,Nilles:2020gvu}.

The superpotential could be expanded in powers of the supermultiplets $\varphi^{(I)}$ as follows:
\begin{equation}
\mathcal{W}=\sum_n Y_{I_1...I_n}(\tau)~ \varphi^{(I_1)}... \varphi^{(I_n)}\,.
\end{equation}
The $n$-th order term is invariant provided the functions $Y_{I_1...I_n}(\tau)$ obey:
\begin{equation}
\label{eq:Ytras-auto}Y_{I_1...I_n}(\gamma\tau)=j(\gamma,\tau)^{k_Y(n)} \rho^{(Y)}(\gamma)~Y_{I_1...I_n}(\tau)\,.
\end{equation}
Each term of the superpotential should be invariant under the action of $\Gamma$ and its total weight should be vanishing such that $k_Y(n)$ and $\rho^{(Y)}$ have to fulfill
\begin{eqnarray}
\nonumber&&k_Y(n)+k_{I_1}+....+k_{I_n}=0\,,~~~\rho^{(Y)}\times \rho^{{(I_1)}}\times ... \times \rho^{{(I_n)}}\supset\mathbf{1}\,,
\end{eqnarray}
where $\mathbf{1}$ denotes the invariant singlet of the finite group $\Gamma_{finite}$. When we restrict to transformation $\gamma$ of the group $\Gamma_{normal}$ in eq.~\eqref{eq:Ytras-auto}, we obtain:
\begin{equation}
Y_{I_1...I_n}(\gamma\tau)=j(\gamma,\tau)^{k_Y(n)}~Y_{I_1...I_n}(\tau)\,,\quad \gamma\in \Gamma_{normal}\,.
\end{equation}
Thus the function
\begin{equation}
\Psi(g)=j(g,\tau_0)^{-k_Y(n)}Y_{I_1...I_n}(g\tau_0)
\end{equation}
is an automorphic form for $G$, $K$ and $\Gamma_{normal}$. The automorphic form $\Psi(g)$ is a smooth complex function and it satisfies
\begin{eqnarray}
\nonumber \Psi(\gamma \textsf{g})&=&\Psi(\textsf{g})\,,~\quad ~\gamma\in \Gamma_{normal}~~\,,\\
\Psi(gh)&=&j(h,\tau_0)^{-1}~\Psi(\textsf{g})\,,\quad h\in K\,.
\end{eqnarray}
Moreover, the automorphic form $\Psi(g)$ is required to be an eigenfunction of all the Casimir operators of $G$ and it should satisfy moderate growth condition~\cite{Borel,Borel2,klingen1990introductory}. The symplectic modular invariant theory in section~\ref{subsec:symplectic-mod-theory} corresponds to the following choice
\begin{equation}
G=Sp(2g,\mathbb{R})\,,~K=Sp(2g,\mathbb{R})\cap O(2g,\mathbb{R})\,,~\tau_0=i \mathds{1}_g,~\Gamma=Sp(2g,\mathbb{Z})\,,~\Gamma_{normal}=\Gamma_g(n)\,,
\end{equation}
and the automorphy factor $j(\gamma,\tau)=\det(C\tau+D)$. We can also choose other non-compact Hermitian Symmetric Space as multi-dimensional moduli space which are summarized in table~\ref{tab:HSS-coset}, and the symplectic modular group is exactly the type $\mathbf{III}_{g}$.

\begin{table}[th!]
\centering
\begin{tabular}{|c|c|c|c|}\hline\hline
Type & Group $G$ & Compact subgroup $K$ & $\dim_{\mathbb{C}}G/K$  \\ \hline
$\mathbf{I}_{m,n}$ & $U(m,n)$ & $U(m)\times U(n)$ & $mn$   \\ \hline
$\mathbf{II}_{m}$ & $SO^*(2m)$ & $U(m)$ & $\frac{1}{2}m(m-1)$    \\ \hline
 $\mathbf{III}_{m}$ & $Sp(2m)$ & $U(m)$ & $\frac{1}{2}m(m+1)$    \\ \hline
 $\mathbf{IV}_{m}$ & $SO(m,2)$ & $SO(m)\times SO(2)$ & $m$    \\ \hline
$\mathbf{V}$ &  $E_{6,-14}$ & $SO(10)\times SO(2)$ & $16$  \\ \hline
$\mathbf{VI}$ &  $E_{7,-25}$ & $E_6\times U(1)$ & $27$   \\ \hline \hline
\end{tabular}
\caption{\label{tab:HSS-coset} Irreducible hermitian symmetric manifolds of noncompact type and their complex dimension~\cite{Ding:2020zxw}.}
\end{table}

\section{Fermion mass hierarchies and texture zeros from modular symmetry}

The modular symmetry is quite predictive by minimizing the symmetry breaking sector, nevertheless it does not yet provide a convincing explanation of the charged lepton masses. Usually the mass hierarchy is achieved by hand
by introducing one parameter for each charged lepton species. The values of the free parameters are adjusted to reproduce the charged lepton masses, which are not predicted, but just fitted as in the Standard Model. In this section, we shall review how the fermion mass hierarchies can be naturally reproduced in the framework of modular symmetry.

\subsection{Weighton mechanism}

Froggatt-Nielsen (FN) mechanism is a well-known approach to understand the quark and lepton mass hierarchies~\cite{Froggatt:1978nt}.
The FN mechanism assumes an additional $U(1)_{FN}$ flavor symmetry under which the quarks and leptons carry various charges. The $U(1)_{FN}$ flavor symmetry is spontaneously broken by the vacuum expectation value (VEV) of a new ``flavon'' field $\theta$, where $\theta$ is a neutral scalar under the SM but carries one unit of $U(1)_{FN}$ charge. In the FN mechanism, the fermion Yukawa couplings (except the top quark Yukawa coupling) are forbidden at the renormalisable level due to the $U(1)_{FN}$ symmetry. The small effective Yukawa couplings then originate from non-renormalisable contact operators where the FN charges of fermions are compensated by powers of $\theta$, leading to suppression by powers of the small ratio $\langle \theta \rangle /M_{FN}$, where $\langle \theta \rangle$ and $M$ denote the VEV of $\theta$ and the cut-off scale of the contact interaction respectively.

The fermion mass hierarchies can be naturally reproduced in the framework of modular symmetry through the so called weighton mechanism~\cite{King:2020qaj}. The mechanism is analogous to the FN mechanism, but without requiring any Abelian symmetry to be introduced, nor any SM singlet flavon to break it. The modular weights of fermion fields play the role of FN charges, and a SM singlet field $\phi$ with $k_{\phi}=1$ (i.e. weight $-1$). The field $\phi$ plays the role of a flavon and it is called a ``weighton''. This mechanism can be illustrated with the benchmark modular invariant model for leptons in~\cite{Feruglio:2017spp}. The representation and weight assignments matter fields are assumed to be~\cite{Feruglio:2017spp}:
\begin{equation}
L\sim\left(\mathbf{3}, 1\right)\,,~~e^{c}\sim\left(\mathbf{1}, 1\right)\,,~~\mu^{c}\sim\left(\mathbf{1}'', 1\right)\,,~~\tau^c\sim\left(\mathbf{1}', 1\right)\,,~~N^c\sim\left(\mathbf{3}, 1\right)\,,~~H_{u,d}\sim\left(\mathbf{1}, 0\right)\,,
\end{equation}
where the first numbers in the parentheses denote the transformation under $A_4$ modular symmetry and the second numbers are the modular weight. Thus the superpotential for the charged lepton Yukawa couplings read as
\begin{eqnarray}
\nonumber\mathcal{W}_e
&=&\alpha e^c (LY^{(2)}_{\mathbf{3}})_{\mathbf{1}}H_d
+\beta \mu^c (LY^{(2)}_{\mathbf{3}})_{\mathbf{1}'}H_d + \gamma \tau^c (LY^{(2)}_{\mathbf{3}})_{\mathbf{1}''}H_d\\
\nonumber
&=&\alpha e^c(L_1 Y_1+L_2 Y_3+L_3 Y_2)H_d+\beta \mu^c (L_3 Y_3+L_1 Y_2+L_2 Y_1)H_d\\
&&~+\gamma \tau^c (L_2 Y_2+L_3 Y_1+L_1 Y_3)H_d\,,
\label{eq:We-Ferugllio}
\end{eqnarray}
where $Y_{1,2,3}(\tau)$ are the three independent level 3 and weight 2 modular forms given in Eq.~\eqref{eq:MF-level-3-w2}. Consequently the charged lepton Yukawa coupling matrix takes the following form:
\begin{equation}
\mathcal{Y}_e =\begin{pmatrix}
\alpha\,Y_1 ~&~\alpha\,Y_3 ~&~\alpha\,Y_2 \\
\beta\, Y_2 ~&~\beta\, Y_1  ~&~\beta\, Y_3 \\
\gamma\, Y_3~&~ \gamma\, Y_2~&~\gamma\, Y_1
\end{pmatrix}\,.
\label{eq:ye-Ferugllio}
\end{equation}
For $\tau=i\infty$ the modular triplet has a simple vaccum alignment
\begin{equation}
\left(Y_1, Y_2, Y_3\right)|_{\tau=i\infty}=\left(1, 0, 0\right)\,,
\end{equation}
which can be seen from Eq.~\eqref{eq:Y1-Y2-Y3-Feruglio}. Hence the charged lepton Yukawa is diagonal in the limit $\tau=i\infty$,
\begin{equation}
\mathcal{Y}_e|_{\tau=i\infty}=\begin{pmatrix}
\alpha  ~&~ 0 ~&~ 0 \\
0 ~&~\beta  ~&~ 0 \\
0~&~ 0~&~\gamma
\end{pmatrix}\,,
\end{equation}
which implies
\begin{equation}
\label{eq:me-mu-mtau-ratio}m_e:m_{\mu}:m_{\tau}=\alpha :\beta :\gamma\,.
\end{equation}
Hierarchial values of the three couplings $\alpha$, $\beta$, $\gamma$ are necessary to accommodate the observed charged lepton masses. In detailed numerical analysis, the best fit values of modulus and the free couplings are found to be~\cite{Ding:2019zxk}
\begin{equation}
\langle\tau\rangle=0.0386+2.23i\,,~~\beta/\alpha=207.908\,,~~\gamma/\alpha=3673.38\,,
\end{equation}
which is in agreement with the estimate of Eq.~\eqref{eq:me-mu-mtau-ratio}.

This model can be recast in natural form by introducing a single weighton  $\phi$ which is a SM and $A_4$ singlet with $k_{\phi}=1$. The representation assignments of lepton fields are kept intact while their modular weights are changed as follows~\cite{King:2020qaj},
\begin{eqnarray}
\nonumber&& L\sim\left(\mathbf{3}, 1\right)\,,~~e^{c}\sim\left(\mathbf{1}, -3\right)\,,~~\mu^{c}\sim\left(\mathbf{1}'', -1\right)\,,~~\tau^c\sim\left(\mathbf{1}', 0\right)\,,\\
&& N^c\sim\left(\mathbf{3}, 1\right)\,,~~H_{u,d}\sim\left(\mathbf{1}, 0\right)\,,~~\phi\sim\left(\mathbf{1}, 1\right)\,.
\end{eqnarray}
Then the superpotential for the charged lepton masses takes the form
\begin{eqnarray}
\mathcal{W}_e
&=&\alpha_e \left(\frac{\phi}{\Lambda}\right)^4 e^c (LY^{(2)}_{\mathbf{3}})_{\mathbf{1}}H_d
+\beta_e  \left(\frac{\phi}{\Lambda}\right)^2 \mu^c (LY^{(2)}_{\mathbf{3}})_{\mathbf{1}'}H_d + \gamma_e  \frac{\phi}{\Lambda} \tau^c (LY^{(2)}_{\mathbf{3}})_{\mathbf{1}''}H_d\,,
\end{eqnarray}
which gives rise to a charged lepton Yukawa coupling matrix similar to Eq.~\eqref{eq:ye-Ferugllio}, except that it involves
powers of $\phi$ controlling the hierarchies,
\begin{equation}
\mathcal{Y}_e =\begin{pmatrix}
\alpha_e  \tilde{\phi}^4\,Y_1 ~&~\alpha_e \tilde{\phi}^4 \,Y_3 ~&~\alpha_e  \tilde{\phi}^4 \,Y_2 \\
\beta_e  \tilde{\phi}^2 Y_2 ~&~\beta_e  \tilde{\phi}^2 Y_1  ~&~\beta_e  \tilde{\phi}^2 Y_3 \\
 \gamma_e  \tilde{\phi} Y_3 ~&~ \gamma_e  \tilde{\phi} Y_2~&~\gamma_e  \tilde{\phi} Y_1
\end{pmatrix},~~~\tilde{\phi}\equiv\frac{\langle\phi\rangle}{\Lambda}\,,
 \label{eq:ye-King}
\end{equation}
where $\langle\phi\rangle$ denote the VEV of the weighton $\phi$ and $\Lambda$ is a dimensionful cut-off flavor scale. Similar to the usual driving field mechanism familiar from flavon models~\cite{Altarelli:2005yx}, the weighton  may be driven by a leading order superpotential term\footnote{A $U(1)_R$ symmetry in necessary in the driving field mechanism, the driving superfield $\chi$ carry two unit $R$ charges, the weighton $\phi$ and Higgs superfields have $R=0$ and the matter superfields have $R=1$. }
\begin{equation}
\mathcal{W}_{driv}= \chi\left(Y^{(4)}_{\mathbf{1}}\frac{\phi^4}{M_{fl}^2} -M^2\right),
\label{eq:driving-King}
\end{equation}
where $\chi$ is an $A_4$ singlet driving superfield with zero modular weight,
while $M$ is a free dimensionful mass scale. The F-flatness condition
$F_{\chi}=\frac{\partial W_{driv}}{\partial \chi}=0$ applied to Eq.~\eqref{eq:driving-King} then drives a weighton vev, $\langle \phi \rangle = (M^2 M^2_{fl}/Y^{(4)}_{\mathbf{1}})^{1/4}$. In the limit $\tau\rightarrow i\infty$, the above Yukawa coupling matrix $\mathcal{Y}_e$ is diagonal and the charged lepton mass ratios are
\begin{equation}
\label{eq:me-mu-mtau-ratio-2}m_e:m_{\mu}:m_{\tau}=\alpha_e\tilde{\phi}^4:\beta_e \tilde{\phi}^2 :\gamma_e\tilde{\phi}\,.
\end{equation}
It is remarkable that the electron, muon and tau masses are suppressed by $\tilde{\phi}^4$, $\tilde{\phi}^2$ and $\tilde{\phi}$ respectively. Assuming order one coefficients $\alpha_e,\beta_e, \gamma_e\sim\mathcal{O}(1)$, the charged lepton mass hierarchies can be naturally produced for the small parameter $\tilde{\phi}\approx1/15$. Furthermore, there will be additional terms corresponding to higher weight modular forms, $Y^{(4)}_{\mathbf{3}}$,
compensated by extra powers of weighton fields $\phi$, which will give corrections to the charged lepton superpotential~\cite{King:2020qaj},
\begin{equation}
\Delta \mathcal{W}_e=\alpha'_e \left(\frac{\phi}{\Lambda}\right)^6 e^c (LY^{(4)}_{\mathbf{3}})_{\mathbf{1}}H_d
+\beta'_e \mu^c \left(\frac{\phi}{\Lambda}\right)^4 (LY^{(4)}_{\mathbf{3}})_{\mathbf{1}'}H_d + \gamma'_e \tau^c \left(\frac{\phi}{\Lambda}\right)^3 (LY^{(4)}_{\mathbf{3}})_{\mathbf{1}''}H_d\,,
\end{equation}
which are suppressed by $\tilde{\phi}^2$ with respect to the leading superpotential in Eq.~\eqref{eq:ye-King}. The weighton mechanism can also be applied to explain the quark mass hierarchies~\cite{King:2020qaj,Kuranaga:2021ujd}.

\subsection{Mass hierarchies, large lepton mixing and residual modular symmetry }

As shown in section~\ref{subsec:fixed-points-SL2Z}, the modular group $SL(2, \mathds{Z})$ only has three independent fixed points $\tau_S=i$, $\tau_{ST}=e^{2\pi i/3}=\omega$ and $\tau_T=i\infty$ which are left invariant under the action of $S$, $ST$ and $T$. Since certain residual symmetry listed in table~\ref{tab:stabilizer-modular} is preserved at the fixed points, the modular forms has definite alignments at the modular fixed points. For example, the weight 2 and level 3 modular forms $Y_{1,2,3}(\tau)$ in Eq.~\eqref{eq:Y1-Y2-Y3-Feruglio} in the basis of Eq.~\eqref{eq:rho-S-T-A4} have the following alignments at fixed points:
\begin{eqnarray}
\nonumber \left(Y_1, Y_2, Y_3\right)\Big|_{\tau=i\infty}&=&\left(1, 0, 0\right)\,,\\
\nonumber\left(Y_1, Y_2, Y_3\right)\Big|_{\tau=i}
&=&Y_1(i)\left(1, 1-\sqrt{3}, -2+\sqrt{3}\right)\,, \\
\left(Y_1, Y_2, Y_3\right)\Big|_{\tau=\omega}&=&Y_1(\omega)\left(1, \omega, -\omega^2/2\right)\,,
\end{eqnarray}
with $Y_1(i)\simeq1.023$ and $Y_1(\omega)\simeq0.949$. As a consequence, the residual modular symmetry group would enforce some entries of the fermion mass matrix to be vanishing in proper representation basis of the finite modular group. As the value of the modulus $\tau$ moves away from the fixed points, these entries will generically become non-zero. The magnitudes of such residual-symmetry-breaking entries will be controlled by the size of the departure of $\tau$ from the fixed points and by the field transformation properties under the residual symmetry group~\cite{Okada:2020ukr,Feruglio:2021dte,Novichkov:2021evw,Feruglio:2022kea,Feruglio:2023mii}. This is shown in what follows.

After electroweak symmetry breaking, the fermion mass term can be generally written as\footnote{For Majorana fermions, $\psi$ and $\psi^c$ are identical and this implies $k^c=k$ and $\rho^c=\rho$. }
\begin{equation}
\mathcal{L}_{m}=\psi^c_iM_{ij}(\tau)\psi_j\,.
\end{equation}
The matter superfields $\psi^{(c)}_i$ can be arranged into a column vector $\psi^{(c)}$ transforming under the finite modular symmetry
\begin{equation}
\psi\xrightarrow{\gamma}\, (c \tau + d)^{-k} \rho(\gamma) \,\psi \,, ~~~ \psi^c \,\xrightarrow{\gamma}\, (c \tau + d)^{-k^c} \rho^c(\gamma)\, \psi^c \,,
\end{equation}
where $\rho(\gamma)$ and $\rho^c(\gamma)$ are representations of the finite modular group $\Gamma_N$ or $\Gamma'_N$. Hence each entry $M(\tau)_{ij}$ is a modular form of level $N$ and weight $K \equiv k+k^c$, and modular invariance requires the mass matrix $M(\tau)$ to transform as
\begin{equation}
\label{eq:mass-matrix-mod-trans}
M(\tau)\, \xrightarrow{\gamma}\, M(\gamma \tau) = (c \tau + d)^K \rho^c(\gamma)^{*} M(\tau) \rho(\gamma)^{\dagger} \,.
\end{equation}
One can use this transformation rule to constrain the form of the mass matrix $M(\tau)$ by taking $\tau$ to be close to the modular fixed points and setting $\gamma$ to the residual symmetry generator.

\begin{itemize}

\item{$\tau\rightarrow i\infty$}

In modular symmetry models with $\tau$ in the vicinity of $i\infty$, it is convenient to choose the $T$-diagonal basis $\rho^{(c)}(T) =\text{diag} (\rho^{(c)}_i)$. The multiplication rule $T^{N}=1$ in Eq.~(\ref{eq:GammaN-rules}, \ref{eq:GammaNp-rules}) implies generally $(\rho^c_i \rho_j)^{*} = \zeta^{p_{ij}}$ with $\zeta=e^{2\pi i/N}$ and $0 \leq p_{ij} < N$. Then Eq.~\eqref{eq:mass-matrix-mod-trans} implies that the entries $M_{ij}$ can be expanded as~\cite{Novichkov:2021evw}:
\begin{equation}
  \label{eq:series-T}
  M_{ij}(\tau) = a_0\, q^{p_{ij}}_N + a_1\, q^{p_{ij}+N}_N + a_2\, q^{p_{ij}+2N} + \ldots \,,~~~q_N\equiv \exp \left( 2 \pi i \tau / N \right)~~q_N=e^{2\pi i\tau/N}
\end{equation}
in the vicinity of $i\infty$. Hence $M_{ij}$ is of order $\mathcal{O}(\epsilon^{p_{ij}})$ with $\epsilon=|q_N|=e^{-2\pi\texttt{Im}(\tau)/N}$.

\item{$\tau\rightarrow\omega$ }

For the analysis of models where $\tau$ is in the vicinity of $\tau_{ST}=\omega$,  it is convenient to switch to the basis where the product $ST$ is represented by a diagonal matrix, i.e.,
$\rho^{(c)}(ST) =\text{diag} (\rho^{(c)}_i)$, where the diagonal elements are power of $\omega$ because of $(ST)^3=1$. Depending on the transformation matrices $\rho^{(c)}(ST)$ and the modular weights $k$ and $k^c$, it turns out that the entries $M_{ij}\sim \mathcal{O}(\epsilon^{l_{ij}})$ for $\omega^{K}\rho^{c}_i\rho_j=\omega^{l_{ij}}$ with $l_{ij}=0, 1, 2$, where $\epsilon=\left|\frac{\tau-\omega}{\tau-\omega^2}\right|$~\cite{Novichkov:2021evw}.

\item{$\tau\rightarrow i$}

In this case, it is convenient to switch to the basis where the modular generator $S$ is represented by a diagonal matrix with $\rho^{(c)}(S) =\text{diag}(\rho^{(c)}_i)$, where the diagonal elements are powers of $i$ because of $S^4=1$. It is found that the entries $M_{ij}\sim \mathcal{O}(\epsilon^{n_{ij}})$ for $i^{K}\rho^{c}_i\rho_j=(-1)^{n_{ij}}$ with $n_{ij}=0, 1$, where $\epsilon=\left|\frac{\tau-i}{\tau+i}\right|$~\cite{Novichkov:2021evw}.

\end{itemize}

In short, each entry $M_{ij}$ the fermion mass matrix is of order $\mathcal{O}(\epsilon^l)$, where $\epsilon$ parameterises the
deviation of $\tau$ from the fixed points, and the power $l$ can be extracted from products of factors which correspond to representations
of the residual symmetry group. It is notable that the value of $l$ only depends on the representations of the matter fields. The fermion mass matrices in the vicinity of modular fixed points have been analyzed for the homogeneous finite modular groups $\Gamma'_N$ of levels $N\leq 5$, it is found that hierarchical spectra can be obtained for a small list of representation pairs~\cite{Novichkov:2021evw}, the most promising of which are collected in table~\ref{tab:good_patterns}. Here the coupling constants in modular flavor models are assumed to be of the same order of magnitude, and the modular forms should be properly normalized~\cite{Petcov:2023fwh}. Furthermore, it is claimed that the modular invariant models exhibit a universal behavior around the fixed points, and the scaling of fermion masses and mixing parameters is independent of the details of the theory~\cite{Feruglio:2022kea,Feruglio:2023mii}.

\begin{table}[hptb!]
\small
\centering
\renewcommand{\arraystretch}{1.5}
\begin{tabular}{|cccll|}\hline\hline
$N$ ~&~ $\Gamma'_N \;(\text{or}~ \Gamma_N)$ ~&~ \texttt{Mass ratios} ~&~ \texttt{Modulus} $\tau$ ~&~ \texttt{Viable} $\mathbf{r} \otimes \mathbf{r}^c$ \\ \hline
2 ~&~ $S_3$ ~&~ $(1,\epsilon,\epsilon^2)$
~&~ $\tau \simeq \omega$
~&~ $[\mathbf{2}\oplus\mathbf{1}^{(\prime)}] \otimes [\mathbf{1}\oplus\mathbf{1}^{(\prime)}\oplus\mathbf{1}']$
\\[2mm]\hline

\multirow{3}{*}{3}~&~\multirow{3}{*}{$\Gamma'_3=T'\;(~\text{or}~ A_4)$} ~&~ \multirow{3}{*}{$(1,\epsilon,\epsilon^2)$} ~&~ $\tau \simeq \omega$  ~&~ $[\mathbf{1}_a\oplus\mathbf{1}_a\oplus\mathbf{1}_a'] \otimes [\mathbf{1}_b\oplus\mathbf{1}_b\oplus\mathbf{1}_b'']$
\\[1mm]
~&~ ~&~ ~&~ \multirow{2}{*}{$\tau \simeq i \infty$} ~&~$[\mathbf{1}_a\oplus\mathbf{1}_a\oplus\mathbf{1}_a'] \otimes [\mathbf{1}_b\oplus\mathbf{1}_b\oplus\mathbf{1}_b'']$\\[-2mm]
~&~ ~&~ ~&~ ~&~ $ \text{with}~\mathbf{1}_a \neq (\mathbf{1}_b)^*$\\[2mm] \hline

\multirow{6}{*}{4} ~&~ \multirow{6}{*}{$S_4'\;(\;\text{or}~ S_4)$} ~&~ \multirow{2}{*}{$(1,\epsilon,\epsilon^2)$}
~&~ \multirow{2}{*}{$\tau \simeq \omega$} ~&~ $[\mathbf{3}_a \text{, or }
\mathbf{2}\oplus\mathbf{1}^{(\prime)} \text{, or }
\mathbf{\hat{2}}\oplus\mathbf{\hat{1}}^{(\prime)}]$
\\[-2mm]
~&~ ~&~ ~&~ ~&~~ $\otimes [\mathbf{1}_b\oplus\mathbf{1}_b\oplus\mathbf{1}_b']$ \\[1mm]
~&~ ~&~ \multirow{4}{*}{$(1,\epsilon,\epsilon^3)$} ~&~ \multirow{4}{*}{$\tau \simeq i \infty$} ~&~ $\mathbf{3}\hphantom{'} \otimes [\mathbf{2}\oplus\mathbf{1} \text{, or }
\mathbf{1}\oplus\mathbf{1}\oplus\mathbf{1}']$,
\\[-1mm]
~&~ ~&~ ~&~ ~&~~$\mathbf{3}' \otimes [\mathbf{2}\oplus\mathbf{1}' \text{, or }
\mathbf{1}\oplus\mathbf{1}'\oplus\mathbf{1}'],$\\[-1mm]
~&~ ~&~ ~&~ ~&~~$\mathbf{\hat{3}}' \otimes [\mathbf{\hat{2}}\oplus\mathbf{\hat{1}} \text{, or }
\mathbf{\hat{1}}\oplus\mathbf{\hat{1}}\oplus\mathbf{\hat{1}}'],$ \\[-1mm]

~&~ ~&~ ~&~ ~&~ $\mathbf{\hat{3}}\hphantom{'} \otimes [\mathbf{\hat{2}}\oplus\mathbf{\hat{1}'} \text{, or }
\mathbf{\hat{1}}\oplus\mathbf{\hat{1}}'\oplus\mathbf{\hat{1}}']$
\\[2mm] \hline

5 ~&~ $A_5'\;(\;\text{or}~ A_5)$ ~&~ $(1,\epsilon,\epsilon^4)$ ~&~ $\tau \simeq i \infty$ ~~&~ $\mathbf{3}\otimes\mathbf{3}'$ \\ \hline\hline
\end{tabular}
\caption{\label{tab:good_patterns} Hierarchical mass patterns which can be realized in the vicinity of symmetric points~\cite{Novichkov:2021evw}. These patterns are unaffected by the exchange \(\mathbf{r}\leftrightarrow \mathbf{r}^c\) and may only be viable for certain weights, where $\mathbf{r}$ and $\mathbf{r}^c$ are representation of $\psi$ and $\psi^c$ respectively. Subscripts run over irreps of a certain dimension, and \(\mathbf{1}_a'''=\mathbf{1}_a\) for \(N=3\), while \(\mathbf{1}_a''=\mathbf{1}_a\) for \(N=4\). Primes in parenthesis are uncorrelated. Notice the representations without a hat have a direct correspondence with $\Gamma_N$ representations, while the hatted representations are specific to $\Gamma'_N$.  }
\end{table}

The three generations of left-handed lepton fields are usually assigned to a triplet of flavor symmetry in order to produce the large lepton mixing angles. In the symmetry limit with $\tau$ at the fixed points, the charged lepton and neutrino sectors would share the same residual symmetry so that the lepton mixing matrix would be block diagonal~\cite{Ding:2019gof} if the lepton masses are non-degenerate. Hence the triplet assignment of left-handed leptons is disfavored in the vicinity of modular fixed points. There are only four cases in which the deviation of $\tau$ from the fixed point can generate hierarchical charged-lepton masses together with the viable lepton mixing~\cite{Reyimuaji:2018xvs,Novichkov:2021evw},
\begin{enumerate}[label=(\roman*)]
\item $L \leadsto 1 \oplus 1 \oplus 1$, $E^c \leadsto 1 \oplus r$, where $1$ is some real singlet of the flavour symmetry, and $r$ is some (possibly reducible) representation such that $r \not\supset 1$;
\item $L\leadsto\mathbf{1} \oplus \mathbf{1} \oplus \mathbf{1}^{*}$, $E^c \leadsto \mathbf{1}^* \oplus r$, where $\mathbf{1}$ is some complex singlet of the flavour symmetry, $\mathbf{1}^{*}$ is its conjugate, and $r$ is some (possibly reducible) representation such that $r\not\supset \mathbf{1}, \mathbf{1}^{*}$;
\item all charged-lepton masses vanish in the symmetric limit, i.e.~the corresponding hierarchical pattern involves only positive powers of $\epsilon$, e.g. $(\epsilon, \epsilon^2, \epsilon^3)$. The small tau lepton mass in comparison to the electroweak scale could be explained in this case, and the three left-handed leptons can be a triplet of the modular group;
\item all light neutrino masses vanish in the symmetric limit, i.e.~$L$ decomposes into three (possibly identical) complex singlets none of which are conjugated to each other.
\end{enumerate}

Motivated by the model independent and universal behaviors around the modular fixed points, a concrete model with the $S'_4$ modular symmetry has been constructed, and the light neutrino masses are generated by the type-I seesaw mechanism~\cite{Novichkov:2021evw}. The modular transformations of the lepton fields are assumed to be
\begin{equation}
L_1, L_2 \sim (\mathbf{\hat{1}}, 2),\quad L_3 \sim (\mathbf{\hat{1}'}, 2),\quad E^c \sim (\mathbf{\hat{3}}, 4),\quad N^c \sim (\mathbf{3}', 1)\,,
\end{equation}
where the first numbers in the parenthesis denote the transformations under $S'_4$ and the second numbers are the modular weights. The corresponding superpotential for lepton mass reads~\cite{Novichkov:2021evw}:
\begin{equation}
\label{eq:S4_omega_W}
\begin{aligned}
\mathcal{W} &= \left[
\alpha_1 \left( Y^{(4,6)}_{\mathbf{3'}, 1} E^c L_1 \right)_{\mathbf{1}} +
\alpha_2 \left( Y^{(4,6)}_{\mathbf{3'}, 2} E^c L_1 \right)_{\mathbf{1}} \right. \\
&+ \left.\alpha_3 \left( Y^{(4,6)}_{\mathbf{3'}, 1} E^c L_2 \right)_{\mathbf{1}} +\alpha_4 \left( Y^{(4,6)}_{\mathbf{3'}, 2} E^c L_2 \right)_{\mathbf{1}} +\alpha_5 \left( Y^{(4,6)}_{\mathbf{3}} E^c L_3 \right)_{\mathbf{1}} \right] H_d \\
&+\left[g_1 \left( Y^{(4,3)}_{\mathbf{\hat{3}}} N^c L_1 \right)_{\mathbf{1}} +g_2 \left( Y^{(4,3)}_{\mathbf{\hat{3}}} N^c L_2 \right)_{\mathbf{1}}+g_3 \left( Y^{(4,3)}_{\mathbf{\hat{3}'}} N^c L_3 \right)_{\mathbf{1}} \right] H_u \\
&+\Lambda \left( Y^{(4,2)}_{\mathbf{2}} (N^c)^2 \right)_{\mathbf{1}} \,.
\end{aligned}
\end{equation}
Since $L_1$ and $L_2$ are indistinguishable, one of the constants $\alpha_i$, with $i = 1, \dots, 4$, are effectively not an independent parameter and can be set to zero by a suitable rotation without loss of generality. We choose to set $\alpha_2 = 0$. All the coupling constants are enforced to be real by the gCP symmetry, and consequently the model depends on 9 effective parameters including $\text{Re}\,\tau$ and $\text{Im}\,\tau$. The best fit value of the modulus is found to be $\tau\simeq-0.496+0.877i$ which is close to the fixed point $\omega$, the charged lepton mass hierarchies are produced without fine-tuning~\cite{Novichkov:2021evw}, and the leptonic Dirac CP violation phase is predicted to be around $\pi$. See~\cite{Novichkov:2021evw}
for more numerical results. The closeness of the modulus to the modular symmetry fixed points can also help to explain the quark mass hierarchies~\cite{Petcov:2023vws,deMedeirosVarzielas:2023crv}. Furthermore, it is shown that the modular symmetry models exhibit a universal behavior in the vicinity of fixed points, independent of details of the models such as the modular weights and representation assignments of matter field under the finite modular groups~\cite{Feruglio:2022koo,Feruglio:2023mii}.

\subsection{Texture zeros and modular symmetry}

Texture zero is an interesting attempt to understand the patterns of fermion masses and flavor mixing, it assumes that some entries of the fermion mass matrices are vanishing in order to reduce the number of free parameters~\cite{Fritzsch:1977za,Weinberg:1977hb,Wilczek:1977uh}.  A typical example is the Fritzsch-type quark mass matrices\cite{Fritzsch:1977vd,Fritzsch:1979zq}:
\begin{equation}
M_u=\begin{pmatrix}
0  ~& A_u  ~& 0 \\
A^{*}_u  ~& 0 ~& B_u \\
0  ~& B^{*}_u  ~& C_u
\end{pmatrix},~~~~M_d=\begin{pmatrix}
0  ~& A_d  ~& 0 \\
A^{*}_d  ~& 0 ~& B_d \\
0  ~& B^{*}_d  ~& C_d
\end{pmatrix}\,.
\end{equation}
It relates the Cabibbo angle $\theta_{\rm C}$ to the ratio between the down and strange quark masses via
\begin{equation}
\tan\theta_{\rm C}\approx\sqrt{\frac{m_d}{m_s}}\,,
\end{equation}
which is the Gatto-Sartori relation~\cite{Gatto:1968ss}. As regard the texture zero of lepton sector, in the basis of diagonal charged lepton mass matrix, the Majorana neutrino mass matrix can have at most two zero entries and only seven out of the total fifteen two-zero textures are
compatible with current experimental data at $3\sigma$ level~\cite{Frampton:2002yf,Xing:2002ta,Fritzsch:2011qv,Meloni:2014yea},
\begin{eqnarray}
\nonumber&&\mathbf{A}_1: \begin{pmatrix}
0 &0 & \times \\
0 & \times & \times \\
\times & \times & \times
\end{pmatrix},~~~\mathbf{A}_2:\begin{pmatrix}
0 & \times & 0 \\
\times & \times & \times \\
0  & \times & \times
\end{pmatrix},~~~\mathbf{B}_1: \begin{pmatrix}
\times & \times & 0 \\
\times & 0 & \times \\
0 & \times & \times
\end{pmatrix},~~~\mathbf{B}_2: \begin{pmatrix}
\times & 0 & \times \\
0 & \times & \times\\
\times & \times & 0
\end{pmatrix}\,,\\
&&\mathbf{B}_3: \begin{pmatrix}
\times & 0  & \times\\
0 & 0 & \times \\
\times & \times & \times
\end{pmatrix},~~~\mathbf{B}_4: \begin{pmatrix}
\times & \times & 0 \\
\times & \times & \times \\
0  & \times & 0
\end{pmatrix},~~~\mathbf{C}: \begin{pmatrix}
\times & \times & \times \\
\times & 0 & \times \\
\times & \times & 0
\end{pmatrix}\,,
\end{eqnarray}
where $\times$ denotes an arbitrary complex number. The phenomenology of texture zeros in both quark and lepton sectors have been widely studied in literature, see Refs.~\cite{Fritzsch:1999ee,Gupta:2012fsl} for reviews. Systematical and complete analyses of all possible texture zeros have been carried out for both quark~\cite{Ludl:2015lta} and lepton mass matrices~\cite{Ludl:2014axa}.
It is found that the predictivity of pure texture zero models is quite weak and the predictive mass matrices need relations among the non-zero matrix elements. The most straightforward way to impose vanishing Yukawa couplings is by enforcing them with Abelian flavour symmetries~\cite{Grimus:2004hf,GonzalezFelipe:2014zjk,Camara:2020efq}. In such case, the non-zero entries of the fermion mass matrices are uncorrelated since the three generations of matter fields are not linked by the Abelian symmetry group.

One can freely assign the modular weights of matter fields in the bottom-up approach of modular flavor symmetry, thus the modular symmetry has the merits of both abelian flavor symmetry and discrete non-abelian flavor symmetry. Moreover, it is known that modular forms in certain representations of $\Gamma'_N$ (or $\Gamma_N$) are absent at lower weights. In addition, the even weight modular forms at level $N$ are in the irreducible representations of $\Gamma_N$~\cite{Feruglio:2017spp} while the odd weight modular forms are in the irreducible representations specific to $\Gamma'_N$~\cite{Liu:2019khw}. The structure of modular forms makes the modular symmetry to be a natural framework producing texture zeros of fermion mass matrices. Moreover, the nonzero entries are related by modular symmetry so that modular symmetry can considerably improve the predictive power of texture zero. The possible texture zeros of quark and lepton mass matrices have been systematically analyzed in the context of $T'$ modular symmetry~\cite{Lu:2019vgm,Ding:2022aoe}. The possible texture zero patterns of the charged fermions mass matrices and neutrino mass matrix in the $T'$ modular symmetry are summarized in table~\ref{tab:classes_charged_lepton_mass_matrix} and table~\ref{tab:classes_Majorana_neutrino_mass_matrix} respectively.

It was shown that the two-zero textures of the Majorana neutrino mass matrix can be reproduced from the $A_{4}$ modular group~\cite{Zhang:2019ngf}, where all matter fields were assigned to $A_4$ singlets and thus correlations among nonzero entries of lepton mass matrices are lost.  Moreover, the vanishing entries of fermion mass matrices can naturally appear~\cite{Ding:2019gof} if the modulus $\tau$ is at certain modular fixed point listed in table~\ref{tab:stabilizer-modular}. The nearest neighbor interaction form of the quark mass matrices can be obtained at the fixed point $\tau=\omega$ in the $A_4$ modular flavor symmetry by taking account multi-Higgs fields~\cite{Kikuchi:2022svo}.

\begin{table}[t!]
 \centering
\begin{tabular}{|l|} \hline\hline
$\mathcal{C}_{1}^{(1)}:\left( \begin{array}{ccc} 0 & \times & \times \\
\times & \times & \times \\
\times & \times & \times \end{array} \right)$ \\ \hline
$\mathcal{C}_{2}^{(1)}:\left( \begin{array}{ccc} 0 & \times & \times \\
\times & 0 & \times \\
\times & \times & \times \end{array} \right)\,,~~
\mathcal{C}_{2}^{(2)}:\left( \begin{array}{ccc} \times & \times & \times \\
\times & \times & \times \\
0 & 0 & \times \end{array} \right)\,,~~
\mathcal{C}_{2}^{(3)}:\left( \begin{array}{ccc} \times & \times & 0 \\
\times & \times & 0 \\
\times & \times & \times \end{array} \right)$ \\ \hline
$\mathcal{C}_{3}^{(1)}: \left( \begin{array}{ccc} \times & \times & 0 \\
\times & \times & \times \\
0 & 0 & \times \end{array} \right)\,,~~
\mathcal{C}_{3}^{(2)}:\left( \begin{array}{ccc} \times & 0 & \times \\
\times & 0 & \times \\
0 & \times & \times \end{array} \right)\,,~~
\mathcal{C}_{3}^{(3)}:\left( \begin{array}{ccc} 0 & \times & \times \\
\times & 0 & \times \\
\times & \times & 0 \end{array} \right)$ \\ \hline
$\mathcal{C}_{4}^{(1)}:\left( \begin{array}{ccc} \times & \times & 0 \\
\times & \times & 0 \\
0 & 0 & \times \end{array} \right)\,,~~
\mathcal{C}_{4}^{(2)}:\left( \begin{array}{ccc} 0 & \times & 0 \\
\times & 0 & 0 \\
\times & \times & \times \end{array} \right)\,,~~
\mathcal{C}_{4}^{(3)}:\left( \begin{array}{ccc} 0 & \times & \times \\
\times & 0 & \times \\
 0 & 0 & \times \end{array} \right)$  \\ \hline
 $\mathcal{C}_{6}^{(1)}:\left( \begin{array}{ccc} 0 & \times & 0 \\
\times & 0 & 0 \\
 0 & 0 & \times \end{array} \right)$ \\ \hline \hline
\end{tabular}
\caption{\label{tab:classes_charged_lepton_mass_matrix} Texture-zero classification for the (rank-3) charged fermions mass matrices which can be realised from $T'$ modular symmetry, up to row and column permutations. In the notation of Grimus et al~\cite{Ludl:2014axa}, $\mathcal{C}_{3}^{(1)}=3_{1}^{(l)}$, $\mathcal{C}_{3}^{(2)}=3_{2}^{(l)}$, $\mathcal{C}_{3}^{(3)}=3_{3}^{(l)}$,  $\mathcal{C}_{4}^{(1)}=4_{4}^{(l)}$, $\mathcal{C}_{4}^{(2)}=4_{1}^{(l)}$, $\mathcal{C}_{4}^{(3)}=4_{2}^{(l)}$, $\mathcal{C}_{6}^{(1)}=6_{1}^{(l)}$. }
\end{table}

\begin{table}[hptb!]
\centering
\begin{tabular}{|l|} \hline\hline
$\mathcal{W}_{1}^{(1)}:\left( \begin{array}{ccc} \times & \times & \times \\
\times & \times & \times \\
\times & \times & 0 \end{array} \right)\,,\quad
\mathcal{W}_{1}^{(2)}:\left( \begin{array}{ccc} \times & 0 & \times \\
 0 & \times & \times \\
\times & \times & \times \end{array} \right)$ \\ \hline
$\mathcal{W}_{2}^{(1)}:\left( \begin{array}{ccc} 0 & \times & \times \\
\times & 0 & \times \\
\times & \times & \times \end{array} \right)\,,\quad
\mathcal{W}_{2}^{(2)}:\left( \begin{array}{ccc} \times & \times & 0 \\
\times & \times & 0 \\
0 & 0 & \times \end{array} \right)$ \\ \hline
$\mathcal{W}_{3}^{(1)}:\left( \begin{array}{ccc} 0 & 0 & \times \\
 0 & 0 & \times \\
\times & \times & \times \end{array} \right)\,,\quad
\mathcal{W}_{3}^{(2)}:\left( \begin{array}{ccc} \times & \times & 0 \\
\times & \times & 0 \\
 0 & 0 & 0 \end{array} \right)$ \\ \hline
$\mathcal{W}_{4}^{(1)}:\left( \begin{array}{ccc} 0 & 0 & \times \\
 0 & 0 & \times \\
\times & \times & 0 \end{array} \right)\,,\quad
\mathcal{W}_{4}^{(2)}:\left( \begin{array}{ccc} 0 & \times & 0 \\
\times & \times & 0 \\
 0 & 0 & 0 \end{array} \right)\,,\quad \mathcal{W}_{4}^{(3)}\left(\begin{array}{ccc} \times & 0 & 0 \\
0 & \times & 0 \\
0 & 0 & 0 \end{array} \right)\,,\quad \mathcal{W}_{4}^{(4)}:\left( \begin{array}{ccc} \times & 0 & 0 \\
 0 & 0 & \times \\
 0 & \times & 0 \end{array} \right)$ \\ \hline
$\mathcal{W}_{5}^{(1)}:\left( \begin{array}{ccc} 0 & \times & 0 \\
\times & 0 & 0 \\
0 & 0 & 0 \end{array} \right)$ \\ \hline\hline
\end{tabular}
\caption{\label{tab:classes_Majorana_neutrino_mass_matrix}Texture-zero patterns for the Majorana neutrino mass matrix $M_\nu$ which can be realised by $T'$ modular symmetry, up to row and column permutations. Notice that only $\mathcal{W}_{1}^{(1)}$\,, $\mathcal{W}_{2}^{(2)}$\,, $\mathcal{W}_{3}^{(1)}$\,, $\mathcal{W}_{3}^{(2)}$\,, $\mathcal{W}_{4}^{(1)}$ and $\mathcal{W}_{4}^{(4)}$ can be obtained if neutrino masses are described by the Weinberg operator. On the other hand, all the above textures except $\mathcal{W}_{4}^{(4)}$ can be achieved if neutrino masses are generated via the minimal type I seesaw mechanism. The above texture zeros of neutrino mass matrices have been discussed in~\cite{Ludl:2014axa}, see Ref.~~\cite{Ding:2022aoe} for  the notation correspondence between ours and those of~\cite{Ludl:2014axa}.  }
\end{table}

\section{Examples of modular models}
In this section we apply the formalism of the previous sections to some examples of concrete models. We begin with some general model building considerations and an overview of models in the current literature.
We then discuss a general class of models of leptons based on $A_4$ before turning to minimal models of leptons (and quarks) based on $S'_4$ with a single modulus field. Finally we discuss a model of leptons based on three $S_4$ groups with three moduli fields, the Littlest Modular Seesaw model, so called because it is highly predictive.

\subsection{General model building considerations and overview of models}
In general, the building blocks of modular flavor symmetry models are modular forms of level $N$ and matter supermultiplets $\varphi$ which transform in representations $\rho_{\varphi}$ of a finite modular group $\Gamma_N$ (or $\Gamma'_N$) and carry certain modular weights $k_{\varphi}$. When considering a particular model, one has to specify both $\rho_{\varphi}$ and $k_{\varphi}$ of each matter field\footnote{The representation $\rho_{\varphi}$ and weight $k_{\varphi}$ of matter fields can be freely assigned in bottom-up models, while they could possibly be fixed in the top-down approach~\cite{Nilles:2020nnc,Nilles:2020kgo,Nilles:2020tdp,Baur:2020jwc,Nilles:2020gvu}.},  then it is straightforward to write out the modular invariant superpotential for fermion masses by using the modular forms of level $N$ listed in Appendix~\ref{app:Gamma3-MF3}.

Many modular flavor symmetry models for lepton masses and mixing, based on different choices of level, modular weights and representations of matter fields have been proposed. Moreover, the modular symmetry has been also extended to the quark sector to describe the flavor structure of both quarks and leptons with a common modulus $\tau$. A summary of the modular invariant models in the current literature is given in table~\ref{tab:modular-models-summary}.

In modular invariant models for leptons, the three generations of lepton doublets are usually assigned to be an irreducible triplet of the finite modular group $\Gamma_N$ or $\Gamma'_N$ in order to minimize the number of free parameters and the K\"ahler potential is assumed to be the minimal form otherwise additional free parameters and sizable corrections would be invoked~\cite{Chen:2019ewa,Lu:2019vgm}. The superpotential of charged leptons usually depends only 3 dimensionless parameters. The latter are adjusted to reproduce
the charged lepton masses. All the remaining physical observables such as neutrino masses and lepton mixing angles and CP violation phases are usualy determined by the VEV of modulus $\tau$ and additional two (or three) free couplings. The gCP symmetry could enforce the couplings to be real in certain representation basis and the predictive power of the model would be improved   further. Hence the modular invariant models could involve a small number of free parameters, and they can predict not only the lepton mixing but also lepton masses. Moreover, modular symmetry has been used to address the strong CP problem~\cite{Feruglio:2023uof,Ahn:2023iqa}, and it is also exploited to new physics beyond SM~\cite{Nomura:2019jxj,Okada:2020oxh,Okada:2020dmb,Zhang:2021olk,Tanimoto:2021ehw,Kobayashi:2021ajl,Kobayashi:2021uam,Kobayashi:2021pav,Behera:2022wco,Kobayashi:2022jvy,Mishra:2023cjc,Abe:2023ylh,Nomura:2023kwz}. The moduli in modular flavor symmetric models may possibly play the role of inflaton~\cite{Gunji:2022xig,Abe:2023ylh}.

\begin{table}[hptb]
\centering
\resizebox{1.0\textwidth}{!}{
\begin{tabular}{|c|c|c|c|c|} \hline\hline
& $\Gamma_N/\Gamma'_N$ & leptons alone & quarks alone & leptons $\&$ quarks \\ \hline
$N=2$ &  $S_3$  & \cite{Kobayashi:2018vbk,Okada:2019xqk,Meloni:2023aru} & \cite{Kobayashi:2018wkl}& --- \\ \hline

\multirow{2}{*}{$N=3$} & $A_4$ & \cite{Feruglio:2017spp,Kobayashi:2018vbk,Kobayashi:2018scp,Criado:2018thu,Kobayashi:2018wkl,Novichkov:2018yse,Nomura:2019yft,Ding:2019zxk,Nomura:2019lnr,Asaka:2019vev,Ding:2019gof,Zhang:2019ngf,Nomura:2019xsb,Kobayashi:2019gtp,Wang:2019xbo,Ding:2020yen,Aoki:2020eqf,Asaka:2020tmo,
Okada:2020brs,Kashav:2021zir,Okada:2021qdf,deMedeirosVarzielas:2021pug,Kang:2022psa,Abbas:2022slb,Devi:2023vpe,CentellesChulia:2023zhu,Nomura:2023usj,Kumar:2023moh,Kobayashi:2023qzt} & \cite{Okada:2018yrn,Petcov:2023vws,Petcov:2022fjf,Kikuchi:2023jap}  & \cite{Okada:2019uoy,King:2020qaj,Abbas:2020qzc,Okada:2020rjb,Yao:2020qyy,Nomura:2021yjb} \\ \cline{2-5}

 & $T'$ & \cite{Liu:2019khw,Okada:2022kee,Ding:2022aoe}  & --- & \cite{Lu:2019vgm,Benes:2022bbg} \\\hline

\multirow{2}{*}{$N=4$} & $S_4$ & \cite{Ding:2019gof,Penedo:2018nmg,Novichkov:2018ovf,deMedeirosVarzielas:2019cyj,Kobayashi:2019mna,King:2019vhv,Criado:2019tzk,Wang:2019ovr,Wang:2020dbp,Nomura:2021ewm} & --- &  \cite{Qu:2021jdy} \\ \cline{2-5}

& $S'_4$ & \cite{Novichkov:2020eep,Ding:2022nzn} & \cite{Abe:2023ilq} & \cite{Liu:2020akv,Abe:2023qmr} \\\hline

\multirow{2}{*}{$N=5$} & $A_5$ & \cite{Novichkov:2018nkm,Ding:2019xna,Criado:2019tzk,deMedeirosVarzielas:2022ihu} & --- & ---  \\ \cline{2-5}

& $A'_5$ & \cite{Wang:2020lxk,Wang:2021mkw,Behera:2021eut,Behera:2022wco}  & --- & \cite{Yao:2020zml}\\\hline

\multirow{2}{*}{$N=6$} & $\Gamma_6\cong S_3\times A_4$ & --- & \cite{Kikuchi:2023cap} & ---  \\ \cline{2-5}

& $\Gamma'_6\cong S_3\times T'$ & \cite{Li:2021buv}  & --- & --- \\\hline

\multirow{2}{*}{$N=7$} & $\Gamma_7$ & \cite{Ding:2020msi} & --- & ---  \\ \cline{2-5}

& $\Gamma'_7$ & ---  & --- & --- \\\hline\hline

\end{tabular}}
\caption{\label{tab:modular-models-summary} Summary of modular invariant models based on SM gauge group.  }
\end{table}

\subsection{Overview of $A_4$ modular models }

The study of modular symmetry initiated from the modular group $A_4$~\cite{Feruglio:2017spp}, it is instructive to recapitulate the $A_4$ modular invariant lepton model since it comprises all the essential features of the modular symmetry models. The model is realized at level $N=3$, and light neutrinos get their masses via the type I see-saw mechanism. Both the left-handed lepton fields $L$ and the right-handed neutrino fields $N^{c}$ are assumed to be triplet of $A_4$, and the three right-handed charged leptons $e^{c}$, $\mu^{c}$ and $\tau^c$ are assigned to transform as $A_4$ singlets $\mathbf{1}$, $\mathbf{1}''$ and $\mathbf{1}'$ respectively. All the lepton fields carry one unit of modular weight, and the Higgs fields $H_{u,d}$ are invariant under $A_4$ with zero modular weight, i.e.
\begin{equation}
L\sim\left(\mathbf{3}, 1\right)\,,~~e^{c}\sim\left(\mathbf{1}, 1\right)\,,~~\mu^{c}\sim\left(\mathbf{1}'', 1\right)\,,~~\tau^c\sim\left(\mathbf{1}', 1\right)\,,~~N^c\sim\left(\mathbf{3}, 1\right)\,,~~H_{u,d}\sim\left(\mathbf{1}, 0\right)\,,
\end{equation}
where the first numbers in the parentheses denote the transformation under $A_4$ modular symmetry and the second numbers are the modular weight. In a standard notation, the superpotential for the charged lepton masses is given by:
\begin{equation}
\mathcal{W}_e
=\alpha e^c(LY^{(2)}_{\mathbf{3}})_{\mathbf{1}}H_d
+\beta \mu^c(LY^{(2)}_{\mathbf{3}})_{\mathbf{1}'}H_d + \gamma \tau^c(LY^{(2)}_{\mathbf{3}})_{\mathbf{1}''}H_d\,,
\label{eq:We-A4-Feruglio}
\end{equation}
where $Y^{(2)}_{\mathbf{3}}(\tau)=(Y_1, Y_2, Y_3)^{T}$ denotes the irreducible triplet of level-3 weight-2 modular forms in Eq.~\eqref{eq:MF-level-3-w2}. The couplings $\alpha$, $\beta$ and $\gamma$ can be made real through redefinition of the fields $e^{c}$, $\mu^c$, $\tau^c$. Using the contraction rules of two $A_4$ triplets in Eq.~\eqref{eq:3x3-rules-A4}, we can read out the charged lepton mass matrix as follow,
\begin{equation}
\label{eq:Me-A4} M_e =\begin{pmatrix}
\alpha Y_1(\tau) ~&\alpha Y_3(\tau) ~&\alpha Y_2(\tau) \\
\beta Y_2(\tau) ~&\beta Y_1(\tau)  ~&\beta Y_3(\tau) \\
 \gamma Y_3(\tau) ~& \gamma Y_2(\tau) ~& \gamma Y_1(\tau)
 \end{pmatrix} v_d\,.
\end{equation}
The superpotential relevant to neutrino masses is:
\be
\mathcal{W}_\nu
=g_1((N^c\,L)_{\mathbf{3}_S}Y^{(2)}_{\mathbf{3}})_\mathbf{1}H_u+g_2((N^c\,L)_{\mathbf{3}_A}Y^{(2)}_{\mathbf{3}})_\mathbf{1}H_u
+\frac{1}{2}\Lambda_L (\left(N^c N^c\right)_\mathbf{3_S}Y^{(2)}_{\mathbf{3}})_\mathbf{1}\,,
\label{wnu}
\ee
where the phase of $g_1$ can be absorbed into the left-handed lepton field $L$, while $g_2$ is complex. The Dirac neutrino mass matrix $m_D$ and heavy Majorana neutrino mass matrix $m_N$ take the following form
\begin{eqnarray}
\nonumber&&\qquad\quad~~~ M_N = \begin{pmatrix}
2Y_1(\tau) ~&~ -Y_3(\tau) ~&~ -Y_2(\tau) \\
 -Y_3(\tau) ~&~ 2Y_2(\tau)  ~&~ -Y_1(\tau)  \\
 -Y_2(\tau) ~&~ -Y_1(\tau) ~&~2Y_3(\tau)
\end{pmatrix}\Lambda_L\,,\\
\label{mnuss}&&
M_D =\begin{pmatrix}
2g_1Y_1(\tau)        ~&~  (-g_1+g_2)Y_3(\tau) ~&~ (-g_1-g_2)Y_2(\tau) \\
(-g_1-g_2)Y_3(\tau)  ~&~     2g_1Y_2(\tau)    ~&~ (-g_1+g_2)Y_1(\tau)  \\
 (-g_1+g_2)Y_2(\tau) ~&~ (-g_1-g_2)Y_1(\tau)  ~&~ 2g_1Y_3(\tau)
\end{pmatrix}v_{u}\,.
\end{eqnarray}
The light neutrino mass matrix is $M_\nu= -M_D^T M_N^{-1} M_D $.
Charged lepton masses can be reproduced by adjusting the parameters $\alpha$, $\beta$ and $\gamma$, while neutrino masses and the lepton mass matrix depend also on additional five parameters: one overall scale $|g_1|^2v^2_u/\Lambda_L$, the complex combination $g_2/g_1$ and the $\tau$ VEV. This model can accommodate both normal ordering and inverted ordering neutrino mass spectrum. The best fit values of the free parameters as well as the neutrino masses and mixing parameters at the best fit points are shown in table~\ref{fitresults-A4-Feruglio}.

\begin{table}[hptb!]
\centering
\resizebox{1.0\textwidth}{!}{
\begin{tabular}{|c|c|c|c|c|c|c|c|c|c|c|}\hline\hline
\multicolumn{11}{|c|}{Goodness of fit against NuFit 5.2 values without SK atmospheric data} \\\hline
\multirow{4}{*}{NO} & $\chi_{\text{min}}^{2}$ & ${\tt Re}(\tau)$ & ${\tt Im}(\tau)$ & $|g_2/g_1|$ & ${\tt arg}(g_2/g_1)$ & $\frac{|g_1|^2 v_u^2}{\Lambda_L}({\rm eV})$ & $\alpha v_d({\rm MeV})$ & $\beta/\alpha$ & $\gamma/\alpha$ & \\ \cline{2-11}

& $1.921$ & $0.049$ & $2.264$ & $1.109$ & $1.899$ & $0.049$ & $0.360$ & $210.767$ & $3581.720$ &  \\ \cline{2-11}
&$\sin^2\theta_{12}$&$\sin^2\theta_{13}$&$\sin^2\theta_{23}$&$\delta_{CP}/\pi$&$\alpha_{21}/\pi$&$\alpha_{31}/\pi$ & $m_1({\rm eV})$&$m_2({\rm eV})$&$m_3({\rm eV})$& $|m_{\beta\beta}|({\rm eV})$ \\
\cline{2-11}
&$0.303$&$0.0220$&$0.575$&$1.416$&$0.023$&$1.014$ & $0.0985$&$0.0988$&$0.1105$ & $0.0985$\\
  \hline
 \multirow{4}{*}{IO} & $\chi_{\text{min}}^{2}$ & ${\tt Re}(\tau)$ & ${\tt Im}(\tau)$ & $|g_2/g_1|$ & ${\tt arg}(g_2/g_1)$ & $\frac{|g_1|^2 v_u^2}{\Lambda_L}({\rm eV})$ & $\alpha v_d({\rm MeV})$ & $\beta/\alpha$ & $\gamma/\alpha$ & \\
  \cline{2-11}
 & $1.129$ & $0.498$ & $1.141$ & $0.968$ & $0.022$ & $0.028$ & $68.517$ & $16.527$ & $4.826\times 10^{-3}$ &  \\
  \cline{2-11}
&$\sin^2\theta_{12}$&$\sin^2\theta_{13}$&$\sin^2\theta_{23}$&$\delta_{CP}/\pi$&$\alpha_{21}/\pi$&$\alpha_{31}/\pi$ & $m_1({\rm eV})$&$m_2({\rm eV})$&$m_3({\rm eV})$& $|m_{\beta\beta}|({\rm eV})$ \\
\cline{2-11}
&$0.305$&$0.0222 $&$0.575$&$1.745$&$0.728$&$0.813$ & $0.0638$&$0.0644$&$0.0405$&$0.0332$\\
  \hline
  \multicolumn{11}{|c|}{Goodness of fit against NuFit 5.2 values with SK atmospheric data} \\
  \hline
  \multirow{4}{*}{NO} & $\chi_{\text{min}}^{2}$ & ${\tt Re}(\tau)$ & ${\tt Im}(\tau)$ & $|g_2/g_1|$ & ${\tt arg}(g_2/g_1)$ & $\frac{|g_1|^2 v_u^2}{\Lambda_L}({\rm eV})$ & $\alpha v_d({\rm MeV})$ & $\beta/\alpha$ & $\gamma/\alpha$ & \\
  \cline{2-11}
 & $0.129$ & $0.284$ & $1.871$ & $1.123$ & $4.651$ & $0.032$ & $0.362$ & $3544.019$ & $209.654$ &  \\
  \cline{2-11}
&$\sin^2\theta_{12}$&$\sin^2\theta_{13}$&$\sin^2\theta_{23}$&$\delta_{CP}/\pi$&$\alpha_{21}/\pi$&$\alpha_{31}/\pi$ & $m_1({\rm eV})$&$m_2({\rm eV})$&$m_3({\rm eV})$& $|m_{\beta\beta}|({\rm eV})$ \\
\cline{2-11}
&$0.303$&$0.0222 $&$0.445$&$1.298$&$1.944$&$0.933$ & $0.0635$&$0.0641$&$0.0809$&$0.0629$\\
  \hline
  \multirow{4}{*}{IO} & $\chi_{\text{min}}^{2}$ & ${\tt Re}(\tau)$ & ${\tt Im}(\tau)$ & $|g_2/g_1|$ & ${\tt arg}(g_2/g_1)$ & $\frac{|g_1|^2 v_u^2}{\Lambda_L}({\rm eV})$ & $\alpha v_d({\rm MeV})$ & $\beta/\alpha$ & $\gamma/\alpha$ & \\
  \cline{2-11}
 & $2.793$ & $0.498$ & $1.139$ & $0.971$ & $1.018$ & $0.028$ & $68.431$ & $16.530$ & $4.826\times 10^{-3}$ &  \\
  \cline{2-11}
&$\sin^2\theta_{12}$&$\sin^2\theta_{13}$&$\sin^2\theta_{23}$&$\delta_{CP}/\pi$&$\alpha_{21}/\pi$&$\alpha_{31}/\pi$ & $m_1({\rm eV})$&$m_2({\rm eV})$&$m_3({\rm eV})$& $|m_{\beta\beta}|({\rm eV})$ \\
\cline{2-11}
&$0.307$&$0.0222 $&$0.564$&$1.731$&$0.728$&$0.812$ & $0.0636$&$0.0642$&$0.0405$&$0.0331$\\
  \hline\hline
\end{tabular}}
\caption{\label{fitresults-A4-Feruglio} The best fit values of free couplings and lepton masses and lepton mixing parameters in the original $A_4$ modular model of Feruglio~\cite{Feruglio:2017spp}, we use the latest global fit results of NuFIT v5.2 without/with SK atmospheric data~\cite{Esteban:2020cvm}.}
\end{table}

In the above original $A_4$ modular model~\cite{Feruglio:2017spp}, the right-handed lepton fields $e^{c}$, $\mu^{c}$ and $\tau^c$ have the same modular weight and they transform as different $A_4$ singlets. In fact, we can distinguish $e^{c}$, $\mu^{c}$ and $\tau^c$ through the combination of modular weights and representation assignments under $A_4$. If two or all of $e^{c}$, $\mu^{c}$ and $\tau^c$ are assigned to the same $A_4$ singlet representation, they could be distinguished by different modular weights. The left-handed lepton fields are assigned to be $A_4$ triplet in order to produce the large lepton mixing angles. Guided by the principle of simplicity, we use the lower weight modular forms of level 3 as much as possible. Then the superpotential for charged lepton masses can take the following ten possible forms~\cite{Ding:2019zxk}:
\begin{itemize}[labelindent=-1.8em, leftmargin=1.8em]
\item{$(e^c, \mu^c, \tau^c)\sim(\mathbf{1}, \mathbf{1}, \mathbf{1})$, $(k_{e^c}+k_L, k_{\mu^c}+k_L, k_{\tau^c}+k_L)=(2, 4, 6)$}

The modular invariant superpotential for charged lepton Yukawa couplings is given by
\begin{eqnarray}
\nonumber \mathcal{W}_e&=&\alpha e^c(LY^{(2)}_{\mathbf{3}})_\mathbf{1}H_d
+\beta \mu^c(LY^{(4)}_{\mathbf{3}})_\mathbf{1}H_d + \gamma_1 \tau^c(LY^{(6)}_{\mathbf{3}I})_\mathbf{1}H_d + \gamma_2 \tau^c(LY^{(6)}_{\mathbf{3}II})_\mathbf{1}H_d\\
\nonumber&=&\alpha e^c\left(L_1 Y_1+L_2 Y_3+L_3 Y_2\right)H_d\\
\nonumber&&+\beta \mu^c\Big[L_1(Y^2_1-Y_2Y_3)\,+\,L_2(Y^2_2-Y_1Y_3)+L_3(Y^2_3-Y_1Y_2)\Big]H_d\\
\nonumber&&+\gamma_1 \tau^c\Big[L_1(Y^3_1+2Y_1Y_2Y_3)\,+\,L_2(Y^2_1Y_3+2Y^2_3Y_2)\,+\,L_3(Y^2_1Y_2+2Y^2_2Y_3)\Big]H_d\\
\label{eq:We-A4-1}
&&+\gamma_2 \tau^c\Big[L_1(Y^3_3+2Y_1Y_2Y_3)\,+\,L_2(Y^2_3Y_2+2Y^2_2Y_1)\,+\,L_3(Y^2_3Y_1+2Y^2_1Y_2)\Big]H_d\,,~~~~~~
\end{eqnarray}
where $Y^{(4)}_{\mathbf{3}}$ and $Y^{(6)}_{\mathbf{3}I}$, $Y^{(6)}_{\mathbf{3}II}$ are weight 4 and weight 6 modular forms at level 3 respectively, and they are given in Eqs.~(\ref{eq:MF-w4l3}, \ref{eq:MF-w6}).

\item{$(e^c, \mu^c, \tau^c)\sim(\mathbf{1}', \mathbf{1}', \mathbf{1}')$, $(k_{e^c}+k_L, k_{\mu^c}+k_L, k_{\tau^c}+k_L)=(2, 4, 6)$}
\begin{eqnarray}
\nonumber\mathcal{W}_e&=&\alpha e^c(LY^{(2)}_{\mathbf{3}})_{\mathbf{1}''}H_d+\beta\mu^c(LY^{(4)}_{\mathbf{3}})_{\mathbf{1}''}H_d +\gamma_1\tau^c(LY^{(6)}_{\mathbf{3}I})_{\mathbf{1}''}H_d + \gamma_2 \tau^c(LY^{(6)}_{\mathbf{3}II})_{\mathbf{1}''}H_d\\
\nonumber&=&\alpha e^c(L_1 Y_3+L_2 Y_2+L_3 Y_1)H_d\\
\nonumber&&+\beta\mu^c\Big[L_1(Y^2_2-Y_1Y_3)+L_2(Y^2_3-Y_1Y_2)+L_3(Y^2_1-Y_2Y_3)\Big]H_d\\
\nonumber&&+\gamma_1\tau^c\Big[L_1(Y^2_1Y_3+2Y^2_3Y_2)+L_2(Y^2_1Y_2+2Y^2_2Y_3)+L_3(Y^3_1+2Y_1Y_2Y_3)\Big]H_d\\
\label{eq:We-A4-2}&&+\gamma_2\tau^c\Big[L_1(Y^2_3Y_2+2Y^2_2Y_1)+L_2(Y^2_3Y_1+2Y^2_1Y_2)+L_3(Y^3_3+2Y_1Y_2Y_3)\Big]H_d\,.~~~~~~
\end{eqnarray}

\item{$(e^c, \mu^c, \tau^c)\sim(\mathbf{1}'', \mathbf{1}'', \mathbf{1}'')$, $(k_{e^c}+k_L, k_{\mu^c}+k_L, k_{\tau^c}+k_L)=(2, 4, 6)$}
\begin{eqnarray}
\nonumber \mathcal{W}_e
&=&\alpha e^c(LY^{(2)}_{\mathbf{3}})_\mathbf{1'}H_d
+\beta \mu^c(LY^{(4)}_{\mathbf{3}})_\mathbf{1'}H_d + \gamma_1 \tau^c(LY^{(6)}_{\mathbf{3}I})_\mathbf{1'}H_d + \gamma_2 \tau^c(LY^{(6)}_{\mathbf{3}II})_\mathbf{1'}H_d\\
\nonumber &=&\alpha e^c(L_1Y_2+L_2Y_1+L_3Y_3)H_d\\
\nonumber&&+\beta\mu^c\Big[L_1(Y^2_3-Y_1Y_2)+L_2(Y^2_1-Y_2Y_3)+L_3(Y^2_2-Y_1Y_3)\Big]H_d\\
\nonumber&&+\gamma_1\tau^c\Big[L_1(Y^2_1Y_2+2Y^2_2Y_3)+L_2(Y^3_1+2Y_1Y_2Y_3)+L_3(Y^2_1Y_3+2Y^2_3Y_2)\Big]H_d\\
&&+\gamma_2\tau^c\Big[L_1(Y^2_3Y_1+2Y^2_1Y_2)+L_2(Y^3_3+2Y_1Y_2Y_3)+L_3(Y^2_3Y_2+2Y^2_2Y_1)\Big]H_d\,.~~~~~~
\label{eq:We-A4-3}
\end{eqnarray}

\item{$(e^c, \mu^c, \tau^c)\sim(\mathbf{1}, \mathbf{1}, \mathbf{1}')$, $(k_{e^c}+k_L, k_{\mu^c}+k_L, k_{\tau^c}+k_L)=(2, 4, 2)$}
\begin{eqnarray}
\nonumber\mathcal{W}_e&=&\alpha e^c(LY^{(2)}_{\mathbf{3}})_\mathbf{1}H_d
+\beta \mu^c(L\,Y^{(4)}_{\mathbf{3}})_\mathbf{1}H_d + \gamma \tau^c(LY^{(2)}_{\mathbf{3}})_{\mathbf{1}''}H_d\\
\nonumber&=&\alpha e^c(L_1 Y_1+L_2 Y_3+L_3 Y_2)H_d+\beta \mu^c\Big[L_1(Y^2_1-Y_2Y_3)+L_2(Y^2_2-Y_1Y_3)\\
&&+L_3(Y^2_3-Y_1Y_2)\Big]H_d+\gamma \tau^c(L_1 Y_3+L_2Y_2+L_3Y_1)H_d\,.
\label{eq:We-A4-4}
\end{eqnarray}

\item{$(e^c, \mu^c, \tau^c)\sim(\mathbf{1}, \mathbf{1}, \mathbf{1}'')$, $(k_{e^c}+k_L, k_{\mu^c}+k_L, k_{\tau^c}+k_L)=(2, 4, 2)$}
\begin{eqnarray}
\nonumber\mathcal{W}_e&=&\alpha e^c(LY^{(2)}_{\mathbf{3}})_\mathbf{1}H_d
+\beta \mu^c(L\,Y^{(4)}_{\mathbf{3}})_\mathbf{1}H_d + \gamma \tau^c(LY^{(2)}_{\mathbf{3}})_\mathbf{1'}H_d\\
\nonumber&=&\alpha e^c(L_1 Y_1+L_2 Y_3+L_3 Y_2)H_d+\beta \mu^c\Big[L_1(Y^2_1-Y_2Y_3)+L_2(Y^2_2-Y_1Y_3)\\
&&+L_3(Y^2_3-Y_1Y_2)\Big]H_d+\gamma \tau^c(L_1 Y_2+L_2 Y_1+L_3 Y_3)H_d\,.
\label{eq:We-A4-5}
\end{eqnarray}

\item{$(e^c, \mu^c, \tau^c)\sim(\mathbf{1}', \mathbf{1}', \mathbf{1})$, $(k_{e^c}+k_L, k_{\mu^c}+k_L, k_{\tau^c}+k_L)=(2, 4, 2)$}
\begin{eqnarray}
\nonumber\mathcal{W}_e&=&\alpha e^c(LY^{(2)}_{\mathbf{3}})_{\mathbf{1}''}H_d
+\beta \mu^c(LY^{(4)}_{\mathbf{3}})_{\mathbf{1}''}H_d + \gamma \tau^c(LY^{(2)}_{\mathbf{3}})_\mathbf{1}H_d\\
\nonumber&=&\alpha e^c(L_1 Y_3+L_2 Y_2+L_3 Y_1)H_d+\beta \mu^c\Big[L_1(Y^2_2-Y_1Y_3)+L_2(Y^2_3-Y_1Y_2)\\
&&+L_3(Y^2_1-Y_2Y_3)\Big]H_d +\gamma \tau^c(L_1 Y_1+L_2 Y_3+L_3 Y_2)H_d\,.
\label{eq:We-A4-6}
\end{eqnarray}

\item{$(e^c, \mu^c, \tau^c)\sim(\mathbf{1}', \mathbf{1}', \mathbf{1}'')$, $(k_{e^c}+k_L, k_{\mu^c}+k_L, k_{\tau^c}+k_L)=(2, 4, 2)$}
\begin{eqnarray}
\nonumber\mathcal{W}_e&=&\alpha e^c(LY^{(2)}_{\mathbf{3}})_{\mathbf{1}''}H_d
+\beta \mu^c(L\,Y^{(4)}_{\mathbf{3}})_{\mathbf{1}''}H_d + \gamma \tau^c(LY^{(2)}_{\mathbf{3}})_{\mathbf{1}'}H_d\\
\nonumber&=&\alpha e^c(L_1 Y_3+L_2 Y_2+L_3 Y_1)H_d+\beta \mu^c\Big[L_1(Y^2_2-Y_1Y_3)+L_2(Y^2_3-Y_1Y_2)\\
&&+L_3(Y^2_1-Y_2Y_3)\Big]H_d+\gamma \tau^c(L_1 Y_2+L_2 Y_1+L_3 Y_3)H_d\,.
\label{eq:We-A4-7}
\end{eqnarray}

\item{$(e^c, \mu^c, \tau^c)\sim(\mathbf{1}'', \mathbf{1}'', \mathbf{1})$, $(k_{e^c}+k_L, k_{\mu^c}+k_L, k_{\tau^c}+k_L)=(2, 4, 2)$}
\begin{eqnarray}
\nonumber\mathcal{W}_e&=&\alpha e^c(LY^{(2)}_{\mathbf{3}})_{\mathbf{1}'}H_d
+\beta \mu^c(LY^{(4)}_{\mathbf{3}})_{\mathbf{1}'}H_d + \gamma \tau^c(LY^{(2)}_{\mathbf{3}})_\mathbf{1}H_d\\
\nonumber&=&\alpha e^c(L_1 Y_2+L_2 Y_1+L_3 Y_3)H_d+\beta \mu^c\Big[L_1(Y^2_3-Y_1Y_2)+L_2(Y^2_1-Y_2Y_3)\\
&&+L_3(Y^2_2-Y_1Y_3)\Big]H_d+\gamma \tau^c(L_1 Y_1+L_2 Y_3+L_3 Y_2)H_d\,.
\label{eq:We-A4-8}
\end{eqnarray}

\item{$(e^c, \mu^c, \tau^c)\sim(\mathbf{1}'', \mathbf{1}'', \mathbf{1}')$, $(k_{e^c}+k_L, k_{\mu^c}+k_L, k_{\tau^c}+k_L)=(2, 4, 2)$}
\begin{eqnarray}
\nonumber\mathcal{W}_e&=&\alpha e^c(LY^{(2)}_{\mathbf{3}})_{\mathbf{1}'}H_d
+\beta \mu^c(LY^{(4)}_{\mathbf{3}})_{\mathbf{1}'}H_d + \gamma \tau^c(LY^{(2)}_{\mathbf{3}})_{\mathbf{1}''}H_d\\
\nonumber&=&\alpha e^c(L_1 Y_2+L_2Y_1+L_3 Y_3)H_d+\beta\mu^c\Big[L_1(Y^2_3-Y_1Y_2)+L_2(Y^2_1-Y_2Y_3)\\
&&+L_3(Y^2_2-Y_1Y_3)\Big]H_d+\gamma \tau^c(L_1 Y_3+L_2 Y_2+L_3 Y_1)H_d\,.
\label{eq:We-A4-9}
\end{eqnarray}

\item{$(e^c, \mu^c, \tau^c)\sim(\mathbf{1}, \mathbf{1}'', \mathbf{1}')$, $(k_{e^c}+k_L, k_{\mu^c}+k_L, k_{\tau^c}+k_L)=(2, 2, 2)$}
\begin{eqnarray}
\nonumber\mathcal{W}_e&=&\alpha e^c(LY^{(2)}_{\mathbf{3}})_{\mathbf{1}}H_d
+\beta \mu^c(LY^{(2)}_{\mathbf{3}})_{\mathbf{1}'}H_d + \gamma \tau^c(LY^{(2)}_{\mathbf{3}})_{\mathbf{1}''}H_d\\
\nonumber&=&\alpha e^c(L_1 Y_1+L_2 Y_3+L_3 Y_2)H_d+\beta \mu^c(L_1 Y_2+L_2 Y_1+L_3 Y_3)H_d\\
&&+\gamma \tau^c(L_1 Y_3+L_2 Y_2+L_3 Y_1)H_d\,,
\label{eq:We-A4-10}
\end{eqnarray}
which is exactly charged lepton sector of the Feruglio's $A_4$ modular model~\cite{Feruglio:2017spp}.

\end{itemize}

If the right-handed neutrinos $N^c$ are assigned to be a triplet of $A_4$, the superpotential of neutrino masses can only take three possible forms in the scenarios with the level 3 and weight 2 modular form~\cite{Ding:2019zxk}:
\begin{itemize}[labelindent=-1.8em, leftmargin=1.8em]
\item{ $(k_{N^c}, k_L)=(0, 2)$}

The neutrino superpotential invariant under the modular symmetry takes the following form:
\begin{eqnarray}
\nonumber \mathcal{W}_\nu&=&g_1((N^c\,L)_{\mathbf{3}_S}Y^{(2)}_{\mathbf{3}})_\mathbf{1}H_u+g_2((N^cL)_{\mathbf{3}_A}Y^{(2)}_{\mathbf{3}})_\mathbf{1}H_u\,
+\Lambda \left(N^c N^c\right)_\mathbf{1}\\
\nonumber&=&g_1\Big[(2N^c_1\,L_1\,-N^c_2\,L_3-N^c_3\,L_2)Y_1+(2N^c_3\,L_3\,-N^c_1\,L_2-N^c_2\,L_1)Y_3 \\
\nonumber&& +(2N^c_2L_2-N^c_1L_3-N^c_3L_1)Y_2\Big]H_u+g_2\Big[(N^c_2L_3-N^c_3L_2)Y_1\\
&& +(N^c_1L_2-N^c_2L_1)Y_3+(N^c_3L_1-N^c_1L_3)Y_2\Big]H_u+\Lambda(N^c_1N^c_1+2N^c_2N^c_3)\,.~~~
\label{eq:Wnu-A4-s1}
\end{eqnarray}

\item{ $(k_{N^c}, k_L)=(1, 1)$}

\begin{eqnarray}
\nonumber\mathcal{W}_\nu
&=&g_1((N^c\,L)_{\mathbf{3}_S}Y^{(2)}_{\mathbf{3}})_\mathbf{1}H_u+g_2((N^c\,L)_{\mathbf{3}_A}Y^{(2)}_{\mathbf{3}})_\mathbf{1}H_u
+\Lambda (\left(N^c N^c\right)_\mathbf{3_S}Y)_\mathbf{1}\\
\nonumber&=&g_1\Big[(2N^c_1L_1-N^c_2L_3-N^c_3L_2)Y_1+(2N^c_3L_3-N^c_1L_2-N^c_2L_1)Y_3\\
\nonumber&&+(2N^c_2L_2-N^c_1L_3-N^c_3L_1)Y_2\Big]H_u+g_2\Big[(N^c_2L_3-N^c_3L_2)Y_1+(N^c_1L_2-N^c_2L_1)Y_3\\
\nonumber&&+(N^c_3L_1-N^c_1L_3)Y_2\Big]H_u+2\Lambda\Big[(N^c_1N^c_1-N^c_2N^c_3)Y_1+(N^c_3N^c_3-N^c_1N^c_2)Y_3\\
&&+(N^c_2N^c_2-N^c_1N^c_3)Y_2\Big]\,.
\label{eq:Wnu-A4-s2}
\end{eqnarray}

\item{ $(k_{N^c}, k_L)=(1, -1)$}

\begin{eqnarray}
\nonumber \mathcal{W}_\nu
&=&g((N^c\,L)_\mathbf{1}H_u+\Lambda (\left(N^c\, N^c\right)_{\mathbf{3}_S}Y^{(2)}_{\mathbf{3}})_\mathbf{1}\\
\nonumber&=&g(N^c_1\,L_1\,+N^c_2\,L_3+N^c_3\,L_2)H_u+2\Lambda\Big[(N^c_1N^c_1-N^c_2N^c_3)Y_1 \\
&& +(N^c_3N^c_3-N^c_1N^c_2)Y_3+(N^c_2N^c_2-N^c_1N^c_3)Y_2\Big]\,.
\label{eq:Wnu-A4-s3}
\end{eqnarray}

\end{itemize}

Combining the charged lepton sector with the neutrino sector, we have 30 simple $A_4$ modular symmetry model for lepton, and the experimental data  can be accommodated for the first two cases of the neutrino sector. We summarize the phenomenologically viable models in table~\ref{tab:viable-A4-models}, where the predictions for the lightest neutrino masses are presented in the last two columns for the viable models. Notice that the $\chi^2$ function could have more than one local minima at which different predictions for neutrino masses and mixing parameters are reached. A full classification of the $A_4$ modular models with gCP is performed in~\cite{Yao:2020qyy}, and the minimal model can describe the experimental lepton data with only 7 real free parameters.

\begin{table}[hptb]
\centering
\begin{tabular}{|c|c|c|c|c|c|} \hline\hline
$(\rho_{e^c}, \rho_{\mu^c}, \rho_{\tau^c})$  & $k_L$  & $(k_{e^c}, k_{\mu^c}, k_{\tau^c})$ & $k_{N^c}$  & NO ($m_{1}/\text{meV}$) & IO ($m_{3}/\text{meV}$) \\ \hline
\multirow{2}{*}{$(\mathbf{1}, \mathbf{1}, \mathbf{1})$} &  2 &  $(0, 2, 4)$ & 0 & \ding{52} $(5.58)$ & \ding{52} $(0.43)$ \\ \cline{2-6}
&  1 &  $(1, 3, 5)$ & 1 & \ding{52} $(11.94)$ & \ding{52} $(25.46)$\\\hline
\multirow{2}{*}{$(\mathbf{1}', \mathbf{1}', \mathbf{1}')$} &  2 &  $(0, 2, 4)$ & 0 & \ding{52} $(5.53)$ & \ding{52} $(1.53)$ \\ \cline{2-6}
&  1 &  $(1, 3, 5)$ & 1 & \ding{52} $(2.38)$ & \ding{52} $(10.35)$\\\hline
\multirow{2}{*}{$(\mathbf{1}'', \mathbf{1}'', \mathbf{1}'')$} &  2 &  $(0, 2, 4)$ & 0 & \ding{52} $(11.32)$ & \ding{52} $(3.99)$ \\ \cline{2-6}
 &  1 &  $(1, 3, 5)$ & 1 & \ding{52} $(3.02)$ & \ding{52} $(10.98)$\\\hline
\multirow{2}{*}{$(\mathbf{1}, \mathbf{1}, \mathbf{1}')$}  & 2  & $(0, 2, 0)$ & 0 & \ding{56} $(-)$ & \ding{52} $(61.89)$ \\ \cline{2-6}
&  1 &  $(1, 3, 1)$ & 1 & \ding{56} $(-)$ & \ding{56} $(-)$\\\hline
\multirow{2}{*}{$(\mathbf{1}, \mathbf{1}, \mathbf{1}'')$}   & 2  & $(0, 2, 0)$ & 0 & \ding{56} $(-)$ & \ding{52} $(50.55)$ \\ \cline{2-6}
&  1 &  $(1, 3, 1)$ & 1 & \ding{52} $(40.14)$ & \ding{52} $(25.37)$\\\hline
\multirow{2}{*}{$(\mathbf{1}', \mathbf{1}', \mathbf{1})$}   & 2  & $(0, 2, 0)$ & 0 & \ding{56} $(-)$ & \ding{52} $(40.65)$ \\ \cline{2-6}
&  1 &  $(1, 3, 1)$ & 1 & \ding{52} $(6.66)$ & \ding{52} $(25.36)$\\\hline
\multirow{2}{*}{$(\mathbf{1}', \mathbf{1}', \mathbf{1}'')$}   & 2  & $(0, 2, 0)$ & 0 & \ding{52} $(42.26)$ & \ding{52} $(53.67)$ \\ \cline{2-6}
&  1 &  $(1, 3, 1)$ & 1 & \ding{52} $(28.91)$ & \ding{52} $(10.17)$\\\hline
\multirow{2}{*}{$(\mathbf{1}'', \mathbf{1}'', \mathbf{1})$}  & 2  & $(0, 2, 0)$ & 0 & \ding{56} $(-)$ & \ding{52} $(22.78)$ \\ \cline{2-6}
&  1 &  $(1, 3, 1)$ & 1 & \ding{52} $(6.15)$ & \ding{52} $(40.45)$\\\hline
\multirow{2}{*}{$(\mathbf{1}'', \mathbf{1}'', \mathbf{1}')$}   & 2  & $(0, 2, 0)$ & 0 & \ding{52} $(42.36)$ & \ding{52} $(44.84)$ \\ \cline{2-6}
&  1 &  $(1, 3, 1)$ & 1 & \ding{52} $(78.76)$ & \ding{56} $(-)$\\\hline
\multirow{2}{*}{$(\mathbf{1}, \mathbf{1}'', \mathbf{1}')$} & 2  & $(0, 2, 0)$ & 0 & \ding{52} $(42.39)$ & \ding{52} $(48.58)$ \\ \cline{2-6}
&  1 &  $(1, 3, 1)$ & 1 & \ding{52} $(98.50)$ & \ding{52} $(40.54)$\\\hline\hline
\end{tabular}
\caption{\label{tab:viable-A4-models}Summary of $A_4$ modular symmetry models for leptons. In the last two columns, we use ``\ding{52}'' (``\ding{56}'') denote the models which are (not) compatible with the global fitting results of NuFIT v5.2 without SK atmospheric data~\cite{Esteban:2020cvm} at $3\sigma$ level, and the preditions for the lightest neutrino mass are presented for viable models. The model in the last line is exactly the Feruglio $A_4$ modular models discussed earlier. }
\end{table}

\subsection{Minimal $S'_4$ modular model for leptons}

This model is based on the finite modular symmetry $S'_4$~\cite{Ding:2022nzn}, the group theory of $S'_4$ and the explicit expressions of the modular forms of level 4 are listed in Appendix~\ref{app:group-MF-N4}. The neutrino masses are generated by type-I seesaw mechanism with three right-handed neutrinos $N^c=(N^c_1, N^c_2, N^c_3)^T$. In this model, both the left-handed lepton fields $L$ and right-handed neutrinos $N^c$ are assumed to transform as $\mathbf{3}$ under $S'_4$,  the first two generations of the right-handed charged leptons $E^c_{D}=(E^c_1, E^c_2)^{T}$ are assigned to a doublet $\mathbf{\widehat{2}}$ of $S'_4$, and the third right-handed charged leptons $E^c_3$ is a $S'_4$ singlet $\mathbf{\widehat{1}'}$. The representation and weight assignments of the lepton fields are given by:
\begin{eqnarray}
\nonumber&&\rho_{E^c}=\mathbf{\widehat{2}}\oplus\mathbf{\widehat{1}'},~~~ \rho_{L} = \mathbf{3},~~~\rho_{N^c}=\mathbf{3},~~~\rho_{H_u}=\rho_{H_d}=\mathbf{1}\,,\\
&& k_{E_{1,2,3}^c}=4,~~~k_{N^c}=1,~~~k_{L}=-1,~~~k_{H_u}=k_{H_d}=0\,.
\end{eqnarray}
This model can be revised into a supergravity version by simply revising the modular weights of matter fields as $k_{E_{1,2,3}^c}=9/2$, $k_{N^c}=3/2$, $k_{L}=-1/2$~\cite{Ding:2022nzn}.

Then one can read off the modular invariant superpotential for the lepton masses as follow:
\begin{align}
\nonumber \mathcal{W}_e &= \alpha \left( E_D^c L Y^{(3)}_{\mathbf{\widehat{3}'}}\right)_{\mathbf{1}} H_d +\beta \left( E_D^c L Y^{(3)}_{\mathbf{\widehat{3}}}\right)_{\mathbf{1}} H_d + \gamma \left(E_3^c L Y^{(3)}_{\mathbf{\widehat{3}}}\right)_{\mathbf{1}} H_d\,, \\
\mathcal{W}_\nu &=  g_1 \left(N^c L\right)_{\mathbf{1}} H_u + \Lambda \left( (N^c N^c)_{\mathbf{2},s} Y^{(2)}_{\mathbf{2}}\right)_{\mathbf{1}}\,.
\end{align}
We include gCP symmetry in this model, and consequently all coupling constants are enforced to be real since both modular generators $S$ and $T$ are represented by symmetric and unitary matrices and all Clebsch-Gordan coefficients are real in the chosen basis of Appendix~\ref{app:group-MF-N4}. Note that $\beta$ would be a complex number and the phases of others can be absorbed into fields if gCP is not considered. Using the contraction rules of $S'_4$ group listed in Appendix~\ref{app:group-MF-N4}, we find the lepton mass matrices are
\begin{align}
\label{eq:S4DClept1}
\nonumber&M_e=  \begin{pmatrix}
2\alpha Y^{(3)}_{\mathbf{\widehat{3}'},1} ~&~ -\alpha Y^{(3)}_{\mathbf{\widehat{3}'},3}+\sqrt{3}\beta Y^{(3)}_{\mathbf{\widehat{3}},2}   ~&~  -\alpha Y^{(3)}_{\mathbf{\widehat{3}'},2}+\sqrt{3}\beta Y^{(3)}_{\mathbf{\widehat{3}},3}  \\
-2\beta Y^{(3)}_{\mathbf{\widehat{3}},1} ~&~\sqrt{3}\alpha Y^{(3)}_{\mathbf{\widehat{3}'},2}+\beta Y^{(3)}_{\mathbf{\widehat{3}},3}  ~&~\sqrt{3}\alpha Y^{(3)}_{\mathbf{\widehat{3}'},3}+\beta Y^{(3)}_{\mathbf{\widehat{3}},2} \\
 \gamma Y^{(3)}_{\mathbf{\widehat{3}},1} ~&~ \gamma Y^{(3)}_{\mathbf{\widehat{3}},3} ~&~ \gamma Y^{(3)}_{\mathbf{\widehat{3}},2}
\end{pmatrix}v_d  \,, \\
&M_D=g\begin{pmatrix}
1  ~& 0 ~& 0 \\
0  ~& 0 ~& 1 \\
0  ~& 1 ~& 0
\end{pmatrix}v_u\,,\quad M_N= \Lambda \begin{pmatrix}
2 Y^{(2)}_{\mathbf{2},1} ~& 0 ~& 0  \\
0 ~&\sqrt{3} Y^{(2)}_{\mathbf{2},2} ~& -Y^{(2)}_{\mathbf{2},1} \\
0 ~& -Y^{(2)}_{\mathbf{2},1} ~&\sqrt{3} Y^{(2)}_{\mathbf{2},2}
\end{pmatrix}\,,
\end{align}
where $v_d=\langle H_d\rangle$ and $v_u=\langle H_u\rangle$ are VEVs of Higgs fields. Applying the famous seesaw formula, we can obtain the light neutrino mass matrix as follow
\begin{equation}
M_\nu= -M_D^T M_N^{-1} M_D = \frac{g^2v_u^2}{\Lambda}\begin{pmatrix}
-\frac{1}{2 Y^{(2)}_{\mathbf{2},1}} ~&~ 0 ~&~ 0  \\
0 ~&~ \frac{\sqrt{3} \; Y^{(2)}_{\mathbf{2},2}}{Y^{(2)2}_{\mathbf{2},1}-3Y^{(2)2}_{\mathbf{2},2}} ~&~ \frac{Y^{(2)}_{\mathbf{2},1}}{Y^{(2)2}_{\mathbf{2},1}-3Y^{(2)2}_{\mathbf{2},2}} \\
0 ~&~ \frac{Y^{(2)}_{\mathbf{2},1}}{Y^{(2)2}_{\mathbf{2},1}-3Y^{(2)2}_{\mathbf{2},2}} ~&~ \frac{\sqrt{3}\; Y^{(2)}_{\mathbf{2},2}}{Y^{(2)2}_{\mathbf{2},1}-3Y^{(2)2}_{\mathbf{2},2}}
\end{pmatrix}\,,
\end{equation}
which contributes to the atmospheric mixing angle. It is remarkable that the light neutrino masses are fixed by the modulus $\tau$ up to an overall scale $g^2v^2_u/\Lambda$,
\begin{equation}
\label{eq:NeutrinoMass}
m_1=\frac{1}{|2Y^{(2)}_{\mathbf{2},1}|} \frac{g^2v_u^2}{\Lambda}\,,\quad m_2=\frac{1}{|Y^{(2)}_{\mathbf{2},1}-\sqrt{3}Y^{(2)}_{\mathbf{2},2}|}\frac{g^2v_u^2}{\Lambda}\,,\quad
m_3=\frac{1}{|Y^{(2)}_{\mathbf{2},1}+\sqrt{3}Y^{(2)}_{\mathbf{2},2}|}\frac{g^2v_u^2}{\Lambda}\,.
\end{equation}
In this model, the lepton mass matrices as well as flavor observables only depend on four constants $\alpha$, $\beta$, $\gamma$, $g^2/\Lambda$ plus the complex modulus $\tau$. It is the minimal modular invariant model in the present literature as far as we know. We find that the experimental data of lepton masses and flavor mixing can be accommodated only if the neutrino mass is normal ordering. The best-fit values of the input parameters are determined to be\footnote{The determinant of the charged lepton mass matrix is
\begin{eqnarray*}
\det[M_e(\tau)]=-96\sqrt{6}v_d^3\gamma(\beta^2-3\alpha^2)\eta^{18}(\tau)\,.
\end{eqnarray*}
Consequently the small electron mass can be naturally reproduced for $\beta\approx\pm\sqrt{3}\alpha$. This is the reason why the best fit value $\beta/\alpha= 1.73048$ which is quite close to $\sqrt{3}$.}
\begin{equation}
\begin{gathered}
\label{eq:best-fit values}
\langle \tau \rangle =-0.193773+1.08321i\,,\quad\beta/\alpha= 1.73048\,,\quad \gamma/\alpha=0.27031\,,\\
\alpha v_d = 244.621~ \text{MeV}\,,\quad g^2v^2_u/\Lambda= 29.0744~\text{meV}\,,
\end{gathered}
\end{equation}
where $\alpha v_d$ and $g^2v^2_u/\Lambda$ are fixed by the measured values of the electron mass and the solar mass squared splitting $\Delta m^2_{21}$ respectively~\cite{Esteban:2020cvm}. At this best fit point, the lepton masses and flavor mixing parameters are given by
\begin{eqnarray}
\nonumber&\sin^2\theta_{12}=0.328920\,,~~ \sin^2\theta_{13}=0.0218499\,,~~ \sin^2\theta_{23}=0.506956\,,~~ \delta_{CP}=1.34256\pi\,,\\
\nonumber&\alpha_{21}= 1.32868\pi \,,~~\alpha_{31}= 0.544383\pi \,,~~ m_e/m_\mu=0.00472633 ,~~m_\mu/m_\tau=0.0587566\,,  \\
\nonumber &m_1=14.4007~\text{meV}\,,~~ m_2 =16.7803~\text{meV}\,,~~ m_3 =51.7755~\text{meV}\,,\\
&m_\beta=16.8907~\text{meV} \,,~~ m_{\beta\beta}=9.25333~\text{meV}\,,
\end{eqnarray}
which are within the $3\sigma$ intervals of the latest global fit NuFIT v5.1 without SK atmospheric data~\cite{Esteban:2020cvm}. The charged lepton mass ratios are compatible with their renormalization group (RG) running values at the GUT scale $2\times 10^{16}$~GeV, where $M_{\text{SUSY}}=1~\mathrm{TeV}\,,\tan\beta=5$ is taken as a benchmark~\cite{Antusch:2013jca}. Here $m_\beta$ is the effective neutrino mass probed by direct kinematic search in tritium beta decay and $m_{\beta\beta}$ is the effective mass in neutrinoless double beta decay. Our predictions for these two effective neutrino masses are far below the present experimental bounds $m_{\beta}<0.8$ eV from KATRIN~\cite{KATRIN:2021uub} and $m_{\beta\beta}\leq (36-156)$ meV from KamLAND-Zen~\cite{KamLAND-Zen:2022tow} at $90\%$ confidence level. Moreover, the neutrino mass sum is predicted to be $m_1+m_2+m_3=82.9565$ meV which is compatible with the upper limit of Planck $\sum_i m_i < 120$ meV~\cite{Planck:2018vyg}.

\begin{figure}[t!]
\centering
\includegraphics[width=0.65\textwidth]{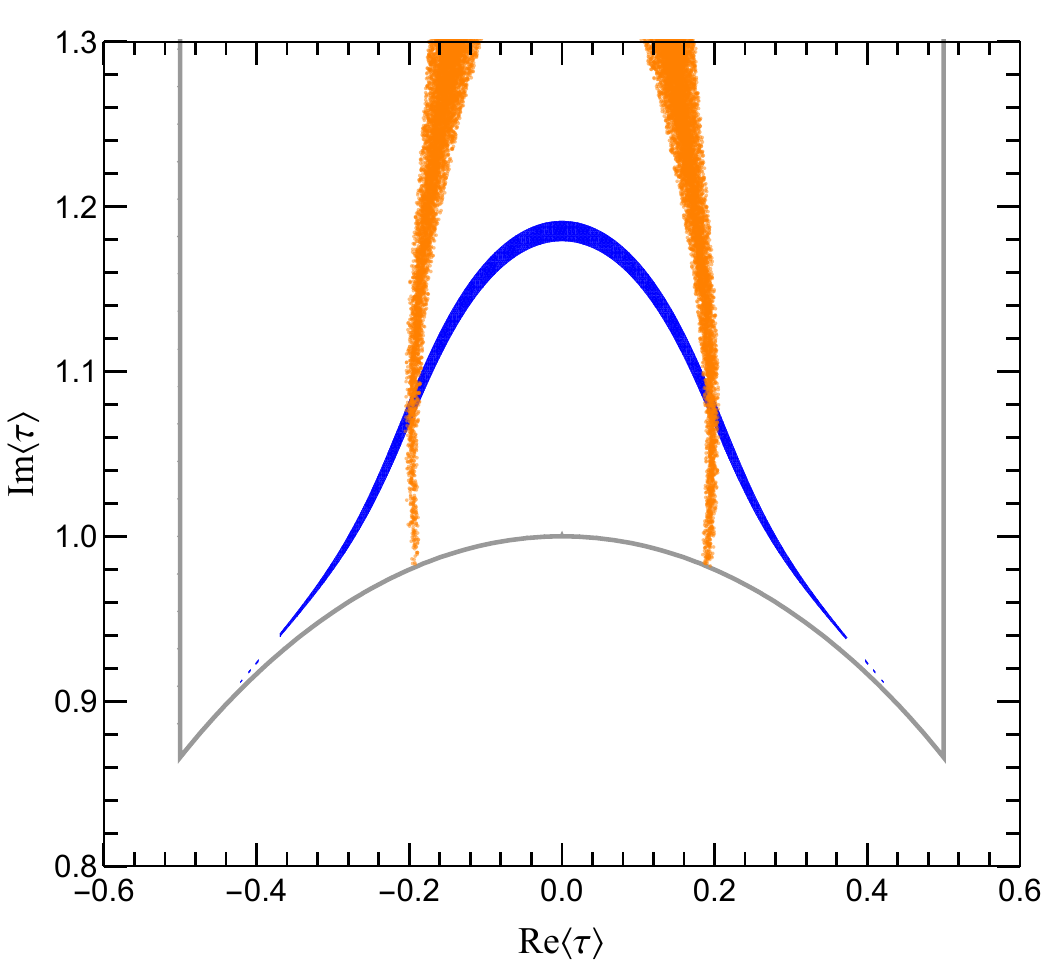}
\caption{The region of modulus $\tau$ compatible with experimental data, where the gray line is the boundary of the fundamental domain. The blue region represents the feasible range of $\langle\tau\rangle$ compatible with the data $\Delta m^2_{21}/\Delta m^2_{31}$ of the neutrino mass squared difference~\cite{Esteban:2020cvm}. The orange area denotes  the viable region of $\langle\tau\rangle$ limited only by the measured values of the charged lepton mass ratios and the reactor mixing angle $\theta_{13}$~\cite{Antusch:2013jca,Esteban:2020cvm}. }
\label{fig:tauRegion}
\end{figure}

It is notable that the light neutrino masses only depend on the complex modulus $\tau$ up to the overall scale $g^2v^2_u/\Lambda$. Therefore the experimental data of $\Delta m^2_{21}/\Delta m^2_{31}$ can efficiently  constrain the range of $\langle\tau\rangle$, where $\Delta m^2_{21}\equiv m^2_2-m^2_1$ and $\Delta m^2_{31}\equiv m^2_3-m^2_1$ are the solar and atmospheric neutrino mass squared differences respectively, as is shown in the blue region of figure~\ref{fig:tauRegion}. After further including the precisely measured values of the reactor angle $\theta_{13}$ and the charged lepton mass ratios $m_e/m_{\mu}$, $m_{\mu}/m_{\tau}$, the modulus VEV $\langle\tau\rangle$ would be limited in two small regions around $-0.19+1.08i$ and $0.19+1.08i$.

\begin{figure}[hptb!]
\centering
\includegraphics[width=1.0\textwidth]{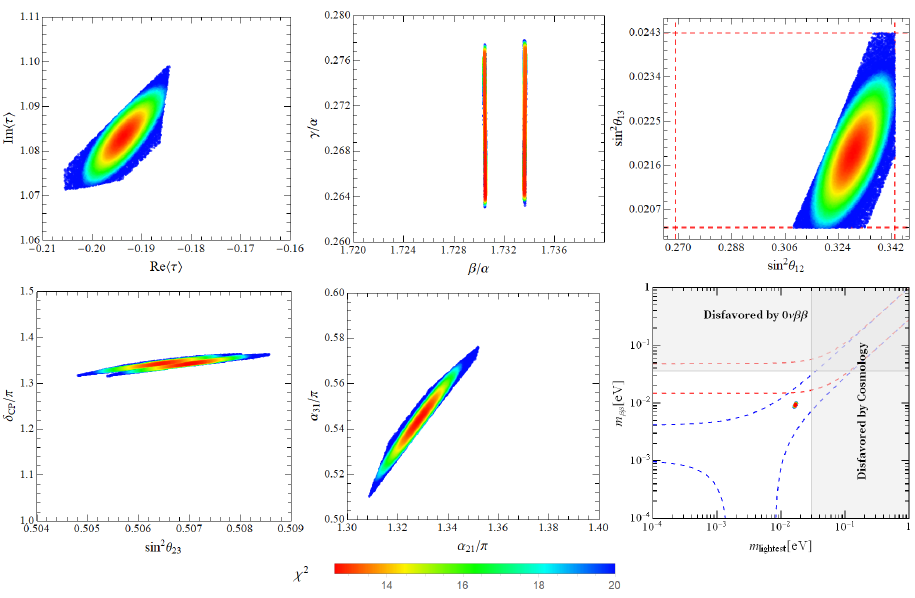}
\caption{The predicted correlations between the input free parameters, neutrino mixing angles, and CP violation phases in the minimal model. The plots only display the points that can reproduce the charged lepton masses, $\Delta m^2_{21}$, $\Delta m^2_{31}$ and all the three lepton mixing angles at $3\sigma$ level~\cite{Esteban:2020cvm}. In the top-right panel, the red dashed lines are the $3\sigma$ bounds of the mixing angles. In the bottom right panel for $m_{\beta\beta}$, the blue (red) dashed lines represent the most general allowed regions for normal ordering (inverted ordering) neutrino mass spectrum, where the neutrino oscillation parameter are varied within their $3\sigma$ ranges. Moreover, the vertical grey exclusion band stands for the most radical upper bound $\sum_i m_i<0.12~\text{eV}$ from Planck~\cite{Planck:2018vyg}. The horizontal grey band represents the present upper limit $m_{\beta\beta}\leq (36-156)$ meV from KamLAND-Zen~\cite{KamLAND-Zen:2022tow}.}
\label{fig:correlations}
\end{figure}

Furthermore, we explore the parameter space of this minimal model. Requiring the three charged lepton masses $m_{e, \mu, \tau}$, the three lepton mixing angles $\theta_{12}$, $\theta_{13}$, $\theta_{23}$ and the neutrino squared
mass splittings $\Delta m^2_{21}$ and $\Delta m^2_{31}$ to lie in the experimentally allowed $3\sigma$ regions~\cite{Esteban:2020cvm}, the
correlations between the free parameters and flavor observables are shown in figure~\ref{fig:correlations}. We see that the atmospheric neutrino mixing angle $\sin^2\theta_{23}$ is predicted to lie in the range of $[0.504, 0.510]$ which is in the second octant, the Dirac CP violation phase $\delta_{CP}$ is in the narrow interval $[1.316\pi, 1.364\pi]$. These predictions for $\theta_{23}$ and $\delta_{CP}$ could be tested in future long baseline neutrino experiments DUNE~\cite{DUNE:2015lol} and T2HK~\cite{Hyper-KamiokandeProto-:2015xww}. Moreover, the neutrino mixing angles $\sin^2\theta_{12}$ and $\sin^2\theta_{13}$ also show a certain correlation, this feature is expected to be tested at JUNO~\cite{JUNO:2015sjr} which can measure the solar angle $\theta_{12}$ with sub-percent precision. Moreover, the Majorana CP violation phases are also found to lie in quite narrow regions $\alpha_{21}\in [1.309\pi, 1.352\pi]$ and $\alpha_{31}\in[0.510\pi, 0.576\pi]$. Consequently we have a definite prediction for the effective Majorana mass $m_{\beta\beta}$ in the interval $[8.543~\text{meV}, 10.010~\text{meV}]$ which is within the sensitivity of future ton-scale neutrinoless double beta decay experiments.

\subsection{Extension of $S'_4$ modular model to quark sector without GUTs}

As summarized in table~\ref{tab:modular-models-summary}, the modular symmetry is explored to explain the lepton masses and mixing in most case. In a similar way, the modular symmetry can also help to understand the quark mass hierarchies and CKM mixing matrix. It is highly nontrivial to address the experimental data of both quark and lepton sectors with a single modular symmetry in terms of a small number of free parameters. We see that 16 real input parameters~\cite{Lu:2019vgm} or more~\cite{Okada:2018yrn,Okada:2019uoy,Okada:2020rjb} are required to describe the masses and mixing patterns of quark and lepton in the models based on level $N=3$ modular group. In the following, we shall present a  $S'_4$ modular symmetry model which can explain the flavor structure of quarks and leptons simultaneously for a common value of modulus $\tau$, and this model involves only 15 real free parameters including the real and imaginary parts of the modulus $\tau$~\cite{Liu:2020akv}.

This model is formulated in the framework of minimal supersymmetric standard model (MSSM), and the gCP symmetry is imposed so that all coupling constants are real in our working basis. The light neutrino masses are assumed to be generated by the type-I seesaw mechanism. The three generations of left-handed lepton doublets $L$ and right-handed neutrinos $N^c$ are assigned to two triplets $\mathbf{3}$ of $S'_4$, while the right-handed charged leptons $E_1^c$, $E_2^c$ and $E_3^c$ transform as $\mathbf{1}$, $\mathbf{1}$ and $\mathbf{\hat{1}'}$ respectively. The modular weights of lepton fields are chosen as $k_{N^c}=0$, $k_L=2$, $k_{E_1^c}=2$, $k_{E_2^c}=0$ and $k_{E_3^c}=1$. Then the modular invariant superpotential for lepton masses is given by
\begin{eqnarray}
\nonumber && \mathcal{W}_{e}=\alpha_e (E^{c}_{1}L Y_{\mathbf{3}}^{(4)} )_{\mathbf{1}}H_{d}+\beta_e (E^{c}_{2}L Y_{\mathbf{3}}^{(2)} )_{\mathbf{1}}H_{d} +\gamma_e (E^{c}_{3}L  Y_{\mathbf{\hat{3}}}^{(3)} )_{\mathbf{1}}H_{d}\,,\\
&& \mathcal{W}_{\nu}=g_1( N^{c} LY_{\mathbf{2}}^{(2)})_{\mathbf{1}}H_{u}+g_2(N^{c} L Y_{\mathbf{3}}^{(2)})_{\mathbf{1}}H_{u}+\Lambda (N^{c}N^{c})_{\mathbf{1}}\,,
\label{eq:unification-S4p}
\end{eqnarray}
from which we can read out the lepton mass matrices as follows,
\begin{eqnarray}
\label{eq:Ml-lepton-quark}
\nonumber&& M_{e}=\begin{pmatrix}
\alpha_e  Y_{\mathbf{3}, 1}^{(4)} ~&~\alpha_e Y_{\mathbf{3}, 3}^{(4)} ~&~ \alpha_e  Y_{\mathbf{3}, 2}^{(4)}\\
\beta_e  Y_{\mathbf{3}, 1}^{(2)} ~&~\beta_e Y_{\mathbf{3}, 3}^{(2)} ~&~\beta_e Y_{\mathbf{3}, 2}^{(2)} \\
\gamma_e  Y_{\mathbf{\hat{3}}, 1}^{(3)} ~&~ \gamma_e Y_{\mathbf{\hat{3}}, 3}^{(3)} ~&~ \gamma_e  Y_{\mathbf{\hat{3}}, 2}^{(3)}
\end{pmatrix}v_{d}\,,~\quad~ M_{N}=\begin{pmatrix}
 1 ~&~ 0 ~&~ 0 \\
 0 ~&~ 0 ~&~ 1 \\
 0 ~&~ 1 ~&~ 0
\end{pmatrix}\Lambda\,, \\
&& M_D=\begin{pmatrix}
2g_1 Y_{\mathbf{2}, 1}^{(2)}  ~& g_2Y_{\mathbf{3}, 3}^{(2)}  ~&  -g_2Y_{\mathbf{3}, 2}^{(2)} \\
-g_2Y_{\mathbf{3}, 3}^{(2)}  ~&\sqrt{3}\,g_1 Y_{\mathbf{2}, 2}^{(2)} ~& -g_1 Y_{\mathbf{2}, 1}^{(2)}+g_2Y_{\mathbf{3}, 1}^{(2)} \\
g_2Y_{\mathbf{3}, 2}^{(2)}  ~& -g_1 Y_{\mathbf{2}, 1}^{(2)}-g_2Y_{\mathbf{3}, 1}^{(2)}  ~&\sqrt{3}\,g_1 Y_{\mathbf{2}, 2}^{(2)}
\end{pmatrix}v_{u}\,.
\end{eqnarray}
The phases of $\alpha_e$, $\beta_e$, $\gamma_e$, $g_1$ and $\Lambda$ can be absorbed by lepton fields, and the coupling $g_2$ is complex (real) without (with) gCP. As regards the representation assignment of quark fields, we assign the left-handed quarks $Q$ to be triplet $\mathbf{3}$ of $S'_4$, $u^c$, $c^c$ and $t^c$ transform as $\mathbf{1}$, $\mathbf{1}$ and $\mathbf{\hat{1}'}$ respectively under $S'_4$, the down type quarks $d^c,\,s^c,\,b^c$ are $S'_4$ singlets $\mathbf{1}',\,\mathbf{\hat{1}},\,\mathbf{\hat{1}'}$ respectively.
Although both $u^c$ and $c^c$ are invariant under $S'_4$, their modular weights are different. The modular weights of quark fields are chosen to satisfy the relations $k_{Q}=4-k_{u^c}=6-k_{c^c}=3-k_{t^c}=4-k_{d^c}=5-k_{s^c}=5-k_{b^c}$. We can read off the superpotential of the quark sector as follows,
\begin{eqnarray}
\nonumber&& \mathcal{W}_{u}=\alpha_{u}(u^{c}Q Y_{\mathbf{3}}^{(4)} )_{\mathbf{1}}H_{u}+\beta_{u}(c^{c}QY_{\mathbf{3}I}^{(6)} )_{\mathbf{1}}H_{u}+\gamma_{u}( c^{c}QY_{\mathbf{3}II}^{(6)} )_{\mathbf{1}}H_{u}+\delta_{u}(t^{c}QY_{\mathbf{\hat{3}}}^{(3)})_{\mathbf{1}}H_{u}\,,\\
&& \mathcal{W}_{d}=\alpha_{d}(d^{c}Q Y_{\mathbf{3}'}^{(4)} )_{\mathbf{1}}H_{d}+\beta_{d}(s^{c} QY_{\mathbf{\hat{3}'}I}^{(5)} )_{\mathbf{1}}H_{d}+\gamma_{d}(s^{c} QY_{\mathbf{\hat{3}'}II}^{(5)})_{\mathbf{1}}H_{d}+\delta_{d}(b^{c}Q Y_{\mathbf{\hat{3}}}^{(5)})_{\mathbf{1}}H_{d} \,.\quad~~~
\label{eq:Wq-quark-lepton}
\end{eqnarray}
Then the quark mass matrices are given by
\begin{eqnarray}
\label{eq:Mq-quark-lepton}
\nonumber&& M_{u}=\begin{pmatrix}
\alpha_{u} Y_{\mathbf{3}, 1}^{(4)} ~&~\alpha_{u} Y_{\mathbf{3}, 3}^{(4)} ~&~\alpha_{u} Y_{\mathbf{3}, 2}^{(4)} \\
\beta_{u} Y_{\mathbf{3}I, 1}^{(6)}+\gamma_{u} Y_{\mathbf{3}II, 1}^{(6)} ~&~\beta_{u} Y_{\mathbf{3}I, 3}^{(6)}+\gamma_{u} Y_{\mathbf{3}II, 3}^{(6)} ~&~\beta_{u} Y_{\mathbf{3}I, 2}^{(6)}+\gamma_{u} Y_{\mathbf{3}II, 2}^{(6)} \\
 \delta_{u} Y_{\mathbf{\hat{3}}, 1}^{(3)} ~&~ \delta_{u} Y_{\mathbf{\hat{3}}, 3}^{(3)} ~&~ \delta_{u} Y_{\mathbf{\hat{3}}, 2}^{(3)} \\
\end{pmatrix}v_{u}\,, \\[0.1in]
&& M_{d}=\begin{pmatrix}
\alpha_{d} Y_{\mathbf{3}', 1}^{(4)} ~&~\alpha_{d} Y_{\mathbf{3}', 3}^{(4)} ~&~\alpha_{d} Y_{\mathbf{3}', 2}^{(4)} \\
\beta_{d} Y_{\mathbf{\hat{3}'}I, 1}^{(5)}+\gamma_{d} Y_{\mathbf{\hat{3}'}II, 1}^{(5)} ~&~\beta_{d} Y_{\mathbf{\hat{3}'}I, 3}^{(5)}+\gamma_{d} Y_{\mathbf{\hat{3}'}II, 3}^{(5)} ~&~\beta_{d} Y_{\mathbf{\hat{3}'}I, 2}^{(5)}+\gamma_{d} Y_{\mathbf{\hat{3}'}II, 2}^{(5)} \\
\delta_{d} Y_{\mathbf{\hat{3}}, 1}^{(5)} ~&~ \delta_{d} Y_{\mathbf{\hat{3}}, 3}^{(5)} ~&~ \delta_{d} Y_{\mathbf{\hat{3}}, 2}^{(5)}
\end{pmatrix}v_{d}\,.
\end{eqnarray}
The parameters $\alpha_{u,d}$, $\beta_{u,d}$ and $\delta_{u,d}$ can be made real and positive by field redefinition while $\gamma_{u}$ and $\gamma_{d}$ are complex. If the gCP symmetry is imposed on the model, all couplings are constrained to be real, and $\gamma_{u}$ and $\gamma_{d}$ are either positive or negative.

It is notable that the model has less free parameters  than the number of observable quantities including the masses and mixing parameters of quarks and leptons. We perform a comprehensive numerical scan over the parameter space, we find that good agreement with experimental data can be achieved for the following value of the modulus $\tau$ common to quark and lepton sectors,
\begin{equation}
\langle\tau\rangle=-0.2111+1.5201i\,,
\end{equation}
which is mainly determined by the quark masses and CKM mixing parameters. Given this value of $\tau$, the charged lepton masses can be reproduced by adjusting $\alpha_e$, $\beta_{e}$ and $\gamma_{e}$, only two real parameters $g_1^2v_u^2/\Lambda$ and $g_2/g_1$ describe the nine neutrino observables: three neutrino masses, three neutrino mixing angles and three CP violating phases. The best fit values of the free parameters of both lepton and quark sectors are found to be
\begin{eqnarray}
\nonumber&&\hskip-0.3in\beta_u/\alpha_u=1711.9384\,,~~ \gamma_{u}/\alpha_u=25512.4486\,,~~\delta_{u}/\alpha_u=90.5459\,,\\
\nonumber&&\hskip-0.3in \alpha_u v_u= 2.9637\times 10^{-3}~\text{GeV}\,,~~\beta_d/\alpha_d=234.6646\,,~~\gamma_{d}/\alpha_d=235.3665\,,\\
\nonumber&&\hskip-0.3in  \delta_{d}/\alpha_d=4.0687\,,~~\alpha_d v_d= 4.5279\times 10^{-3}~\text{GeV}\,,~~\beta_{e}/\alpha_e=1.8514\times 10^{-3}\,,\\
&& \hskip-0.3in  \gamma_{e}/\alpha_e=0.0622\,,~~ g_2/g_1=-1.3682\,,~\alpha_e v_d=1.2485~\text{GeV}\,,~g_1^2v^2_u/\Lambda= 8.1856 ~\text{meV}\,.
\end{eqnarray}
The masses and mixing parameters of quarks and leptons are predicted to be
\begin{eqnarray}
\nonumber&&\theta^q_{12}=0.22751\,,~~ \theta^q_{13}=0.003379\,,~~ \theta^q_{23}=0.038885\,,~~ \delta^q_{CP}=75.5572^{\circ}\,,\\
\nonumber&&m_u/m_c=0.001908\,,~~ m_c/m_t=0.002724\,,~~ m_d/m_s=0.050120\,,~~ m_s/m_b=0.017707\,,\\
\nonumber&&\sin^2\theta^l_{12}=0.35020\,,~~ \sin^2\theta^l_{13}=0.02179\,,~~ \sin^2\theta^l_{23}=0.56443\,,\\
\nonumber&&\delta^l_{CP}=266.3662^{\circ}\,,~~\alpha_{21}=1.1476\pi\,,~~\alpha_{31}=0.1515\pi\,,\\
\nonumber&&m_1=3.5234 ~\text{meV}\,,~~ m_2=9.3013 ~\text{meV}\,,~~ m_3=50.0901 ~\text{meV}\,,\\
\label{eq:num-quark-lepton-S4p}&&\sum_im_i=62.9149 ~\text{meV}\,,~~ |m_{\beta\beta}|=2.5405~\text{meV}\,.
\end{eqnarray}
All the observables are in the experimentally preferred regions except that the solar mixing angle $\theta^l_{12}$ is slightly larger than its $3\sigma$ upper bound. The sum of neutrino masses is determined to be 62.9149 meV, this is compatible with the latest bound $\sum_im_i<120$ meV at $95\%$ confidence level from Planck~\cite{Planck:2018vyg}.

Recently a more predictive model for quark and lepton has been constructed with the modular binary octahedral symmetry $2O$~\footnote{As shown in table~\ref{tab:NorSubgroupSL2Z}, the binary octahedral group $2O$ belong to the general finite modular group and its group ID is [48, 28]. The group $2O$ is the Schur cover of $S_4$ of ``$-$'' type but distinct
from $S'_4$.}, and the experimental data can be accommodated by using only 14 real free parameters~\cite{Ding:2023ydy}. It is the minimal modular invariant model for quarks and lepton at present.

\subsection{\label{subsec:modular-littlest-seesaw}Littlest seesaw model with modular $S_4^3$ symmetry and three moduli}

As explained in section~\ref{subsec:fixed-points-SL2Z} and summarized in table~\ref{tab:stabilizer-modular}, there are only three modular symmetry fixed points $\tau_0=i$, $e^{2\pi i/3}$ and $i\infty$ in the fundamental domain, which are invariant under the action of $\gamma_0=S$, $ST$ and $T$ respectively. At the fixed point $\tau_0$ fulfilling $\gamma_0\tau_0=\tau_0$ in Eq.~\eqref{eq:gamm0-tau0}, using Eq.~\eqref{eq:MF-decomp} and setting $\gamma=\gamma_0$ and $\tau=\tau_0$,  we find that the weight $k$ modular multiplet $Y^{(k)}_{\mathbf{r}}(\tau)$ fulfills the following relation,
\begin{equation}
\label{eq:MF_align}\rho_\mathbf{r}(\gamma_0)Y^{(k)}_\mathbf{r}(\tau_0)=J^{-k}(\gamma_0,\tau_0)Y^{(k)}_\mathbf{r}(\tau_0)\,,
\end{equation}
where $J(\gamma, \tau)=c\tau+d$ is the automorphy factor defined in Eq.~\eqref{eq:J-automorphy-fac}. One can straightforwardly obtain
\begin{equation}
J(\gamma_0,\tau_0)=\left\{\begin{array}{cc}
-i\,, ~&~ \gamma_0=S,~~\tau_0=i \\
\omega^2\,, ~&~ \gamma_0=ST,~~\tau_0=\omega \\
1\,,  ~&~\gamma_0=T,~~\tau_0=i\infty
\end{array}
\right.\,.
\end{equation}
Hence the modular multiplets $Y^{(k)}_\mathbf{r}(\tau_0)$ at the fixed point $\tau_0$ is actually the eigenvector of the representation matrix $\rho_\mathbf{r}(\gamma_0)$ with eigenvalue $J^{-k}(\gamma_0,\tau_0)$.
As a consequence, the alignment of $Y^{(k)}_\mathbf{r}(\tau_0)$ can be easily fixed once the presentation basis is fixed.

Moreover, there are infinite fixed points in the upper half plane with
$\tau_f=\gamma'\tau_0$, $\gamma_f=\gamma'\gamma_0\gamma'^{-1}$ satisfying $\gamma_f\tau_f=\tau_f$ for any element $\gamma'\in\overline{\Gamma}$, as shown in Eq.~\eqref{eq:fp-general}. The value of the modular form at the fixed point $\gamma_f$ satisfies\footnote{After some algebra calculation,  we can check that the modular multiplet $Y^{(k)}_\mathbf{r}(\tau_f)$ also fulfills the following property
\begin{eqnarray*}
\rho_\mathbf{r}(\gamma_f)Y^{(k)}_\mathbf{r}(\tau_f) = J^{-k}(\gamma_0,\tau_0)Y^{(k)}_\mathbf{r}(\tau_f)\,.
\end{eqnarray*}}
\begin{equation}
\label{eq:Y-tauf}Y^{(k)}_\mathbf{r}(\tau_f)= Y^{(k)}_\mathbf{r}(\gamma'\tau_0) = J^k(\gamma', \tau_0)\rho_\mathbf{r}(\gamma')Y^{(k)}_\mathbf{r}(\tau_0)\,.
\end{equation}
Therefore the alignment of the modular multiplet $Y^{(k)}_\mathbf{r}(\tau_f)$ is proportional to $\rho_\mathbf{r}(\gamma')Y^{(k)}_\mathbf{r}(\tau_0)$.
If $\gamma'\in\Gamma(N)$, then $\rho_\mathbf{r}(\gamma')=\mathds{1}$ so that we have only a finite number of independent directions of $Y^{k)}_\mathbf{r}(\tau_f)$ although there are infinite numbers of fixed points $\tau_f$. It is sufficient to only consider these fixed points $\tau_f$ with $\gamma'\in\Gamma_N (\Gamma'_N)$. We give some fixed points and the corresponding alignments of the triplet modular forms of level 4 in table~\ref{tab:MF-tauf-alignment}, see Ref.~\cite{Ding:2019gof} for the complete results for all the nonequivalent fixed points.

\begin{table}[t!]
\begin{center}
\resizebox{1.0\textwidth}{!}{
\begin{tabular}{|c|c|c|c|c|}\hline\hline
 $\tau$ & \multicolumn{2}{c|}{$Y^{(2)}_\mathbf{3}(\tau)$, $Y^{(6)}_\mathbf{3,I}(\tau)$} & $Y^{(4)}_\mathbf{3}(\tau)$, $Y^{(6)}_\mathbf{3'}(\tau)$ & $Y^{(4)}_\mathbf{3'}(\tau)$, $Y^{(6)}_\mathbf{3,II}(\tau)$ \\
\hline\hline
 $i$ & \multicolumn{2}{c|}{$(1,1+\sqrt{6},1-\sqrt{6})$} & $(1,-\frac{1}{2},-\frac{1}{2})$ & $(1,1-\sqrt{\frac{3}{2}},1+\sqrt{\frac{3}{2}})$  \\
\hline
 $i+1$ & \multicolumn{2}{c|}{$(1,-\frac{\omega}{3}(1+i\sqrt{2}),-\frac{\omega^2}{3}(1+i\sqrt{2}))$} & $(0,1,-\omega)$ & $(1,\frac{i\omega}{\sqrt{2}},\frac{i\omega^2}{\sqrt{2}})$  \\
\hline
$i+2$ & \multicolumn{2}{c|}{$(1,\frac{1}{3}(-1+i\sqrt{2}),\frac{1}{3}(-1+i\sqrt{2}))$} & $(0,1,-1)$ & $(1,-\frac{i}{\sqrt{2}},-\frac{i}{\sqrt{2}})$  \\
\hline
 $i+3$ & \multicolumn{2}{c|}{$(1,\omega(1+\sqrt{6}),\omega(1-\sqrt{6}))$} & $(1,-\frac{\omega}{2},-\frac{\omega^2}{2})$ & $(1,\omega(1-\sqrt{\frac{3}{2})},\omega^2(1+\sqrt{\frac{3}{2}}))$ \\
\hline\hline
$\tau$ & $Y^{(2)}_\mathbf{3}(\tau)$ & $Y^{(4)}_\mathbf{3}(\tau)$,$Y^{(4)}_\mathbf{3'}(\tau)$ & $Y^{(6)}_\mathbf{3,II}(\tau)$,$Y^{(6)}_\mathbf{3'}(\tau)$ & $Y^{(6)}_\mathbf{3,I}(\tau)$\\ \hline \hline
 $\omega$ & $(0,1,0)$ & $(0,0,1)$ & $(1,0,0)$ & \multirow{7}*{$(0,0,0)$}\\
\cline{1-4}
 $\omega+1$ & $(1,1,-\frac{1}{2})$ & $(1,-\frac{1}{2},1)$& $(1,-2,-2)$  &  \\
\cline{1-4}
 $\omega+2$ & $(1,-\frac{\omega^2}{2},\omega)$ & $(1,\omega^2,-\frac{\omega}{2})$ & $(1,-2\omega^2,-2\omega)$& \\
\cline{1-4}
 $\omega+3$ & $(1,\omega,-\frac{\omega^2}{2})$ & $(1,-\frac{\omega}{2},\omega^2)$ & $(1,-2\omega,-2\omega^2)$ & \\
\cline{1-4}
 $\rho/\sqrt{3}$ & $(1,-\frac{\omega}{2},\omega^{2})$ & $(1,\omega,-\frac{\omega^{2}}{2})$ & $(1,-2\omega,-2\omega^{2})$ & \\
 \cline{1-4}
  $\rho/\sqrt{3}+1$ & $(0,0,1)$ & $(0,1,0)$ & $(1,0,0)$ & \\
 \cline{1-4}
   $\rho/\sqrt{3}+2$ & $(1,-\frac{1}{2},1)$ & $(1,1,-\frac{1}{2})$ & $(1,-2,-2)$ & \\
 \cline{1-4}
 $\rho/\sqrt{3}+3$ & $(1,\omega^2,-\frac{\omega}{2})$ & $(1,-\frac{\omega^2}{2},\omega)$ &$(1,-2\omega^2,-2\omega)$ & \\
\hline\hline
\end{tabular} }
\caption{\label{tab:MF-tauf-alignment} The alignments of triplet modular forms $Y_{\mathbf{3}, \mathbf{3'}}(\tau)$ of level 4 up to weight 6 with the available fixed moduli in orbifolds, where $\rho =e^{i\pi/6}$. We have ignored the overall constant appearing in each alignment.}
\end{center}
\end{table}

These alignments at the fixed points could give a rich phenomenology of neutrino mixing in the framework of tri-direct modular approach~\cite{Ding:2019gof}. In the following, we shall present a Littlest seesaw model with multiple modular symmetry $S^A_4\times S^B_4\times S^C_4$~\cite{deMedeirosVarzielas:2022fbw}. In the following, we adopt the $S_4$ basis of Ref.~\cite{Ding:2019gof} where the triplet representations $\mathbf{3}$ and $\mathbf{3}'$ correspond to $\mathbf{3}'$ and $\mathbf{3}$ of~\cite{deMedeirosVarzielas:2022fbw} respectively. The two columns of the Dirac neutrino mass matrix are triplet modular forms of level 4 with the following alignment at the fixed points\footnote{The second column could also chosen as $\tau_B =\frac{8+i}{5}$, $Y_{ \mathbf{3}}^{(2)} (\tau_B ) = (1, 1 -\sqrt{6}, 1 +\sqrt{6})$.},
\begin{eqnarray}
\nonumber \tau_A = i+2 &:& \quad  Y_{\mathbf{3}}^{(4)} (\tau_A ) = (0, 1, -1) \,, \\
\label{eq:neutrino-Mod-Littlest}\tau_B =i &:& \quad Y_{ \mathbf{3}}^{(2)} (\tau_B ) = (1, 1 +\sqrt{6}, 1 -\sqrt{6}) \,.
\end{eqnarray}
The charged lepton Yukawa coupling matrix is enforced to be diagonal through the weights $2$, $4$, and $6$  modular forms in the triplet representation
$\mathbf{3}$ of $S_4$ at $\tau_C=\omega$:
\begin{equation}
\label{eq:Ye-Mod-Littlest}\tau_C = \omega ~~:~~ Y_{\mathbf{3}}^{(2)} (\tau_C ) = (0, 1, 0), ~~Y_{ \mathbf{3}}^{(4)} (\tau_C ) = (0, 0, 1),~~ Y^{(6)}_\mathbf{3,II} (\tau_C ) = (1, 0, 0)\,.
\end{equation}
Notice that the above fixed points as well as the vacuum alignments in Eqs.~(\ref{eq:neutrino-Mod-Littlest}, \ref{eq:Ye-Mod-Littlest}) are preferred by orbifold constructions~\cite{deAnda:2023udh}.

\begin{table}[hptb]
\centering
\begin{tabular}{| l | c c c c c c|}
\hline \hline
Fields & $S_4^A$ & $S_4^B$ & $S_4^C$ & \!$k_A$\! & \!$k_B$\! & \!$k_C$\!\\
\hline \hline
$L$ & $\mathbf{1}$ & $\mathbf{1}$ & $\mathbf{3}$ & 0 & 0 & 0\\
$e^c$ & $\mathbf{1}$ & $\mathbf{1}$ & $\mathbf{1}$ & 0 & 0 & \!$-6$\! \\
$\mu^c$ & $\mathbf{1}$ & $\mathbf{1}$ & $\mathbf{1}$ & 0 & 0 & \!$-4$\! \\
$\tau^c$ & $\mathbf{1}$ & $\mathbf{1}$ & $\mathbf{1}$ & 0 & 0 & \!$-2$\! \\
$N_A^c$ & $\mathbf{1}$ & $\mathbf{1}$ & $\mathbf{1}$ & $-4$ & 0 & 0 \\
$N_B^c$ & $\mathbf{1}$ & $\mathbf{1}$ & $\mathbf{1}$ & 0 & $-2$ & 0 \\
\hline
$\Phi_{AC}$ & $\mathbf{3}$ & $\mathbf{1}$ & $\mathbf{3}$ & 0 & 0 & 0 \\
$\Phi_{BC}$ & $\mathbf{1}$ & $\mathbf{3}$ & $\mathbf{3}$ & 0 & 0 & 0 \\
\hline \hline
\end{tabular}~~
\begin{tabular}{| l | c c c c c c|}
\hline \hline
Yuk/Mass &$S_4^A$ & $S_4^B$ & $S_4^C$ & $k_A$ & $k_B$ & $k_C$\\
\hline \hline
$Y_e(\tau_C)$ & $\mathbf{1}$ & $\mathbf{1}$ & $\mathbf{3}$ & 0 & 0 & $6$ \\
$Y_\mu(\tau_C)$ & $\mathbf{1}$ & $\mathbf{1}$ & $\mathbf{3}$ & 0 & 0 & $4$ \\
$Y_\tau(\tau_C)$ & $\mathbf{1}$ & $\mathbf{1}$ & $\mathbf{3}$ & 0 & 0 & $2$ \\
$Y_A(\tau_A)$ & $\mathbf{3}$ & $\mathbf{1}$ & $\mathbf{1}$ & $4$ & 0 & 0 \\
$Y_B(\tau_B)$ & $\mathbf{1}$ & $\mathbf{3}$ & $\mathbf{1}$ & 0 & $2$ & 0 \\\hline
$M_A(\tau_A)$ & $\mathbf{1}$ & $\mathbf{1}$ & $\mathbf{1}$ & $8$ & 0 & 0 \\
$M_B(\tau_B)$ & $\mathbf{1}$ & $\mathbf{1}$ & $\mathbf{1}$ & 0 & $4$ & 0
\\
\hline \hline
\end{tabular}
\caption{\label{tab:modular-Littlest-seesaw-model}Transformation properties of fields and modular forms (Yuk/Mass) under the modular symmetries $S^A_4\times S^B_4\times S^C_4$ in the Littlest modular seesaw model~\cite{deMedeirosVarzielas:2022fbw}.}
\end{table}

The transformation of the matter fields and the relevant modular forms under the finite modular symmetry group $S^A_4\times S^B_4\times S^C_4$ are listed in table~\ref{tab:modular-Littlest-seesaw-model}, and the $SU(2)$ doublets $H_{u,d}$ are assumed to  transform trivially under all flavour symmetries~\cite{deMedeirosVarzielas:2022fbw}. The superpotential in the  charged lepton and neutrino sectors is given by
\begin{eqnarray}\label{eq:superpot}
\mathcal{W}_\ell &=&\left[y_e L Y_e(\tau_C) e^c + y_{\mu} L Y_\mu(\tau_C) \mu^c + y_{\tau} L Y_\tau(\tau_C) \tau^c \right] H_d \nonumber \\
\nonumber &&+ \frac{1}{\Lambda}\left[y_{a} L \Phi_{AC} Y_A(\tau_A) N_A^c + y_{s} L \Phi_{BC} Y_B(\tau_B) N_B^c \right] H_u \\
&&+ \frac{1}{2} M_A(\tau_A) N_A^c N_A^c + \frac{1}{2} M_B(\tau_B) N_B^c N_B^c\,.
\end{eqnarray}
Given the specific shapes of the modular forms in Eq.~\eqref{eq:Ye-Mod-Littlest},  we obtain a diagonal charged lepton mass matrix for $\tau_C=\omega$:
\begin{equation}
M_e= \begin{pmatrix}
 y_e ~&~ 0 ~&~ 0 \\
 0 ~&~ y_\mu ~&~ 0 \\
 0 ~&~ 0 ~&~ y_\tau
\end{pmatrix}v_d\,.
\end{equation}
The mass hierarchies of charged leptons are not addressed in this model, and it can be naturally explained through the weighton mechanism without affecting the remaining predictions of the model~\cite{deMedeirosVarzielas:2022fbw}.

\begin{figure}[t!]
\centering
\begin{tabular}{cc}
\includegraphics[width=0.58\linewidth]{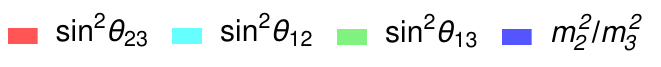}\\
\includegraphics[width=0.88\linewidth]{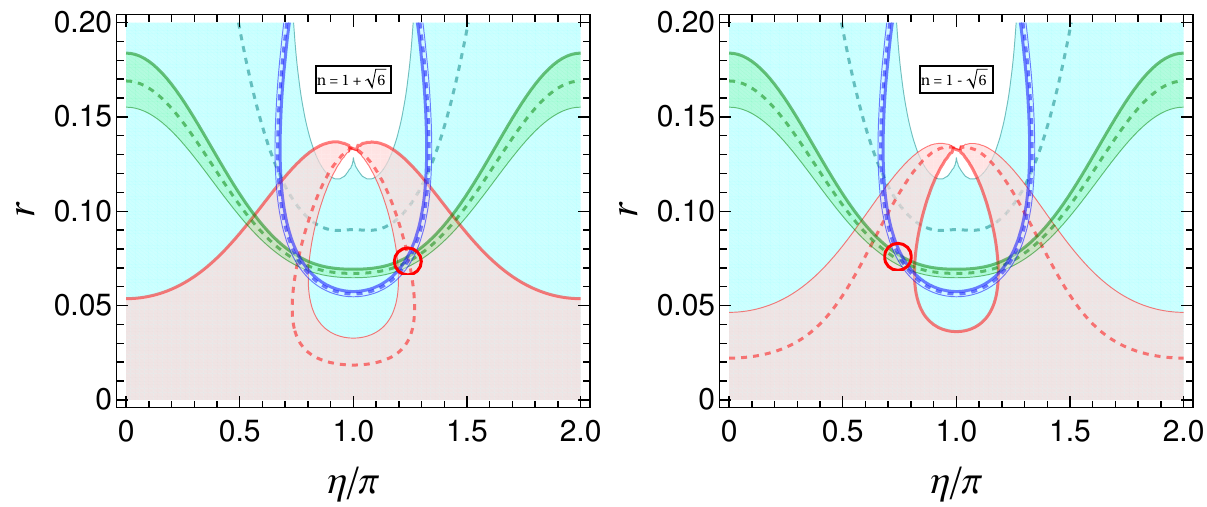}
\end{tabular}
\caption{\label{fig:Modular-LSS-with-SK} The contour plots of $\sin^2\theta_{12}$, $\sin^2\theta_{13}$, $\sin^{2}\theta_{23}$ and $m^2_{2}/m^2_{3}$ in the $\eta/\pi-r$ plane for the modular Littlest seesaw (left) and the flipped version (right) with $n=1+\sqrt{6}$. The cyan, red, green and blue areas denote the $3\sigma$ regions of $\sin^{2}\theta_{23}$, $\sin^{2}\theta_{13}$ and $m_{2}^{2}/m_{3}^{2}$ respectively. The solid lines denote the $3\sigma$ upper bounds, the thin lines denote the $3\sigma$ lower bounds
and the dashed lines refer to their best fit values, as adopted from NuFIT 5.2 with SK data~\cite{Esteban:2020cvm}. The red circle indicates the best fit region.}
\end{figure}

Regarding the neutrino sector, $M_A(\tau_A)$ and $M_B(\tau_B)$ are singlet modular forms in the modular space $S_4^A$, $S_4^B$ with weights $k_A =8$ and $k_B=4$, respectively. The term $N_A^c N_B^c$ is forbidden by the symmetries, since it transforms non-trivially under both $S_4^{A}$ and under $S_4^B$ and there are no one-dimensional modular forms of weight 2 at level 4. Thus the RH neutrino mass matrix is diagonal with
\begin{equation}
M_N=\begin{pmatrix} M_A(\tau_A) ~&~ 0 \\
0 ~&~ M_B(\tau_B) \end{pmatrix}\,.
\end{equation}
The neutrino Yukawa couplings involve the  bi-triplets flavons $\Phi_{AC}$ and $\Phi_{BC}$. As shown in~\cite{deMedeirosVarzielas:2019cyj}, the VEVs of $\Phi_{AC}$ and $\Phi_{BC}$ can take the following forms
\begin{eqnarray} \label{eq:bi-triplet-vev}
\langle \Phi_{AC} \rangle_{i\alpha} = v_{AC} (P_{23})_{i\alpha}\,,~~
\langle \Phi_{BC} \rangle_{m\alpha} = v_{BC} (P_{23})_{m\alpha}\,.
\end{eqnarray}
Here again, $P_{23}$ represents the (2,3) row/column-switching transformation matrix, and $\alpha=1,2,3$ corresponds the entries of the triplet of $S_4^C$, while $i=1,2,3$ ($m=1,2,3$) corresponds to those of $S_4^A$ ($S_4^B$).
The VEV in Eq.~\eqref{eq:bi-triplet-vev} breaks three modular $S_4$'s to a single modular $S_4$ symmetry, and it can be achieved through the standard driving field method, see~\cite{deMedeirosVarzielas:2019cyj} for details.
Choosing the specific fixed points $\tau_A=i+2$, $\tau_B=i$ for the two remaining moduli fields,  we can achieve a flipped CSD($n$) structure with $n=1-\sqrt{6}$~\cite{Ding:2019gof,deMedeirosVarzielas:2019cyj}
 and the Dirac neutrino mass matrix is\footnote{Notice that a CSD($n$) structure with the second and third entries of the second column interchanged
would be achieved if the second modulus is fixed to $\tau_B=(8+i)/5$, and the corresponding value of $n$ is $n=1+\sqrt6\simeq 3.45$.}:
\begin{equation}
\label{eq:Dirac-Modular-LSS}
M_D=\begin{pmatrix}
0 ~&~   b\\
a ~&~  b\left(1-\sqrt{6}\right) \\
-a ~&~ b\left(1+\sqrt{6}\right)
\end{pmatrix} v_u\,,
\end{equation}
with $a=-y_a v_{AC}/\Lambda$, $b=y_s v_{BC}/\Lambda$. The light neutrino mass matrix given by the type-I seesaw mechanism is determined to be
\begin{eqnarray}
\label{eq:mnu-modular-LSS}M_{\nu}=m_a\left(
\begin{array}{ccc}
 0 ~& 0 ~& 0 \\
 0 ~& 1 ~& -1 \\
 0 ~& -1 ~& 1 \\
\end{array}
\right)+m_s e^{i\eta}\left(
\begin{array}{ccc}
 1 ~& 1-\sqrt{6} ~& 1+\sqrt{6} \\
 1-\sqrt{6} ~& 7-2\sqrt{6} ~& -5 \\
 1+\sqrt{6} ~& -5 ~& 7+2\sqrt{6} \\
\end{array}
\right)\,,
\end{eqnarray}
where $m_a=|a|^2v^2_u/M_A$, $m_b=|b|^2v^2_u/M_B$ and $\eta$ is the relative phase of the two terms.  It is notable that the neutrino mass matrix of Eq.~\eqref{eq:mnu-modular-LSS} only depends on three free parameters $m_a$, $m_s$ and $\eta$ are involved in the neutrino mass matrix. One can check that the column vector $\left(2, -1, -1\right)^{T}$ is an eigenvector of $m_{\nu}$ with vanishing eigenvalue and it could only be first column of the lepton mixing matrix to be compatible with the experimental data~\cite{Esteban:2020cvm}. Therefore the neutrino mass spectrum is normal ordering, and lightest neutrino is massless $m_1=0$, and the lepton mixing matrix is of the following form
\begin{equation}
U_{PMNS}=\begin{pmatrix}
\sqrt{\frac{2}{3}}  ~& -  ~&~  -  \\
-\frac{1}{\sqrt{6}}  ~&  -  ~&~  -   \\
-\frac{1}{\sqrt{6}}  ~&  -  ~&~  -
\end{pmatrix}\,,
\end{equation}
where the first column is in common with the tri-bimaximal mixing pattern. As a consequence, the following sum rules arise,
\begin{equation}
3\cos^2\theta_{12}\cos^2\theta_{13}=2,\quad \cos\delta_{CP}=-\frac{\cot2\theta_{23}(1-5\sin^2\theta_{13})}{2\sin\theta_{13}\sqrt{2-6\sin^2\theta_{13}}}\,.
\end{equation}
The model is quite predictive and the light neutrino mass matrix in Eq.~\eqref{eq:mnu-modular-LSS} only depends on three real parameters $m_a$, $m_s$ and $\eta$. We plot the contour plots of $\sin^2\theta_{13}$, $\sin^2\theta_{12}$, $\sin^2\theta_{23}$ and $m^2_2/m^2_3$ in the plane $r$ versus $\eta/\pi$ in figure~\ref{fig:Modular-LSS-with-SK} where $r=m_s/m_a$.
We see that the experimental data of $\sin^2\theta_{23}$ and $m^2_2/m^2_3$ leave only two small allowed parameter regions in the plane $r-\eta$. These two small regions lead to the same predictions for lepton mixing angles while the CP violation phases are of opposite sign. Thus only the one labelled in red circle in figure~\ref{fig:Modular-LSS-with-SK} is preferred by the present value of $\delta_{CP}$~\cite{Esteban:2020cvm}. Once we have the values of $r$ and $\eta$, thanks to the high predictivity of the model we can derive all the physical parameters and we can test them against the observed values. The regions of free parameters and lepton mixing parameters allowed the experimental data at $3\sigma$ level listed in table~\ref{tab:bf-modular-LSS}. Furthermore, this model has been extended to $SU(5)$ Grand Unified Theory (GUT)~\cite{deMedeirosVarzielas:2023ujt}.

\begin{table}[hptb!]
\centering
\resizebox{1.00\textwidth}{!}{
\begin{tabular}{|c|c|c|c|c|c|}
  \hline \hline
\multicolumn{3}{|c|}{Modular Littlest seesaw}  & \multicolumn{3}{c|}{Flipped modular Littlest seesaw}  \\ \hline\hline
& \text{bf} & allowed ranges & &  \text{bf} & allowed ranges\\
  \hline
 $\eta/\pi$ & 1.240  & $[1.197,1.276]$   & $\eta/\pi$& $0.742$ & $[0.725,0.806]$ \\
  \cline{1-3}
  \cline{4-6}
 $r$ & 0.0734 & $[0.0684,0.0786]$   &  $r$ & 0.0758 & $[0.0683,0.0786]$  \\
  \cline{1-3}
  \cline{4-6}
$\sin^{2}\theta_{13}$ & 0.0223 & $[0.0205,0.0240]$  &  $\sin^{2}\theta_{13}$ & 0.0231 & $[0.0205,0.0240]$  \\
  \cline{1-3}
  \cline{4-6}
 $\sin^{2}\theta_{12}$ & 0.318 & $[0.317,0.319]$ & $\sin^{2}\theta_{12}$ & 0.318 & $[0.317,0.319]$  \\
  \cline{1-3}
  \cline{4-6}
 $\sin^{2}\theta_{23}$ & 0.447 & $[0.408,0.483]$  &  $\sin^{2}\theta_{23}$ & 0.535 & $[0.517,0.595]$  \\
  \cline{1-3}
  \cline{4-6}
$\delta_{CP}/\pi$ &$-0.575$  &$[-0.640,-0.522]$ & $\delta_{CP}/\pi$ & $-0.452$ & $[-0.478,-0.354]$  \\
  \cline{1-3}
  \cline{4-6}
$\beta/\pi$ &  $0.474$  &$[0.408,0.555]$  &  $\beta/\pi$ &  $-0.441$  & $[-0.562,-0.409]$ \\
  \cline{1-3}
  \cline{4-6}
$m_{2}^{2}/m_{3}^{2}$ & 0.0297 & $[0.0270,0.0321]$   & $m_{2}^{2}/m_{3}^{2}$ &  0.0283 & $[0.0270,0.0321]$ \\
  \hline \hline
\end{tabular}}
\caption{\label{tab:bf-modular-LSS} The best fit values and the allowed region of the free parameters $\eta$ and $r$ as well as lepton mixing parameters, where all the lepton mixing angles and neutrino mass squared differences are required to lie in $3\sigma$ regions of experimental data~\cite{Esteban:2020cvm}. The neutrino mass matrix of the flipped modular Littlest seesaw is given by Eq.~\eqref{eq:mnu-modular-LSS}. The second and third entries of the second column of $M_D$ in Eq.~\eqref{eq:Dirac-Modular-LSS} are permutated in the modular Littlest seesaw.}
\end{table}

\section{Modular GUT}

\begin{figure}[t!]
\centering
\includegraphics[width=0.45\textwidth]{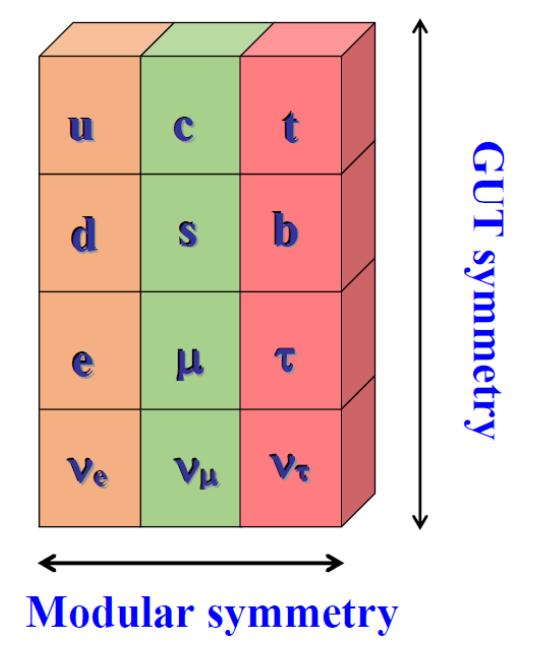}
\caption{The diagrammatic sketch combining modular symmetry with GUTs.   }
\label{fig:GUT-modular}
\end{figure}

It is known that the electromagnetic and weak interactions are unified in the SM, and the SM involves three different gauge groups $SU(3)\times SU(2)\times U(1)$ with independent gauge couplings. Grand Unified Theories (GUTs) are well motivated extensions of the SM, and it unifies the electroweak and strong interactions into a single force. In general, the gauge group of GUTs should include the SM gauge group as a subgroup. The quark and lepton fields within a generation are usually assigned to very few gauge multiplets of GUTs, and consequently GUTs allow to connect the lepton mass matrices with the quark mass matrices. However, the matter fields multiplets in different generations transform in the same way under GUTs consequently the gauge symmetry of GUTs has some constraint on the Yukawa couplings, but can not explain the hierarchies of different Yukawa couplings. In order to understand the different patterns of quark and lepton mixing, one usually imposes flavor symmetry which relates matter fields of three generations so that the strong connection between the elements of quark and lepton mass matrices. For the case of traditional flavor symmetry, flavons and complicated vacuum alignment are also required in the context of GUTs, see Ref.~\cite{King:2017guk} for a review. As shown in previous sections, the modular flavor symmetry could simplify the vacuum alignment problem significantly. Thus we implement the modular symmetry in GUTs to address the mass hierarchies and mixing patterns of quarks and leptons, this paradigm is shown in figure~\ref{fig:GUT-modular}. In the following, we shall focus on $SU(5)$ and $SO(10)$ GUTs, before discussing flipped $SU(5)_F\times U(1)_X$.

\begin{table}[hptb]
\centering
\begin{tabular}{|c|c|c|c|c|} \hline\hline
Level & $\Gamma_N$($\Gamma'_N$)  & $SU(5)$ & $SU(5)_F\times U(1)_X$ & $SO(10)$ \\   \hline

$N=2$ &  $S_3$ & \cite{Kobayashi:2019rzp,Du:2020ylx} & ---& --- \\ \hline

\multirow{2}{*}{$N=3$} & $A_4$  & \cite{deAnda:2018ecu,Chen:2021zty} & \cite{Charalampous:2021gmf,Du:2022lij}  &  \cite{Ding:2021eva,Ding:2022bzs}  \\ \cline{2-4}
 & $T'$ & --- &  --- & --- \\\hline

\multirow{2}{*}{$N=4$} & $S_4$  & \cite{Zhao:2021jxg,King:2021fhl,Ding:2021zbg,deMedeirosVarzielas:2023ujt}  & --- &  --- \\ \cline{2-4}
& $S'_4$  & --- &  --- &  --- \\\hline

\multirow{2}{*}{$N=5$} & $A_5$ & --- &  ---  &  --- \\ \cline{2-4}
& $A'_5$ & --- &  --- &  --- \\\hline

\multirow{2}{*}{$N=6$} & $\Gamma_6\cong S_3\times A_4$  & --- &  --- &  --- \\ \cline{2-4}
& $\Gamma'_6\cong S_3\times T'$   & \cite{Abe:2023dvr} & --- &  --- \\ \hline

\multirow{2}{*}{$N=7$} & $\Gamma_7$  & --- &  --- &  --- \\ \cline{2-4}
& $\Gamma'_7$ & ---  & --- &  ---\\\hline\hline
\end{tabular}
\caption{\label{tab:modular-models-GUT-summary} Summary of modular invariant models based on GUTs.  }
\end{table}

\subsection{Modular SU(5)}

\begin{figure}[hptb!]
\centering
\includegraphics[width=0.99\textwidth]{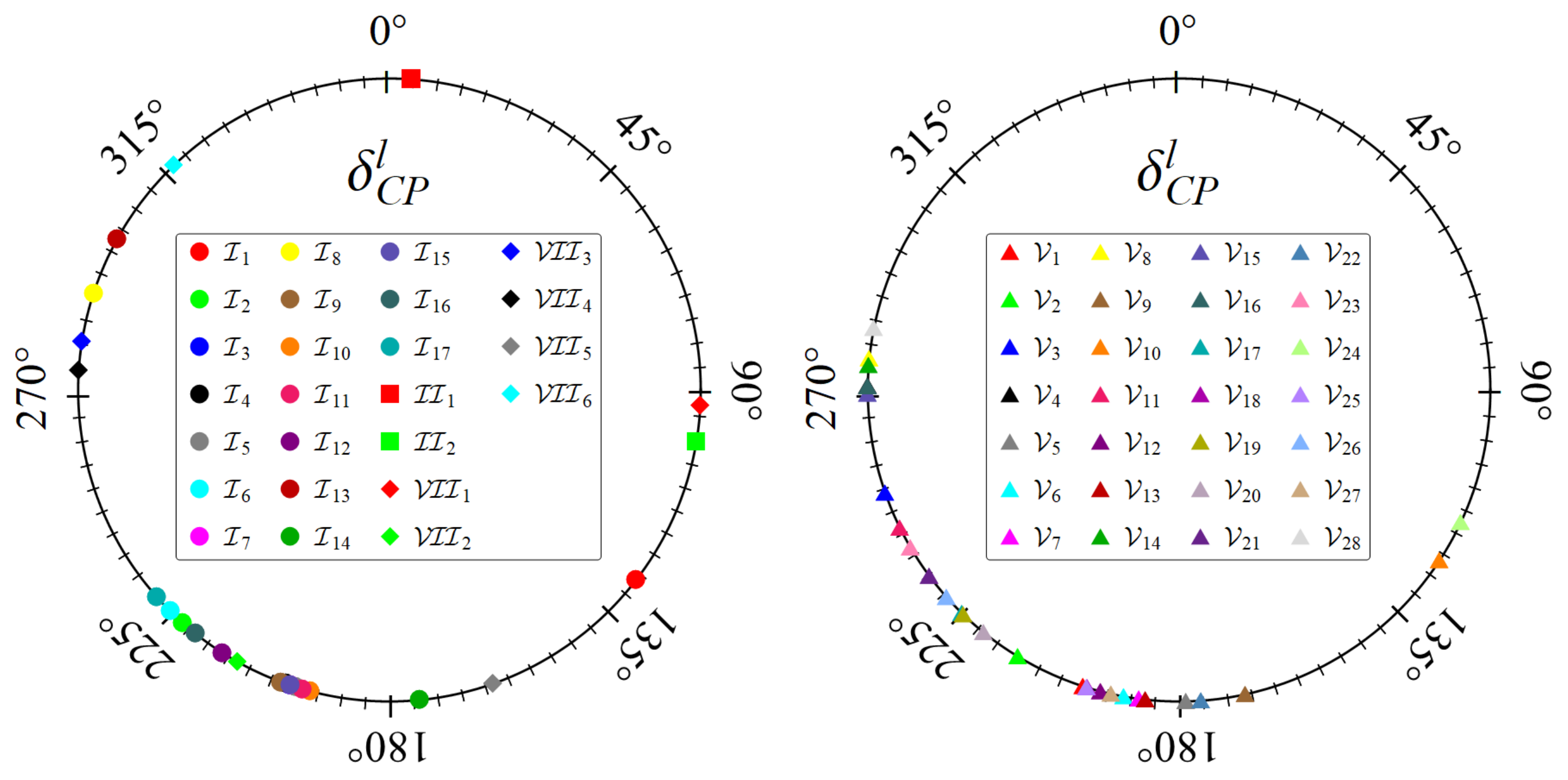}
\caption{\label{fig:deltaCP-SU5}The predictions for the Dirac CP phase $\delta^l_{CP}$ in the viable $SU(5)$ models based on $A_4$ modular symmetry, see \cite{Chen:2021zty} for details of the model. }
\end{figure}

\begin{figure}[hptb!]
\centering
\includegraphics[width=0.8\textwidth]{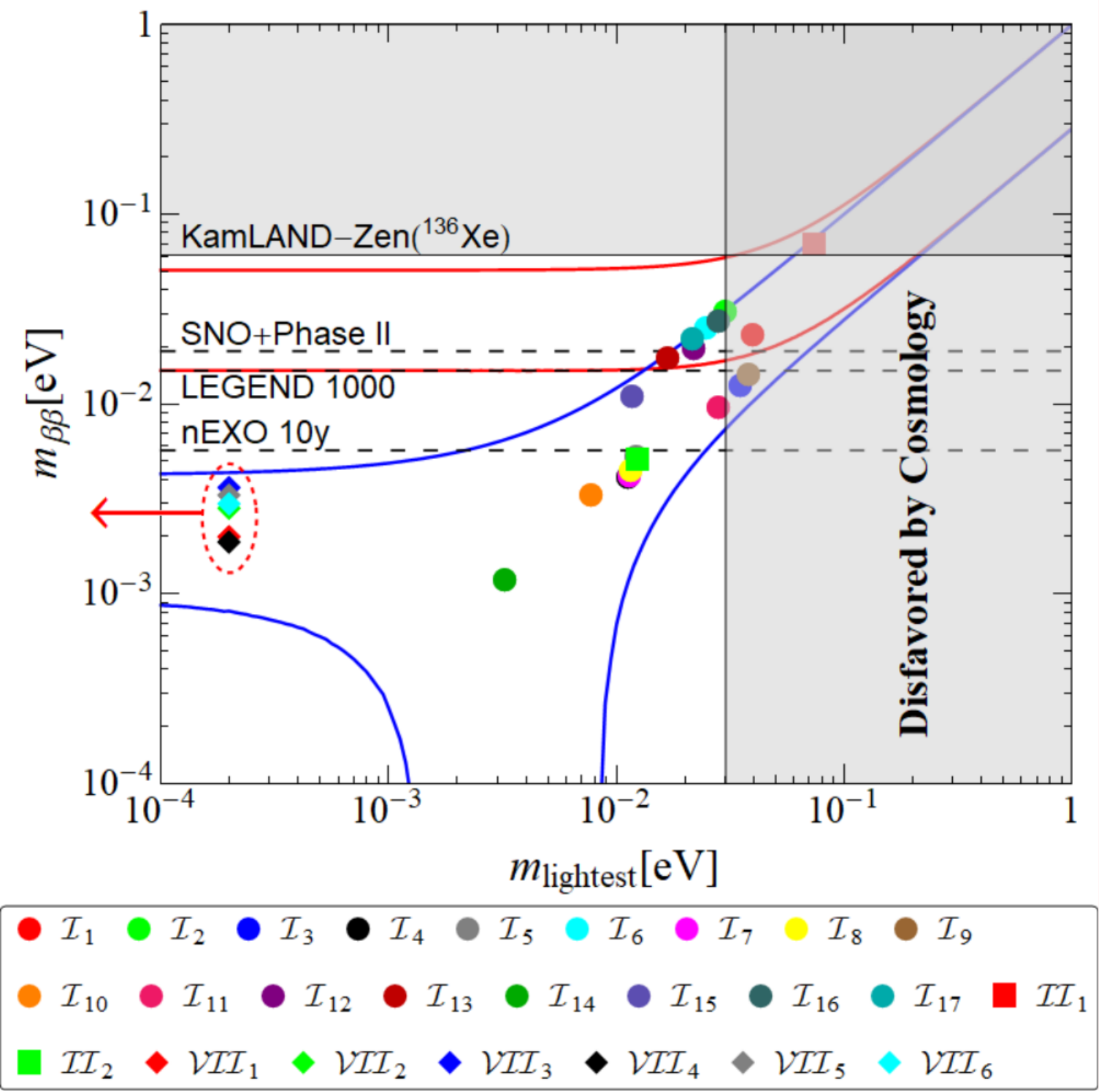}
\caption{\label{fig:mbb-SU5-part1}The predictions for the effective mass of the neutrinoless double beta decay in the viable $SU(5)$ models based on $A_4$ modular symmetry. }
\end{figure}

\begin{figure}[hptb!]
\centering
\includegraphics[width=0.8\textwidth]{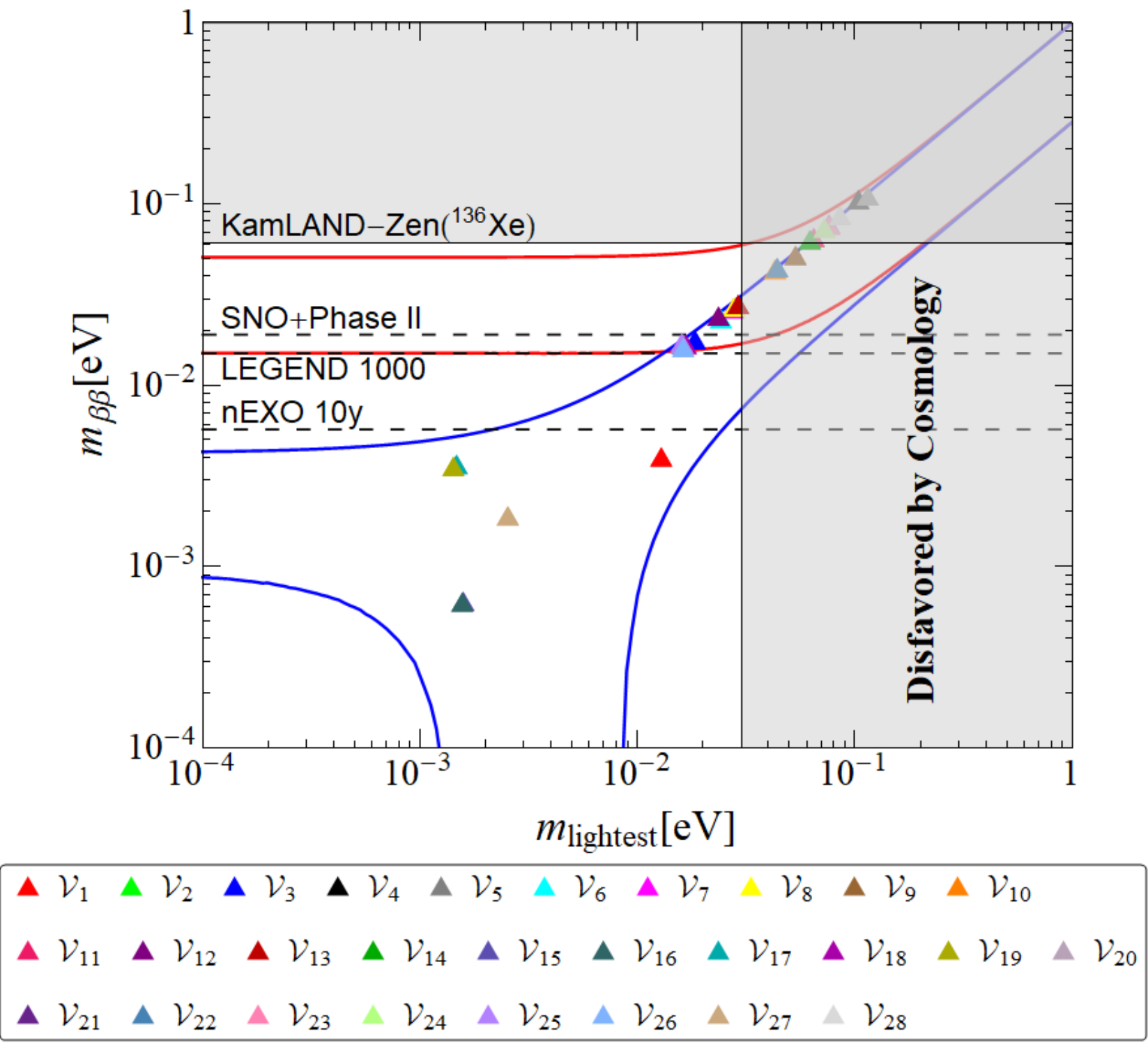}
\caption{\label{fig:mbb-SU5-part2} The continuum of figure~\ref{fig:mbb-SU5-part1}. }
\end{figure}

The minimal extension of the SM to GUT is based on the $SU(5)$ gauge group~\cite{Georgi:1974sy}, the strong, weak and electromagnetic coupling constants merge into a single coupling constant at high energy of order $10^{15}\sim10^{16}$ GeV. The observed differences in the couplings at low energy are induced by the renormalization group evolution effect. In $SU(5)$ GUT, all the 15 left-handed quark and lepton fields within each family are embedded into the $\overline{\mathbf{5}}$-plet $F$ and $\mathbf{10}$-plet $T$:
\begin{eqnarray}
F=\begin{pmatrix}
d^c_r\\
d^c_g\\
d^c_b\\
e\\
-\nu
\end{pmatrix}\,,~~~
T=\frac{1}{\sqrt{2}}\begin{pmatrix}
0~&u^c_b~&-u^c_g~&-u_r~&-d_r\\
-u^c_b~&0~&u^c_r~&-u_g~&-d_g\\
u^c_g~&-u^c_r~&0~&-u_b~&-d_b\\
u_r~&u_g~&u_b~&0~&e^c\\
d_r~&d_g~&d_b~&-e^c~&0
\end{pmatrix}\,,
\label{eq:su5multiplets}
\end{eqnarray}
where the superscript $c$ denotes the charge conjugated fields. The neutrinos are massless in the minimal $SU(5)$ GUT, we introduce the right-handed neutrinos $\nu^c$ to generate light neutrino masses through the type-I seesaw mechanism. Note that $\nu^c$ are $SU(5)$ singlets. The $SU(5)$ gauge symmetry is spontaneously broken down to the SM gauge group $SU(3)_C\times SU(2)\times U(1)_Y$ by the VEV of the Higgs $H_{\mathbf{24}}$ in the adjoint representation of $SU(5)$. In the minimal $SU(5)$ model, two Higgs multiplets $H_{\mathbf{5}}$ and $H_{\overline{\mathbf{5}}}$ in the fundamental representation $\mathbf{5}$ and antifundamental representation $\overline{\mathbf{5}}$ of $SU(5)$ further break the SM gauge symmetry into $SU(3)_C\times U(1)_{\text{EM}}$. The charged leptons and down-type quarks are in the same GUT multiplets, the down type quark mass matrix is the transpose of the charged lepton mass matrix in the minimal $SU(5)$ such that the masses of the charged leptons and down quarks would be identical. The observed mass difference of down quarks and charged leptons can be explained by introducing the Higgs multiplet $H_{\mathbf{\overline{45}}}$~\cite{Georgi:1979df}. The corresponding contributions to the charged lepton and down quark mass matrices are different by factor of $-3$. If some modular flavor symmetry $\Gamma_N$ or $\Gamma'_N$ is imposed in $SU(5)$ GUT, the three generations of matter fields $F$, $T$ and $N$ would be assigned to transform as irreducible representations of the finite modular groups. Then the most general superpotential for fermion masses can be written as,
\begin{equation}
\label{eq:superpotential-SU5}\mathcal{W}=\nu^c\nu^c f_M(Y) + \nu^cFH_{5} f_N(Y)+FTH_{\overline{5}}f_{D}(Y)+FTH_{\overline{45}}f'_{D}(Y)+TTH_{5}f_{U}(Y)\,,
\end{equation}
where $f_M(Y)$, $f_N(Y)$, $f_{D}(Y)$, $f'_{D}(Y)$ and $f_{U}(Y)$ are functions of modular forms, and they are fixed by modular invariance. The interplays of the $SU(5)$ GUT and the modular symmetry $S_3$~\cite{Kobayashi:2019rzp,Du:2020ylx}, $A_4$~\cite{deAnda:2018ecu,Chen:2021zty}, $S_4$~\cite{Zhao:2021jxg,King:2021fhl,Ding:2021zbg} and $\Gamma'_6\cong S_3\times T'$~\cite{Abe:2023dvr} have been studied. A comprehensive analysis of $SU(5)$ model with $A_4$ modular symmetry has been performed in~\cite{Chen:2021zty}, the models with two right-handed neutrinos are found to be predictive. The predictions of the phenomenological viable models for the lepton Dirac CP violation phase and the effective mass of the neutrinoless double decay are displayed in figure~\ref{fig:deltaCP-SU5} and figures~\ref{fig:mbb-SU5-part1}, \ref{fig:mbb-SU5-part2} respectively.

\subsection{Modular SO(10)}

The $SO(10)$ GUT theory embeds all SM fermions of a generation plus a right-handed neutrino into the $\mathbf{16}$ spinor representation of $SO(10)$~\cite{Fritzsch:1974nn}. The tensor product of two spinor representations gives
\begin{equation}
\mathbf{16} \otimes \mathbf{16} = \mathbf{10}_{S} \oplus \mathbf{120}_{A} \oplus \mathbf{126}_{S}\,,
\end{equation}
where the subscripts $S$ and $A$ denote the symmetric and antisymmetric parts of the tensor products respectively in flavor space. Therefore the renormalizable Yukawa interactions for the quark and lepton masses requires the Higgs multiplets in the $SO(10)$ representations $\mathbf{10}$, $\mathbf{120}$ and $\overline{\mathbf{126}}$, and the superpotential of the renormalizable Yukawa couplings is of the following form:
\begin{equation}
\label{eq:Yukawa-SO10}\mathcal{W}_Y=\mathcal{Y}_{ij}^{10} \psi_{i} \psi_{j}H+\mathcal{Y}_{ij}^{\overline{126}}  \psi_{i} \psi_{j}\overline{\Delta}+ \mathcal{Y}_{ij}^{120}\psi_{i} \psi_{j} \Sigma\,,
\end{equation}
where $i, j=1, 2, 3$ are generation indices, $\psi$ refers to the matter fields in the spinor representation $\mathbf{16}$, $H$, $\overline{\Delta}$ and $\Sigma$ denote the Higgs fields in the representations $\mathbf{10}$, $\overline{\mathbf{126}}$ and $\mathbf{120}$ respectively. Moreover, $\mathcal{Y}^{10}$, $\mathcal{Y}^{\overline{126}}$ and $\mathcal{Y}^{120}$ are $3\times3$ Yukawa matrices in flavor space. The $SO(10)$ gauge symmetry enforces that $\mathcal{Y}^{10}$ and $\mathcal{Y}^{\overline{126}}$ are symmetric while $\mathcal{Y}^{120}$ is antisymmetric, i.e.
\begin{equation}
\mathcal{Y}_{ij}^{10} = \mathcal{Y}_{ji}^{10},\quad \mathcal{Y}_{ij}^{\overline{126}} = \mathcal{Y}_{ji}^{\overline{126}}, \quad
\mathcal{Y}_{ij}^{120} = -\mathcal{Y}_{ji}^{120}\,.
\end{equation}
The rank of the $SO(10)$ group is equal to five which is one more than SM gauge group, consequently there are a plenty of ways breaking $SO(10)$ to SM. The  decomposition of the above Higgs multiplets under the SM gauge symmetry $SU(3)_C\times SU(2)_L\times U(1)_Y$ which are relevant to  fermion masses are
\begin{eqnarray}
\nonumber&&\mathbf{10}\supset(\mathbf{1}, \mathbf{2}, 1/2)\oplus(\mathbf{1}, \mathbf{2}, -1/2)\equiv \Phi^{10}_d\oplus\Phi^{10}_u\,,\\
\nonumber&&\mathbf{120}\supset(\mathbf{1}, \mathbf{2}, 1/2)\oplus(\mathbf{1}, \mathbf{2}, -1/2)\oplus(\mathbf{1}, \mathbf{2}, 1/2)\oplus(\mathbf{1}, \mathbf{2}, -1/2)\equiv \Phi^{120}_d\oplus\Phi^{120}_u\oplus\Phi'^{120}_d\oplus\Phi'^{120}_u\,,\\
\label{eq:decomposition-SO10}&&\mathbf{\overline{126}}\supset(\mathbf{1}, \mathbf{2}, 1/2)\oplus(\mathbf{1}, \mathbf{2}, -1/2)\oplus(\mathbf{1}, \mathbf{1}, 0)\oplus(\mathbf{1}, \mathbf{3}, 1)\equiv \Phi^{\overline{126}}_d\oplus\Phi^{\overline{126}}_u\oplus \overline{\Delta}_R\oplus \overline{\Delta}_L\,.
\end{eqnarray}
We see that the $SO(10)$ Higgs fileds $H$, $\overline{\Delta}$ and $\Sigma$ have eight $SU(2)_L$ doublets $\Phi^{10}_d$, $\Phi^{10}_u$, $\Phi^{120}_d$, $\Phi^{120}_u$, $\Phi'^{120}_d$, $\Phi'^{120}_u$, $\Phi^{\overline{126}}_d$ $\Phi^{\overline{126}}_u$ carrying the same quantum numbers as the Higgs doublets $H_u$ and $H_d$. They can all develop VEVs and thus generate masses of quarks and charged leptons. The light Higgs boson observed at the LHC, is in general a linear combination of these $SU(2)_L$ doublets above. Moreover, the VEVs of the neutral components of the $SU(2)_L$ singlet $\overline{\Delta}_R$ and triplet $\overline{\Delta}_L$ generate the Majorana masses of the heavy right-handed neutrinos and light left-handed neutrinos respectively.  Decomposing the Yukawa superpotential in Eq.~\eqref{eq:Yukawa-SO10}, the quark and lepton mass matrices are determined to be
\begin{eqnarray}
\nonumber&&M_u=v^{10}_u\mathcal{Y}^{10}+v^{\overline{126}}_u\mathcal{Y}^{\overline{126}}+(v^{120}_u+v'^{120}_u)\mathcal{Y}^{120}\,,\\
\nonumber&&M_d=v^{10}_d\mathcal{Y}^{10}+v^{\overline{126}}_d\mathcal{Y}^{\overline{126}}+(v^{120}_d+v'^{120}_d)\mathcal{Y}^{120}\,,\\
\nonumber&&M_{e}=v^{10}_d\mathcal{Y}^{10}-3v^{\overline{126}}_d\mathcal{Y}^{\overline{126}}+(v^{120}_d-3v'^{120}_d)\mathcal{Y}^{120}\,,\\
\nonumber&&M_D=v^{10}_u\mathcal{Y}^{10}-3v^{\overline{126}}_u\mathcal{Y}^{\overline{126}}+(v^{120}_u-3v'^{120}_u)\mathcal{Y}^{120}\,,\\
\nonumber&&M_R=v^{\overline{126}}_R\mathcal{Y}^{\overline{126}}\\ \label{eq:mass-matrices-SO10}
&&M_L=v^{\overline{126}}_L\mathcal{Y}^{\overline{126}}\,,
\end{eqnarray}
where $M_D$, $M_R$ and $M_L$ are the Dirac and Majorana neutrino mass matrices in type I and type II seesaw mechanism, and the VEVs are defined as
\begin{eqnarray}
\nonumber&&v^{10}_{u,d}=\langle\Phi^{10}_{u,d}\rangle\,,~~~v^{120}_{u,d}=\langle\Phi^{120}_{u,d}\rangle\,,~~~v'^{120}_{u,d}=\langle\Phi'^{120}_{u,d}\rangle\,,\\
&&v^{\overline{126}}_{u,d}=\langle\Phi^{\overline{126}}_{u,d}\rangle\,,~~~v^{\overline{126}}_R=\langle\overline{\Delta}_R\rangle\,,~~~v^{\overline{126}}_L=\langle\overline{\Delta}_L\rangle\,.
\end{eqnarray}
Hence the $SO(10)$ GUT naturally predicts seesaw mechanism generating neutrino mass, and the light neutrino mass matrix is given by
\begin{equation}
\label{eq:mnu-I-II-SO10}M_{\nu}=M_{L}-M_DM^{-1}_RM^T_D\,,
\end{equation}
which is identical with Eq.~\eqref{eq:TypIIMassMatrix}. From the expressions of the fermion mass matrices in Eq.~\eqref{eq:mass-matrices-SO10}, we see that all fermion mass matrices are proportional to the same Yukawa matrix if only one of the Higgs fields $H$, $\overline{\Delta}$, $\Sigma$ is employed. As a result, both quark and lepton mixing matrices would be a three-dimensional unit matrix. Hence at least two Higgs fields are necessary for realistic fermion spectrum and flavor mixing, and the non-vanishing neutrino masses requires the $\mathbf{126}$ dimensional Higgs $\overline{\Delta}$ must be present. The minimal choice for the second Higgs is that the $\mathbf{10}$ dimensional multiplet $H$, this is the so-called minimal $SO(10)$ GUT.

\subsubsection{$SO(10)$ GUT with $A_4$ modular symmetry}

If some modular symmetry $\Gamma_N$ or $\Gamma'_N$ is implemented in $SO(10)$ GUT, the different elements of the Yukawa matrices $\mathcal{Y}^{10}$, $\mathcal{Y}^{\overline{126}}$ and $\mathcal{Y}^{120}$ would be correlated. Since $A_4$ is the smallest finite modular group with three dimensional irreducible representation, we will take the $A_4$ modular symmetry to explain the idea of $SO(10)$ modular GUT. The three generations of matter fields $\psi_{1,2,3}$ are assumed to transform as a triplet $\mathbf{3}$ under $A_4$ modular symmetry\footnote{If the matter fields were assigned to be $A_4$ singlets, the effect of $A_4$ modular symmetry is equivalent to an Abelian flavour symmetry and consequently the correlations among nonzero entries of lepton mass matrices are lost. Moreover, the modular forms appearing in the lepton mass matrices can be absorbed into the Yukawa coupling constants.}, and its modular weight is denoted as $k_F$. All the three Higgs multiplets $H$, $\overline{\Delta}$ and $\Sigma$ are assumed to be invariant under $A_4$, thus the known $SO(10)$ breaking patterns would be kept intact even in the presence of modular symmetry. The modular weights of $H$, $\overline{\Delta}$ and $\Sigma$ are denoted as $k_{10}$, $k_{\overline{126}}$ and $k_{120}$ respectively, and one can set $k_{10}=0$ without loss of generality. Hence the modular invariant superpotential of $SO(10)$ can be written as~\cite{Ding:2021eva}
\begin{eqnarray}
\nonumber\mathcal{W}_{Y}&=&\sum_{\mathbf{r}_a}
\alpha_a\left((\psi\psi H)_{\mathbf{r}'_a}Y_{\mathbf{r}_a}^{(2k_F+k_{10})}(\tau)\right)_{\mathbf{1}}+\sum_{\mathbf{r}_c}
\beta_b\left((\psi\psi \Sigma)_{\mathbf{r}'_b}Y_{\mathbf{r}_b}^{(2k_F+k_{120})}(\tau)\right)_{\mathbf{1}}\\ \label{eq:W_Y}
&&~~~+\sum_{\mathbf{r}_c}\gamma_c\left((\psi\psi \overline{\Delta})_{\mathbf{r}'_c}Y_{\mathbf{r}_c}^{(2k_F+k_{\overline{126}})}(\tau)\right)_{\mathbf{1}}\,,
\end{eqnarray}
where the $A_4$ representations $\mathbf{r}_{a,b,c}$ and $\mathbf{r}'_{a,b,c}$ fulfill $\mathbf{r}'_a\otimes \mathbf{r}_a=\mathbf{r}'_b\otimes \mathbf{r}_b=\mathbf{r}'_c\otimes \mathbf{r}_c=\mathbf{1}$. We have denoted $\alpha_a\equiv\alpha_{\mathbf{r}_a}^{(2k_F+k_{10})}$,
$\beta_b\equiv\beta_{\mathbf{r}_b}^{(2k_F+k_{120})}$, $\gamma_c\equiv \gamma_{\mathbf{r}_c}^{(2k_F+k_{\overline{126}})}$ for simplicity of notation. We see that the modular symmetry enforces the Yukawa couplings $\mathcal{Y}_{ab}^{10}$,  $\mathcal{Y}_{ab}^{\overline{126}}$ and $\mathcal{Y}_{ab}^{120}$ in Eq.~\eqref{eq:Yukawa-SO10} are modular forms of level 3. It is remarkable that the $SO(10)$  modular models are completely specified by the modular weights of matter fields and Higgs fields. With the tensor products of the $A_4$ group listed in Eq.~\eqref{eq:3x3-rules-A4}, one can read out of the Yukawa coupling of $\mathcal{Y}^{10}$ as follow
\begin{eqnarray}
\nonumber\mathcal{Y}^{10}&=&\alpha_1Y_{\mathbf1}^{(2k_F+k_{10})}(\tau)\left(
\begin{matrix}
 1 ~& 0 ~& 0 \\
 0 ~& 0 ~& 1 \\
 0 ~& 1 ~& 0 \\
\end{matrix}
\right)+\alpha_2Y_{\mathbf1'}^{(2k_F+k_{10})}(\tau)\left(
\begin{matrix}
 0 ~& 0 ~& 1 \\
 0 ~& 1 ~& 0 \\
 1 ~& 0 ~& 0 \\
\end{matrix}
\right)\\
\nonumber &&+\alpha_3Y_{\mathbf1''}^{(2k_F+k_{10})}(\tau)\left(
\begin{matrix}
 0 ~& 1 ~& 0 \\
 1 ~& 0 ~& 0 \\
 0 ~& 0 ~& 1 \\
\end{matrix}
\right)+\alpha_S\left(\begin{matrix}
 2Y^{(2k_F+k_{10})}_{{{\mathbf3}},1}(\tau) ~&~ -Y^{(2k_F+k_{10})}_{{{\mathbf3}},3}(\tau)  ~&~ -Y^{(2k_F+k_{10})}_{{{\mathbf3}},2}(\tau) \\
 -Y^{(2k_F+k_{10})}_{{{\mathbf3}},3}(\tau) ~&~ 2Y^{(2k_F+k_{10})}_{{{\mathbf3}},2}(\tau)  ~&~ -Y^{(2k_F+k_{10})}_{{{\mathbf3}},1}(\tau) \\
 -Y^{(2k_F+k_{10})}_{{{\mathbf3}},2}(\tau) ~&~ -Y^{(2k_F+k_{10})}_{{{\mathbf3}},1}(\tau)  ~&~ 2Y^{(2k_F+k_{10})}_{{{\mathbf3}},3}(\tau)
\end{matrix}
\right)\\
&&+\alpha_A\left(\begin{matrix}
 0 ~&~ -Y^{(2k_F+k_{10})}_{{{\mathbf3}},3}(\tau)  ~&~ Y^{(2k_F+k_{10})}_{{{\mathbf3}},2}(\tau) \\
 Y^{(2k_F+k_{10})}_{{{\mathbf3}},3}(\tau) ~&~ 0  ~&~ -Y^{(2k_F+k_{10})}_{{{\mathbf3}},1}(\tau) \\
 -Y^{(2k_F+k_{10})}_{{{\mathbf3}},2}(\tau) ~&~ Y^{(2k_F+k_{10})}_{{{\mathbf3}},1}(\tau)  ~&~ 0
\end{matrix}
\right)\,,
\end{eqnarray}
where all the linearly independent modular forms at weight $2k_F+k_{10}$ should be considered. The Yukawa matrix $\mathcal{Y}^{\overline{126}}$ is of a similar form with $\alpha_a$ and $k_{10}$ replaced by $\gamma_a$ and $k_{\overline{126}}$ respectively. However, the Yukawa coupling $\mathcal{Y}^{120}$ is antisymmetric and it is determined to be of the following form
\begin{eqnarray}
\nonumber\mathcal{Y}^{120}&=&\beta_A\left(\begin{matrix}
 0 ~&~ -Y^{(2k_F+k_{120})}_{{{\mathbf3}},3}(\tau)  ~&~ Y^{(2k_F+k_{120})}_{{{\mathbf3}},2}(\tau) \\
 Y^{(2k_F+k_{120})}_{{{\mathbf3}},3}(\tau) ~&~ 0  ~&~ -Y^{(2k_F+k_{120})}_{{{\mathbf3}},1}(\tau) \\
 -Y^{(2k_F+k_{120})}_{{{\mathbf3}},2}(\tau) ~&~ Y^{(2k_F+k_{120})}_{{{\mathbf3}},1}(\tau)  ~&~ 0
\end{matrix}
\right)\,.
\end{eqnarray}
It was found that the experimental data of fermion masses and mixing can be accommodated in both the minimal $SO(10)$ GUT and the extension including the $\mathbf{120}$ dimensional Higgs field $\Sigma$~\cite{Ding:2021eva}. The right-handed neutrinos naturally appear in $SO(10)$ GUT, the mass and Yukawa coupling of the right-handed neutrinos are closely related to those of quarks and charged leptons. It is notable that the matter-antimatter asymmetry of the Universe could also be generated through leptogenesis in the modular $SO(10)\times A_4$ models~\cite{Ding:2022bzs}.

\subsection{Modular flipped $SU(5)$}

The Flipped $SU(5)$ GUT has been proposed long time ago~\cite{Barr:1981qv,Derendinger:1983aj} as an alternative symmetry breaking pattern of the $SO(10)$ gauge group. It is based on the gauge group $SU(5)\times U(1)_X$. Flipped $SU(5)$ GUT can easily be derived naturally from weakly-coupled string theory. Flipped $SU(5)$ GUT can naturally solve the notorious doublet-triplet splitting problem via the missing partner mechanism~\cite{Antoniadis:1987dx} to and efficiently suppress the proton decay rate~\cite{Derendinger:1983aj}. The flipped $SU(5)$ differ from the Georgi-Glashow $SU(5)$ in the manner in which the SM particle fields are embedded into representations of the group. In contrast to Eq.~\eqref{eq:su5multiplets}, the matter representations in the flipped $SU(5)$ model are
\begin{equation}
\widetilde{F}= \begin{pmatrix} u_r^c \\ u_g^c \\ u_b^c \\ e \\ -\nu \end{pmatrix}\sim(\mathbf{\bar{5}}, -3), ~~~ \widetilde{T}=\begin{pmatrix} 0 & d_b^c & -d_g^c & u_r & d_r \\ -d_b^c & 0 & d_r^c & u_g & d_g \\ d_g^c & -d_r^c & 0 & u_b & d_b \\ -u_r & -u_g & -u_b & 0 & \nu^c \\ -d_r & -d_g & -d_b & -\nu^c & 0 \end{pmatrix}\sim(\mathbf{10}, 1),
~~~ e^c\sim(\mathbf{1}, 5)\,,
 \label{eq:flippedSU5fields}
\end{equation}
which flips the position of $u^{c}\leftrightarrow d^c$, $e^c\leftrightarrow\nu^c$ in the regular $SU(5)$ representations while the remaining unaltered. Notice that the first numbers in bold denote the $SU(5)$ representation while the second numbers stand for the $U(1)_X$ charges in Eq.~\eqref{eq:flippedSU5fields}. As opposed to the Georgi-Glashow $SU(5)$ which is broken to the SM gauge group via a Higgs multiplet in the $\mathbf{24}$-dimensional representation, the flipped $SU(5)$ doesn't require Higgs fields in large dimensional representations. The GUT and electroweak symmetry breaking is driven by the following Higgs fields
\begin{equation}
H\sim(\mathbf{10}, 1),~~~\overline{H}\sim(\overline{\mathbf{10}}, -1),~~~h\sim(\mathbf{5}, -2),~~~\overline{h}\sim(\overline{\mathbf{5}}, 2)\,.
\end{equation}
The superpotential for the Yukawa couplings of quarks and charged leptons is given by
\begin{equation}
\mathcal{W}\supset y^{D}_{ij}\widetilde{T}_i\widetilde{T}_jh+y^{U}_{ij}\widetilde{T}_i\widetilde{F}_j\overline{h}+y^{E}_{ij}\widetilde{F}_ie^c_j h\,,
\end{equation}
where $y^{D}_{ij}$, $y^{U}_{ij}$ and $y^{E}_{ij}$ are the Yukawa coupling matrices of down-type quarks, up-type quarks and charged lepton respectively. In comparison with the Georgi-Glashow $SU(5)$, the down quark and charged lepton Yukawa couplings are not related in flipped $SU(5)$ and they can be adjusted separately. The neutrino masses can be generated through the double seesaw mechanism by introducing three additional superfields $S_{1,2,3}\sim(\mathbf{1}, 0)$ which are invariant under $SU(5)\times U(1)_X$. The superpotential relevant for neutrino masses is given by
\begin{equation}
\mathcal{W}\supset y^{U}_{ij}\widetilde{T}_i\widetilde{F}_j\overline{h}+y^{S}_{ij}\widetilde{T}_iS_j\overline{H}+\frac{1}{2}(M_S)_{ij}S_iS_j\,.
\end{equation}
The Yukawa couplings are not fixed by the GUT gauge symmetry. In order to understand the flavor puzzle of SM from GUT, flavor symmetry acting in flavor space is generally necessary to obtain certain flavor pattern at the GUT scale. One could use the modular symmetry to avoid the complicated vacuum alignment, the three generations of chiral matter fields of the flipped $SU(5)$ would be arranged to be multiplets of $\Gamma_N$ (or $\Gamma'_N$) with certain modular weights. Then all the Yukawa couplings $y^{D}_{ij}$, $y^{U}_{ij}$, $y^{E}_{ij}$ and $(M_S)_{ij}$ would be modular forms of level $N$~\cite{Charalampous:2021gmf,Du:2022lij}, the quark and lepton mass matrices would depend on a smaller number of free parameters.

\section{Top-down approaches}
This review is devoted to bottom-up approaches to modular symmetry, and its application to flavor models. However, there are certain aspects of modular symmetry which arise from top-down considerations but which are relevant to model building and which should therefore also be considered in this review. In this section, after briefly introducing the role of modular symmetry in string theory, we turn to a couple of these issues, including eclectic flavor symmetry and moduli stabilisation.

\subsection{Modular symmetry from String theory }

In the framework of superstring theory and extra dimensions, orbifold compactifications of two extra dimensions are often done on a torus~\cite{Ferrara:1989bc,Ferrara:1989qb} (for a review see~\cite{DHoker:2022dxx}) and in superstring theory, the single lattice vector which describes the torus (in the convention that the other lattice vector has unit length and lies along the real axis) is promoted to the status of a field, called the modulus field $\tau$, where its vacuum expectation value (VEV) fixes the geometry of the torus~\cite{Ishiguro:2021ccl,Cremades:2004wa,Ishiguro:2020tmo}.
Upon compactifation of the six extra dimensions the modular group or its congruence subgroup emerges naturally in the four-dimensional effective theory. The origin of the modular group in particular is attributed to the specific geometry of the compactification manifold.

From the top-down perspective, the superstring theory in 10 dimensions is a promising framework of unifying all four fundamental interactions. The consistency of the theory requires the six-dimensional extra compact space besides our familiar four-dimensional spacetime. It is remarkable that the non-Abelian discrete flavor symmetry such as $D_4$ and $\Delta(54)$ can arise in certain compactification schemes~\cite{Kobayashi:2004ya,Kobayashi:2006wq,Abe:2009vi}. Moreover, the string duality transformations generate the modular symmetry. The matter fields transform nontrivially under the modular symmetry, consequently the modular symmetry could constrain the flavor structure of quarks and leptons and it enforces the Yukawa couplings to be modular forms~\cite{Feruglio:2017spp}.
In bottom-up models with modular symmetry alone, the finite modular groups $\Gamma_N$ and $\Gamma'_N$ ($N=2, 3, 4, 5, 6, 7$) play the role of flavor symmetry, More generally the finite modular groups can be expressed as the quotient groups of $SL(2, \mathbb{Z})$ over its normal subgroups with finite index~\cite{Liu:2021gwa}.

There has been considerable effort devoted to studying modular symmetry arising from orbifolds in top-down Heterotic string constructions~\cite{Baur:2019kwi,Nilles:2021glx}. However there is a limited choice of consistent 10d toroidal orbifolds consistent with supersymmetry, for example all such orbifolds based on $(T^2)^3/(Z_N\times Z_M)$ have been classified~\cite{Fischer:2012qj}.
One of these examples, $(T^2)^3/(Z_4\times Z_2)$, has been recently studied in a bottom-up approach to modular symmetry involving three moduli at their fixed points $i, i+2, \omega$, leading to a minimal predictive (littlest) seesaw model of leptons~\cite{deAnda:2023udh}, as discussed earlier in the review.

Apart from Heterotic string theory, modular symmetry is also present in other types of string theory such as Type I, Type IIA, Type IIB, and possibily F-theory. In such explicit constructions of modular family symmetries from string theory, one typically finds very constrained modular weights assignments defined by the compactification~\cite{Kikuchi:2023clx}. For example,
for Type IIB on magnetised branes one typically has $k=-1/2$, while in Type IIA on intersecting branes $k=-1/2 + \text{shift}$ where the shift depends on intersecting numbers and does not alter selection rules~\cite{Kobayashi:2020hoc}, as compared to heterotic string theory on a
Calabi-Yao manifold where the weights are $k=-1$ for untwisted and $k=-n/2$ for twisted depending on details of compactification~\cite{Nilles:2021glx}.

In general, top-down approaches to Heterotic string theory suggest that the finite modular symmetry will typically be accompanied by a non-commuting flavour symmetry leading to so called Eclectic flavour symmetry~\cite{Nilles:2020nnc,Nilles:2020kgo,Nilles:2020tdp,Baur:2020jwc,Ding:2023ynd}, as we discuss in the following subsection.

\subsection{Eclectic flavor symmetry }

Motivated by top-down constructions, the idea of eclectic flavor group (EFG) has been proposed in~\cite{Nilles:2020nnc,Nilles:2020kgo}, and it has been developed in a series of papers~\cite{Baur:2019kwi,Baur:2019iai,Nilles:2020nnc,Nilles:2020kgo,Nilles:2020tdp,Nilles:2020gvu,Baur:2021mtl}. The eclectic flavor group is a maximal extension of the traditional flavor group by finite modular group, and certain consistency conditions have to be fulfilled in order to consistently combine modular symmetry with traditional flavor symmetry.

Under the action of a traditional flavor transformation $g$ or a modular transformation $\gamma$, the complex modulus $\tau$ and a generic matter field multiplet $\psi$ transform as follow~\cite{Feruglio:2017spp,Liu:2019khw}
\begin{equation}
\left\{\begin{array}{l}
\tau \stackrel{g}{\longrightarrow}\tau, \quad \psi\stackrel{g}{\longrightarrow}\rho(g)\psi,~~~~g\in G_{\text{fl}}\,, \\
\tau\stackrel{\gamma}{\longrightarrow}\gamma\tau\equiv\frac{a\tau+b}{c\tau+d},~~~\psi\stackrel{\gamma}{\longrightarrow}(c\tau+d)^{-k_{\psi}}\rho(\gamma)\psi,~~~\gamma=\begin{pmatrix}
a  &  b\\
c  & d
\end{pmatrix}\in\Gamma \,,
\end{array}\right.
\end{equation}
where $k_{\psi}$ is the modular weight of the matter field multiplet $\psi$, and $\rho(g)$ and $\rho(\gamma)$ are unitary representations of traditional flavor group $\mathcal{G}_{\text{fl}}$ and the finite modular group $\Gamma_{N}$ or $\Gamma^\prime_{N}$, respectively. We see that the flavor symmetry transformation leaves the modulus $\tau$ invariant. Consequently the traditional flavor symmetry and modular symmetry are distinguished  by their action on the modulus $\tau$. Let us first perform a modular transformation $\gamma\in\Gamma$, subsequently a traditional flavor transformation $g$ and last perform the inverse modular transformation  $\gamma^{-1}$. Since the modulus $\tau$ is invariant under this chain of transformations, the resulting transformation should be another traditional flavor symmetry transformation $g'$~\cite{Nilles:2020nnc}, i.e.    \begin{equation}\label{eq:consistency-cond-EFG}
\rho(\gamma) \rho(g) \rho^{-1}(\gamma)=\rho(g^\prime), \qquad g,g^\prime\in \mathcal{G}_{\text{fl}}, \quad \gamma \in \Gamma \,.
\end{equation}
This is the consistency condition between traditional flavor group and finite modular group. Eq.~\eqref{eq:consistency-cond-EFG} implies that the modular transformation $\gamma$ maps the traditional group element $g$ into another element $g^\prime$. Therefore Eq.~\eqref{eq:consistency-cond-EFG} defines a automorphism of the traditional flavor group $\mathcal{G}_{\text{fl}}$, i.e.
\begin{equation}\label{eq:consistency-cond2-EFG}
\rho(\gamma) \rho(g) \rho^{-1}(\gamma)=\rho(u_{\gamma}(g)) \,, \qquad \forall g\in G_{f}\,,
\end{equation}
where $u_{\gamma}$ is an automorphism $u_{\gamma}:\mathcal{G}_{\text{fl}}\rightarrow \mathcal{G}_{\text{fl}}$ with $u_{\gamma}(g)=g'$. If $u_{\gamma}$ is the trivial identity automorphism with $u_{\gamma}=1$ for any modular transformation $\gamma$ so that $u_{\gamma}(g)=g$, the modular symmetry transformation would commute with the flavor symmetry transformation and consequently the eclectic flavor group is the direct product $G_f\times \Gamma_N$ or $G_f\times \Gamma'_N$~\cite{Nilles:2020nnc}. It is called the quasi-eclectic flavor symmetry~\cite{Chen:2021prl}.

Since the finite modular groups $\Gamma_N$ and $\Gamma^\prime_{N}$ can be generated by the two generators $S$ and $T$, it is sufficient to impose the consistency condition in Eq.~\eqref{eq:consistency-cond2-EFG} on the two outer automorphisms $u_{S}$ and $u_{T}$
\begin{equation}
\label{eq:ST_Cons}
\rho(S)\,\rho(g)\,\rho^{-1}(S)= \rho(u_{S}(g)), \qquad \rho(T)\,\rho(g)\,\rho^{-1}(T)~=~ \rho(u_{T}(g))\,,
\end{equation}
where $\rho(S)$ and $\rho(T)$ are matrix representations of the two automorphisms $u_{S}$ and $u_{T}$ respectively, and they should satisfy the multiplication rules of the finite modular group $\Gamma_N$ or $\Gamma^\prime_N$ in Eqs.~(\ref{eq:GammaN-rules}, \ref{eq:GammaNp-rules}). In other words, the outer automorphisms $u_{S}$ and $u_{T}$ should also satisfy the multiplication rules of the finite modular group $\Gamma_N$ or $\Gamma^\prime_N$:
\begin{equation}\label{eq:uS_uT_rules}
\left(u_{S}\right)^{N_s} =\left(u_{T}\right)^N =\left(u_{S} \circ u_{T}\right)^3=1,  \qquad \left(u_{S}\right)^2  \circ u_{T} = u_{T} \circ \left(u_{S}\right)^2\,,
\end{equation}
with $N_s=4$ for $\Gamma'_N$ and $N_s=2$ for $\Gamma_N$.

The EFG can be consistently combined with the gCP symmetry. As explained in section~\ref{subsec:gCP-modular-symmetry}, a new generator $K_{*}$ corresponding to gCP transformation could be introduced and the modular group  $\Gamma\cong SL(2,\mathbb{Z})$ is enhanced to  $\Gamma^*\cong GL(2,\mathbb{Z})$~\cite{Novichkov:2019sqv}:
\begin{equation}
\Gamma^{*}= \Big\{ \tau \stackrel{S}{\longrightarrow} -1/\tau, ~~ \tau \stackrel{T}{\longrightarrow} \tau + 1, ~~  \tau \stackrel{K_*}{\longrightarrow} -\tau^{*} \Big\}\,.
\end{equation}
The gCP transformation $K_{*}$ acts on the modulus $\tau$ and the matter field as follows,
\begin{equation}\label{eq:K*_to_matter}
\tau\stackrel{K_{*}}{\longrightarrow}-\bar{\tau},\quad	\psi(x)\stackrel{K_{*}}{\longrightarrow} \rho(K_*)[\psi^{\dagger}(t, -{\bm x})]^T\,,
\end{equation}
where the gCP transformation $\rho(K_*)$ is a unitary matrix, and the obvious action of CP on the spinor indices is omitted for the case of $\psi$ being spinor. Requiring that the gCP transformation $K_{*}$ be of order 2 with $(K_{*})^2=1$, we can obtain
\begin{equation}\label{eq:rho_K*_order2}
		\rho^*(K_*)= \rho^{-1}(K_*)\,.
\end{equation}
The gCP transformation has to be compatible with both the traditional flavor symmetry and the finite modular group, and its allowed form is strongly constrained by the corresponding restricted consistency conditions. The consistency between the EFG and gCP symmetry requires the following consistency conditions have to be satisfied~\cite{Novichkov:2019sqv,Ding:2021iqp}:
\begin{eqnarray}
\label{eq:con_uK} &\rho(K_*)\rho^*(g)\rho^{-1} (K_*)= \rho(u_{{K}_*}(g))\,, \qquad \forall g\in \mathcal{G}_{\text{fl}}\,,\\
\label{eq:consistency_K*_mod} &\rho(K_*)\rho^{*}(S)\rho^{-1}(K_*)=\rho^{-1}(S),\qquad 	\rho(K_*)\rho^{*}(T)\rho^{-1}(K_*)=\rho^{-1}(T)\,,
\end{eqnarray}
where $u_{{K}_*}$ is an automorphism of the traditional flavor symmetry group $\mathcal{G}_{\text{fl}}$. It is sufficient to consider the element $g$ being the generators of $\mathcal{G}_{\text{fl}}$, and one can fix the explicit form of the gCP transformation $\rho(K_*)$ up to an overall irrelevant phase by solving Eqs.~(\ref{eq:con_uK},\ref{eq:consistency_K*_mod}). Hence the automorphism $u_{K_*}$ of the traditional flavor group $\mathcal{G}_{\text{fl}}$ should satisfy the following relations:
\begin{equation}\label{eq:auto_Kstar_rules}
(u_{{K}_*})^2=1, \qquad  u_{{K}_*}\circ u_{S}\circ u_{{K}_*}=u_{{S}}^{-1}, \qquad  u_{{K}_*}\circ u_{T}\circ u_{{K}_*}=u_{{T}}^{-1}\,.
\end{equation}

\begin{table}[t!]
\center
\begin{tabular}{|cc|c|l|c|c|}
\hline\hline
flavor group    & GAP     & $\mathrm{Aut}(\mathcal{G}_\mathrm{fl})$ & \multicolumn{2}{c|}{finite modular}    & eclectic flavor\\
$\mathcal{G}_\mathrm{fl}$ & ID     &                                         & \multicolumn{2}{c|}{groups}   & group\\ \hline\hline
$Q_8$  & [ 8, 4 ]   & $S_4$   & without $\mathcal{CP}$ & $S_3$               & $\mathrm{GL}(2,3)$\\ \cline{4-6}
  &      &     & with $\mathcal{CP}$    & --     & --\\ \hline
$\mathbbm{Z}_3\times \mathbbm{Z}_3$     & [ 9, 2 ]   & $\mathrm{GL}(2,3)$    & without $\mathcal{CP}$ & $S_3$    & $\Delta(54)$\\ \cline{4-6}
 &   &    & with $\mathcal{CP}$    & $S_3\times\mathbbm{Z}_2$   & [108, 17]\\ \hline
$A_4$   & [ 12, 3 ]  & $S_4$  & without $\mathcal{CP}$ & $S_3$  & $S_4$\\
  &    &     &      & $S_4$    & $S_4$\\ \cline{4-6}
  &    &    & with $\mathcal{CP}$    & --    & --\\ \hline
$T'$   & [ 24, 3 ]  & $S_4$   & without $\mathcal{CP}$ & $S_3$  & $\mathrm{GL}(2,3)$\\ \cline{4-6}
   &    &  & with $\mathcal{CP}$    & --     & --\\ \hline
$\Delta(27)$    & [ 27, 3 ]  & [ 432, 734 ]   & without $\mathcal{CP}$ & $S_3$    & $\Delta(54)$\\
 &   &   &    & $T'$   & $\Omega(1)$\\ \cline{4-6}
 &   &   & with $\mathcal{CP}$    & $S_3\times\mathbbm{Z}_2$    & [108, 17]\\
 &   &   &     & $\mathrm{GL}(2,3)$  & [1296, 2891]\\ \hline
$\Delta(54)$ & [ 54, 8  ] & [ 432, 734 ]    & without $\mathcal{CP}$ & $T'$  & $\Omega(1)$\\ \cline{4-6}
 &   &   & with $\mathcal{CP}$   & $\mathrm{GL}(2,3)$  & [1296, 2891]\\
\hline\hline
\end{tabular}
\caption{Examples of traditional flavor groups, their extensions by finite modular groups and the resulting eclectic flavor groups. Here $\mathrm{Aut}(\mathcal{G}_\mathrm{fl})$ denotes the group of automorphisms of the traditional flavor group $\mathcal{G}_\mathrm{fl}$. This table is taken from Ref.~\cite{Nilles:2020nnc}.
\label{tab:EFG-candidates}}
\end{table}

The EFG extensions of some popular traditional flavor symmetry group are listed in table~\ref{tab:EFG-candidates}. The scheme of eclectic flavor group is more predictive than the finite modular group and the traditional flavor group alone, and it combines the advantages of both approaches so that both superpotential and K\"ahler potential would be severely restricted by EFG. Usually only the minimal K\"ahler potential is adopted in concrete modular models. However, the K\"ahler potential is less constrained by modular symmetry alone, and the most general K\"ahler potential compatible with modular symmetry is of the following form~\cite{Chen:2019ewa,Lu:2019vgm},
\begin{eqnarray}
\label{eq:kahler-MFS}\mathcal{K}=(-i\tau+i\bar{\tau})^{-k_{\psi}}\left(\psi^{\dagger}\psi\right)_{\mathbf{1}}+\sum_{n,\bm{r_1},\bm{r_2}} c^{(n,\bm{r_1},\bm{r_2})} (-i \tau+ i \bar \tau)^{-k_{\psi}+n}
\left(\psi^{\dagger} Y^{(n)\dagger}_{\bm{r_{1}}} Y^{(n)}_{\bm{r_2}} \psi  \right)_{\bm{1}}\,,
\end{eqnarray}
where $\psi$ stands for a generic matter field multiple with modular weight $k_{\psi}$, and $ Y^{(n)}_{\bm{r}}$ is a weight $n\neq0$ modular form in the representation $\bm{r}$ of finite modular group. Usually only the first term of Eq.~\eqref{eq:kahler-MFS} is taken, this is the so-called minimal K\"ahler potential. However, we see that the K\"ahler potential generally has a lot of terms and the couplings $c^{(n,\bm{r_1},\bm{r_2})}$ are not suppressed. Consequently these additional terms can be as important as the first term, and thus the predictive power of modular flavor symmetry would be reduced~\cite{Chen:2019ewa}. How to control the K\"ahler potential is an open question of the modular flavor symmetry approach. The presence of traditional flavor symmetry $\mathcal{G}_\mathrm{fl}$ leads to strong constraints on both the modular transformations of matter fields and the K\"ahler potential. As a result, the minimal K\"ahler potential can be reproduced and the higher order corrections suppressed by powers of $\langle\Phi\rangle/\Lambda$, where $\Lambda$ denotes the cutoff scale and $\langle\Phi\rangle$ represents the vacuum expectation value (VEV) of flavons breaking the traditional flavor symmetry. See Refs.~\cite{Baur:2022hma,Ding:2023ynd,Li:2023dvm} for model construction with EFG.

\subsection{Modulus stabilization}

A long standing problem of string theory is that of finding a successful potential for the moduli fields, one which stabilizes them at a global minimum which corresponds to a de Sitter (dS) vacuum, namely one with a (slightly) positive cosmological constant. The theoretical framework for discussing the issue of moduli stabilisation is that of effective supergravity theories.

For example one approach to modulus stabilisation in Type IIB string theory was that in~\cite{Giddings:2001yu, Gukov:1999ya, Curio:2000sc, Ashok:2003gk, Denef:2004cf}. More recently, in the framework of modular flavour symmetry a 3-form flux in Type IIB models has been considered~\cite{Ishiguro:2020tmo} based on complex structure moduli with a $T^6_{}/(Z^{}_2 \times Z^\prime_2)$ orbifold. The minima of the moduli fields are found to be clustered around the left cusp $\tau = \omega$ in the fundamental domain. Another example involving non-perturbative effects was studied in~\cite{Kobayashi:2019xvz,Kobayashi:2019uyt,Kikuchi:2023uqo}.

In Heterotic string theory, it is traditional to consider non-perturbative effect arising from gaugino condensation~\cite{Dine:1985rz, Nilles:1982ik, Ferrara:1982qs}. The potential for the dilaton, and the K\"ahler and complex structure moduli is flat at tree level, but threshold corrections~\cite{Kaplunovsky:1987rp, Dixon:1990pc,Antoniadis:1991fh, Antoniadis:1992rq} or worldsheet instantons can modify the potential leading to minima~\cite{Cicoli:2013rwa}. The effect of modular symmetries on the K\"ahler moduli has also been considered~\cite{Font:1990nt, Gonzalo:2018guu}, leading to anti-de Sitter (AdS) vacua with negative cosmological constant. A similar approach was followed in~\cite{Novichkov:2022wvg} with minima close to the stabiliser $\tau = \omega$.

Recently~\cite{Knapp-Perez:2023nty} showed that the AdS vacua can be uplifted to become dS vacua by the effects of matter in the superpotential~\cite{Lebedev:2006qq, Lebedev:2006qc} following the Kachru-Kallosh-Linde-Trivedi (KKLT) scenario~\cite{Kachru:2003aw}.

Alternatively, it is possible to realise dS vacua even without introducing the matter superpotential~\cite{Leedom:2022zdm} using non-perturbative corrections to the dilaton K\"ahler potential proposed by Shenker~\cite{Shenker:1990uf}. This leads to metastable dS vacua at $\tau = {\rm i}$ and $\omega$ in some regions of multiple moduli parameter space~\cite{King:2023snq}. The 1-loop correction may lead to the slight deviation from the tree level results of modulus VEV~\cite{Kobayashi:2023spx}.

\section{Summary and conclusion}

In this review we have systematically reviewed the formalism and applications of modular symmetry to flavour models in general and neutrino mass and mixing models in particular, from the bottom-up point of view.
We began with a survey of neutrino mass and lepton mixing, including a brief history of neutrino mass and mixing in the period of most rapid developments, 1998-2012, then summarised what we have learned, what we still don't know, and why the new paradigm demanded by neutrino physics cannot be consistent with the Standard Model, before mentioning a number of possible mechanisms for the origin of neutrino mass, focussing on various Majorana seesaw mechanisms. We then showed how the PMNS lepton mixing matrix may be constructed and parameterised, and how its parameters are constrained by the latest global fits, before discussing the flavour puzzle of the Standard Model. We then gave a very brief reminder of neutrino mass and mixing models strategies without modular symmetry, starting with
some simple patterns of lepton mixing, then introducing family (or flavour) symmetries which can reproduce these structures using the direct approach. This was followed by a discussion of the semi-direct and tri-direct CP approaches, where the simple patterns of lepton mixing receive the necessary corrections to make them phenomenologically viable whilst still maintaining some degree of predictivity due to a subgroup of the family symmetry being preserved in either the charged lepton sector or the neutrino sector, leading to phenomenologically testable sum rules.

We then turned to the main subject of this review, namely a pedagogical introduction to modular symmetry as a candidate for family symmetry, from the bottom-up point of view. Unlike usual flavour symmetries, which act linearly on flavon fields, modular symmetry acts non-linearly on the modulus field, $\tau$, which transforms into $(a \tau + b)/(c\tau +d)$, where $a,b,c,d$ are integers with $ac-bd=1$. Such symmetry transformations form a Special Linear group of $2\times 2$ matrices whose elements are integers $\mathbb{Z}$, hence the name $SL(2,\mathbb{Z})$. In 6d theories, the modulus field, $\tau$, defines the shape of the compactified 2d torus, and modular symmetry is the group of symmetry transformations on $\tau$ which leaves the torus invariant. The modulus field $\tau$, is constrained to lie in the upper half of the complex plane, in a fundamental domain which excludes both a unit circle centred on the origin and real parts of magnitude greater than $0.5$, which means that modular symmetry is always necessarily broken. Although there is no value of $\tau$ which preserves the modular group, there are some values which preserve part of the symmetry, and these are called fixed points which preserve a residual symmetry or stabilizer. For example $\tau = i$ is invariant under the transformation $\tau \rightarrow 1/(-\tau )$, corresponding to a modular transformation with $a=0,b=1,c=-1,d=0$, which is identified with the $2\times 2$ matrix generator $S$. We show that there are only three inequivalent fixed points namely $\tau = i, \omega$ and $i\infty$ which play a special role both in string theory, and in phenomenologically.

After the introduction to the above full modular group,
$SL(2,\mathbb{Z})$ we introduced its projective subgroup $PSL(2,\mathbb{Z})$ in which transformations related by an overall minus sign are identified (since $a,b,c,d$ all change sign which cancels in the ratio), corresponding to the faithful group action on the modulus field $\tau$. We also introduce their respective principal congruence subgroups,  $\Gamma(N)$ and $\overline{\Gamma}(N)$, of level $N=1,2,3,\ldots$, which essentially consist of unit matrices mod $N$. Although all the above groups are infinite, finite discrete subgroups emerge by taking the quotient of the modular groups with their principal congruence subgroups, namely
$\Gamma'_N\equiv SL(2,\mathbb{Z})/\Gamma(N)$ and $\Gamma_N\equiv PSL(2,\mathbb{Z})/\overline{\Gamma}(N)$.

The resulting finite groups (which are essentially the interesting finite parts of the modular groups, once the
uninteresting infinite sets of matrices related to unit matrices mod $N$ have been removed) which emerge
are well known from their applications to flavour model building, as  discussed earlier.
For example
$\Gamma_{N=3,4,5}$ are isomorphic to $A_4,S_4,A_5$, respectively, while $\Gamma'_{N=3,4,5}$ are isomorphic to the respective double cover groups  $T',S'_4,A'_5$. Modular symmetry not only provides a theoretical origin for such well known flavour groups, which all contain triplet representations and can accommodate three families of fermion representations, but the Yukawa couplings are generally modular forms, which are mathematical functions of $\tau$, which depend on its modular weight, and transform as a representation of the finite modular group.
In this framework a Yukawa operator involving fields whose modular weights do not add up to zero, is therefore permitted if the Yukawa coupling is a modular form which carries a weight which balances that of the fields, where the Yukawa coupling itself may transform as a representation of $\Gamma_N$ or $\Gamma'_N$, obviating the need for flavon fields.

We have derived the modular forms of level $N$ and even weight $k$, using a range of techniques, and showed that they form representations of the finite modular group $\Gamma_N$, considering the derivation for $N=3$ in detail as an example. We then extended the discussion to include the double cover groups $\Gamma'_N$ with integer modular weights, and the metaplectic covers with rational modular weights. We also introduced the vector valued modular forms, based on a new formalism which is more general than that of quotients of principal congruence subgroups, and allows more general finite modular groups beyond $\Gamma_N$ or $\Gamma'_N$, and their corresponding modular forms, to be identified.
In an extensive Appendix, we give the irreducible representations, Clebsch-Gordon coefficients
and modular forms of $\Gamma'_N$ and $\Gamma_N$ explicitly for the levels $N=2, 3, 4, 5$.

The interplay between modular symmetry and generalized CP symmetry was then discussed, where we showed that the modulus field should transform under CP as $\tau \rightarrow -\tau^*$, where the minus sign is necessary in order for the modulus field to remain in the upper half of the complex plane. We derived the CP transformation on matter multiplets (which are chiral supermultiplets). We also highlighted the special choice of the symmetric basis of representation matrices, in which the generalised CP transformation on the matter fields become just the canonical CP transformations consisting of unit matrices and complex conjugation, which is always possible for all examples so far studied. The modular forms themselves also transform under CP, which we show to be identical to a corresponding matter field with a similar representation and weight.
In the symmetric basis the Clebsch-Gordan coefficients and all coupling constants are real.
We noted that $\tau$ has CP invariant lines all along the imaginary axis, so that such values would not violate CP, and also showed the other lines along the boundary of the fundamental domain respect certain combinations of residual modular symmetry and CP.
For a single modulus on a CP invariant line, we showed that this would lead to all physical CP phases in the lepton sector being trivial,
and hence to break CP requires some deviation from these lines. However this conclusion is not true when multiple moduli are considered.

In general, compactification of extra dimensions generally leads to a number of moduli $\tau_i$ with $i=1,2,\ldots$,
rather than just the single modulus field
$\tau$ as often assumed in bottom-up models, and so it is necessary to generalise the above formalism to the case of more than one modulus field. The geometrically simplest case is that of factorizable tori, which
allows a straightforward generalisation of the formalism to the
case of multiple moduli. The case of non-factorizable or symplectic multiple moduli, is considered by extending the modular group from $SL(2,\mathbb{Z})$ to the Siegel group $Sp(2g,\mathbb{Z})$, where we discussed modular forms appropriate to this case, with the simple case of factorizable tori emerging as a limiting case. In practice, $g=2$ is already quite complicated, where the reduced finite Siegel modular group arises which is more manageable, but still awaits phenomenological exploitation. By contrast, the simple multiple factorizable tori case, also commonly assumed in string theory compactifications, provides a straightforward generalisation with many phenomenological applications already in the literature.

Modular strategies for understanding fermion mass hierarchies were discussed, including the weighton mechanism and appealing to small deviations from the fixed points. The weighton mechanism is analogous to the Froggatt-Nielsen mechanism but without requiring any Abelian symmetry to be introduced. The modular weights play the role of FN charges, and a singlet field with non-zero modular weight plays the role of the FN flavon field, and is called a weighton. Lepton mixing may be successfully addressed in this approach. Alternatively, a second strategy relies on the observation that modular forms may contain zeros at the fixed points. Consider for example the weight 2 level 3 triplet modular form in the standard basis $(Y_1,Y_2,Y_3)=(1,0,0)$ at $\tau = i\infty$. Under small deviations from this fixed point, the modular form becomes
$(Y_1,Y_2,Y_3)=(1,\epsilon,\epsilon^2)$, where $\epsilon \ll 1$, allowing charged lepton hierarchies to be addressed. The requirement of successful lepton mixing is not easy to achieve in this approach, although a successful example has been given. Another approach to reducing the number of free parameters is the idea of texture zeroes in the mass matrices, which arise naturally in the framework of modular symmetry since modular forms in certain representations of $\Gamma_N$ or $\Gamma'_N$ are absent at lower weights, with the modular symmetry providing extra predictive power for the non-zero elements.

A variety of examples of modular models exist in the literature based on the SM gauge group and different finite modular groups of levels
$N=2-7$. As an example, we gave a detailed overview of a general class of $N=3$ lepton models based on $A_4$ with a single modulus field, as in the original Feruglio seesaw model which we showed can still provide a good fit to the data, and more generally assigning the $A_4$ modular weights and representations of the leptons in all possible ways, and tabulating the phenomenologically successful cases. We then presented a minimal phenomenologically successful model of leptons with a single modulus, based on $S'_4$ with leptons carrying specific modular weights in specific representations, leading to a six parameter fit to the lepton data. Moving to multiple moduli, we then discussed a highly predictive model based on three $S_4$ groups, with the three moduli located at the fixed points $i,i+2$ and $\omega$, capable of reproducing the Littlest Seesaw model, with the lepton data reproduced with only six free parameters (three in the neutrino sector and three required to fit the charged lepton masses). There are also modular models in the literature capable of explaining both leptons and quarks with a single modulus, and we discussed such an example with spontaneously broken CP symmetry.

We then extended the discussion to include Grand Unified Theories (GUTs) based on modular $SU(5)$, $SO(10)$ and flipped $SU(5)_F\times U(1)$. Most examples concern $SU(5)$, where we summarise all phenomenologically successful $A_4$ models in terms of the representations and modular weights of the fields, including right-handed neutrinos to provide a type-I seesaw mechanism. We also discuss an analysis of $SO(10)$ models with three Higgs multiplets, where the right-handed neutrinos are are a necessary ingredient, but the resulting seesaw mechanism may be type-I or type-II dominated. The three Higgs multiplets of the $SO(10)$ models, namely
$\mathbf{10}$, $\overline{\mathbf{126}}$ and $\mathbf{120}$, include the latter two rather large representations which are not readily available in most string constructions. This motivates the study of more string friendly groups such as flipped $SU(5)_F\times U(1)$ where the symmetry breaking and fermion mass structure can be achieved with smaller Higgs representations, and these types of models have also been discussed in the framework of modular symmetry. In the modular invariant models, the exchange of moduli between electrons and neutrinos can induce a non-standard neutrino interactions which can leads to a shift of the neutrino mass matrix in a region with non-vanishing electron number density~\cite{Ding:2020yen}. Hence the presence of moduli can potentially be tested in neutrino oscillation experiments~\cite{Ding:2020yen}.

Finally we briefly mentioned some issues related to top-down approaches based on string theory, including eclectic flavour symmetry and
moduli stabilisation. Although this review is mainly concerned with bottom-up considerations, it is appropriate to recall that the whole subject of modular symmetry is very much inspired by string theory in extra dimensions, and it is important to have this in mind when considering bottom-up models. In particular, six dimensional theories play a crucial role in the interpretation of the modulus field $\tau$, and in general one must expect moduli fields for ten dimensional theories, with three moduli fields for the simplest three-tori compactifications. It is somewhat sobering to realise that many realistic string compactifications lead to exclusively matter fields with half-integer modular weights, as compared the rich choice of modular weights usually assumed in bottom-up models. Another observation is that realistic Heterotic string constructions tend to yield traditional flavour symmetries together with, but not commuting with, the finite modular symmetries, leading to so-called eclectic symmetry models. On the one hand this is very unfortunate, since such symmetries can only be broken at the expense of re-introducing flavon fields (which should be present as massless modes of the string construction), which are the very fields that we wished to be rid of in the beginning.  On the other hand this is fortunate, since the presence of the traditional (non-modular) flavour symmetry controls the corrections to the K\"ahler potential which could not be controlled by modular symmetry alone. This is because the insertion of the dimensionless Yukawa coupling modular forms in a particular representation leads to unsuppressed corrections, while the
while the analogous corrections with traditional flavour symmetry involve powers of flavon insertions suppressed by the mass dimensions of the flavon fields. Moreover, realistic eclectic symmetry models, in which the full symmetry is realised as the automorphism of the chosen flavour symmetry, tend to be quite constrained and predictive, containing the advantages of both approaches. Finally we discussed the issue of moduli stabilisation, which is a long standing issue in string theory, and very much related to the discussion of fixed points, with $i$ and $\omega$, or regions close to these fixed points being preferred, while $\infty$ is considered somewhat unphysical from this point of view.

In conclusion, we have reviewed the idea of finite modular symmetry, motivated from string theory and extra dimensions, and its application to bottom-up flavour models in general and neutrino mass and lepton mixing models in particular. We have seen how the dimensionless Yukawa couplings become dynamical objects, which transform as modular forms, mathematical functions controlled by a single modulus field $\tau$, and which also transform as representations of the finite modular symmetry, rather like flavon fields. The promise of being able to have a well motivated modular origin for the non-Abelian discrete family or flavour symmetry, which is broken without the use of arbitrary flavon fields has led to considerable interest in this approach, and we have summarised the latest formalism and applications of this approach. The string motivations however also lead us to eclectic flavour symmetry which re-introduces traditional favour symmetry and flavons, together with the finite modular symmetry, which may help to control the K\"ahler potential and lead to a highly constrained framework. However, in the end, the bottom-up model builder cannot help but feel that we have come full circle and are back again at the beginning where we started with flavons being present once more. We are reminded of the quote by T.S. Eliot: ``We shall not cease from exploration and the end of all our exploring will be to arrive where we started and know the place for the first time.'' From this point of view, the study of modular symmetry has certainly enriched our understanding of the flavour problem, even if it cannot claim to be the final answer.

\section*{Acknowledgements}

We are grateful to Jun-Nan Lu and Xiang-Gan Liu for their kind helps in numerical calculations and plotting some of the figures. We acknowledge Peng Chen, Francisco J. de Anda, Ivo de Medeiros Varzielas, Ferruccio Feruglio, Simon J.D. King, Miguel Levy, Cai-Chang Li, Xiang-Gan Liu, Jun-Nan Lu, Bu-Yao Qu, Xin Wang, Chang-Yuan Yao and Ye-Ling Zhou for fruitful collaboration on modular symmetry.  GJD is supported by the National Natural Science Foundation of China under Grant Nos.~11975224, 11835013, 12375104. SFK acknowledges the STFC Consolidated Grant ST/L000296/1 and the European Union's Horizon 2020 Research and Innovation programme under Marie Sk\l{}odowska-Curie grant agreement HIDDeN European ITN project (H2020-MSCA-ITN-2019//860881-HIDDeN).

\section*{Appendix}

\begin{appendix}

\section{\label{app:derivation-Ferruccio} Derivation of Eq.~\eqref{eq:Ferruccio1} }

The well-known modular transformation is
\begin{equation}
\tau\rightarrow \tau'=\gamma\tau=\frac{a\tau+b}{c\tau+d},~~\text{with}~~ ad-bc=1\,,
\end{equation}
which implies
\begin{eqnarray}
&&\frac{d\tau'}{d\tau}=\frac{a(c\tau+d)-c(a\tau+b)}{(c\tau+d)^2}=\frac{ad-bc}{(c\tau+d)^2}=\frac{1}{(c\tau+d)^2}\\
\label{eq:dtau-datau'}&&\frac{d\tau}{d\tau'}=\frac{1}{\frac{d\tau'}{d\tau}}=(c\tau+d)^2\,.
\end{eqnarray}
Ref.~\cite{Feruglio:2017spp} starts from a function $f(\tau)$ which transforms under modular transformation in the following way,
\begin{equation}
f(\tau)\rightarrow f(\tau')=e^{i\alpha}(c\tau+d)^kf(\tau)\,,
\end{equation}
which is exactly Eq.~\eqref{eq:f-phase}. Then we can see that the derivative of the $\log f(\tau)$ transforms as
\begin{eqnarray}
\frac{d}{d\tau}\log f(\tau)\rightarrow \frac{d}{d\tau'}\log f(\tau')&=& \frac{d\tau}{d\tau'}\frac{d}{d\tau}\log f(\tau')\\
&=&\frac{d\tau}{d\tau'}\frac{d}{d\tau}\log\left[e^{i\alpha}(c\tau+d)^kf(\tau)\right]\\
&=&\frac{d\tau}{d\tau'}\frac{d}{d\tau}\left[\log e^{i\alpha}+\log(c\tau+d)^k+\log f(\tau)\right] \\
&=&\frac{d\tau}{d\tau'}\left[\frac{kc}{c\tau+d}+\frac{d}{d\tau}\log f(\tau)\right]\\
&=&(c\tau+d)^2\left[\frac{kc}{c\tau+d}+\frac{d}{d\tau}\log f(\tau)\right]\\
&=&(c\tau+d)^2\frac{d}{d\tau}\log f(\tau)+kc(c\tau+d)\,.
\end{eqnarray}
We have used Eq.~\eqref{eq:dtau-datau'} in above.

\section{\label{app:tau-gCP-transform} Derive the gCP transformation of the modulus $\tau$ from consistency condition  }

As shown in Eqs.~(\ref{eq:varphi-mod-trans}, \ref{eq:varphi-gCP-trans}), a chiral supermultiplets $\varphi$ transforms under modular symmetry and gCP  as follows,
\begin{eqnarray}
\nonumber \text{modular symmetry}&:&\varphi(x)\stackrel{\gamma}{\longrightarrow} (c\tau+d)^{-k_{\varphi}}\rho_{\mathbf{r}}(\gamma)\varphi(x)\,,~~~\gamma=\begin{pmatrix}
a ~& b \\
c ~& d
\end{pmatrix}\in\Gamma\,, \\
\text{gCP}&:&\varphi(x)\stackrel{\mathcal{CP}}{\longrightarrow} X_{\mathbf{r}}\overline{\varphi}(x_{\mathcal{P}})
\end{eqnarray}
In order to combine generalized CP symmetry consistently with the modular symmetry consistently, as explain in section~\ref{ch:05-gCP}, the following consistency condition of Eq.~\eqref{eq:consistency-cond1} has to be satisfied:
\begin{equation}
X_{\mathbf{r}} \rho^*_\mathbf{r}(\gamma) X_{\mathbf{r}}^{-1} =\left(\frac{c_{\gamma'}\tau+d_{\gamma'}}{c\tau^*_{\mathcal{CP}^{-1}}+d}\right)^{-k_{\varphi}} \rho_\mathbf{r}(\gamma'),~~~~\gamma'=\begin{pmatrix}
a_{\gamma'}  &  b_{\gamma'} \\
c_{\gamma'}  &  d_{\gamma'}
\end{pmatrix}\in\Gamma\,.
\end{equation}
Since the matrices $X_{\mathbf{r}}$, $\rho^*_\mathbf{r}(\gamma)$ and $\rho_\mathbf{r}(\gamma')$ are independent of the moduli $\tau$, the overall factor has to be a constant:
\begin{equation}
\label{eq:CP-cons1}
\frac{c_{\gamma'}\tau+d_{\gamma'}}{c\tau^*_{\mathcal{CP}^{-1}}+d}=\lambda_{\gamma}\,,
\end{equation}
where $\lambda_{\gamma}$ is a complex parameter with $|\lambda_{\gamma}|=1$ because both $X_{\mathbf{r}}$ and $\rho_{\mathbf{r}}$ and unitary matrices. In general, the value of $\lambda_{\gamma}$ depends on $\gamma$. The constraint in Eq.~\eqref{eq:CP-cons1} allows to determine the gCP transformation of the modulus $\tau$~\cite{Novichkov:2019sqv}. Applying the above condition of Eq.~\eqref{eq:CP-cons1} to the generator $\gamma=S$, we obtain
\begin{equation}
\label{eq:gCP-der1}c_{S'}\tau + d_{S'}=-\lambda_S\tau^{*}_{~\mathcal{CP}^{-1}}\,,
\end{equation}
which leads to
\begin{eqnarray}
\nonumber&&\tau^{*}_{~\mathcal{CP}^{-1}}=-\frac{1}{\lambda_S}\left(c_{S'}\tau + d_{S'}\right)=-\lambda^{*}_S\left(c_{S'}\tau + d_{S'}\right)\,,\\
\label{eq:tau-CP-1}&&\tau_{\mathcal{CP}} =-\frac{1}{c_{S'}} \left(\lambda_S\tau^\ast+d_{S'}\right)\,.
\end{eqnarray}
Then we proceed to consider the consistency chain $\mathcal{CP}\rightarrow T\rightarrow\mathcal{CP}^{-1}$:
\begin{equation}
\tau \xrightarrow{\mathcal{CP}}~ -\frac{1}{c_{S'}} \left(\lambda_S\tau^\ast+d_{S'}\right) \xrightarrow{T}~ -\frac{1}{c_{S'}} \left[\lambda_S\left(\tau^\ast+1\right)+d_{S'}\right]\xrightarrow{\mathcal{CP}^{-1}} \tau - \frac{\lambda_S}{c_{S'}} \,,
\end{equation}
where the CP transformation of $\tau$ in Eq.~\eqref{eq:tau-CP-1} has been used. The resulting transformation should be a modular transformation, consequently the combination $\lambda_S/c_{S'}$ should be a real integer.
Since $|\lambda_S|=1$ and $c_{S'}$ is an integer so we should have
\begin{equation}
\label{eq:lc}\lambda_S=\pm1\,,\qquad c'_S=\pm1\,,
\end{equation}
which implies that $d_{S'}/c_{S'}$ is an integer as well.  Furthermore the requirement $\texttt{Im}\tau_{\mathcal{CP}}>0$ entails $\lambda_S = c_{S'} =\pm1$. As a consequence, the $\mathcal{CP}$ transformation rule of the complex modulus $\tau$ in Eq.~\eqref{eq:tau-CP-1} is
\begin{equation}
\tau \,\xrightarrow{\mathcal{CP}}\,, n-\tau^{*}\,,
\label{eq:taucp_shift}
\end{equation}
where $n=d_{S'}/c_{S'}$ is some integer. Let us consider the composition $\mathcal{CP}'=\mathcal{CP}\circ T^{n}$ of the CP transformation $\mathcal{CP}$ and the modular transformation $T^{n}$,
\begin{equation}
\label{eq:tau-CPp}\tau \xrightarrow{\mathcal{CP}} -\tau^* +n \xrightarrow{T^n}-\tau^{*}\,.
\end{equation}
This composition transformation $\mathcal{CP}'$ acts on the chiral superfield $\varphi(x)$ as
\begin{equation}
\varphi(x)\xrightarrow{{\cal CP}} X_{\mathbf{r}}\overline{\varphi}(x_{\mathcal{P}}) \xrightarrow{T^n}  X_{\mathbf{r}}\rho^*_\mathbf{r}(T^n)\overline{\varphi}(x_{\mathcal{P}})=X'_{\mathbf{r}}\overline{\varphi}(x_{\mathcal{P}}),~~~X'_{\mathbf{r}}\equiv X_{\mathbf{r}}\rho^*_\mathbf{r}(T^n)\,.
\end{equation}
We see that $X'_{\mathbf{r}}$ the same properties as the original CP transformation up to a redefinition of $X_{\mathbf{r}}$.
As shown in Eq.~\eqref{eq:tau-CPp},
\begin{equation}
\label{eq:tau-gCP-petcov}\tau \xrightarrow{\mathcal{CP}'}-\tau^{*} \,.
\end{equation}
Thus we derive exactly the same gCP transformation rule of $\tau$ as Eq.~\eqref{eq:linearCP}. It is obvious that the modulus $\tau$ is invariant under two successive CP transformation,i.e.
\begin{equation}
\tau \xrightarrow{\mathcal{CP}'}  -\tau^* \xrightarrow{\mathcal{CP}'} \tau\,,
\end{equation}
such that $\mathcal{CP}'^2=1$ is satisfied. Without loss of generality, we could choose the generators of the full symmetry group to be $\mathcal{CP}'$, $S$, $T$ instead of $\mathcal{CP}$, $S$, $T$. With a slight abuse of notation, in this paper we shall denote $\mathcal{CP}'$ as $\mathcal{CP}$ and $X'_{\mathbf{r}}$ as $X_{\mathbf{r}}$ for notation simplicity.

\section{\label{app:Gamma3-MF3} Finite modular groups $\Gamma'_N$ and $\Gamma_N$ and modular forms of level $N$  }

In this appendix, we shall give the irreducible representations and the Clebsch-Gordon of $\Gamma'_N$ and $\Gamma_N$ for the level $N=2, 3, 4, 5$. The definitions of $\Gamma_N$ and $\Gamma'_N$ are given in Eq.~\eqref{eq:Gamma_N-def} and Eq.~\eqref{eq:Gammap_N-def} respectively.
Both group generators $S$ and $T$ would be represented by symmetric matrices in all irreducible representations, consequently the gCP transformation would reduce to the canonical CP transformation with $X_{\bm{r}}=\mathds{1}_{\bm{r}}$~\cite{Novichkov:2019sqv,Ding:2021iqp}, as shown in section~\ref{subsec:gcp-transfm-matter}. We shall also report the explicit expressions of the modular forms at level $N$.

\subsection{\label{app:group-MF-N2}$N=2$}

The groups $\Gamma'_2$ and $\Gamma_2$ are isomorphic to the $S_3$ group which is the symmetry group of a regular triangle. $S_{3}$ can be generated by $S$ and $T$ fulfilling the following relations~\cite{Ishimori:2010au}
\begin{equation}
	S^2=T^2=(ST)^3=1\,.
\end{equation}
The six elements of $S_{3}$ are divided to three conjugacy classes
\begin{equation}
	1C_1=\left\{1 \right\}, \quad 3C_2=\{S,T,TST\},\quad  2C_3=\{ST,TS\}\,.
\end{equation}
where $nC_k$ denotes a conjugacy class with $n$ elements and the  subscript $k$ is the order of the elements. The finite modular group $S_{3}$ has two singlet representations $\bm{1}$ and $\bm{1^\prime}$, and one double representation $\bm{2}$. The modular generators $S$ and $T$ are represented by
\begin{eqnarray}
\label{eq:S3-irre} \begin{array}{cccc}
\bm{1}:~~ & S=1, ~&~ T=1\,, \\
\bm{1^\prime}:~~ & S=1, & T=-1\,, \\
\bm{2}:~~ &~ S=-\frac{1}{2}
\left(\begin{array}{cc}	1 ~&~ \sqrt{3} \\
\sqrt{3} &~ -1 \\
\end{array}\right) , & T=\left(\begin{array}{cc}
1 &~ 0 \\
0 &~ -1 \\
\end{array}\right) \,.
\end{array}
\end{eqnarray}
The Kronecker products between different irreducible representations can be obtained from the character table
\begin{equation}\label{eq:S3_KP}
\bm{1}\otimes \bm{1^\prime}=\bm{1^{\prime}} , \quad  \bm{1^\prime}\otimes \bm{2}=\bm{2}, \quad \bm{2}\otimes \bm{2}= \bm{1}\oplus \bm{1^{\prime}}\oplus \bm{2}\,,
\end{equation}
For the tensor product of the singlet $\bm{1^{\prime}}$ with a doublet $\bm{2}$, we have
\begin{equation}
\bm{1^\prime}\otimes\bm{2}=\bm{2},~~~\text{with}~~~\bm{2}=\alpha\left(\begin{array}{c}
\beta_1 \\
-\beta_2
\end{array}\right)\,.
\end{equation}
The contraction rules for two $S_3$ doublets are found to be
\begin{eqnarray}
\begin{array}{lll}
\bm{2}\otimes\bm{2}=\bm{1}\oplus\bm{1^\prime}\oplus\bm{2},& \qquad\qquad &
\text{with}\qquad \left\{\begin{array}{l}
\bm{1}\;=\alpha_1\beta_1+\alpha_2\beta_2  \\
\bm{1^\prime}=\alpha_1\beta_2-\alpha_2\beta_1  \\
\bm{2}\;=\left(\begin{array}{c}
\alpha_2\beta_2-\alpha_1\beta_1  \\
\alpha_1\beta_2+\alpha_2\beta_1
\end{array}\right)
\end{array}\right. \\
\\[-8pt]
\end{array}
\end{eqnarray}
As shown in Eq.~\eqref{eq:vanishing_cond}, there is no odd weight modular form of level 2. From the dimension formula of Eq.~\eqref{eq:dime_N2}, we know that the linear space of weight $2k$ and level 2 modular forms has dimension $k+1$. At weight 2, there are two linearly independent modular forms of level 2, and they can be arranged into a doublet of $S_3$~\cite{Kobayashi:2018vbk}:
\begin{equation}
Y^{(2)}_{\mathbf{2}}=\begin{pmatrix}
Y_1(\tau) \\
Y_2(\tau)
\end{pmatrix}
\end{equation}
with
\begin{eqnarray}
Y_1(\tau) &=&\frac{i}{4\pi}\left[ \frac{\eta'(\tau/2)}{\eta(\tau/2)}  +\frac{\eta'((\tau +1)/2)}{\eta((\tau+1)/2)}- 8\frac{\eta'(2\tau)}{\eta(2\tau)}  \right] ,\nonumber \\
Y_2(\tau) &=& \frac{\sqrt{3}i}{4\pi}\left[ \frac{\eta'(\tau/2)}{\eta(\tau/2)}  -\frac{\eta'((\tau +1)/2)}{\eta((\tau+1)/2)}   \right]  ,
\end{eqnarray}
The $q$-expansion of the modular forms $Y_{1,2}(\tau)$ is given by
\begin{eqnarray}
Y_1(\tau) &=& 1/8+3 q+3 q^2+12 q^3+3 q^4+18 q^5+12 q^6+24 q^7+3 q^8+39 q^9+18 q^{10} \cdots \,,\nonumber \\
Y_2(\tau) &=& \sqrt 3 q^{1/2} (1+4 q+6 q^2+8 q^3+13 q^4+12 q^5+14 q^6+24 q^7+18 q^8+20 q^9 \cdots)~~\quad\,.
\end{eqnarray}
The higher weight modular forms of level 2 are the product of $Y_1$ and $Y_2$ as follows:
\begin{eqnarray}
\nonumber&&Y^{(4)}_{\bm{1}}=\left(Y^{(2)}_{\bm{2}}Y^{(2)}_{\bm{2}}\right)_{\bm{1}}=Y_1^2+Y_2^2\,,~~Y^{(4)}_{\bm{2}}
=\left(Y^{(2)}_{\bm{2}}Y^{(2)}_{\bm{2}}\right)_{\bm{2}}
=\left(\begin{array}{c}Y_2^2-Y_1^2\\
2 Y_1 Y_2\end{array}\right)\,, \\
\nonumber &&Y^{(6)}_{\bm{1}}
=\left(Y^{(2)}_{\bm{2}}Y^{(4)}_{\bm{2}}\right)_{\bm{1}} =3Y_1Y^2_2-Y^3_1\,,~~~Y^{(6)}_{\bm{1^{\prime}}}=\left(Y^{(2)}_{\bm{2}}Y^{(4)}_{\bm{2}}\right)_{\bm{1^{\prime}}}=3Y^2_1Y_2-Y^3_2\,,\\
\label{eq:Yw2to8}&&Y^{(6)}_{\bm{2}}=\left(Y^{(2)}_{\bm{2}}Y^{(4)}_{\bm{1}}\right)_{\bm{2}}=(Y^2_1+Y^2_2)\left(\begin{array}{c}Y_1\\
Y_2\end{array}\right)\,.
\end{eqnarray}
The modular form multiplets of level 2 are summarized in table~\ref{tab:MF-L5-W6}.

\subsection{\label{app:group-MF-N3} $N=3$}

The $\Gamma'_3\cong T'$ group is the double covering of the tetrahedral group $\Gamma_3\cong A_4$. All the elements of $T'$ can be generated by three generators $S$, $T$ and $R$ which obey the following relations~\footnote{Alternatively the $T'$ group can be generated by $S$ and $T$ fulfilling $S^4=T^3=(ST)^3=1$, $S^2T=TS^2$.}~\cite{Liu:2019khw}:
\begin{equation}
S^{2}=R,~~ (ST)^{3}=T^{3}=R^{2}=1,~~RT = TR\,.
\end{equation}
The generator $R$ commutes with all elements of the group, and the center of $T^{\prime}$ is the $Z_2$ subgroup generated by $R$. The group $A_4$ can be reproduced by setting $R=1$. The 24 elements of $T'$ group belong to 7 conjugacy classes:
\begin{eqnarray}
\nonumber 1 C_1:&&\,1\,, \\
\nonumber 1 C_2:&&\,R\,, \\
\nonumber 6 C_4:&&\,S,\,T^{-1}ST,\,TST^{-1},\,SR,\,T^{-1}STR,\,TST^{-1}R \,, \\
\nonumber 4 C_6:&&\,TR,\,TSR,\,STR,\,T^{-1}ST^{-1}R \,, \\
\nonumber 4 C_3:&&\,T^{-1},\,ST^{-1}R,\,T^{-1}SR,\,TSTR \,, \\
\nonumber 4 C'_3:&&\,T,\,TS,\,ST,\,T^{-1}ST^{-1} \,, \\
\label{T'conj_class} 4C'_6:&&\,ST^{-1},\,T^{-1}S,\,TST,\,T^{-1}R \,.
\end{eqnarray}
The $T'$ group has a triplet representation $\mathbf{3}$ and three singlets representations $\mathbf{1}$, $\mathbf{1}'$ and $\mathbf{1}''$ in common with $A_4$. In addition, it has three two-dimensional spinor representations $\mathbf{\widehat{2}}$, $\mathbf{\widehat{2}}'$ and $\mathbf{\widehat{2}}''$. In our working basis, the generators $S$, and $T$ are represented by the following symmetric and unitary matrices:
\begin{eqnarray}
\label{eq:irr-Tp} \begin{array}{cccc}
\mathbf{1:} & S=1, ~&~ T=1\,, \\
\mathbf{1}':& S=1, & T=\omega\,, \\
\mathbf{1}'': & S=1, & T=\omega^{2} \,,\\
\mathbf{\widehat{2}:} ~&~ S=-\frac{i}{\sqrt{3}}
\begin{pmatrix}
1  ~& \sqrt{2}  \\
\sqrt{2}  ~&  -1
\end{pmatrix}, & T=\left(\begin{array}{cc}
\omega & 0 \\
0 & 1 \\
\end{array}\right) \,,\\
\mathbf{\widehat{2}}': ~&~ S=-\frac{i}{\sqrt{3}}
\begin{pmatrix}
1  ~& \sqrt{2}  \\
\sqrt{2}  ~&  -1
\end{pmatrix}, & T=\left(\begin{array}{cc}
\omega^{2} & 0 \\
0 & \omega \\
\end{array}\right) \,,\\
\mathbf{\widehat{2}}'': ~&~ S=-\frac{i}{\sqrt{3}}
\begin{pmatrix}
1  ~& \sqrt{2}  \\
\sqrt{2}  ~&  -1
\end{pmatrix}, & T=\left(\begin{array}{cc}
1 & 0 \\
0 & \omega^{2} \\
\end{array}\right) \,,\\
\mathbf{3:} & S=\frac{1}{3}\left(\begin{array}{ccc}
-1&2&2\\
2&-1&2\\
2&2&-1
\end{array}\right),
 ~&~ T=\left(
\begin{array}{ccc}
1~&0~&0\\
0~&\omega~&0\\
0~&0~&\omega^2
\end{array}\right) \,,
\end{array}
\end{eqnarray}
with $\omega=e^{i2\pi/3}$. Notice the generator $R=1$ in the unhatted representations and $R=-1$ in the unhatted representations. Obviously the representation matrices of $S$ and $T$ are unitary and symmetric in all irreducible representations. Notice that the two-dimensional representation matrices are related to those of Refs.~\cite{Liu:2019khw,Lu:2019vgm} by a similarity transformation, while the remaining ones are the same. The Kronecker products between different irreducible representations of $T'$ are given by
\begin{eqnarray}
\nonumber&&\mathbf{1}^a\otimes \mathbf{r}^b = \mathbf{r}^b \otimes \mathbf{1}^a= \mathbf{r}^{a+b~(\text{mod}~3)},~~~~~{\rm for}~~\mathbf{r} = \mathbf{1}, \mathbf{\widehat{2}}\,,\\
\nonumber&&\mathbf{1}^a \otimes \mathbf{3} = \mathbf{3}\otimes \mathbf{1}^a=\mathbf{3}\,,\\
\nonumber&&\mathbf{\widehat{2}}^a \otimes \mathbf{\widehat{2}}^b=\mathbf{3}\oplus \mathbf{1}^{a+b+1~(\text{mod}~3)}\,, \\
\nonumber&&\mathbf{\widehat{2}}^a \otimes \mathbf{3} =\mathbf{3} \otimes \mathbf{\widehat{2}}^a = \mathbf{\widehat{2}}\oplus \mathbf{\widehat{2}}'\oplus \mathbf{\widehat{2}}''\,, \\
\label{eq:mult}&&\mathbf{3} \otimes \mathbf{3} = \mathbf{3}_S \oplus \mathbf{3}_A \oplus \mathbf{1} \oplus \mathbf{1}' \oplus \mathbf{1}''\,,
\end{eqnarray}
where $a,b=0, 1, 2$ and we have denoted $\mathbf{1}\equiv\mathbf{1}^0$, $\mathbf{1}'\equiv\mathbf{1}^{1}$, $\mathbf{1}''\equiv\mathbf{1}^{2}$ for singlet representations and $\mathbf{\widehat{2}}\equiv\mathbf{\widehat{2}}^0$, $\mathbf{\widehat{2}}'\equiv\mathbf{\widehat{2}}^{1}$, $\mathbf{\widehat{2}}''\equiv\mathbf{\widehat{2}}^{2}$ for the doublet representations. The notations $\mathbf{3}_S$ and $\mathbf{3}_A$ stand for the symmetric and antisymmetric triplet combinations respectively. In the following, we report the Clebsch-Gordon (CG) coefficients of the $T'$ group in the chosen basis. We shall use $\alpha_i$ ($\beta_i $) to denote the elements of the first (second) representation of the product.
\begin{eqnarray}
\mathbf{1}^a\otimes\mathbf{1}^b &=& \mathbf{1}^{a+b~(\text{mod}~3)} \sim\alpha\beta \,, \\
\mathbf{1}^a\otimes\mathbf{\widehat{2}}^b &=& \mathbf{\widehat{2}}^{a+b~(\text{mod}~3)} \sim \left(\begin{array}{c}\alpha\beta_1\\
\alpha\beta_2 \\ \end{array}\right) \,,\\
  \mathbf{1}\otimes\mathbf{3} &=& \mathbf{3} \sim \left(\begin{array}{c}\alpha\beta_1\\
\alpha\beta_2 \\
\alpha\beta_3 \end{array}\right) \,, \\
\mathbf{1}'\otimes\mathbf{3} &=& \mathbf{3} \sim \left(\begin{array}{c}\alpha\beta_3\\
\alpha\beta_1 \\
\alpha\beta_2 \end{array}\right) \,, \\
\mathbf{1}''\otimes\mathbf{3} &=& \mathbf{3} \sim \left(\begin{array}{c}\alpha\beta_2\\
\alpha\beta_3 \\
\alpha\beta_1 \end{array}\right) \,,\\
\mathbf{\widehat{2}}\otimes\mathbf{\widehat{2}}=\mathbf{\widehat{2}}'\otimes \mathbf{\widehat{2}}''=\mathbf{1}'\oplus\mathbf{3} ~&\text{with}&~\left\{
\begin{array}{l}
\mathbf{1}'\sim\alpha_1\beta_2-\alpha_2\beta_1 \\ [0.1in]
\mathbf{3}\sim
\left(\begin{array}{c}\alpha_2\beta_2 \\
\frac{1}{\sqrt{2}}(\alpha_1\beta_2+\alpha_2\beta_1)  \\
-\alpha_1\beta_1 \end{array}\right)
\end{array}
\right. \\
\mathbf{\widehat{2}}\otimes\mathbf{\widehat{2}}'=\mathbf{\widehat{2}}''\otimes\mathbf{\widehat{2}}''=\mathbf{1}''\oplus\mathbf{3}~&\text{with}&~\left\{
\begin{array}{l}
\mathbf{1}''\sim\alpha_1\beta_2-\alpha_2\beta_1\\ [0.1in]
\mathbf{3}\sim
 \left(\begin{array}{c} -\alpha_1\beta_1\\
\alpha_2\beta_2 \\
 \frac{1}{\sqrt{2}}(\alpha_1\beta_2+\alpha_2\beta_1) \end{array}\right)
\end{array}
\right. \\
\mathbf{\widehat{2}}\otimes\mathbf{\widehat{2}}''=\mathbf{\widehat{2}}'\otimes\mathbf{\widehat{2}}'=\mathbf{1}\oplus \mathbf{3} ~&\text{with}&~\left\{
\begin{array}{l}
\mathbf{1} \sim\alpha_1\beta_2-\alpha_2\beta_1 \\ [0.1in]
\mathbf{3}\sim
\left(\begin{array}{c}  \frac{1}{\sqrt{2}}(\alpha_1\beta_2+\alpha_2\beta_1)\\
 -\alpha_1\beta_1\\
\alpha_2\beta_2  \end{array}\right)
\end{array}
\right. \\
\mathbf{\widehat{2}}\otimes\mathbf{3}=\mathbf{\widehat{2}}\oplus\mathbf{\widehat{2}}'\oplus\mathbf{\widehat{2}}'' ~&\text{with}&~\left\{
\begin{array}{l}
\mathbf{\widehat{2}}\sim
 \left(\begin{array}{c}\alpha_1\beta_1+\sqrt{2}\alpha_2\beta_2 \\
 -\alpha_2\beta_1+\sqrt{2}\alpha_1\beta_3 \end{array}\right)  \\ [0.1in]
\mathbf{\widehat{2}}'\sim
\left(\begin{array}{c}\alpha_1\beta_2+\sqrt{2}\alpha_2\beta_3 \\
-\alpha_2\beta_2+\sqrt{2}\alpha_1\beta_1  \end{array}\right) \\ [0.1in]
\mathbf{\widehat{2}}''\sim
\left(\begin{array}{c}\alpha_1\beta_3+\sqrt{2}\alpha_2\beta_1 \\
-\alpha_2\beta_3+\sqrt{2}\alpha_1\beta_2  \end{array}\right) \\ [0.1in]
\end{array}
\right. \\
\mathbf{\widehat{2}}'\otimes\mathbf{3}=\mathbf{\widehat{2}}\oplus\mathbf{\widehat{2}}'\oplus\mathbf{\widehat{2}}'' ~&\text{with}&~\left\{
\begin{array}{l}
\mathbf{\widehat{2}}\sim
\left(\begin{array}{c}\alpha_1\beta_3+\sqrt{2}\alpha_2\beta_1 \\
-\alpha_2\beta_3+\sqrt{2}\alpha_1\beta_2  \end{array}\right)  \\ [0.1in]
\mathbf{\widehat{2}}'\sim
 \left(\begin{array}{c}\alpha_1\beta_1+\sqrt{2}\alpha_2\beta_2 \\
 -\alpha_2\beta_1+\sqrt{2}\alpha_1\beta_3 \end{array}\right) \\ [0.1in]
\mathbf{\widehat{2}}''\sim
\left(\begin{array}{c}\alpha_1\beta_2+\sqrt{2}\alpha_2\beta_3 \\
-\alpha_2\beta_2+\sqrt{2}\alpha_1\beta_1  \end{array}\right) \\ [0.1in]
\end{array}
\right. \\
\mathbf{\widehat{2}}''\otimes\mathbf{3}=\mathbf{\widehat{2}}\oplus\mathbf{\widehat{2}}'\oplus\mathbf{\widehat{2}}'' ~&\text{with}&~\left\{
\begin{array}{l}
\mathbf{\widehat{2}}\sim
\left(\begin{array}{c}\alpha_1\beta_2+\sqrt{2}\alpha_2\beta_3 \\
-\alpha_2\beta_2+\sqrt{2}\alpha_1\beta_1  \end{array}\right)  \\ [0.1in]
\mathbf{\widehat{2}}'\sim
\left(\begin{array}{c}\alpha_1\beta_3+\sqrt{2}\alpha_2\beta_1 \\
-\alpha_2\beta_3+\sqrt{2}\alpha_1\beta_2  \end{array}\right) \\ [0.1in]
\mathbf{\widehat{2}}''\sim
 \left(\begin{array}{c}\alpha_1\beta_1+\sqrt{2}\alpha_2\beta_2 \\
 -\alpha_2\beta_1+\sqrt{2}\alpha_1\beta_3 \end{array}\right) \\ [0.1in]
\end{array}
\right. \\
\label{eq:3x3-rules-A4}\mathbf{3}\otimes\mathbf{3}=\mathbf{3}_S\oplus\mathbf{3}_A\oplus\mathbf{1}\oplus\mathbf{1}'\oplus\mathbf{1}'' ~&\text{with}&~\left\{
\begin{array}{l}
\mathbf{3}_S\sim
 \left(\begin{array}{c} 2\alpha_1\beta_1 -\alpha_2\beta_3 -\alpha_3\beta_2 \\
 2\alpha_3\beta_3 -\alpha_1\beta_2 -\alpha_2\beta_1  \\
 2\alpha_2\beta_2 -\alpha_1\beta_3 -\alpha_3\beta_1 \end{array}\right) \\ [0.1in]
\mathbf{3}_A\sim
 \left(\begin{array}{c}\alpha_2\beta_3 -\alpha_3\beta_2 \\
\alpha_1\beta_2 -\alpha_2\beta_1  \\
\alpha_3\beta_1 -\alpha_1\beta_3 \end{array}\right) \\ [0.1in]
\mathbf{1} \sim\alpha_1\beta_1 +\alpha_2\beta_3 +\alpha_3\beta_2 \\ [0.1in]
\mathbf{1}' \sim\alpha_3\beta_3 +\alpha_1\beta_2 +\alpha_2\beta_1 \\ [0.1in]
\mathbf{1}'' \sim\alpha_2\beta_2 +\alpha_1\beta_3 +\alpha_3\beta_1 \\ [0.1in]
\end{array}
\right.
\end{eqnarray}
The linear space spanned by level 3 and weight $k$ modular forms has dimension $k+1$, and a general vector $\mathcal{M}_{k}(\Gamma(3))$ in the linear space can be explicitly constructed by using the Dedekind eta function $\eta(\tau)$:
\begin{equation}
\mathcal{M}_{k}(\Gamma(3))=\sum_{m+n=k;\\ m,n\ge0} c_{mn} \frac{\eta^{3m}(3\tau)\eta^{3n}(\tau /3 )}{\eta^{m+n}(\tau)}=\sum_{m+n=k;\, m,n\ge0}c_{mn} \left[\frac{\eta^3(3\tau)}{\eta(\tau)}\right]^m\left[\frac{\eta^{3}(\tau/3)}{\eta(\tau)}\right]^n\,,
\end{equation}
where $c_{mn}$ are general complex coefficients and $\eta(\tau)$ is the Dedekind eta-function defined in Eq.~\eqref{eq:eta-func}. There are two linearly independent weight 1 and level 3 modular forms which can be arranged in a $T'$ doublet:
\begin{equation}
\label{eq:Y1-Y2}
Y^{(1)}_{\mathbf{\widehat{2}}}(\tau)=\begin{pmatrix}
Y_1(\tau) \\
Y_2(\tau)
\end{pmatrix}\,,
\end{equation}
with
\begin{equation}
Y_1(\tau)=\sqrt{2}\,\frac{\eta^{3}(3\tau)}{\eta(\tau)},~~~~ Y_2(\tau)=-\frac{\eta^{3}(3\tau)}{\eta(\tau)}-\frac{1}{3}\frac{\eta^{3}(\tau / 3)}{\eta(\tau)}\,.
\end{equation}
The $q$-expansion of $Y_{1,2}(\tau)$ reads
\begin{eqnarray}
\nonumber Y_1(\tau)&=&\sqrt{2}\, q^{1/3} \left(1+q +2 q^2+2 q^4+q^5 + 2q^6+q^8+2q^9+2q^{10}+2q^{12} +\ldots\right) ,\\
\label{eq:q_exp_weigh1}Y_2(\tau)&=& -1/3 -2 q - 2 q^3 - 2 q^4- 4 q^7 - 2 q^9 -2q^{12}-4q^{13}+\ldots\,.
\label{q-expansion}
\end{eqnarray}
Notice that $Y_1(\tau)$ and $Y_2(\tau)$ are algebraically independent. The level-3 modular form of integer weight are polynomials of degree $k$ in $Y_1(\tau)$ and $Y_2(\tau)$, and higher-weight modular forms can be constructed from tensor product of lower-weight ones.

To be more specific, the weight-2 modular forms can be generated from the tensor products of two $Y^{(1)}_{\mathbf{\widehat{2}}}$,
\begin{equation}
Y^{(2)}_{\mathbf{3}}=\left(Y^{(1)}_{\mathbf{\widehat{2}}}Y^{(1)}_{\mathbf{\widehat{2}}}\right)_{\mathbf{3}}=\left(Y^2_2,\quad\sqrt{2}Y_1Y_2,\quad -Y^2_1 \right)^T\,.
\label{eq:modfvec1-N3}
\end{equation}
Then we can use the weight 1 and weight 2 modular forms to construct the weight 3 modular forms,
\begin{equation}
\begin{aligned}
&Y^{(3)}_{\mathbf{\widehat{2}}}=\left(Y^{(1)}_{\mathbf{\widehat{2}}}Y^{(2)}_{\mathbf{3}}\right)_{\mathbf{\widehat{2}}}= \left(3Y_1Y^2_2 ,\quad -\sqrt{2}Y^3_1-Y^3_2 \right)^T,\\
&Y^{(3)}_{\mathbf{\widehat{2}}''}=\left(Y^{(1)}_{\mathbf{\widehat{2}}}Y^{(2)}_{\mathbf{3}}\right)_{\mathbf{\widehat{2}}''}=\left(-Y^3_1+\sqrt{2}Y^3_2,\quad 3Y_2Y^2_1 \right)^T \,.
\end{aligned}
\end{equation}
At weight $k=4$, we find five independent modular forms which can be arranged into two singlets $\mathbf{1}$ and $\mathbf{1'}$ and a triplet of $T'$,
\begin{equation}
\begin{aligned}
&\hskip-0.12in Y^{(4)}_{\mathbf{3}}=\left(Y^{(1)}_{\mathbf{\widehat{2}}}Y^{(3)}_{\mathbf{\widehat{2}}}\right)_{\mathbf{3}} = \left(-\sqrt{2}Y^3_1Y_2-Y^4_2,~ -Y^4_1+\sqrt{2}Y_1Y^3_2, ~ -3Y^2_1Y^2_2 \right)^T,\\
&\hskip-0.12in Y^{(4)}_{\mathbf{1'}}=\left(Y^{(1)}_{\mathbf{\widehat{2}}}Y^{(3)}_{\mathbf{\widehat{2}}}\right)_{\mathbf{1'}}=-\sqrt{2}Y^4_1-4Y_1Y^3_2 \,, \\
&\hskip-0.12in Y^{(4)}_{\mathbf{1}}=\left(Y^{(1)}_{\mathbf{\widehat{2}}}Y^{(3)}_{\mathbf{\widehat{2}}''}\right)_{\mathbf{1}}=4Y^3_1Y_2-\sqrt{2}Y^4_2 \,.
\end{aligned}
\end{equation}
Notice that the contraction $\left(Y^{(1)}_{\mathbf{\widehat{2}}}Y^{(3)}_{\mathbf{\widehat{2}}''}\right)_{\mathbf{3}}=-Y^{(4)}_{\mathbf{3}}$.
The above weight 4 modular forms can also be obtained from the tensor products $\left(Y^{(2)}_{\mathbf{3}}Y^{(2)}_{\mathbf{3}}\right)_{\mathbf{3}}$, $\left(Y^{(2)}_{\mathbf{3}}Y^{(2)}_{\mathbf{3}}\right)_{\mathbf{1'}}$, $\left(Y^{(2)}_{\mathbf{3}}Y^{(2)}_{\mathbf{3}}\right)_{\mathbf{1}}$ with $Y^{(2)}_{\mathbf{3}}$ in Eq.~\eqref{eq:modfvec1-N3}. In particular, using Eqs.~(\ref{eq:3x3-rules-A4}, \ref{eq:modfvec1-N3}) it is obvious to see the constraint
\begin{equation}
\left(Y^{(2)}_{\mathbf{3}}Y^{(2)}_{\mathbf{3}}\right)_{\mathbf{1''}}=0\,,
\end{equation}
which was previously obtained from the $q$-expansion in~\cite{Feruglio:2017spp}. Similarly, the independent weight 5 modular forms can be constructed from the tensor products of weight 1 and weight 4 modular forms as follows,
\begin{equation}
\begin{aligned}
&\hskip-0.15in Y^{(5)}_{\mathbf{\widehat{2}}}=\left(Y^{(1)}_{\mathbf{\widehat{2}}}Y^{(4)}_{\mathbf{3}}\right)_{\mathbf{\widehat{2}}}= \left[-2\sqrt{2}Y_{1}^{3}Y_{2}+Y_{2}^{4}\right]\left(Y_{1},\quad Y_{2}\right)^T,\\
&\hskip-0.15in Y^{(5)}_{\mathbf{\widehat{2}}'}=\left(Y^{(1)}_{\mathbf{\widehat{2}}}Y^{(4)}_{\mathbf{3}}\right)_{\mathbf{\widehat{2}}'}=\left[-Y_{1}^{4}-2\sqrt{2}Y_{1}Y_{2}^{3}\right]\left(Y_{1},\quad Y_{2}\right)^T,\\
&\hskip-0.15in Y^{(5)}_{\mathbf{\widehat{2}}''}=\left(Y^{(1)}_{\mathbf{\widehat{2}}}Y^{(4)}_{\mathbf{3}}\right)_{\mathbf{\widehat{2}}''}=\left(5Y^3_1Y^2_2-\sqrt{2}Y^5_2,~ -\sqrt{2}Y^5_1+5Y^2_1Y^3_2 \right)^T \,.
\end{aligned}
\end{equation}
Notice that $Y^{(5)}_{\mathbf{\widehat{2}}}$ and $Y^{(5)}_{\mathbf{\widehat{2}}'}$ are proportional to the contractions $Y^{(1)}_{\mathbf{\widehat{2}}}Y^{(4)}_{\mathbf{1}}$ and $Y^{(1)}_{\mathbf{\widehat{2}}}Y^{(4)}_{\mathbf{1'}}$ respectively. Finally, the linearly independent weight 6 modular forms of level 3 can be decomposed into one singlet $\mathbf{1}$ and two triplets $\mathbf{3}$ under $T'$,
\begin{eqnarray}
\nonumber Y^{(6)}_{\mathbf{3}I}&=& \left(Y^{(1)}_{\mathbf{\widehat{2}}}Y^{(5)}_{\mathbf{\widehat{2}}}\right)_{\mathbf{3}} = \left[-2\sqrt{2}Y_{1}^{3}Y_{2}+Y_{2}^{4}\right] \left( Y_{2}^{2},\quad\sqrt{2}Y_{1}Y_{2},\quad -Y_{1}^{2}\right)^{T},\\
\nonumber Y^{(6)}_{\mathbf{3}II}&=&\left(Y^{(1)}_{\mathbf{\widehat{2}}}Y^{(5)}_{\mathbf{\widehat{2}}'}\right)_{\mathbf{3}}=\left[-Y_{1}^{4}-2\sqrt{2}Y_{1}Y_{2}^{3}\right] \left( -Y_{1}^{2},\quad Y_{2}^{2},\quad\sqrt{2}Y_{1}Y_{2}\right)^{T},\\
Y^{(6)}_{\mathbf{1}}&=&\left(Y^{(1)}_{\mathbf{\widehat{2}}}Y^{(5)}_{\mathbf{\widehat{2}}''}\right)_{\mathbf{1}}=\sqrt{2}Y^6_2-\sqrt{2}Y^6_1+10Y^3_1Y^3_2 \,.
\label{eq:modfvec2}
\end{eqnarray}
We summarize the level 3 modular forms up to weight 6 in table~\ref{tab:MF-L5-W6}.

\begin{table}[htbp]
\centering
\begin{tabular}{|c|c|c|c|c|c|c|c|}\hline\hline
\diagbox{$k$}{$N$}~&~2~&~3 ~&~4~&~5 \\ \hline
1 ~&~ --- ~&~ $Y^{(1)}_{\mathbf{\widehat{2}}}$ ~&~ $Y_{\mathbf{\widehat{3}'}}^{(1)}$  ~&~ $Y^{(1)}_{\mathbf{\widehat{6}}}$ \\\hline

2 ~&~$Y^{(2)}_{\mathbf{2}}$ ~&~ $Y^{(2)}_{\mathbf{3}}$ ~&~ $Y_{\mathbf{2}}^{(2)}, Y_{\mathbf{3}}^{(2)}$ ~&~ $Y_{\mathbf{3}}^{(2)},  Y_{\mathbf{3'}}^{(2)}, Y_{\mathbf{5}}^{(2)}$ \\\hline

3 ~&~ --- ~&~ $Y^{(3)}_{\mathbf{\widehat{2}}}, Y^{(3)}_{\mathbf{\widehat{2}}''}$ ~&~ $Y_{\mathbf{\widehat{1}'}}^{(3)}, Y_{\mathbf{\widehat{3}}}^{(3)}, Y_{\mathbf{\widehat{3}'}}^{(3)}$ ~&~ $Y_{\mathbf{\widehat{4}'}}^{(3)}, Y_{\mathbf{\widehat{6}}I}^{(3)}, Y_{\mathbf{\widehat{6}}II}^{(3)}$ \\\hline

4 ~&~$Y^{(4)}_{\bm{1}},  Y^{(4)}_{\bm{2}}$ ~&~ $Y^{(4)}_{\mathbf{1}}, Y^{(4)}_{\mathbf{1}'}, Y^{(4)}_{\mathbf{3}}$ ~&~ $\left.\begin{array}{l}
Y_{\mathbf{1}}^{(4)}, Y_{\mathbf{2}}^{(4)}, Y_{\mathbf{3}}^{(4)}, \\ Y_{\mathbf{3'}}^{(4)}
\end{array}
\right.$ ~&~ $\left.\begin{array}{l}
Y_{\mathbf{1}}^{(4)}, Y_{\mathbf{3}}^{(4)}, Y_{\mathbf{3'}}^{(4)}, \\ Y_{\mathbf{4}}^{(4)}, Y_{\mathbf{5}I}^{(4)}, Y_{\mathbf{5}II}^{(4)}
\end{array}
\right.$ \\\hline

5~&~ --- ~&~  $Y^{(5)}_{\mathbf{\widehat{2}}}, Y^{(5)}_{\mathbf{\widehat{2}}'}, Y^{(5)}_{\mathbf{\widehat{2}}''}$ ~&~ $\left.\begin{array}{l}
Y_{\mathbf{\widehat{2}}}^{(5)}, Y_{\mathbf{\widehat{3}}}^{(5)}, Y_{\mathbf{\widehat{3}'}I}^{(5)}, \\
Y_{\mathbf{\widehat{3}'}II}^{(5)}
\end{array}
\right.$ ~&~ $\left.\begin{array}{l}
Y_{\mathbf{\widehat{2}}}^{(5)}, Y_{\mathbf{\widehat{2}'}}^{(5)}, Y_{\mathbf{\widehat{4}'}}^{(5)},\\
Y_{\mathbf{\widehat{6}}I}^{(5)}, Y_{\mathbf{\widehat{6}}II}^{(5)}, Y_{\mathbf{\widehat{6}}III}^{(5)}
\end{array}
\right.$ \\\hline

6~&~ $Y^{(6)}_{\bm{1}}, Y^{(6)}_{\bm{1^{\prime}}}, Y^{(6)}_{\bm{2}}$ ~&~ $Y^{(6)}_{\mathbf{1}}, Y^{(6)}_{\mathbf{3}I}, Y^{(6)}_{\mathbf{3}II}$ ~&~ $\left.\begin{array}{l}
Y_{\mathbf{1}}^{(6)}, Y_{\mathbf{1'}}^{(6)}, Y_{\mathbf{2}}^{(6)}, \\ Y_{\mathbf{3}I}^{(6)},Y_{\mathbf{3}II}^{(6)},Y_{\mathbf{3'}}^{(6)}
\end{array}
\right.$ ~&~ $\left.\begin{array}{l}
Y_{\mathbf{1}}^{(6)}, Y_{\mathbf{3}I}^{(6)}, Y_{\mathbf{3}II}^{(6)}, \\ Y_{\mathbf{3'}I}^{(6)}, Y_{\mathbf{3'}II}^{(6)}, Y_{\mathbf{4}I}^{(6)}, \\ Y_{\mathbf{4}II}^{(6)}, Y_{\mathbf{5}I}^{(6)}, Y_{\mathbf{5}II}^{(6)}
\end{array}
\right.$ \\
\hline\hline
\end{tabular}
\caption{\label{tab:MF-L5-W6}Summary of modular forms up to weight 6 and level 5. Note that the odd weight modular forms are in the hatted representations of $\Gamma'_N$ while the even weight modular forms are in the unhatted representations of $\Gamma'_N$.  }
\end{table}

\subsection{\label{app:group-MF-N4} $N=4$}

The homogeneous finite modular group $\Gamma'_4\cong S'_4$ is the double covering of $\Gamma_4\cong S_4$ which is the symmetry group of a cube. It has 48 elements, and it can be generated by three generators $S$, $T$ and $R$ obeying the following relations:
\begin{equation}
S^2=R,\quad (ST)^3=T^4=R^2=1,\quad TR=RT\,.
\end{equation}
The group \texttt{ID} of $S'_4$ in \texttt{GAP}~\cite{GAP} is [48, 30]. Notice that $S_4$ is not a subgroup of $S'_4$, it is isomorphic to the quotient group of $S'_4$ over $Z_2^{R}$, i.e., $S_4\cong S'_4/Z_2^{R}$, where $Z_2^{R}=\{1,R\}$ is the center and a normal subgroup of $S'_4$. Hence the group $S_4$ can be reproduced by setting $R=1$. Note that $S'_4$ is isomorphic to the semidirect product of $A_4$ with $Z_4$, namely $S'_4 \cong A_4 \rtimes Z_4$. Therefore it can be expressed in terms of another set of generators $s$, $t$ and $r$ satisfying the relations:
\begin{equation}
s^2=(st)^3=t^3=1,\quad r^4=1,\quad r s r^{-1}=s, \quad r t r^{-1} = (st)^2\,,
\end{equation}
where $s$ and $t$ generate a $A_4$ subgroup, $r$ generates a $Z_4$ subgroup, and the last two relations define the semidirect product ``$\rtimes$''. The generators $s,t$ and $r$ are related to $S$, $T$ and $R$ by
\begin{align}
\nonumber & s = T^2 R,\quad t=(ST)^2, \quad r=T \,,\\
& S= t^2 r^3, \quad T=r, \quad R=r^2 s\,.
\end{align}
All the elements of $S_4^{\prime}$ group can be divided into 10 conjugacy classes:
\begin{eqnarray}
\nonumber 1C_1&=&\{1\}\,, \\
\nonumber 1C_2&=&\{R\}=(1C_1)\cdot R\,, \\
\nonumber 3C_2&=&\left\{T^2,~ST^2S^3,~(ST^2)^2\right\}\,, \\
\nonumber3C_2'&=&\left\{T^2R,~ST^2S,~(ST^2)^2R\right\}=(3C_2)\cdot R\,, \\
\nonumber 8C_3&=&\left\{ST,~TS,~(ST)^2,~(TS)^2,~TS^3T^2,~T^2ST^3,~T^2S^3T,~T^3ST^2\right\}\,, \\
\nonumber6C_4&=&\left\{S,~TST^3,~T^2ST^2,~T^3ST,~TST^2S^3,~ST^2S^3T\right\}\,, \\
\nonumber 6C_4'&=&\left\{T,~ST^2,~T^2S,~T^3S^2,~TST,~STS^3\right\}\,, \\
\nonumber 6C_4''&=&\left\{SR,~TST^3R,~T^2ST^2R,~T^3STR,~TST^2S,~ST^2S^3TR\right\}=(6C_4)\cdot R\,, \\
\nonumber 6C_4'''&=&\left\{TR,~ST^2R,~T^2SR,~T^3,~TSTR,~STS\right\}=(6C_4')\cdot R \,, \\
\nonumber8C_6&=&\left\{STR,~TSR,~(ST)^2R,~(TS)^2R,~TS^3T^2R,~T^2ST^3R,~\right.\\
  && \left.  T^2S^3TR,~T^3ST^2R\right\}=(8C_3)\cdot R \,.
\end{eqnarray}
Note that one half of these conjugacy classes can be written as the product of the other half with $R$. The group $S'_4$ has four singlet representations $\mathbf{1},\mathbf{1}^{\prime},\mathbf{\widehat{1}}$ and $\mathbf{\widehat{1}}^{\prime}$, two doublet representations $\mathbf{2}$ and $\mathbf{\widehat{2}}$, and four triplet representations $\mathbf{3},\mathbf{3}^{\prime},\mathbf{\widehat{3}}$ and $\mathbf{\widehat{3}}^{\prime}$. The representation matrices of the generators $S$ and $T$ are given by
\begin{eqnarray}
\label{eq:irr-S4p} \begin{array}{cccc}
\mathbf{1}: & S=1, ~&~ T=1\,, \\
\mathbf{1}':& S=-1, & T=-1\,, \\
\mathbf{\widehat{1}} : & S=i, & T=-i \,,\\
\mathbf{\widehat{1}^{\prime}} : & S=-i, & T=i \,,\\
\mathbf{2}: & S=\dfrac{1}{2}\begin{pmatrix}
  -1 ~&\sqrt{3} \\
\sqrt{3} ~& 1 \\
\end{pmatrix}, ~&~ T=\begin{pmatrix}
 1 ~& 0 \\
 0 ~& -1 \\
\end{pmatrix}\,, \\
\mathbf{\widehat{2}}: & S=\dfrac{i}{2}\begin{pmatrix}
 -1 ~&\sqrt{3} \\
\sqrt{3} ~& 1 \\
\end{pmatrix}, ~&~ T=-i\begin{pmatrix}
 1 ~& 0 \\
 0 ~& -1 \\
\end{pmatrix}\,, \\
\mathbf{3}: & S=\dfrac{1}{2}\begin{pmatrix}
 0 ~&\sqrt{2} ~&\sqrt{2} \\
\sqrt{2} ~& -1 ~& 1 \\
\sqrt{2} ~& 1 ~& -1
\end{pmatrix}, ~&~ T=\begin{pmatrix}
 1 ~& 0 ~& 0 \\
 0 ~& i ~& 0 \\
 0 ~& 0 ~& -i
\end{pmatrix}\,, \\
\mathbf{3^{\prime}}: & S=-\dfrac{1}{2}\begin{pmatrix}
 0 ~&\sqrt{2} ~&\sqrt{2} \\
\sqrt{2} ~& -1 ~& 1 \\
\sqrt{2} ~& 1 ~& -1
\end{pmatrix}, ~&~ T=-\begin{pmatrix}
 1 ~& 0 ~& 0 \\
 0 ~& i ~& 0 \\
 0 ~& 0 ~& -i
\end{pmatrix}\,, \\
\mathbf{\widehat{3}}: & S=\dfrac{i}{2}\begin{pmatrix}
 0 ~&\sqrt{2} ~&\sqrt{2} \\
\sqrt{2} ~& -1 ~& 1 \\
\sqrt{2} ~& 1 ~& -1
\end{pmatrix}, ~&~ T=-i\begin{pmatrix}
 1 ~& 0 ~& 0 \\
 0 ~& i ~& 0 \\
 0 ~& 0 ~& -i
\end{pmatrix}\,, \\
\mathbf{\widehat{3}^{\prime}}: & S=-\dfrac{i}{2}\begin{pmatrix}
 0 ~&\sqrt{2} ~&\sqrt{2} \\
\sqrt{2} ~& -1 ~& 1 \\
\sqrt{2} ~& 1 ~& -1
\end{pmatrix}, ~&~ T=i\begin{pmatrix}
 1 ~& 0 ~& 0 \\
 0 ~& i ~& 0 \\
 0 ~& 0 ~& -i
\end{pmatrix}\,.
\end{array}
\end{eqnarray}
In the unhatted representations $\mathbf{1}$, $\mathbf{1}^{\prime}$, $\mathbf{2}$, $\mathbf{3}$ and $\mathbf{3}^{\prime}$, the generator $R=\mathds{1}$ is an identity matrix so that $S'_4$ can not be distinguished from $S_4$ in these representations. We have the generator $R=-\mathds{1}$ in the hatted representations $\mathbf{\widehat{1}}$, $\mathbf{\widehat{1}}^{\prime}$, $\mathbf{\widehat{2}}$, $\mathbf{\widehat{3}}$ and $\mathbf{\widehat{3}}^{\prime}$, these representations are novel and specific to $S'_4$. The character table of $S_4^{\prime}$ can be straightforwardly obtained from the representation matrices in Eq.~\eqref{eq:irr-S4p}, as shown in table~\ref{tab:character-tab-S4prime}. The Kronecker products between different irreducible representations $S'_4$ read as follows:
\begin{eqnarray}
\nonumber&&\mathbf{1} \otimes \mathbf{1} = \mathbf{1}^{\prime} \otimes \mathbf{1}^{\prime} = \mathbf{\widehat{1}} \otimes \mathbf{\widehat{1}}^{\prime} = \mathbf{1}, \quad \mathbf{1} \otimes \mathbf{\widehat{1}} = \mathbf{1}^{\prime} \otimes \mathbf{\widehat{1}}^{\prime} = \mathbf{\widehat{1}}  ,\\
\nonumber&&\mathbf{1} \otimes \mathbf{1}^{\prime} = \mathbf{\widehat{1}} \otimes \mathbf{\widehat{1}} = \mathbf{\widehat{1}}^{\prime} \otimes \mathbf{\widehat{1}}^{\prime} = \mathbf{1}^{\prime},\quad \mathbf{1} \otimes \mathbf{\widehat{1}}^{\prime} = \mathbf{1}^{\prime} \otimes \mathbf{\widehat{1}} = \mathbf{\widehat{1}}^{\prime} ,\\
\nonumber&&\mathbf{1} \otimes \mathbf{2} = \mathbf{1}^{\prime} \otimes \mathbf{2} = \mathbf{\widehat{1}} \otimes \mathbf{\widehat{2}} = \mathbf{\widehat{1}}^{\prime} \otimes \mathbf{\widehat{2}} = \mathbf{2},\quad \mathbf{1} \otimes \mathbf{\widehat{2}} = \mathbf{1}^{\prime} \otimes \mathbf{\widehat{2}} = \mathbf{\widehat{1}} \otimes \mathbf{2} = \mathbf{\widehat{1}}^{\prime} \otimes \mathbf{2} = \mathbf{\widehat{2}} ,\\
\nonumber&&\mathbf{1} \otimes \mathbf{3} = \mathbf{1}^{\prime} \otimes \mathbf{3}^{\prime} = \mathbf{\widehat{1}} \otimes \mathbf{\widehat{3}}^{\prime} = \mathbf{\widehat{1}}^{\prime} \otimes \mathbf{\widehat{3}} = \mathbf{3} ,\quad \mathbf{1} \otimes \mathbf{\widehat{3}} = \mathbf{1}^{\prime} \otimes \mathbf{\widehat{3}}^{\prime} = \mathbf{\widehat{1}} \otimes \mathbf{3} = \mathbf{\widehat{1}}^{\prime} \otimes \mathbf{3}^{\prime} = \mathbf{\widehat{3}} ,\\
\nonumber&&\mathbf{1} \otimes \mathbf{3}^{\prime} = \mathbf{1}^{\prime} \otimes \mathbf{3} = \mathbf{\widehat{1}} \otimes \mathbf{\widehat{3}} = \mathbf{\widehat{1}}^{\prime} \otimes \mathbf{\widehat{3}}^{\prime} = \mathbf{3}^{\prime},\quad \mathbf{1} \otimes \mathbf{\widehat{3}}^{\prime} = \mathbf{1}^{\prime} \otimes \mathbf{\widehat{3}} = \mathbf{\widehat{1}} \otimes \mathbf{3}^{\prime} = \mathbf{\widehat{1}}^{\prime} \otimes \mathbf{3} = \mathbf{\widehat{3}}^{\prime} ,\\
\nonumber&&\mathbf{2} \otimes \mathbf{2} = \mathbf{\widehat{2}} \otimes \mathbf{\widehat{2}} = \mathbf{1} \oplus \mathbf{1}^{\prime} \oplus \mathbf{2} ,\quad \mathbf{2} \otimes \mathbf{\widehat{2}} = \mathbf{\widehat{1}} \oplus \mathbf{\widehat{1}}^{\prime} \oplus \mathbf{\widehat{2}} ,\\
\nonumber&&\mathbf{2} \otimes \mathbf{3} = \mathbf{2} \otimes \mathbf{3}^{\prime} = \mathbf{\widehat{2}} \otimes \mathbf{\widehat{3}} = \mathbf{\widehat{2}} \otimes \mathbf{\widehat{3}}^{\prime} = \mathbf{3} \oplus \mathbf{3}^{\prime}, \quad \mathbf{2} \otimes \mathbf{\widehat{3}} = \mathbf{2} \otimes \mathbf{\widehat{3}}^{\prime} = \mathbf{\widehat{2}} \otimes \mathbf{3} = \mathbf{\widehat{2}} \otimes \mathbf{3}^{\prime} = \mathbf{\widehat{3}} \oplus \mathbf{\widehat{3}}^{\prime}  ,\\
\nonumber&&\mathbf{3} \otimes \mathbf{3} = \mathbf{3}^{\prime} \otimes \mathbf{3}^{\prime} = \mathbf{\widehat{3}} \otimes \mathbf{\widehat{3}}^{\prime} = \mathbf{1} \oplus \mathbf{2} \oplus \mathbf{3} \oplus \mathbf{3}^{\prime},\quad \mathbf{3} \otimes \mathbf{\widehat{3}} = \mathbf{3}^{\prime} \otimes \mathbf{\widehat{3}}^{\prime} = \mathbf{\widehat{1}} \oplus \mathbf{\widehat{2}} \oplus \mathbf{\widehat{3}} \oplus \mathbf{\widehat{3}}^{\prime} ,\\
&&\mathbf{3} \otimes \mathbf{3}^{\prime} = \mathbf{\widehat{3}} \otimes \mathbf{\widehat{3}} = \mathbf{\widehat{3}}^{\prime} \otimes \mathbf{\widehat{3}}^{\prime} = \mathbf{1}^{\prime} \oplus \mathbf{2} \oplus \mathbf{3} \oplus \mathbf{3}^{\prime},\quad \mathbf{3} \otimes \mathbf{\widehat{3}}^{\prime} = \mathbf{3}^{\prime} \otimes \mathbf{\widehat{3}} = \mathbf{\widehat{1}}^{\prime} \oplus \mathbf{\widehat{2}} \oplus \mathbf{\widehat{3}} \oplus \mathbf{\widehat{3}}^{\prime} .
\end{eqnarray}

\begin{table}[t!]
\centering
 \begin{tabular}{|c|c|c|c|c|c|c|c|c|c|c|}
 \hline\hline
Classes & $1C_1$ & $1C_2$ & $3C_2$ & $3C_2'$ & $8C_3$ & $6C_4$ & $6C_4'$ & $6C_4''$ & $6C_4'''$ & $8C_6$\\ \hline
$G$ & $1$ & $R$ & $T^2$ & $T^2R$ & $ST$ & $S$ & $T$ & $SR$ & $TR$ & $STR$\\ \hline
$\mathbf{1}$ & $1$ & $1$ & $1$ & $1$ & $1$ & $1$ & $1$ & $1$ & $1$ & $1$\\ \hline
$\mathbf{1^{\prime}}$ & $1$ & $1$ & $1$ & $1$ & $1$ & $-1$ & $-1$ & $-1$ & $-1$ & $1$\\ \hline
$\mathbf{\widehat{1}}$ & $1$ & $-1$ & $-1$ & $1$ & $1$ & $i$ & $-i$ & $-i$ & $i$ & $-1$\\ \hline
$\mathbf{\widehat{1}^{\prime}}$ & $1$ & $-1$ & $-1$ & $1$ & $1$ & $-i$ & $i$ & $i$ & $-i$ & $-1$\\ \hline
$\mathbf{2}$ & $2$ & $2$ & $2$ & $2$ & $-1$ & $0$ & $0$ & $0$ & $0$ & $-1$\\ \hline
$\mathbf{\widehat{2}}$ & $2$ & $-2$ & $-2$ & $2$ & $-1$ & $0$ & $0$ & $0$ & $0$ & $1$\\ \hline
$\mathbf{3}$ & $3$ & $3$ & $-1$ & $-1$ & $0$ & $-1$ & $1$ & $-1$ & $1$ & $0$\\ \hline
$\mathbf{3^{\prime}}$ & $3$ & $3$ & $-1$ & $-1$ & $0$ & $1$ & $-1$ & $1$ & $-1$ & $0$\\ \hline
$\mathbf{\widehat{3}}$ & $3$ & $-3$ & $1$ & $-1$ & $0$ & $-i$ & $-i$ & $i$ & $i$ & $0$\\ \hline
$\mathbf{\widehat{3}^{\prime}}$ & $3$ & $-3$ & $1$ & $-1$ & $0$ & $i$ & $i$ & $-i$ & $-i$ & $0$\\ \hline\hline
\end{tabular}
\caption{\label{tab:character-tab-S4prime}Character table of $S_4^{\prime}$, the representative element of each conjugacy class is given in the second row.}
\end{table}

In the following, we report all CG coefficients of $S'_4$ in the form of $\alpha\otimes\beta$, we use $\alpha_i(\beta_i)$ to denote the component of the left (right) basis vector $\alpha (\beta)$. The notations $\mathbf{I}$, $\mathbf{II}$ and $\mathbf{III}$ stand for singlet, doublet and triplet representations of $S'_4$ respectively.

\begin{itemize}
\item $\mathbf{I}\otimes\mathbf{I}\to\mathbf{I}$\,,
\begin{equation}
\begin{array}{ll}
  \left.\begin{array}{l}
\mathbf{1}\otimes\mathbf{1}\to\mathbf{1_{s}},~~
\mathbf{1}\otimes\mathbf{1'}\to\mathbf{1'}\\
\mathbf{1}\otimes\mathbf{\widehat{1}}\to\mathbf{\widehat{1}},~~~
\mathbf{1}\otimes\mathbf{\widehat{1}'}\to\mathbf{\widehat{1}'}\\
\mathbf{1'}\otimes\mathbf{1'}\to\mathbf{1_{s}},~
\mathbf{1'}\otimes\mathbf{\widehat{1}}\to\mathbf{\widehat{1}'}\\
\mathbf{1'}\otimes\mathbf{\widehat{1}'}\to\mathbf{\widehat{1}},~~
\mathbf{\widehat{1}}\otimes\mathbf{\widehat{1}}\to\mathbf{1'_{s}}\\
\mathbf{\widehat{1}}\otimes\mathbf{\widehat{1}'}\to\mathbf{1},~~~
\mathbf{\widehat{1}'}\otimes\mathbf{\widehat{1}'}\to\mathbf{1'_{s}}
  \end{array}\right\} &~~\text{with}~~
\begin{array}{l}
\mathbf{I} \sim\alpha\beta
\end{array}
\end{array}\nonumber
\end{equation}
\end{itemize}

\begin{itemize}
\item $\mathbf{I}\otimes\mathbf{II}\to\mathbf{II}$\,,
\begin{equation}
\begin{array}{lll}
\begin{array}{c}
    ~  \\ [-1ex]n=0  \\ \\ \\[1ex]n=1 \\ \\
  \end{array} ~~~~&
  \left.\begin{array}{l}
 \mathbf{1}\otimes\mathbf{2}\to\mathbf{2},~~~
\mathbf{1}\otimes\mathbf{\widehat{2}}\to\mathbf{\widehat{2}}\\
\mathbf{\widehat{1}}\otimes\mathbf{2}\to\mathbf{\widehat{2}},~~~
\mathbf{\widehat{1}'}\otimes\mathbf{\widehat{2}}\to\mathbf{2}\\ \\
\mathbf{1'}\otimes\mathbf{2}\to\mathbf{2},~~~
\mathbf{1'}\otimes\mathbf{\widehat{2}}\to\mathbf{\widehat{2}}\\
\mathbf{\widehat{1}}\otimes\mathbf{\widehat{2}}\to\mathbf{2},~~~
\mathbf{\widehat{1}'}\otimes\mathbf{2}\to\mathbf{\widehat{2}}
  \end{array}\right\} &~~\text{with}~~
\begin{array}{l}
\mathbf{II} \sim \alpha M^{(n)}\begin{pmatrix}
\beta_1\\
\beta_2
\end{pmatrix}\\
\end{array}
\end{array}\,,\nonumber
\end{equation}
\end{itemize}
where $M^{(0)}=
\begin{pmatrix}
  1 ~&~ 0\\0 ~&~ 1
\end{pmatrix}
$, $M^{(1)}=
\begin{pmatrix}
  0 ~&~ 1\\-1 ~&~ 0
\end{pmatrix}
$, and the same convention is adopted below.

\begin{itemize}
\item $\mathbf{I}\otimes\mathbf{III}\to\mathbf{III}$\,,
\begin{equation}
\begin{array}{ll}
 \left.\begin{array}{l}
\mathbf{1}\otimes\mathbf{3}\to\mathbf{3},~~~
\mathbf{1}\otimes\mathbf{3'}\to\mathbf{3'}\\
\mathbf{1}\otimes\mathbf{\widehat{3}}\to\mathbf{\widehat{3}},~~~
\mathbf{1}\otimes\mathbf{\widehat{3}'}\to\mathbf{\widehat{3}'}\\
\mathbf{1'}\otimes\mathbf{3}\to\mathbf{3'},~~
\mathbf{1'}\otimes\mathbf{3'}\to\mathbf{3}\\
\mathbf{1'}\otimes\mathbf{\widehat{3}}\to\mathbf{\widehat{3}'},~~
\mathbf{1'}\otimes\mathbf{\widehat{3}'}\to\mathbf{\widehat{3}}\\
\mathbf{\widehat{1}}\otimes\mathbf{3}\to\mathbf{\widehat{3}},~~~\,
\mathbf{\widehat{1}}\otimes\mathbf{3'}\to\mathbf{\widehat{3}'}\\
\mathbf{\widehat{1}}\otimes\mathbf{\widehat{3}}\to\mathbf{3'},~~~
\mathbf{\widehat{1}}\otimes\mathbf{\widehat{3}'}\to\mathbf{3}\\
\mathbf{\widehat{1}'}\otimes\mathbf{3}\to\mathbf{\widehat{3}'},~~\,
\mathbf{\widehat{1}'}\otimes\mathbf{3'}\to\mathbf{\widehat{3}}\\
\mathbf{\widehat{1}'}\otimes\mathbf{\widehat{3}}\to\mathbf{3},~~~
\mathbf{\widehat{1}'}\otimes\mathbf{\widehat{3}'}\to\mathbf{3'}
  \end{array}\right\} &~~\text{with}~~
\begin{array}{l}
\mathbf{III} \sim\alpha\begin{pmatrix}
\beta_1\\
\beta_2\\
\beta_3
\end{pmatrix}
\end{array}
\end{array}\nonumber
\end{equation}
\end{itemize}

\begin{itemize}
\item $\mathbf{II}\otimes\mathbf{II}\to\mathbf{I}_1\oplus\mathbf{I}_2\oplus\mathbf{II}$\,,
\begin{equation}
\begin{array}{lll}
\begin{array}{c}
    ~\\ [-1.1ex]n=0  \\ \\[1ex]n=1 \\
  \end{array} ~~~~&
  \left.\begin{array}{l}
\mathbf{2}\otimes\mathbf{2}\to\mathbf{1'_{a}}\oplus\mathbf{1_{s}}\oplus\mathbf{2_{s}}\\
\mathbf{2}\otimes\mathbf{\widehat{2}}\to\mathbf{\widehat{1}'}\oplus\mathbf{\widehat{1}}\oplus\mathbf{\widehat{2}}\\\\
\mathbf{\widehat{2}}\otimes\mathbf{\widehat{2}}\to\mathbf{1_{a}}\oplus\mathbf{1'_{s}}\oplus\mathbf{2_{s}}
  \end{array}\right\} &~~\text{with}~~~
\begin{array}{l}
\mathbf{I}_1 \sim\alpha_1\beta_2-\alpha_2\beta_1\\[1ex]
\mathbf{I}_2 \sim\alpha_1\beta_1+\alpha_2\beta_2\\[1ex]
\mathbf{II} \sim M^{(n)}\begin{pmatrix}
-\alpha_1\beta_1+\alpha_2\beta_2\\
\alpha_1\beta_2+\alpha_2\beta_1
\end{pmatrix}
\end{array}
\end{array}\nonumber
\end{equation}
\end{itemize}

\begin{itemize}
\item $\mathbf{II}\otimes\mathbf{III}\to\mathbf{III}_1\oplus\mathbf{III}_2$\,,
\begin{equation}
\begin{array}{ll}
 \left.\begin{array}{l}
\mathbf{2}\otimes\mathbf{3}\to\mathbf{3}\oplus\mathbf{3'}\\
\mathbf{2}\otimes\mathbf{3'}\to\mathbf{3'}\oplus\mathbf{3}\\
\mathbf{2}\otimes\mathbf{\widehat{3}}\to\mathbf{\widehat{3}}\oplus\mathbf{\widehat{3}'}\\
\mathbf{2}\otimes\mathbf{\widehat{3}'}\to\mathbf{\widehat{3}'}\oplus\mathbf{\widehat{3}}\\
\mathbf{\widehat{2}}\otimes\mathbf{3}\to\mathbf{\widehat{3}}\oplus\mathbf{\widehat{3}'}\\
\mathbf{\widehat{2}}\otimes\mathbf{3'}\to\mathbf{\widehat{3}'}\oplus\mathbf{\widehat{3}}\\
\mathbf{\widehat{2}}\otimes\mathbf{\widehat{3}}\to\mathbf{3'}\oplus\mathbf{3}\\
\mathbf{\widehat{2}}\otimes\mathbf{\widehat{3}'}\to\mathbf{3}\oplus\mathbf{3'}
  \end{array}\right\} &~~\text{with}~~~
\begin{array}{l}
\mathbf{III}_1 \sim \begin{pmatrix}
2\alpha_1\beta_1\\
-\alpha_1\beta_2+\sqrt{3}\alpha_2\beta_3\\
-\alpha_1\beta_3+\sqrt{3}\alpha_2\beta_2
\end{pmatrix}\\[5ex]
\mathbf{III}_2 \sim \begin{pmatrix}
-2\alpha_2\beta_1\\
\sqrt{3}\alpha_1\beta_3+\alpha_2\beta_2\\
\sqrt{3}\alpha_1\beta_2+\alpha_2\beta_3
\end{pmatrix}
\end{array}
\end{array}\nonumber
\end{equation}
\end{itemize}

\begin{itemize}
\item $\mathbf{III}\otimes\mathbf{III}\to\mathbf{I}\oplus\mathbf{II}\oplus\mathbf{III}_1\oplus\mathbf{III}_2$\,,
\begin{equation}
\begin{array}{lll}
\begin{array}{c}
    ~\\ [-3ex]n=0  \\ \\ \\ \\[3ex]n=1 \\
  \end{array} ~~~~&
  \left.\begin{array}{l}
\mathbf{3}\otimes\mathbf{3}\to\mathbf{1_{s}}\oplus\mathbf{2_{s}}\oplus\mathbf{3_{a}}\oplus\mathbf{3'_{s}}\\
\mathbf{3}\otimes\mathbf{\widehat{3}}\to\mathbf{\widehat{1}}\oplus\mathbf{\widehat{2}}\oplus\mathbf{\widehat{3}}\oplus\mathbf{\widehat{3}'}\\
\mathbf{3'}\otimes\mathbf{3'}\to\mathbf{1_{s}}\oplus\mathbf{2_{s}}\oplus\mathbf{3_{a}}\oplus\mathbf{3'_{s}}\\
\mathbf{3'}\otimes\mathbf{\widehat{3}'}\to\mathbf{\widehat{1}}\oplus\mathbf{\widehat{2}}\oplus\mathbf{\widehat{3}}\oplus\mathbf{\widehat{3}'}\\
\mathbf{\widehat{3}}\otimes\mathbf{\widehat{3}'}\to\mathbf{1}\oplus\mathbf{2}\oplus\mathbf{3}\oplus\mathbf{3'}\\\\
\mathbf{3}\otimes\mathbf{3'}\to\mathbf{1'}\oplus\mathbf{2}\oplus\mathbf{3'}\oplus\mathbf{3}\\
\mathbf{3}\otimes\mathbf{\widehat{3}'}\to\mathbf{\widehat{1}'}\oplus\mathbf{\widehat{2}}\oplus\mathbf{\widehat{3}'}\oplus\mathbf{\widehat{3}}\\
\mathbf{3'}\otimes\mathbf{\widehat{3}}\to\mathbf{\widehat{1}'}\oplus\mathbf{\widehat{2}}\oplus\mathbf{\widehat{3}'}\oplus\mathbf{\widehat{3}}\\
\mathbf{\widehat{3}}\otimes\mathbf{\widehat{3}}\to\mathbf{1'_{s}}\oplus\mathbf{2_{s}}\oplus\mathbf{3'_{a}}\oplus\mathbf{3_{s}}\\
\mathbf{\widehat{3}'}\otimes\mathbf{\widehat{3}'}\to\mathbf{1'_{s}}\oplus\mathbf{2_{s}}\oplus\mathbf{3'_{a}}\oplus\mathbf{3_{s}}
  \end{array}\right\} &~~\text{with}~~~
\begin{array}{l}
\mathbf{I} \sim\alpha_1\beta_1+\alpha_2\beta_3+\alpha_3\beta_2\\[3ex]
\mathbf{II} \sim M^{(n)}\begin{pmatrix}
2\alpha_1\beta_1-\alpha_2\beta_3-\alpha_3\beta_2\\
\sqrt{3}\alpha_2\beta_2+\sqrt{3}\alpha_3\beta_3
\end{pmatrix}\\[3ex]
\mathbf{III}_1 \sim \begin{pmatrix}
\alpha_2\beta_3-\alpha_3\beta_2\\
\alpha_1\beta_2-\alpha_2\beta_1\\
-\alpha_1\beta_3+\alpha_3\beta_1
\end{pmatrix}\\[5ex]
\mathbf{III}_2 \sim \begin{pmatrix}
\alpha_2\beta_2-\alpha_3\beta_3\\
-\alpha_1\beta_3-\alpha_3\beta_1\\
\alpha_1\beta_2+\alpha_2\beta_1
\end{pmatrix}
\end{array}
\end{array}\nonumber
\end{equation}
\end{itemize}

\subsubsection{\label{sec:MF-level4}Integer weight modular forms of level 4}

The linear space spanned by the weight $k$ modular forms at level 4 has dimension $2k+1$, and a general vector $\mathcal{M}_{k}(\Gamma(4))$ in the linear space can be explicitly constructed by using the theta constants~\cite{Liu:2020msy}:
\begin{equation}
\label{eq:Mk-Gamma4}
\mathcal{M}_{k}(\Gamma(4))=\sum_{m+n=2k,\, m,n\ge 0} c_{mn} \theta^{m}_2(\tau)\theta^{n}_3(\tau)\,,
\end{equation}
where $c_{mn}$ are general complex coefficients and the theta constants is defined as
\begin{eqnarray}
\nonumber \theta_2(\tau)&=& \sum_{m\in\mathbb{Z}} e^{2\pi i \tau (m+1/2)^2}=2q^{1/4}(1+q^2+q^6+q^{12}+\dots)\,,\\
\theta_3(\tau)&=&\sum_{m\in\mathbb{Z}} e^{2\pi i \tau m^2}=1+2q+2q^4+2q^9+2q^{16}+\dots \,.
\end{eqnarray}
Hence the weight $k$ modular forms of level 4 can be expressed as the homogeneous polynomials of degree $2k$ in $\theta_{1}$ and $\theta_{2}$. In the following, we report the explicit expressions of the $S'_4$ modular multiplets up to weight 6 in the working basis of Eq.~\eqref{eq:irr-S4p}. We prefer to use $\vartheta_1(\tau)=\theta_3(\tau)$, $\vartheta_2(\tau)=-\theta_2(\tau)$ since $\vartheta_1(\tau)$ and $\vartheta_2(\tau)$ turns out to be one-half weight modular forms and they form a doublet of the metaplectic cover of $S'_4$~\cite{Liu:2020msy}.
\begin{itemize}
\item{$k_Y=1$}
\begin{equation}
Y_{\mathbf{\widehat{3}'}}^{(1)}=\begin{pmatrix}
\sqrt{2}\vartheta_1\vartheta_2\\
-\vartheta_2^2\\
\vartheta_1^2\\
\end{pmatrix}\,.
\end{equation}

\item{$k_Y=2$}

\begin{eqnarray}
\nonumber&&Y_{\mathbf{2}}^{(2)}=\begin{pmatrix}
\vartheta_1^4+\vartheta_2^4\\
-2\sqrt{3}\vartheta_1^2\vartheta_2^2\\
\end{pmatrix}\,,\\
&&Y_{\mathbf{3}}^{(2)}=\begin{pmatrix}
\vartheta_1^4-\vartheta_2^4\\
2\sqrt{2}\vartheta_1^3\vartheta_2\\
2\sqrt{2}\vartheta_1\vartheta_2^3\\
\end{pmatrix}\,.
\end{eqnarray}

\item{$k_Y=3$}

\begin{eqnarray}
\nonumber&&Y_{\mathbf{\widehat{1}'}}^{(3)}=\vartheta_1\vartheta_2\left(\vartheta_1^4-\vartheta_2^4\right)\,, \\
\nonumber&&Y_{\mathbf{\widehat{3}}}^{(3)}=\begin{pmatrix}
4\sqrt{2}\vartheta_1^3\vartheta_2^3\\
\vartheta_1^6+3\vartheta_1^2\vartheta_2^4\\
-\vartheta_2^2\left(3\vartheta_1^4+\vartheta_2^4\right)\\
\end{pmatrix}\,,\\
&&Y_{\mathbf{\widehat{3}'}}^{(3)}=
\begin{pmatrix}
2\sqrt{2}\vartheta_1\vartheta_2\left(\vartheta_1^4+\vartheta_2^4\right)\\
\vartheta_2^6-5\vartheta_1^4\vartheta_2^2\\
5\vartheta_1^2\vartheta_2^4-\vartheta_1^6\\
\end{pmatrix}\,.
\end{eqnarray}

\item{$k_Y=4$}

\begin{eqnarray}
\nonumber&&Y_{\mathbf{1}}^{(4)}=
\vartheta_1^8+14\vartheta_1^4\vartheta_2^4+\vartheta_2^8\,,\\
\nonumber&&Y_{\mathbf{2}}^{(4)}=\begin{pmatrix}
\vartheta_1^8-10\vartheta_1^4\vartheta_2^4+\vartheta_2^8\\
4\sqrt{3}\vartheta_1^2\vartheta_2^2\left(\vartheta_1^4+\vartheta_2^4\right)\\
\end{pmatrix}\,,\\
\nonumber&&Y_{\mathbf{3}}^{(4)}=\begin{pmatrix}
\vartheta_2^8-\vartheta_1^8\\
\sqrt{2}\vartheta_2\left(\vartheta_1^7+7\vartheta_1^3\vartheta_2^4\right)\\
\sqrt{2}\vartheta_1\left(\vartheta_2^7+7\vartheta_1^4\vartheta_2^3\right)\\
\end{pmatrix}\,,\\
&&Y_{\mathbf{3'}}^{(4)}=\vartheta_1\vartheta_2\left(\vartheta_1^4-\vartheta_2^4\right)\begin{pmatrix}
\sqrt{2}\vartheta_1\vartheta_2\\
-\vartheta_2^2\\
\vartheta_1^2\\
\end{pmatrix}\,.
\end{eqnarray}

\item{$k_Y=5$}

\begin{eqnarray}
\nonumber&&Y_{\mathbf{\widehat{2}}}^{(5)}=\vartheta_1\vartheta_2\left(\vartheta_1^4-\vartheta_2^4\right)
\begin{pmatrix}
2\sqrt{3}\vartheta_1^2\vartheta_2^2\\
\vartheta_1^4+\vartheta_2^4\\
\end{pmatrix}\,, \\
\nonumber&&Y_{\mathbf{\widehat{3}}}^{(5)}=
\begin{pmatrix}
-8\sqrt{2}\vartheta_1^3\vartheta_2^3\left(\vartheta_1^4+\vartheta_2^4\right)\\
\vartheta_1^2\left(\vartheta_1^8-14\vartheta_1^4\vartheta_2^4-3\vartheta_2^8\right)\\
\vartheta_2^2\left(3\vartheta_1^8+14\vartheta_1^4\vartheta_2^4-\vartheta_2^8\right)\\
\end{pmatrix}\,, \\
\nonumber&&Y_{\mathbf{\widehat{3}'}I}^{(5)}=
\begin{pmatrix}
2\sqrt{2}\vartheta_1\vartheta_2\left(\vartheta_1^8-10\vartheta_1^4\vartheta_2^4+\vartheta_2^8\right)\\
\vartheta_2^2\left(13\vartheta_1^8+2\vartheta_1^4\vartheta_2^4+\vartheta_2^8\right)\\
-\vartheta_1^2\left(\vartheta_1^8+2\vartheta_1^4\vartheta_2^4+13\vartheta_2^8\right)\\
\end{pmatrix}\,, \\
&&Y_{\mathbf{\widehat{3}'}II}^{(5)}=\left(\vartheta_1^8+14\vartheta_1^4\vartheta_2^4+\vartheta_2^8\right)
\begin{pmatrix}
\sqrt{2}\vartheta_1\vartheta_2\\
-\vartheta_2^2\\
\vartheta_1^2\\
\end{pmatrix}\,.
\end{eqnarray}

\item{$k_Y=6$}

\begin{eqnarray}
\nonumber&&Y_{\mathbf{1}}^{(6)}=
\vartheta_1^{12}-33\vartheta_1^8\vartheta_2^4-33\vartheta_1^4\vartheta_2^8+\vartheta_2^{12}\,,\\
\nonumber&&Y_{\mathbf{1'}}^{(6)}=
\vartheta_1^2\vartheta_2^2\left(\vartheta_1^4-\vartheta_2^4\right)^2 \,, \\
\nonumber&&Y_{\mathbf{2}}^{(6)}=\left(\vartheta_1^8+14\vartheta_1^4\vartheta_2^4+\vartheta_2^8\right)\begin{pmatrix}
\vartheta_1^4+\vartheta_2^4\\
-2\sqrt{3}\vartheta_1^2\vartheta_2^2\\
\end{pmatrix}\,,\\
\nonumber&&Y_{\mathbf{3}I}^{(6)}=\begin{pmatrix}
\vartheta_1^{12}-11\vartheta_1^8\vartheta_2^4+11\vartheta_1^4\vartheta_2^8-\vartheta_2^{12}\\
-\sqrt{2}\vartheta_1^3\vartheta_2\left(\vartheta_1^8-22\vartheta_1^4\vartheta_2^4-11\vartheta_2^8\right)\\
\sqrt{2}\vartheta_1\vartheta_2^3\left(11\vartheta_1^8+22\vartheta_1^4\vartheta_2^4-\vartheta_2^8\right)\\
\end{pmatrix}\,,\\
\nonumber&&Y_{\mathbf{3}II}^{(6)}=\left(\vartheta_1^8+14\vartheta_2^4\vartheta_1^4+\vartheta_2^8\right)
\begin{pmatrix}
\vartheta_1^4-\vartheta_2^4\\
2\sqrt{2}\vartheta_1^3\vartheta_2\\
2\sqrt{2}\vartheta_1\vartheta_2^3\\
\end{pmatrix}\,,\\
&&Y_{\mathbf{3'}}^{(6)}=\vartheta_1\vartheta_2\left(\vartheta_1^4-\vartheta_2^4\right)
\begin{pmatrix}
2\sqrt{2}\vartheta_1\vartheta_2\left(\vartheta_1^4+\vartheta_2^4\right)\\
\vartheta_2^6-5\vartheta_1^4\vartheta_2^2\\
5\vartheta_1^2\vartheta_2^4-\vartheta_1^6\\
\end{pmatrix}\,.
\end{eqnarray}
The above modular forms of level 4 are listed in table~\ref{tab:MF-L5-W6}. The higher weight modular forms can be constructed from the tensor products of the above modular multiplets.

\end{itemize}

\vskip0.2in 

\subsection{\label{app:group-MF-N5} $N=5$}

The finite modular group $\Gamma'_5$ which is the double covering of the icosahedral group $A_5$, can be generated by three generators $S$, $T$ and $R$ which obey the following rules,
\begin{equation}
S^2=R,\quad T^5=(ST)^3=R^2=1,\quad RT=TR\,,
\end{equation}
or equivalently
\begin{equation}
S^4=T^5=(ST)^3=1,\quad S^2T=TS^2\,.
\end{equation}
The group $A_5$ is reproduced by setting $R=1$. The group $A'_5$ has 120 elements which is twice as many elements as $A_5$,
and all the elements can be divided into the 9 conjugacy classes as follows:
\begin{eqnarray}
\nonumber 1C_1&=& \{1\}\,, \\
\nonumber 1C_2&=& \{R\}=(1C_1)\cdot R\,, \\
\nonumber 20C_3&=& \{ST,TS,S^3T^4,T^4S^3,T^2ST^4,T^2S^3T^2,T^3ST^3,T^3S^3T,T^4ST^2,TS^3T^3,\\
\nonumber &&~~~~~ T^2ST^3S,T^3ST^2S,ST^2ST^3,ST^3ST^2,S^3T^3ST,TST^3S^3,S^3T^2ST^4,\\
\nonumber &&~~~~~ T^4ST^2S^3,ST^2ST^2S,TST^3ST\}\,, \\
\nonumber 30C_4&=& \{S,SR,T^2ST^3,T^2ST^3R,T^3ST^2,T^3ST^2R, T^4ST,T^4STR,TST^4,TST^4R,\\
\nonumber &&~~~~~ST^2ST,ST^2STR, TST^2S,TST^2SR,ST^3ST^2S,ST^3ST^2SR,
ST^2ST^3S,ST^2ST^3SR,  \\
\nonumber &&~~~~~ T^2ST^3ST^2, T^2ST^3ST^2R, T^3ST^3ST, T^3ST^3STR, TST^3ST^3, TST^3ST^3R, \\
\nonumber &&~~~~~ ST^2ST^3ST,
ST^2ST^3STR, TST^3ST^2S, TST^3ST^2SR, T^2ST^3ST^2S, T^2ST^3ST^2SR \}\,, \\
\nonumber 12C_5&=& \{T,T^4,T^3S,ST^3,S^3T^2,T^2S^3,T^2ST,TST^2,S^3TS,TS^3T,T^3S^3T^4,T^4S^3T^3\}\,,\\
\nonumber 12C_5'&=& \{T^2,T^3,S^3T^2S,S^3T^3S,T^2ST^2S,ST^2ST^2,T^3ST^3S,ST^3ST^3,T^2ST^3ST,\\
\nonumber &&~~~~~ TST^3ST^2,T^3ST^2ST^4,T^4ST^2ST^3\}\,, \\
\nonumber 20C_6&=& \{STR,TSR,S^3T^4R,T^4S^3R,T^2ST^4R,T^2S^3T^2R,T^3ST^3R,T^3S^3TR,T^4ST^2R,\\
\nonumber &&~~~~~ TS^3T^3R, T^2ST^3SR,T^3ST^2SR,ST^2ST^3R,ST^3ST^2R,S^3T^3STR,TST^3S^3R,\\
\nonumber &&~~~~~ S^3T^2ST^4R, T^4ST^2S^3R,ST^2ST^2SR,TST^3STR \} = (20C_3)\cdot R \,,\\
\nonumber 12C_{10}&=& \{TR,T^4R,T^3SR,ST^3R,S^3T^2R,T^2S^3R,T^2STR,TST^2R,S^3TSR,TS^3TR,\\
\nonumber&&~~~~~ T^3S^3T^4R,T^4S^3T^3R\}=(12C_5)\cdot R\,, \\
\nonumber 12C_{10}'&=&\{T^2R,T^3R,S^3T^2SR,S^3T^3SR,T^2ST^2SR,ST^2ST^2R,T^3ST^3SR,ST^3ST^3R,\\
&&~~~~~ T^2ST^3STR, TST^3ST^2R,T^3ST^2ST^4R,T^4ST^2ST^3R\}=(12C_5')\cdot R\,.
\end{eqnarray}
The number of conjugacy classes is the same as the inequivalent irreducible representations. In addition to the five inequivalent irreducible representations $\mathbf{1}$, $\mathbf{3}$, $\mathbf{3'}$, $\mathbf{4}$, $\mathbf{5}$ of the $A_5$ group, $A'_5$ has four inequivalent representations $\mathbf{\widehat{2}}$, $\mathbf{\widehat{2}'}$, $\mathbf{\widehat{4}'}$ and $\mathbf{\widehat{6}}$. The explicit forms of the generators $S$ and $T$ in each of the irreducible representations are given by
\begin{eqnarray}
\label{eq:irr-A5p} \begin{array}{cccc}
\mathbf{1}: & S=1, ~&~ T=1\,, \\
\mathbf{\widehat{2}}:& S=i\sqrt{\frac{1}{\sqrt{5}\phi}}\begin{pmatrix}
 \phi  & 1 \\
 1 & -\phi  \\
\end{pmatrix}, ~& T=\begin{pmatrix}
 \omega_5^2 & 0 \\
 0 & \omega_5^3 \\
\end{pmatrix}\,, \\
\mathbf{\mathbf{\widehat{2}'}} : & S=i\sqrt{\frac{1}{\sqrt{5}\phi}}\begin{pmatrix}
 1 ~& \phi  \\
 \phi  ~& -1 \\
\end{pmatrix}, ~& T=\begin{pmatrix}
 \omega_5 ~& 0 \\
 0 ~& \omega_5^4 \\
\end{pmatrix} \,,\\
\mathbf{3}: & S=\frac{1}{\sqrt{5}}\begin{pmatrix}
 1 & -\sqrt{2} & -\sqrt{2} \\
 -\sqrt{2} & -\phi  & \frac{1}{\phi } \\
 -\sqrt{2} & \frac{1}{\phi } & -\phi  \\
\end{pmatrix}, ~&~ T=\begin{pmatrix}
 1 & 0 & 0 \\
 0 & \omega _5 & 0 \\
 0 & 0 & \omega _5^4 \\
\end{pmatrix}\,, \\
\mathbf{\mathbf{3}'}: & S=\frac{1}{\sqrt{5}}\begin{pmatrix}
 -1 &\sqrt{2} &\sqrt{2} \\
\sqrt{2} & -\frac{1}{\phi } & \phi  \\
\sqrt{2} & \phi  & -\frac{1}{\phi } \\
\end{pmatrix}, ~&~ T=\begin{pmatrix}
 1 & 0 & 0 \\
 0 & \omega _5^2 & 0 \\
 0 & 0 & \omega _5^3 \\
\end{pmatrix}\,, \\
\mathbf{4}: & S=\frac{1}{\sqrt{5}}\begin{pmatrix}
 1 & \frac{1}{\phi } & \phi  & -1 \\
 \frac{1}{\phi } & -1 & 1 & \phi  \\
 \phi  & 1 & -1 & \frac{1}{\phi } \\
 -1 & \phi  & \frac{1}{\phi } & 1 \\
\end{pmatrix}, ~&~ T=\begin{pmatrix}
 \omega _5 & 0 & 0 & 0 \\
 0 & \omega _5^2 & 0 & 0 \\
 0 & 0 & \omega _5^3 & 0 \\
 0 & 0 & 0 & \omega _5^4 \\
\end{pmatrix}\,,\\
\mathbf{\widehat{4}'}: & S=i\sqrt{\frac{1}{5\sqrt{5}\phi}}\begin{pmatrix}
 -\phi ^2 &\sqrt{3} \phi  & -\sqrt{3} & -\frac{1}{\phi } \\
\sqrt{3} \phi  & \frac{1}{\phi } & -\phi ^2 & -\sqrt{3} \\
 -\sqrt{3} & -\phi ^2 & -\frac{1}{\phi } & -\sqrt{3} \phi  \\
 -\frac{1}{\phi } & -\sqrt{3} & -\sqrt{3} \phi  & \phi ^2 \\
\end{pmatrix}, ~&~ T=\begin{pmatrix}
 \omega _5 & 0 & 0 & 0 \\
 0 & \omega _5^2 & 0 & 0 \\
 0 & 0 & \omega _5^3 & 0 \\
 0 & 0 & 0 & \omega _5^4 \\
\end{pmatrix}\,, \\
\mathbf{5}: & S=\frac{1}{5}\begin{pmatrix}
 -1 &\sqrt{6} &\sqrt{6} &\sqrt{6} &\sqrt{6} \\
\sqrt{6} & \frac{1}{\phi ^2} & -2 \phi  & \frac{2}{\phi } & \phi ^2 \\
\sqrt{6} & -2 \phi  & \phi ^2 & \frac{1}{\phi ^2} & \frac{2}{\phi } \\
\sqrt{6} & \frac{2}{\phi } & \frac{1}{\phi ^2} & \phi ^2 & -2 \phi  \\
\sqrt{6} & \phi ^2 & \frac{2}{\phi } & -2 \phi  & \frac{1}{\phi ^2} \\
\end{pmatrix}, ~&~ T=\begin{pmatrix}
 1 & 0 & 0 & 0 & 0 \\
 0 & \omega _5 & 0 & 0 & 0 \\
 0 & 0 & \omega _5^2 & 0 & 0 \\
 0 & 0 & 0 & \omega _5^3 & 0 \\
 0 & 0 & 0 & 0 & \omega _5^4 \\
\end{pmatrix}\,, \\
\mathbf{\widehat{6}}: & S=i\sqrt{\frac{1}{5\sqrt{5}\phi}}\begin{pmatrix}
 -1 & \phi  & \frac{1}{\phi } &\sqrt{2} \phi  &\sqrt{2} & \phi ^2 \\
 \phi  & 1 & \phi ^2 &\sqrt{2} & -\sqrt{2} \phi  & -\frac{1}{\phi } \\
 \frac{1}{\phi } & \phi ^2 & 1 & -\sqrt{2} &\sqrt{2} \phi  & -\phi  \\
\sqrt{2} \phi  &\sqrt{2} & -\sqrt{2} & -\phi  & -1 &\sqrt{2} \phi  \\
\sqrt{2} & -\sqrt{2} \phi  &\sqrt{2} \phi  & -1 & \phi  &\sqrt{2} \\
 \phi ^2 & -\frac{1}{\phi } & -\phi  &\sqrt{2} \phi  &\sqrt{2} & -1 \\
\end{pmatrix}, ~&~ T=\begin{pmatrix}
 1 & 0 & 0 & 0 & 0 & 0 \\
 0 & 1 & 0 & 0 & 0 & 0 \\
 0 & 0 & \omega _5 & 0 & 0 & 0 \\
 0 & 0 & 0 & \omega _5^2 & 0 & 0 \\
 0 & 0 & 0 & 0 & \omega _5^3 & 0 \\
 0 & 0 & 0 & 0 & 0 & \omega _5^4 \\
\end{pmatrix}\,,
\end{array}
\end{eqnarray}
where $\phi=(1+\sqrt{5})/2$ is the golden ratio, and $\omega_5$ is the quintic unit root $\omega_5=e^{2\pi i/5}$. In the single-valued representations $\mathbf{1}$, $\mathbf{3}$, $\mathbf{3'}$, $\mathbf{4}$ and $\mathbf{5}$, the generator $R$ is represented by the identity matrix $\mathds{1}$, and the elements of $A'_5$ are described by the same set of matrices which represent the elements in $A_5$,  consequently the group $A'_5$ can not be distinguished from the group $A_5$ with these representations. The generator $R$ is $-\mathds{1}$ in the double-valued representations $\mathbf{\widehat{2}}$, $\mathbf{\widehat{2}'}$, $\mathbf{\widehat{4}'}$ and $\mathbf{\widehat{6}}$. Taking trace of the representation matrices, the character table of the  $A'_5$ group can be obtained and shown in table~\ref{tab:character-table}. From the character table, one can straightforwardly calculate the multiplication rules of the irreducible representations as follows,

\begin{table}[hptb!]
  \centering
  \begin{tabular}{|c|c|c|c|c|c|c|c|c|c|c|}
    \hline\hline
    Classes & $1C_1$ & $1C_2$ & $20C_3$ & $30C_4$ & $12C_5$ & $12C_5'$ & $12C_5''$ & $20C_6$ & $12C_{10}$ \\ \hline
    $G$ & $1$ & $R$ & $ST$ & $S$ & $T$ & $T^2$ & $S^3T$ & $RT$ & $T^2R$ \\ \hline
$\mathbf{1}$ & $1$ & $1$ & $1$ & $1$ & $1$ & $1$ & $1$ & $1$ & $1$\\ \hline
$\mathbf{\widehat{2}}$ & $2$ & $-2$ & $-1$ & $0$ & $-\phi$ & $\frac{1}{\phi }$ & $1$ & $\phi$ & $-\frac{1}{\phi }$\\ \hline
$\mathbf{\widehat{2}'}$ & $2$ & $-2$ & $-1$ & $0$ & $\frac{1}{\phi }$ & $-\phi$ & $1$ & $-\frac{1}{\phi }$ & $\phi$\\ \hline
$\mathbf{3}$ & $3$ & $3$ & $0$ & $-1$ & $\phi$ & $-\frac{1}{\phi }$ & $0$ & $\phi$ & $-\frac{1}{\phi }$\\ \hline
$\mathbf{3'}$ & $3$ & $3$ & $0$ & $-1$ & $-\frac{1}{\phi }$ & $\phi$ & $0$ & $-\frac{1}{\phi }$ & $\phi$\\ \hline
$\mathbf{4}$ & $4$ & $4$ & $1$ & $0$ & $-1$ & $-1$ & $1$ & $-1$ & $-1$\\ \hline
$\mathbf{\widehat{4}'}$ & $4$ & $-4$ & $1$ & $0$ & $-1$ & $-1$ & $-1$ & $1$ & $1$\\ \hline
$\mathbf{5}$ & $5$ & $5$ & $-1$ & $1$ & $0$ & $0$ & $-1$ & $0$ & $0$\\ \hline
$\mathbf{\widehat{6}}$ & $6$ & $-6$ & $0$ & $0$ & $1$ & $1$ & $0$ & $-1$ & $-1$ \\ \hline\hline
\end{tabular}
\caption{\label{tab:character-table}Character table of $A'_5$, where $G$ stands for the representative element of each conjugacy class.}
\end{table}

\begin{eqnarray}
\nonumber && \mathbf{\widehat{2}}\otimes \mathbf{\widehat{2}}=\mathbf{1_a}\oplus \mathbf{3_s}\,,\quad \mathbf{\widehat{2}}\otimes \mathbf{\widehat{2}'}=\mathbf{4}\,,\quad \mathbf{\widehat{2}}\otimes \mathbf{3}=\mathbf{\widehat{2}}\oplus \mathbf{\widehat{4}'}\,,\quad \mathbf{\widehat{2}}\otimes \mathbf{3'}=\mathbf{\widehat{2}'}\otimes \mathbf{3}=\mathbf{\widehat{6}}\,,\\
\nonumber &&\mathbf{\widehat{2}}\otimes \mathbf{4}=\mathbf{\widehat{2}'}\oplus \mathbf{\widehat{6}}\,,\quad \mathbf{\widehat{2}}\otimes \mathbf{\widehat{4}'}=\mathbf{3}\oplus \mathbf{5}\,,\quad \mathbf{\widehat{2}}\otimes \mathbf{5}=\mathbf{\widehat{2}'}\otimes \mathbf{5}=\mathbf{\widehat{4}'}\oplus \mathbf{\widehat{6}}\,,\\
\nonumber && \mathbf{\widehat{2}}\otimes \mathbf{\widehat{6}}=\mathbf{3}\otimes \mathbf{4}=\mathbf{3'}\oplus \mathbf{4}\oplus \mathbf{5}\,,\quad \mathbf{\widehat{2}'}\otimes \mathbf{\widehat{2}'}=\mathbf{1_a}\oplus \mathbf{3'_s}\,,\quad \mathbf{\widehat{2}'}\otimes \mathbf{3'}=\mathbf{\widehat{2}'}\oplus \mathbf{\widehat{4}'}\,,\\
\nonumber && \mathbf{\widehat{2}'}\otimes \mathbf{4}=\mathbf{\widehat{2}}\oplus \mathbf{\widehat{6}}\,,\quad \mathbf{\widehat{2}'}\otimes \mathbf{\widehat{4}'}=\mathbf{3'}\oplus \mathbf{5}\,,\quad \mathbf{\widehat{2}'}\otimes \mathbf{\widehat{6}}=\mathbf{3'}\otimes \mathbf{4}=\mathbf{3}\oplus \mathbf{4}\oplus \mathbf{5}\,,\\
\nonumber &&\mathbf{3}\otimes \mathbf{3}=\mathbf{1_s}\oplus \mathbf{3_a}\oplus \mathbf{5_s}\,,\quad \mathbf{3}\otimes \mathbf{3'}=\mathbf{4}\oplus \mathbf{5}\,,\quad \mathbf{3}\otimes \mathbf{\widehat{4}'}=\mathbf{\widehat{2}}\oplus \mathbf{\widehat{4}'}\oplus \mathbf{\widehat{6}}\,,\\
\nonumber && \mathbf{3}\otimes \mathbf{5}=\mathbf{3'}\otimes \mathbf{5}=\mathbf{3}\oplus \mathbf{3'}\oplus \mathbf{4}\oplus \mathbf{5}\,,\quad \mathbf{3}\otimes \mathbf{\widehat{6}}=\mathbf{\widehat{2}'}\oplus \mathbf{\widehat{4}'}\oplus \mathbf{\widehat{6}_1}\oplus \mathbf{\widehat{6}_2}\,,\\
\nonumber &&\mathbf{3'}\otimes \mathbf{3'}=\mathbf{1_s}\oplus \mathbf{3'_a}\oplus \mathbf{5_s}\,,\quad \mathbf{3'}\otimes \mathbf{\widehat{4}'}=\mathbf{\widehat{2}'}\oplus \mathbf{\widehat{4}'}\oplus \mathbf{\widehat{6}}\,,\quad \mathbf{3'}\otimes \mathbf{\widehat{6}}=\mathbf{\widehat{2}}\oplus \mathbf{\widehat{4}'}\oplus \mathbf{\widehat{6}_1}\oplus \mathbf{\widehat{6}_2}\,,\\
\nonumber &&\mathbf{4}\otimes \mathbf{4}=\mathbf{1_s}\oplus \mathbf{3_a}\oplus \mathbf{3'_a}\oplus \mathbf{4_s}\oplus \mathbf{5_s}\,,\quad \mathbf{4}\otimes \mathbf{\widehat{4}'}=\mathbf{\widehat{4}'}\oplus \mathbf{\widehat{6}_1}\oplus \mathbf{\widehat{6}_2}\,,\\
\nonumber && \mathbf{4}\otimes \mathbf{5}=\mathbf{3}\oplus \mathbf{3'}\oplus \mathbf{4}\oplus \mathbf{5_1}\oplus \mathbf{5_2}\,,\quad \mathbf{4}\otimes \mathbf{\widehat{6}}=\mathbf{\widehat{2}}\oplus \mathbf{\widehat{2}'}\oplus \mathbf{\widehat{4}'_1}\oplus \mathbf{\widehat{4}'_2}\oplus \mathbf{\widehat{6}_1}\oplus \mathbf{\widehat{6}_2}\,,\\
\nonumber &&\mathbf{\widehat{4}'}\otimes \mathbf{\widehat{4}'}=\mathbf{1_a}\oplus \mathbf{3_s}\oplus \mathbf{3'_s}\oplus \mathbf{4_s}\oplus \mathbf{5_a}\,,\quad \mathbf{\widehat{4}'}\otimes \mathbf{5}=\mathbf{\widehat{2}}\oplus \mathbf{\widehat{2}'}\oplus \mathbf{\widehat{4}'}\oplus \mathbf{\widehat{6}_1}\oplus \mathbf{\widehat{6}_2}\,,\\
\nonumber &&\mathbf{\widehat{4}'}\otimes \mathbf{\widehat{6}}=\mathbf{3}\oplus \mathbf{3'}\oplus \mathbf{4_1}\oplus \mathbf{4_2}\oplus \mathbf{5_1}\oplus \mathbf{5_2}\,,\\
\nonumber && \mathbf{5}\otimes \mathbf{5}=\mathbf{1_s}\oplus \mathbf{3_a}\oplus \mathbf{3'_a}\oplus \mathbf{4_s}\oplus \mathbf{4_a}\oplus \mathbf{5_{1,s}}\oplus \mathbf{5_{2,s}}\,,\\
\nonumber &&\mathbf{5}\otimes \mathbf{\widehat{6}}=\mathbf{\widehat{2}}\oplus \mathbf{\widehat{2}'}\oplus \mathbf{\widehat{4}'_1}\oplus \mathbf{\widehat{4}'_2}\oplus \mathbf{\widehat{6}_1}\oplus \mathbf{\widehat{6}_2}\oplus \mathbf{\widehat{6}_3}\,,\\
\label{eq:kronecker-Gamma'5}&& \mathbf{\widehat{6}}\otimes \mathbf{\widehat{6}}=\mathbf{1_a}\oplus \mathbf{3_{1,s}}\oplus \mathbf{3_{2,s}}\oplus \mathbf{3'_{1,s}}\oplus \mathbf{3'_{2,s}}\oplus \mathbf{4_s}\oplus \mathbf{4_a}\oplus \mathbf{5_{1,s}}\oplus \mathbf{5_{2,a}}\oplus \mathbf{5_{3,a}}\,,
\end{eqnarray}
where the subscripts $\mathbf{s}$ and $\mathbf{a}$ denote symmetric and antisymmetric combinations respectively. For the product decomposition $\mathbf{3}\otimes \mathbf{\widehat{6}}=\mathbf{\widehat{2}'}\oplus \mathbf{\widehat{4}'}\oplus \mathbf{\widehat{6}_1}\oplus \mathbf{\widehat{6}_2}$, $\mathbf{\widehat{6}_1}$ and $\mathbf{\widehat{6}_2}$ refer to the two sextet representations appearing in the tensor products of $\mathbf{3}$ and $\mathbf{\widehat{6}}$, and similar notations are adopted for other tensor products. We now list the Clebsch-Gordan coefficients which are quite useful in model construction. we use $\alpha_i$ to indicate the elements of the first representation of the product and $\beta_i$ to indicate those of the second representation.

\begin{eqnarray}
 \mathbf{\widehat{2}} \otimes \mathbf{\widehat{2}} = \mathbf{1_a} \oplus \mathbf{3_s} &\text{with}& \left\{
\begin{array}{l}
\mathbf{1_a} \sim\alpha_2\beta_1-\alpha_1\beta_2 \\ [0.1in]
\mathbf{3_s} \sim \begin{pmatrix}
-\alpha_2\beta_1-\alpha_1\beta_2 \\
\sqrt{2}\alpha_2\beta_2 \\
-\sqrt{2}\alpha_1\beta_1 \\
\end{pmatrix}
\end{array}
\right. \\
\mathbf{\widehat{2}} \otimes \mathbf{\widehat{2}'} &=&  \mathbf{4} \sim \begin{pmatrix}
\alpha_1\beta_2 \\
\alpha_2\beta_2 \\
-\alpha_1\beta_1 \\
\alpha_2\beta_1 \\
\end{pmatrix} \,, \\
\mathbf{\widehat{2}} \otimes \mathbf{3} = \mathbf{\widehat{2}} \oplus \mathbf{\widehat{4}'} &\text{with}& \left\{
\begin{array}{l}
\mathbf{\widehat{2}}\sim \begin{pmatrix}
\sqrt{2}\alpha_2\beta_3-\alpha_1\beta_1 \\
\alpha_2\beta_1+\sqrt{2}\alpha_1\beta_2 \\
\end{pmatrix}\\ [0.1in]
\mathbf{\widehat{4}'}\sim
 \begin{pmatrix}
-\sqrt{3}\alpha_1\beta_3 \\
\sqrt{2}\alpha_1\beta_1+\alpha_2\beta_3 \\
\alpha_1\beta_2-\sqrt{2}\alpha_2\beta_1 \\
\sqrt{3}\alpha_2\beta_2 \\
\end{pmatrix}
\end{array}
\right. \\
\mathbf{\widehat{2}} \otimes \mathbf{3'} &=&  \mathbf{\widehat{6}}
 \sim \begin{pmatrix}
-\alpha_1\beta_3 \\
-\alpha_2\beta_2 \\
\alpha_2\beta_3 \\
-\alpha_1\beta_1 \\
-\alpha_2\beta_1 \\
-\alpha_1\beta_2 \\
\end{pmatrix}\,, \\
\mathbf{\widehat{2}} \otimes \mathbf{4} = \mathbf{\widehat{2}'} \oplus \mathbf{\widehat{6}} &\text{with}& \left\{
\begin{array}{l}
\mathbf{\widehat{2}'} \sim \begin{pmatrix}
\alpha_2\beta_3+\alpha_1\beta_4 \\
\alpha_1\beta_2-\alpha_2\beta_1 \\
\end{pmatrix}\\ [0.1in]
\mathbf{\widehat{6}}\sim \begin{pmatrix}
\alpha_1\beta_3-\alpha_2\beta_2 \\
-\alpha_2\beta_2-\alpha_1\beta_3 \\
\alpha_2\beta_3-\alpha_1\beta_4 \\
\sqrt{2}\alpha_2\beta_4 \\
\sqrt{2}\alpha_1\beta_1 \\
\alpha_2\beta_1+\alpha_1\beta_2 \\
\end{pmatrix}
\end{array}
\right.  \\
\mathbf{\widehat{2}} \otimes \mathbf{\widehat{4}'} = \mathbf{3} \oplus \mathbf{5} &\text{with}& \left\{
\begin{array}{l}
\mathbf{3}\sim \begin{pmatrix}
\sqrt{2}\alpha_2\beta_2+\sqrt{2}\alpha_1\beta_3 \\
\alpha_2\beta_3-\sqrt{3}\alpha_1\beta_4 \\
-\sqrt{3}\alpha_2\beta_1-\alpha_1\beta_2 \\
\end{pmatrix}\\ [0.1in]
\mathbf{5} \sim \begin{pmatrix}
\sqrt{2}\alpha_1\beta_3-\sqrt{2}\alpha_2\beta_2 \\
\sqrt{3}\alpha_2\beta_3+\alpha_1\beta_4 \\
2\alpha_2\beta_4 \\
2\alpha_1\beta_1 \\
\sqrt{3}\alpha_1\beta_2-\alpha_2\beta_1 \\
\end{pmatrix}
\end{array}
\right.  \\
\mathbf{\widehat{2}} \otimes \mathbf{5} = \mathbf{\widehat{4}'} \oplus \mathbf{\widehat{6}} &\text{with}& \left\{
\begin{array}{l}
\mathbf{\widehat{4}'} \sim \begin{pmatrix}
2\alpha_2\beta_4+\alpha_1\beta_5 \\
\sqrt{2}\alpha_1\beta_1+\sqrt{3}\alpha_2\beta_5 \\
\sqrt{2}\alpha_2\beta_1-\sqrt{3}\alpha_1\beta_2 \\
\alpha_2\beta_2-2\alpha_1\beta_3 \\
\end{pmatrix}\\ [0.1in]
\mathbf{\widehat{6}} \sim \begin{pmatrix}
-2\alpha_2\beta_3-\alpha_1\beta_4 \\
\alpha_2\beta_3-2\alpha_1\beta_4 \\
2\alpha_1\beta_5-\alpha_2\beta_4 \\
\sqrt{2}\alpha_2\beta_5-\sqrt{3}\alpha_1\beta_1 \\
\sqrt{3}\alpha_2\beta_1+\sqrt{2}\alpha_1\beta_2 \\
-2\alpha_2\beta_2-\alpha_1\beta_3 \\
\end{pmatrix}
\end{array}
\right.  \\
\mathbf{\widehat{2}} \otimes \mathbf{\widehat{6}} = \mathbf{3'} \oplus \mathbf{4} \oplus \mathbf{5} &\text{with}& \left\{
\begin{array}{l}
\mathbf{3'} \sim\begin{pmatrix}
\alpha_2\beta_4-\alpha_1\beta_5 \\
\alpha_2\beta_6-\alpha_1\beta_2 \\
\alpha_2\beta_1+\alpha_1\beta_3 \\
\end{pmatrix}\\ [0.1in]
\mathbf{4} \sim\begin{pmatrix}
\sqrt{2}\alpha_2\beta_5-\alpha_1\beta_6 \\
\alpha_1\beta_1+\alpha_1\beta_2+\alpha_2\beta_6 \\
\alpha_2\beta_1-\alpha_2\beta_2-\alpha_1\beta_3 \\
-\alpha_2\beta_3-\sqrt{2}\alpha_1\beta_4 \\
\end{pmatrix}\\ [0.1in]
\mathbf{5}\sim\begin{pmatrix}
\sqrt{3}\alpha_2\beta_4+\sqrt{3}\alpha_1\beta_5 \\
-\sqrt{2}\alpha_2\beta_5-2\alpha_1\beta_6 \\
\alpha_1\beta_2+\alpha_2\beta_6-2\alpha_1\beta_1 \\
\alpha_2\beta_1+2\alpha_2\beta_2-\alpha_1\beta_3 \\
\sqrt{2}\alpha_1\beta_4-2\alpha_2\beta_3 \\
\end{pmatrix}
\end{array}
\right.  \\
\mathbf{\widehat{2}'} \otimes \mathbf{\widehat{2}'} = \mathbf{1_a} \oplus \mathbf{3'_s} &\text{with}& \left\{
\begin{array}{l}
\mathbf{1_a}\sim\alpha_2\beta_1-\alpha_1\beta_2\\ [0.1in]
\mathbf{3'_s}\sim \begin{pmatrix}
\alpha_2\beta_1+\alpha_1\beta_2 \\
-\sqrt{2}\alpha_1\beta_1 \\
\sqrt{2}\alpha_2\beta_2 \\
\end{pmatrix}
\end{array}
\right.  \\
\mathbf{\widehat{2}'} \otimes \mathbf{3} &=& \mathbf{\widehat{6}}  \sim \begin{pmatrix}
\alpha_2\beta_2-\alpha_1\beta_3 \\
\alpha_2\beta_2+\alpha_1\beta_3 \\
-\sqrt{2}\alpha_1\beta_1 \\
-\sqrt{2}\alpha_1\beta_2 \\
\sqrt{2}\alpha_2\beta_3 \\
\sqrt{2}\alpha_2\beta_1 \\
\end{pmatrix} \,, \\
\mathbf{\widehat{2}'} \otimes \mathbf{3'} = \mathbf{\widehat{2}'} \oplus \mathbf{\widehat{4}'} &\text{with}& \left\{
\begin{array}{l}
\mathbf{\widehat{2}'} \sim\begin{pmatrix}
\alpha_1\beta_1+\sqrt{2}\alpha_2\beta_2 \\
\sqrt{2}\alpha_1\beta_3-\alpha_2\beta_1 \\
\end{pmatrix}\\ [0.1in]
\mathbf{\widehat{4}'}\sim\begin{pmatrix}
\sqrt{2}\alpha_1\beta_1-\alpha_2\beta_2 \\
\sqrt{3}\alpha_2\beta_3 \\
-\sqrt{3}\alpha_1\beta_2 \\
\sqrt{2}\alpha_2\beta_1+\alpha_1\beta_3 \\
\end{pmatrix}
\end{array}
\right.  \\
\mathbf{\widehat{2}'} \otimes \mathbf{4} = \mathbf{\widehat{2}} \oplus \mathbf{\widehat{6}} &\text{with}& \left\{
\begin{array}{l}
\mathbf{\widehat{2}} \sim \begin{pmatrix}
-\alpha_1\beta_1-\alpha_2\beta_3 \\
\alpha_2\beta_4-\alpha_1\beta_2 \\
\end{pmatrix}\\ [0.1in]
\mathbf{\widehat{6}}\sim  \begin{pmatrix}
\sqrt{2}\alpha_2\beta_1 \\
-\sqrt{2}\alpha_1\beta_4 \\
-\sqrt{2}\alpha_2\beta_2 \\
\alpha_1\beta_1-\alpha_2\beta_3 \\
\alpha_1\beta_2+\alpha_2\beta_4 \\
\sqrt{2}\alpha_1\beta_3 \\
\end{pmatrix}
\end{array}
\right.  \\
\mathbf{\widehat{2}'} \otimes \mathbf{\widehat{4}'} = \mathbf{3'} \oplus \mathbf{5} &\text{with}& \left\{
\begin{array}{l}
\mathbf{3'} \sim \begin{pmatrix}
\sqrt{2}\alpha_2\beta_1-\sqrt{2}\alpha_1\beta_4 \\
\alpha_1\beta_1-\sqrt{3}\alpha_2\beta_3 \\
\alpha_2\beta_4-\sqrt{3}\alpha_1\beta_2 \\
\end{pmatrix}\\ [0.1in]
\mathbf{5} \sim\begin{pmatrix}
-\sqrt{2}\alpha_2\beta_1-\sqrt{2}\alpha_1\beta_4 \\
-2\alpha_2\beta_2 \\
-\sqrt{3}\alpha_1\beta_1-\alpha_2\beta_3 \\
\alpha_1\beta_2+\sqrt{3}\alpha_2\beta_4 \\
-2\alpha_1\beta_3 \\
\end{pmatrix}
\end{array}
\right.  \\
\mathbf{\widehat{2}'} \otimes \mathbf{5} = \mathbf{\widehat{4}'} \oplus \mathbf{\widehat{6}} &\text{with}& \left\{
\begin{array}{l}
\mathbf{\widehat{4}'} \sim \begin{pmatrix}
\sqrt{3}\alpha_2\beta_3-\sqrt{2}\alpha_1\beta_1 \\
-2\alpha_1\beta_2-\alpha_2\beta_4 \\
2\alpha_2\beta_5-\alpha_1\beta_3 \\
\sqrt{2}\alpha_2\beta_1+\sqrt{3}\alpha_1\beta_4 \\
\end{pmatrix} \\ [0.1in]
\mathbf{\widehat{6}}\sim \begin{pmatrix}
3\alpha_1\beta_5-\alpha_2\beta_2 \\
3\alpha_2\beta_2+\alpha_1\beta_5 \\
\sqrt{6}\alpha_1\beta_1+2\alpha_2\beta_3 \\
\sqrt{2}\alpha_1\beta_2-2\sqrt{2}\alpha_2\beta_4 \\
2\sqrt{2}\alpha_1\beta_3+\sqrt{2}\alpha_2\beta_5 \\
\sqrt{6}\alpha_2\beta_1-2\alpha_1\beta_4 \\
\end{pmatrix}
\end{array}
\right.  \\
\mathbf{\widehat{2}'} \otimes \mathbf{\widehat{6}} = \mathbf{3} \oplus \mathbf{4} \oplus \mathbf{5} &\text{with}& \left\{
\begin{array}{l}
\mathbf{3} \sim \begin{pmatrix}
\sqrt{2}\alpha_2\beta_3+\sqrt{2}\alpha_1\beta_6 \\
\alpha_1\beta_1+\alpha_1\beta_2+\sqrt{2}\alpha_2\beta_4 \\
\alpha_2\beta_1+\sqrt{2}\alpha_1\beta_5-\alpha_2\beta_2 \\
\end{pmatrix} \\ [0.1in]
\mathbf{4} \sim \begin{pmatrix}
\alpha_2\beta_4-\sqrt{2}\alpha_1\beta_1 \\
\sqrt{2}\alpha_1\beta_3+\alpha_2\beta_5 \\
\alpha_1\beta_4+\sqrt{2}\alpha_2\beta_6 \\
-\sqrt{2}\alpha_2\beta_2-\alpha_1\beta_5 \\
\end{pmatrix}\\ [0.1in]
\mathbf{5} \sim\begin{pmatrix}
\sqrt{6}\alpha_1\beta_6-\sqrt{6}\alpha_2\beta_3 \\
3\alpha_1\beta_2-\alpha_1\beta_1-\sqrt{2}\alpha_2\beta_4 \\
2\alpha_1\beta_3-2\sqrt{2}\alpha_2\beta_5 \\
2\alpha_2\beta_6-2\sqrt{2}\alpha_1\beta_4 \\
\sqrt{2}\alpha_1\beta_5-3\alpha_2\beta_1-\alpha_2\beta_2 \\
\end{pmatrix}
\end{array}
\right.  \\
\mathbf{3} \otimes \mathbf{3} = \mathbf{1_s} \oplus \mathbf{3_a} \oplus \mathbf{5_s} &\text{with}& \left\{
\begin{array}{l}
\mathbf{1_s}\sim\alpha_1\beta_1+\alpha_3\beta_2+\alpha_2\beta_3 \\[0.1in]
\mathbf{3_a} \sim \begin{pmatrix}
\alpha_2\beta_3-\alpha_3\beta_2 \\
\alpha_1\beta_2-\alpha_2\beta_1 \\
\alpha_3\beta_1-\alpha_1\beta_3 \\
\end{pmatrix}\\[0.1in]
\mathbf{5_s} \sim  \begin{pmatrix}
 2\alpha_1\beta_1-\alpha_3\beta_2-\alpha_2\beta_3 \\
 -\sqrt{3}\alpha_2\beta_1-\sqrt{3}\alpha_1\beta_2 \\
\sqrt{6}\alpha_2\beta_2 \\
\sqrt{6}\alpha_3\beta_3 \\
 -\sqrt{3}\alpha_3\beta_1-\sqrt{3}\alpha_1\beta_3 \\
\end{pmatrix}
\end{array}
\right.  \\
\mathbf{3} \otimes \mathbf{3'} = \mathbf{4} \oplus \mathbf{5} &\text{with}& \left\{
\begin{array}{l}
\mathbf{4}\sim \begin{pmatrix}
\sqrt{2}\alpha_2\beta_1+\alpha_3\beta_2 \\
 -\sqrt{2}\alpha_1\beta_2-\alpha_3\beta_3 \\
 -\alpha_2\beta_2-\sqrt{2}\alpha_1\beta_3 \\
\sqrt{2}\alpha_3\beta_1+\alpha_2\beta_3 \\
\end{pmatrix} \\ [0.1in]
\mathbf{5} \sim\begin{pmatrix}
\sqrt{3}\alpha_1\beta_1 \\
\alpha_2\beta_1-\sqrt{2}\alpha_3\beta_2 \\
\alpha_1\beta_2-\sqrt{2}\alpha_3\beta_3 \\
\alpha_1\beta_3-\sqrt{2}\alpha_2\beta_2 \\
\alpha_3\beta_1-\sqrt{2}\alpha_2\beta_3 \\
\end{pmatrix}
\end{array}
\right.  \\
\mathbf{3} \otimes \mathbf{4} = \mathbf{3'} \oplus \mathbf{4} \oplus \mathbf{5} &\text{with}& \left\{
\begin{array}{l}
\mathbf{3'} \sim \begin{pmatrix}
 -\sqrt{2}\alpha_3\beta_1-\sqrt{2}\alpha_2\beta_4 \\
\alpha_3\beta_3+\sqrt{2}\alpha_1\beta_2-\alpha_2\beta_1 \\
\alpha_2\beta_2+\sqrt{2}\alpha_1\beta_3-\alpha_3\beta_4 \\
\end{pmatrix} \\ [0.1in]
\mathbf{4} \sim \begin{pmatrix}
\alpha_1\beta_1-\sqrt{2}\alpha_3\beta_2 \\
 -\sqrt{2}\alpha_2\beta_1-\alpha_1\beta_2 \\
\alpha_1\beta_3+\sqrt{2}\alpha_3\beta_4 \\
\sqrt{2}\alpha_2\beta_3-\alpha_1\beta_4 \\
\end{pmatrix}\\ [0.1in]
\mathbf{5} \sim \begin{pmatrix}
\sqrt{6}\alpha_2\beta_4-\sqrt{6}\alpha_3\beta_1 \\
 2\sqrt{2}\alpha_1\beta_1+2\alpha_3\beta_2 \\
 3\alpha_3\beta_3+\alpha_2\beta_1-\sqrt{2}\alpha_1\beta_2 \\
\sqrt{2}\alpha_1\beta_3-3\alpha_2\beta_2-\alpha_3\beta_4 \\
 -2\alpha_2\beta_3-2\sqrt{2}\alpha_1\beta_4 \\
\end{pmatrix}
\end{array}
\right.  \\
\mathbf{3} \otimes \mathbf{\widehat{4}'} = \mathbf{\widehat{2}} \oplus \mathbf{\widehat{4}'} \oplus \mathbf{\widehat{6}} &\text{with}& \left\{
\begin{array}{l}
\mathbf{\widehat{2}} \sim \begin{pmatrix}
\alpha_3\beta_3+\sqrt{2}\alpha_1\beta_2-\sqrt{3}\alpha_2\beta_1 \\
\alpha_2\beta_2+\sqrt{3}\alpha_3\beta_4-\sqrt{2}\alpha_1\beta_3 \\
\end{pmatrix} \\ [0.1in]
\mathbf{\widehat{4}'} \sim \begin{pmatrix}
-3\alpha_1\beta_1-\sqrt{6}\alpha_3\beta_2 \\
-2\sqrt{2}\alpha_3\beta_3-\sqrt{6}\alpha_2\beta_1-\alpha_1\beta_2 \\
\alpha_1\beta_3+\sqrt{6}\alpha_3\beta_4-2\sqrt{2}\alpha_2\beta_2 \\
\sqrt{6}\alpha_2\beta_3+3\alpha_1\beta_4 \\
\end{pmatrix}\\ [0.1in]
\mathbf{\widehat{6}} \sim \begin{pmatrix}
2\sqrt{2}\alpha_2\beta_4-\sqrt{2}\alpha_3\beta_1 \\
-2\sqrt{2}\alpha_3\beta_1-\sqrt{2}\alpha_2\beta_4 \\
\sqrt{6}\alpha_3\beta_2-2\alpha_1\beta_1 \\
\alpha_2\beta_1+\sqrt{6}\alpha_1\beta_2-\sqrt{3}\alpha_3\beta_3 \\
\sqrt{3}\alpha_2\beta_2+\sqrt{6}\alpha_1\beta_3+\alpha_3\beta_4 \\
\sqrt{6}\alpha_2\beta_3-2\alpha_1\beta_4 \\
\end{pmatrix}
\end{array}
\right.  \\
\left.\begin{array}{c}
\mathbf{3} \otimes \mathbf{5} = \mathbf{3} \oplus \mathbf{3'} \oplus \mathbf{4} \\
\qquad \oplus \mathbf{5}
\end{array}\right. &\text{with}& \left\{
\begin{array}{l}
\mathbf{3}\sim  \begin{pmatrix}
\sqrt{3}\alpha_3\beta_2+\sqrt{3}\alpha_2\beta_5-2\alpha_1\beta_1 \\
\alpha_2\beta_1+\sqrt{3}\alpha_1\beta_2-\sqrt{6}\alpha_3\beta_3 \\
\alpha_3\beta_1+\sqrt{3}\alpha_1\beta_5-\sqrt{6}\alpha_2\beta_4 \\
\end{pmatrix} \\ [0.1in]
\mathbf{3'} \sim \begin{pmatrix}
\sqrt{3}\alpha_1\beta_1+\alpha_3\beta_2+\alpha_2\beta_5 \\
\alpha_1\beta_3-\sqrt{2}\alpha_2\beta_2-\sqrt{2}\alpha_3\beta_4 \\
\alpha_1\beta_4-\sqrt{2}\alpha_2\beta_3-\sqrt{2}\alpha_3\beta_5 \\
\end{pmatrix} \\ [0.1in]
\mathbf{4}\sim \begin{pmatrix}
\alpha_3\beta_3+2\sqrt{2}\alpha_1\beta_2-\sqrt{6}\alpha_2\beta_1 \\
 2\alpha_2\beta_2-\sqrt{2}\alpha_1\beta_3-3\alpha_3\beta_4 \\
 3\alpha_2\beta_3+\sqrt{2}\alpha_1\beta_4-2\alpha_3\beta_5 \\
\sqrt{6}\alpha_3\beta_1-\alpha_2\beta_4-2\sqrt{2}\alpha_1\beta_5 \\
\end{pmatrix}  \\ [0.1in]
\mathbf{5} \sim\begin{pmatrix}
\sqrt{3}\alpha_2\beta_5-\sqrt{3}\alpha_3\beta_2 \\
 -\sqrt{2}\alpha_3\beta_3-\sqrt{3}\alpha_2\beta_1-\alpha_1\beta_2 \\
 -\sqrt{2}\alpha_2\beta_2-2\alpha_1\beta_3 \\
 2\alpha_1\beta_4+\sqrt{2}\alpha_3\beta_5 \\
\sqrt{3}\alpha_3\beta_1+\sqrt{2}\alpha_2\beta_4+\alpha_1\beta_5 \\
\end{pmatrix}
\end{array}
\right.  \\
\left.\begin{array}{c}
\mathbf{3} \otimes \mathbf{\widehat{6}} = \mathbf{\widehat{2}'} \oplus \mathbf{\widehat{4}'} \oplus \mathbf{\widehat{6}_1} \\
\qquad \oplus \mathbf{\widehat{6}_2}
\end{array}\right. &\text{with}& \left\{
\begin{array}{l}
\mathbf{\widehat{2}'} \sim \begin{pmatrix}
\alpha_2\beta_1+\sqrt{2}\alpha_1\beta_3+\sqrt{2}\alpha_3\beta_4-\alpha_2\beta_2 \\
 -\alpha_3\beta_1-\alpha_3\beta_2-\sqrt{2}\alpha_2\beta_5-\sqrt{2}\alpha_1\beta_6 \\
\end{pmatrix}\\ [0.1in]
\mathbf{\widehat{4}'} \sim \begin{pmatrix}
\alpha_3\beta_4-\sqrt{2}\alpha_2\beta_1-2\sqrt{2}\alpha_2\beta_2-2\alpha_1\beta_3 \\
\sqrt{6}\alpha_2\beta_3+\sqrt{6}\alpha_1\beta_4+\sqrt{3}\alpha_3\beta_5 \\
\sqrt{6}\alpha_1\beta_5+\sqrt{6}\alpha_3\beta_6-\sqrt{3}\alpha_2\beta_4 \\
 2\sqrt{2}\alpha_3\beta_1+\alpha_2\beta_5-\sqrt{2}\alpha_3\beta_2-2\alpha_1\beta_6 \\
\end{pmatrix}\\ [0.1in]
\mathbf{\widehat{6}_1} \sim \begin{pmatrix}
 -\alpha_1\beta_1-\sqrt{2}\alpha_3\beta_3 \\
\alpha_1\beta_2+\sqrt{2}\alpha_2\beta_6 \\
\alpha_1\beta_3-\sqrt{2}\alpha_2\beta_1 \\
\sqrt{2}\alpha_3\beta_5-\alpha_1\beta_4 \\
\sqrt{2}\alpha_2\beta_4+\alpha_1\beta_5 \\
\sqrt{2}\alpha_3\beta_2-\alpha_1\beta_6 \\
\end{pmatrix}\\ [0.1in]
\mathbf{\widehat{6}_2} \sim \begin{pmatrix}
\sqrt{2}\alpha_1\beta_2+\alpha_3\beta_3-\alpha_2\beta_6 \\
\sqrt{2}\alpha_1\beta_1-\alpha_3\beta_3-\alpha_2\beta_6 \\
\alpha_2\beta_1-\alpha_2\beta_2-\sqrt{2}\alpha_3\beta_4 \\
\sqrt{2}\alpha_1\beta_4-\sqrt{2}\alpha_2\beta_3 \\
\sqrt{2}\alpha_3\beta_6-\sqrt{2}\alpha_1\beta_5 \\
\sqrt{2}\alpha_2\beta_5-\alpha_3\beta_1-\alpha_3\beta_2 \\
\end{pmatrix}
\end{array}
\right.  \\
\mathbf{3'} \otimes \mathbf{3'} = \mathbf{1_s} \oplus \mathbf{3'_a} \oplus \mathbf{5_s} &\text{with}& \left\{
\begin{array}{l}
\mathbf{1_s} \sim\alpha_1\beta_1+\alpha_3\beta_2+\alpha_2\beta_3\\[0.1in]
\mathbf{3'_a} \sim \begin{pmatrix}
\alpha_2\beta_3-\alpha_3\beta_2 \\
\alpha_1\beta_2-\alpha_2\beta_1 \\
\alpha_3\beta_1-\alpha_1\beta_3 \\
\end{pmatrix}\\[0.1in]
\mathbf{5_s}\sim \begin{pmatrix}
 2\alpha_1\beta_1-\alpha_3\beta_2-\alpha_2\beta_3 \\
\sqrt{6}\alpha_3\beta_3 \\
 -\sqrt{3}\alpha_2\beta_1-\sqrt{3}\alpha_1\beta_2 \\
 -\sqrt{3}\alpha_3\beta_1-\sqrt{3}\alpha_1\beta_3 \\
\sqrt{6}\alpha_2\beta_2 \\
\end{pmatrix}
\end{array}
\right.  \\
\mathbf{3'} \otimes \mathbf{4} = \mathbf{3} \oplus \mathbf{4} \oplus \mathbf{5} &\text{with}& \left\{
\begin{array}{l}
\mathbf{3} \sim \begin{pmatrix}
 -\sqrt{2}\alpha_3\beta_2-\sqrt{2}\alpha_2\beta_3 \\
\sqrt{2}\alpha_1\beta_1+\alpha_2\beta_4-\alpha_3\beta_3 \\
\alpha_3\beta_1+\sqrt{2}\alpha_1\beta_4-\alpha_2\beta_2 \\
\end{pmatrix}\\ [0.1in]
\mathbf{4}\sim \begin{pmatrix}
\alpha_1\beta_1+\sqrt{2}\alpha_3\beta_3 \\
\alpha_1\beta_2-\sqrt{2}\alpha_3\beta_4 \\
\sqrt{2}\alpha_2\beta_1-\alpha_1\beta_3 \\
 -\sqrt{2}\alpha_2\beta_2-\alpha_1\beta_4 \\
\end{pmatrix}\\ [0.1in]
\mathbf{5}\sim \begin{pmatrix}
\sqrt{6}\alpha_2\beta_3-\sqrt{6}\alpha_3\beta_2 \\
\sqrt{2}\alpha_1\beta_1-\alpha_3\beta_3-3\alpha_2\beta_4 \\
 2\sqrt{2}\alpha_1\beta_2+2\alpha_3\beta_4 \\
 -2\alpha_2\beta_1-2\sqrt{2}\alpha_1\beta_3 \\
 3\alpha_3\beta_1+\alpha_2\beta_2-\sqrt{2}\alpha_1\beta_4 \\
\end{pmatrix}
\end{array}
\right.  \\
\mathbf{3'} \otimes \mathbf{\widehat{4}'} = \mathbf{\widehat{2}'} \oplus \mathbf{\widehat{4}'} \oplus \mathbf{\widehat{6}} &\text{with}& \left\{
\begin{array}{l}
\mathbf{\widehat{2}'}\sim \begin{pmatrix}
\sqrt{3}\alpha_3\beta_3-\sqrt{2}\alpha_1\beta_1-\alpha_2\beta_4 \\
\alpha_3\beta_1-\sqrt{3}\alpha_2\beta_2-\sqrt{2}\alpha_1\beta_4 \\
\end{pmatrix}\\ [0.1in]
\mathbf{\widehat{4}'}\sim \begin{pmatrix}
\alpha_1\beta_1+\sqrt{6}\alpha_3\beta_3+2\sqrt{2}\alpha_2\beta_4 \\
\sqrt{6}\alpha_3\beta_4-3\alpha_1\beta_2 \\
\sqrt{6}\alpha_2\beta_1+3\alpha_1\beta_3 \\
2\sqrt{2}\alpha_3\beta_1+\sqrt{6}\alpha_2\beta_2-\alpha_1\beta_4 \\
\end{pmatrix}\\ [0.1in]
\mathbf{\widehat{6}} \sim \begin{pmatrix}
-\alpha_3\beta_2-3\alpha_2\beta_3 \\
3\alpha_3\beta_2-\alpha_2\beta_3 \\
\sqrt{6}\alpha_1\beta_1+\alpha_3\beta_3-\sqrt{3}\alpha_2\beta_4 \\
2\alpha_1\beta_2+\sqrt{6}\alpha_3\beta_4 \\
2\alpha_1\beta_3-\sqrt{6}\alpha_2\beta_1 \\
\sqrt{3}\alpha_3\beta_1+\sqrt{6}\alpha_1\beta_4-\alpha_2\beta_2 \\
\end{pmatrix}
\end{array}
\right.  \\
\left.\begin{array}{c}
\mathbf{3'} \otimes \mathbf{5} = \mathbf{3} \oplus \mathbf{3'} \oplus \mathbf{4} \\
\qquad \oplus \mathbf{5}
\end{array}\right. &\text{with}& \left\{
\begin{array}{l}
\mathbf{3}\sim \begin{pmatrix}
\sqrt{3}\alpha_1\beta_1+\alpha_3\beta_3+\alpha_2\beta_4 \\
\alpha_1\beta_2-\sqrt{2}\alpha_3\beta_4-\sqrt{2}\alpha_2\beta_5 \\
\alpha_1\beta_5-\sqrt{2}\alpha_3\beta_2-\sqrt{2}\alpha_2\beta_3 \\
\end{pmatrix}\\ [0.1in]
\mathbf{3'} \sim \begin{pmatrix}
\sqrt{3}\alpha_3\beta_3+\sqrt{3}\alpha_2\beta_4-2\alpha_1\beta_1 \\
\alpha_2\beta_1+\sqrt{3}\alpha_1\beta_3-\sqrt{6}\alpha_3\beta_5 \\
\alpha_3\beta_1+\sqrt{3}\alpha_1\beta_4-\sqrt{6}\alpha_2\beta_2 \\
\end{pmatrix}\\ [0.1in]
\mathbf{4} \sim \begin{pmatrix}
\sqrt{2}\alpha_1\beta_2+3\alpha_2\beta_5-2\alpha_3\beta_4 \\
 2\sqrt{2}\alpha_1\beta_3+\alpha_3\beta_5-\sqrt{6}\alpha_2\beta_1 \\
\sqrt{6}\alpha_3\beta_1-\alpha_2\beta_2-2\sqrt{2}\alpha_1\beta_4 \\
 2\alpha_2\beta_3-3\alpha_3\beta_2-\sqrt{2}\alpha_1\beta_5 \\
\end{pmatrix}\\ [0.1in]
\mathbf{5} \sim \begin{pmatrix}
\sqrt{3}\alpha_2\beta_4-\sqrt{3}\alpha_3\beta_3 \\
 2\alpha_1\beta_2+\sqrt{2}\alpha_3\beta_4 \\
 -\sqrt{3}\alpha_2\beta_1-\alpha_1\beta_3-\sqrt{2}\alpha_3\beta_5 \\
\sqrt{3}\alpha_3\beta_1+\sqrt{2}\alpha_2\beta_2+\alpha_1\beta_4 \\
 -\sqrt{2}\alpha_2\beta_3-2\alpha_1\beta_5 \\
\end{pmatrix}
\end{array}
\right.  \\
\left.\begin{array}{c}
\mathbf{3'} \otimes \mathbf{\widehat{6}} = \mathbf{\widehat{2}} \oplus \mathbf{\widehat{4}'} \oplus \mathbf{\widehat{6}_1}\\
\qquad \oplus \mathbf{\widehat{6}_2}
\end{array}\right.
 &\text{with}& \left\{
\begin{array}{l}
\mathbf{\widehat{2}}\sim \begin{pmatrix}
\alpha_2\beta_1+\alpha_1\beta_4+\alpha_3\beta_6 \\
\alpha_1\beta_5+\alpha_3\beta_2-\alpha_2\beta_3 \\
\end{pmatrix} \\ [0.1in]
\mathbf{\widehat{4}'} \sim \begin{pmatrix}
\sqrt{6}\alpha_3\beta_5-\sqrt{6}\alpha_1\beta_3-\sqrt{3}\alpha_2\beta_6 \\
\alpha_2\beta_1+\alpha_3\beta_6-3\alpha_2\beta_2-2\alpha_1\beta_4 \\
3\alpha_3\beta_1+\alpha_3\beta_2-\alpha_2\beta_3-2\alpha_1\beta_5 \\
\sqrt{3}\alpha_3\beta_3-\sqrt{6}\alpha_2\beta_4-\sqrt{6}\alpha_1\beta_6 \\
\end{pmatrix}\\ [0.1in]
\mathbf{\widehat{6}_1} \sim \begin{pmatrix}
\alpha_1\beta_1-\alpha_3\beta_4 \\
\alpha_2\beta_5-\alpha_1\beta_2 \\
\alpha_1\beta_3+\alpha_3\beta_5 \\
\alpha_3\beta_6-\alpha_2\beta_1 \\
\alpha_3\beta_2+\alpha_2\beta_3 \\
\alpha_2\beta_4-\alpha_1\beta_6 \\
\end{pmatrix} \\ [0.1in]
\mathbf{\widehat{6}_2} \sim \begin{pmatrix}
\alpha_1\beta_2+\alpha_3\beta_4+\alpha_2\beta_5 \\
\alpha_1\beta_1+\alpha_3\beta_4-\alpha_2\beta_5 \\
\sqrt{2}\alpha_2\beta_6-\alpha_1\beta_3 \\
\alpha_2\beta_1+\alpha_2\beta_2-\alpha_1\beta_4 \\
\alpha_3\beta_1+\alpha_1\beta_5-\alpha_3\beta_2 \\
\sqrt{2}\alpha_3\beta_3+\alpha_1\beta_6 \\
\end{pmatrix}
\end{array}
\right.  \\
\left.\begin{array}{c}
\mathbf{4} \otimes \mathbf{4} = \mathbf{1_s} \oplus \mathbf{3_a} \oplus \mathbf{3'_a}  \\
\qquad \oplus \mathbf{4_s}\oplus \mathbf{5_s}
\end{array}\right. &\text{with}& \left\{
\begin{array}{l}
\mathbf{1_s} \sim\alpha_4\beta_1+\alpha_1\beta_4+\alpha_3\beta_2+\alpha_2\beta_3\\ [0.1in]
\mathbf{3_a} \sim \begin{pmatrix}
\alpha_4\beta_1-\alpha_1\beta_4+\alpha_2\beta_3-\alpha_3\beta_2 \\
\sqrt{2}\alpha_2\beta_4-\sqrt{2}\alpha_4\beta_2 \\
\sqrt{2}\alpha_1\beta_3-\sqrt{2}\alpha_3\beta_1 \\
\end{pmatrix}\\ [0.1in]
\mathbf{3'_a} \sim \begin{pmatrix}
\alpha_2\beta_3-\alpha_3\beta_2+\alpha_1\beta_4-\alpha_4\beta_1 \\
\sqrt{2}\alpha_3\beta_4-\sqrt{2}\alpha_4\beta_3 \\
\sqrt{2}\alpha_1\beta_2-\sqrt{2}\alpha_2\beta_1 \\
\end{pmatrix}\\ [0.1in]
\mathbf{4_s}\sim \begin{pmatrix}
\alpha_3\beta_3+\alpha_4\beta_2+\alpha_2\beta_4 \\
\alpha_1\beta_1+\alpha_4\beta_3+\alpha_3\beta_4 \\
\alpha_4\beta_4+\alpha_2\beta_1+\alpha_1\beta_2 \\
\alpha_2\beta_2+\alpha_3\beta_1+\alpha_1\beta_3 \\
\end{pmatrix}\\ [0.1in]
\mathbf{5_s}\sim\begin{pmatrix}
\sqrt{3}\alpha_4\beta_1+\sqrt{3}\alpha_1\beta_4-\sqrt{3}\alpha_3\beta_2-\sqrt{3}\alpha_2\beta_3 \\
 2\sqrt{2}\alpha_3\beta_3-\sqrt{2}\alpha_4\beta_2-\sqrt{2}\alpha_2\beta_4 \\
\sqrt{2}\alpha_4\beta_3+\sqrt{2}\alpha_3\beta_4-2\sqrt{2}\alpha_1\beta_1 \\
\sqrt{2}\alpha_2\beta_1+\sqrt{2}\alpha_1\beta_2-2\sqrt{2}\alpha_4\beta_4 \\
 2\sqrt{2}\alpha_2\beta_2-\sqrt{2}\alpha_3\beta_1-\sqrt{2}\alpha_1\beta_3 \\
\end{pmatrix}
\end{array}
\right.  \\
\mathbf{4} \otimes \mathbf{\widehat{4}'} = \mathbf{\widehat{4}'} \oplus \mathbf{\widehat{6}_1} \oplus \mathbf{\widehat{6}_2} &\text{with}& \left\{
\begin{array}{l}
\mathbf{\widehat{4}'} \sim \begin{pmatrix}
-\alpha_3\beta_3-\alpha_4\beta_2-\sqrt{3}\alpha_2\beta_4 \\
\alpha_1\beta_1+\sqrt{3}\alpha_4\beta_3-\alpha_3\beta_4 \\
\alpha_4\beta_4+\alpha_2\beta_1-\sqrt{3}\alpha_1\beta_2 \\
\alpha_2\beta_2+\sqrt{3}\alpha_3\beta_1-\alpha_1\beta_3 \\
\end{pmatrix} \\ [0.1in]
\mathbf{\widehat{6}_1} \sim \begin{pmatrix}
-\sqrt{3}\alpha_3\beta_2-\alpha_1\beta_4 \\
\sqrt{3}\alpha_2\beta_3-\alpha_4\beta_1 \\
\alpha_2\beta_4-\sqrt{3}\alpha_4\beta_2 \\
-\sqrt{2}\alpha_1\beta_1-\sqrt{2}\alpha_3\beta_4 \\
\sqrt{2}\alpha_4\beta_4-\sqrt{2}\alpha_2\beta_1 \\
\alpha_3\beta_1+\sqrt{3}\alpha_1\beta_3 \\
\end{pmatrix}\\ [0.1in]
\mathbf{\widehat{6}_2} \sim \begin{pmatrix}
-\sqrt{3}\alpha_4\beta_1-2\alpha_3\beta_2-\alpha_2\beta_3 \\
\sqrt{3}\alpha_1\beta_4+2\alpha_2\beta_3-\alpha_3\beta_2 \\
\sqrt{3}\alpha_2\beta_4-\alpha_4\beta_2-2\alpha_3\beta_3 \\
\sqrt{2}\alpha_4\beta_3-\sqrt{6}\alpha_1\beta_1 \\
\sqrt{2}\alpha_1\beta_2+\sqrt{6}\alpha_4\beta_4 \\
\sqrt{3}\alpha_3\beta_1+\alpha_1\beta_3-2\alpha_2\beta_2 \\
\end{pmatrix}
\end{array}
\right.  \\
\left.\begin{array}{c}
\mathbf{4} \otimes \mathbf{5} = \mathbf{3} \oplus \mathbf{3'} \oplus \mathbf{4}\\
\qquad \oplus \mathbf{5_1} \oplus \mathbf{5_2}
\end{array}\right. &\text{with}& \left\{
\begin{array}{l}
\mathbf{3} \sim \begin{pmatrix}
\sqrt{2}\alpha_3\beta_3+2\sqrt{2}\alpha_1\beta_5-2\sqrt{2}\alpha_4\beta_2-\sqrt{2}\alpha_2\beta_4 \\
2\alpha_2\beta_5+3\alpha_3\beta_4-\alpha_4\beta_3-\sqrt{6}\alpha_1\beta_1 \\
\sqrt{6}\alpha_4\beta_1+\alpha_1\beta_4-2\alpha_3\beta_2-3\alpha_2\beta_3 \\
\end{pmatrix} \\ [0.1in]
\mathbf{3'} \sim \begin{pmatrix}
\sqrt{2}\alpha_1\beta_5+2\sqrt{2}\alpha_2\beta_4-\sqrt{2}\alpha_4\beta_2-2\sqrt{2}\alpha_3\beta_3 \\
2\alpha_4\beta_4+3\alpha_1\beta_2-\sqrt{6}\alpha_2\beta_1-\alpha_3\beta_5 \\
\alpha_2\beta_2+\sqrt{6}\alpha_3\beta_1-2\alpha_1\beta_3-3\alpha_4\beta_5 \\
\end{pmatrix}\\ [0.1in]
\mathbf{4} \sim \begin{pmatrix}
\sqrt{3}\alpha_1\beta_1+\sqrt{2}\alpha_3\beta_4-2\sqrt{2}\alpha_4\beta_3-\sqrt{2}\alpha_2\beta_5 \\
\sqrt{2}\alpha_4\beta_4+2\sqrt{2}\alpha_3\beta_5-\sqrt{3}\alpha_2\beta_1-\sqrt{2}\alpha_1\beta_2 \\
2\sqrt{2}\alpha_2\beta_2+\sqrt{2}\alpha_1\beta_3-\sqrt{3}\alpha_3\beta_1-\sqrt{2}\alpha_4\beta_5 \\
\sqrt{3}\alpha_4\beta_1-2\sqrt{2}\alpha_1\beta_4+\sqrt{2}\alpha_2\beta_3-\sqrt{2}\alpha_3\beta_2 \\
\end{pmatrix} \\ [0.1in]
\mathbf{5_1} \sim \begin{pmatrix}
\sqrt{2}\alpha_1\beta_5+\sqrt{2}\alpha_4\beta_2-\sqrt{2}\alpha_2\beta_4-\sqrt{2}\alpha_3\beta_3 \\
-\sqrt{2}\alpha_1\beta_1-\sqrt{3}\alpha_4\beta_3-\sqrt{3}\alpha_3\beta_4 \\
\sqrt{3}\alpha_3\beta_5+\sqrt{2}\alpha_2\beta_1+\sqrt{3}\alpha_1\beta_2 \\
\sqrt{2}\alpha_3\beta_1+\sqrt{3}\alpha_2\beta_2+\sqrt{3}\alpha_4\beta_5 \\
-\sqrt{3}\alpha_2\beta_3-\sqrt{2}\alpha_4\beta_1-\sqrt{3}\alpha_1\beta_4 \\
\end{pmatrix} \\ [0.1in]
\mathbf{5_2} \sim \begin{pmatrix}
 4\alpha_3\beta_3+2\alpha_1\beta_5+2\alpha_4\beta_2+4\alpha_2\beta_4 \\
 4\alpha_1\beta_1+2\sqrt{6}\alpha_2\beta_5 \\
 2\sqrt{6}\alpha_4\beta_4+2\alpha_2\beta_1-\sqrt{6}\alpha_1\beta_2-\sqrt{6}\alpha_3\beta_5 \\
 2\alpha_3\beta_1+2\sqrt{6}\alpha_1\beta_3-\sqrt{6}\alpha_2\beta_2-\sqrt{6}\alpha_4\beta_5 \\
 4\alpha_4\beta_1+2\sqrt{6}\alpha_3\beta_2 \\
\end{pmatrix}
\end{array}
\right. \quad ~~~~~ \\
\left.\begin{array}{c}
\mathbf{4}\otimes\mathbf{\widehat{6}}=\mathbf{\widehat{2}}\oplus\mathbf{\widehat{2}'}\oplus\mathbf{\widehat{4}'_1} \\
\qquad\qquad\oplus\mathbf{\widehat{4}'_2}\oplus\mathbf{\widehat{6}_1}\oplus\mathbf{\widehat{6}_2}
\end{array}\right. &\text{with}& \left\{
\begin{array}{l}
\mathbf{\widehat{2}}\sim \begin{pmatrix}
\alpha_2\beta_1+\sqrt{2}\alpha_4\beta_5+\alpha_3\beta_6-\alpha_2\beta_2-\alpha_1\beta_3 \\
\sqrt{2}\alpha_1\beta_4+\alpha_4\beta_6+\alpha_2\beta_3-\alpha_3\beta_2-\alpha_3\beta_1 \\
\end{pmatrix}\\ [0.1in]
\mathbf{\widehat{2}'} \sim \begin{pmatrix}
\alpha_4\beta_4+\alpha_3\beta_5+\sqrt{2}\alpha_2\beta_6-\sqrt{2}\alpha_1\beta_2 \\
\sqrt{2}\alpha_4\beta_1+\alpha_1\beta_5-\sqrt{2}\alpha_3\beta_3-\alpha_2\beta_4 \\
\end{pmatrix}\\ [0.1in]
\mathbf{\widehat{4}'_1} \sim \begin{pmatrix}
\alpha_2\beta_6-\alpha_1\beta_2-\sqrt{2}\alpha_4\beta_4-\sqrt{2}\alpha_3\beta_5 \\
-\sqrt{3}\alpha_2\beta_1-\sqrt{3}\alpha_1\beta_3 \\
\sqrt{3}\alpha_3\beta_2+\sqrt{3}\alpha_4\beta_6 \\
\alpha_3\beta_3+\sqrt{2}\alpha_1\beta_5-\alpha_4\beta_1-\sqrt{2}\alpha_2\beta_4 \\
\end{pmatrix}\\ [0.1in]
\mathbf{\widehat{4}'_2} \sim \begin{pmatrix}
\sqrt{3}\alpha_1\beta_1-\sqrt{3}\alpha_1\beta_2-\sqrt{6}\alpha_3\beta_5 \\
\alpha_2\beta_2+2\alpha_3\beta_6-\alpha_2\beta_1-2\alpha_1\beta_3-\sqrt{2}\alpha_4\beta_5 \\
\alpha_3\beta_1+2\alpha_4\beta_6+\alpha_3\beta_2+2\alpha_2\beta_3-\sqrt{2}\alpha_1\beta_4 \\
-\sqrt{3}\alpha_4\beta_1-\sqrt{3}\alpha_4\beta_2-\sqrt{6}\alpha_2\beta_4 \\
\end{pmatrix}\\ [0.1in]
\mathbf{\widehat{6}_1} \sim \begin{pmatrix}
\alpha_2\beta_5-\sqrt{2}\alpha_4\beta_3-\alpha_3\beta_4 \\
-\alpha_3\beta_4-\alpha_2\beta_5-\sqrt{2}\alpha_1\beta_6 \\
\alpha_4\beta_4-\sqrt{2}\alpha_1\beta_1-\alpha_3\beta_5 \\
\alpha_1\beta_3+\alpha_3\beta_6-\alpha_2\beta_1-\alpha_2\beta_2 \\
\alpha_3\beta_1+\alpha_4\beta_6-\alpha_3\beta_2-\alpha_2\beta_3 \\
\alpha_1\beta_5+\alpha_2\beta_4-\sqrt{2}\alpha_4\beta_2 \\
\end{pmatrix}\\ [0.1in]
\mathbf{\widehat{6}_2} \sim \begin{pmatrix}
\alpha_2\beta_5+\sqrt{2}\alpha_4\beta_3+\alpha_3\beta_4-\sqrt{2}\alpha_1\beta_6 \\
\alpha_2\beta_5+\sqrt{2}\alpha_1\beta_6+\sqrt{2}\alpha_4\beta_3-\alpha_3\beta_4 \\
\alpha_3\beta_5-\sqrt{2}\alpha_1\beta_2-\alpha_4\beta_4-\sqrt{2}\alpha_2\beta_6 \\
2\alpha_2\beta_2+\sqrt{2}\alpha_4\beta_5 \\
-2\alpha_3\beta_1-\sqrt{2}\alpha_1\beta_4 \\
\sqrt{2}\alpha_4\beta_1+\sqrt{2}\alpha_3\beta_3-\alpha_2\beta_4-\alpha_1\beta_5 \\
\end{pmatrix}
\end{array}
\right.  \\
\left.\begin{array}{c}
\mathbf{\widehat{4}'} \otimes \mathbf{\widehat{4}'} = \mathbf{1_a} \oplus \mathbf{3_s} \oplus \mathbf{3'_s} \\
\qquad \oplus \mathbf{4_s} \oplus \mathbf{5_a}
\end{array}\right. &\text{with}& \left\{
\begin{array}{l}
\mathbf{1_a} \sim\alpha_4\beta_1-\alpha_1\beta_4+\alpha_3\beta_2-\alpha_2\beta_3\\ [0.1in]
\mathbf{3_s} \sim \begin{pmatrix}
3\alpha_4\beta_1+3\alpha_1\beta_4+\alpha_3\beta_2+\alpha_2\beta_3 \\
2\sqrt{2}\alpha_3\beta_3+\sqrt{6}\alpha_4\beta_2+\sqrt{6}\alpha_2\beta_4 \\
\sqrt{6}\alpha_3\beta_1+\sqrt{6}\alpha_1\beta_3-2\sqrt{2}\alpha_2\beta_2 \\
\end{pmatrix}\\ [0.1in]
\mathbf{3'_s}\sim \begin{pmatrix}
\alpha_4\beta_1+\alpha_1\beta_4-3\alpha_3\beta_2-3\alpha_2\beta_3 \\
\sqrt{6}\alpha_4\beta_3+\sqrt{6}\alpha_3\beta_4-2\sqrt{2}\alpha_1\beta_1 \\
2\sqrt{2}\alpha_4\beta_4-\sqrt{6}\alpha_2\beta_1-\sqrt{6}\alpha_1\beta_2 \\
\end{pmatrix}\\ [0.1in]
\mathbf{4_s} \sim \begin{pmatrix}
\alpha_4\beta_2+\alpha_2\beta_4-\sqrt{3}\alpha_3\beta_3 \\
\sqrt{3}\alpha_1\beta_1+\alpha_4\beta_3+\alpha_3\beta_4 \\
\sqrt{3}\alpha_4\beta_4+\alpha_2\beta_1+\alpha_1\beta_2 \\
-\sqrt{3}\alpha_2\beta_2-\alpha_3\beta_1-\alpha_1\beta_3 \\
\end{pmatrix}\\ [0.1in]
\mathbf{5_a} \sim \begin{pmatrix}
\alpha_4\beta_1-\alpha_1\beta_4+\alpha_2\beta_3-\alpha_3\beta_2 \\
\sqrt{2}\alpha_2\beta_4-\sqrt{2}\alpha_4\beta_2 \\
\sqrt{2}\alpha_4\beta_3-\sqrt{2}\alpha_3\beta_4 \\
\sqrt{2}\alpha_1\beta_2-\sqrt{2}\alpha_2\beta_1 \\
\sqrt{2}\alpha_1\beta_3-\sqrt{2}\alpha_3\beta_1 \\
\end{pmatrix}
\end{array}
\right.  \\
\left.\begin{array}{c}
\mathbf{\widehat{4}'} \otimes \mathbf{5} = \mathbf{\widehat{2}} \oplus \mathbf{\widehat{2}'} \oplus \mathbf{\widehat{4}'} \\
\qquad \oplus \mathbf{\widehat{6}_1} \oplus \mathbf{\widehat{6}_2}
\end{array}\right. &\text{with}& \left\{
\begin{array}{l}
\mathbf{\widehat{2}}\sim \begin{pmatrix}
\sqrt{2}\alpha_2\beta_1+\alpha_1\beta_2-2\alpha_4\beta_4-\sqrt{3}\alpha_3\beta_5 \\
\sqrt{3}\alpha_2\beta_2+\alpha_4\beta_5+\sqrt{2}\alpha_3\beta_1+2\alpha_1\beta_3 \\
\end{pmatrix} \\ [0.1in]
\mathbf{\widehat{2}'} \sim \begin{pmatrix}
\sqrt{3}\alpha_4\beta_3-\alpha_3\beta_4-\sqrt{2}\alpha_1\beta_1-2\alpha_2\beta_5 \\
\sqrt{2}\alpha_4\beta_1+\sqrt{3}\alpha_1\beta_4+2\alpha_3\beta_2-\alpha_2\beta_3 \\
\end{pmatrix} \\ [0.1in]
\mathbf{\widehat{4}'} \sim \begin{pmatrix}
\sqrt{2}\alpha_2\beta_5-\alpha_1\beta_1-\sqrt{2}\alpha_3\beta_4 \\
\sqrt{2}\alpha_4\beta_4+\alpha_2\beta_1+\sqrt{2}\alpha_1\beta_2 \\
\sqrt{2}\alpha_4\beta_5+\alpha_3\beta_1-\sqrt{2}\alpha_1\beta_3 \\
\sqrt{2}\alpha_3\beta_2+\sqrt{2}\alpha_2\beta_3-\alpha_4\beta_1 \\
\end{pmatrix} \\ [0.1in]
\mathbf{\widehat{6}_1} \sim \begin{pmatrix}
\sqrt{2}\alpha_2\beta_4+\sqrt{6}\alpha_1\beta_5-\sqrt{2}\alpha_3\beta_3 \\
-\sqrt{2}\alpha_3\beta_3-\sqrt{6}\alpha_4\beta_2-\sqrt{2}\alpha_2\beta_4 \\
\sqrt{2}\alpha_2\beta_5+\sqrt{6}\alpha_4\beta_3+\sqrt{2}\alpha_3\beta_4 \\
\alpha_3\beta_5+\sqrt{6}\alpha_2\beta_1-\sqrt{3}\alpha_1\beta_2 \\
\alpha_2\beta_2+\sqrt{3}\alpha_4\beta_5-\sqrt{6}\alpha_3\beta_1 \\
\sqrt{6}\alpha_1\beta_4+\sqrt{2}\alpha_2\beta_3-\sqrt{2}\alpha_3\beta_2 \\
\end{pmatrix} \\ [0.1in]
\mathbf{\widehat{6}_2} \sim \begin{pmatrix}
2\alpha_4\beta_2-\sqrt{3}\alpha_2\beta_4-\sqrt{3}\alpha_3\beta_3 \\
\sqrt{3}\alpha_3\beta_3+2\alpha_1\beta_5-\sqrt{3}\alpha_2\beta_4 \\
\sqrt{3}\alpha_3\beta_4-\alpha_4\beta_3-\sqrt{6}\alpha_1\beta_1 \\
\sqrt{2}\alpha_4\beta_4-\sqrt{2}\alpha_1\beta_2-\sqrt{6}\alpha_3\beta_5 \\
\sqrt{2}\alpha_1\beta_3+\sqrt{2}\alpha_4\beta_5-\sqrt{6}\alpha_2\beta_2 \\
\sqrt{3}\alpha_2\beta_3+\sqrt{6}\alpha_4\beta_1-\alpha_1\beta_4 \\
\end{pmatrix}
\end{array}
\right.  \\
\left.\begin{array}{c}
\mathbf{\widehat{4}'} \otimes \mathbf{\widehat{6}} = \mathbf{3} \oplus \mathbf{3'} \oplus \mathbf{4_1} \\
\qquad\qquad \oplus \mathbf{4_2} \oplus \mathbf{5_1} \oplus \mathbf{5_2}
\end{array}\right. &\text{with}& \left\{
\begin{array}{l}
\mathbf{3} \sim \begin{pmatrix}
2\alpha_1\beta_6+\sqrt{6}\alpha_3\beta_4-2\alpha_4\beta_3-\sqrt{6}\alpha_2\beta_5 \\
\sqrt{2}\alpha_1\beta_2+\alpha_4\beta_4+\sqrt{3}\alpha_3\beta_5-2\sqrt{2}\alpha_1\beta_1-\sqrt{6}\alpha_2\beta_6 \\
\sqrt{6}\alpha_3\beta_3+\sqrt{3}\alpha_2\beta_4-2\sqrt{2}\alpha_4\beta_2-\sqrt{2}\alpha_4\beta_1-\alpha_1\beta_5 \\
\end{pmatrix} \\ [0.1in]
\mathbf{3'} \sim \begin{pmatrix}
\sqrt{6}\alpha_4\beta_3+2\alpha_3\beta_4-2\alpha_2\beta_5-\sqrt{6}\alpha_1\beta_6 \\
3\alpha_2\beta_1+\alpha_2\beta_2+\sqrt{3}\alpha_1\beta_3-\sqrt{6}\alpha_4\beta_5-\alpha_3\beta_6 \\
\sqrt{3}\alpha_4\beta_6+3\alpha_3\beta_2-\alpha_2\beta_3-\alpha_3\beta_1-\sqrt{6}\alpha_1\beta_4 \\
\end{pmatrix} \\ [0.1in]
\mathbf{4_1} \sim \begin{pmatrix}
\alpha_1\beta_1-\sqrt{2}\alpha_4\beta_4-\sqrt{3}\alpha_2\beta_6 \\
-\sqrt{3}\alpha_2\beta_2-\alpha_1\beta_3-\sqrt{2}\alpha_4\beta_5 \\
\sqrt{2}\alpha_1\beta_4+\alpha_4\beta_6-\sqrt{3}\alpha_3\beta_1 \\
-\alpha_4\beta_2-\sqrt{3}\alpha_3\beta_3-\sqrt{2}\alpha_1\beta_5 \\
\end{pmatrix} \\ [0.1in]
\mathbf{4_2} \sim \begin{pmatrix}
\sqrt{2}\alpha_3\beta_5-\sqrt{3}\alpha_1\beta_2-\sqrt{6}\alpha_4\beta_4-\alpha_2\beta_6 \\
\alpha_2\beta_1-2\alpha_2\beta_2-\sqrt{3}\alpha_1\beta_3-2\alpha_3\beta_6 \\
\sqrt{3}\alpha_4\beta_6+2\alpha_2\beta_3-\alpha_3\beta_2-2\alpha_3\beta_1 \\
-\sqrt{3}\alpha_4\beta_1-\alpha_3\beta_3-\sqrt{2}\alpha_2\beta_4-\sqrt{6}\alpha_1\beta_5 \\
\end{pmatrix} \\ [0.1in]
\mathbf{5_1}\sim \begin{pmatrix}
\sqrt{6}\alpha_3\beta_4+\sqrt{6}\alpha_2\beta_5 \\
\sqrt{6}\alpha_1\beta_2+\alpha_3\beta_5+\sqrt{2}\alpha_2\beta_6-\sqrt{3}\alpha_4\beta_4 \\
\sqrt{2}\alpha_2\beta_1+\sqrt{2}\alpha_2\beta_2+\sqrt{2}\alpha_3\beta_6-\sqrt{6}\alpha_1\beta_3 \\
\sqrt{2}\alpha_3\beta_1+\sqrt{6}\alpha_4\beta_6-\sqrt{2}\alpha_3\beta_2-\sqrt{2}\alpha_2\beta_3 \\
\sqrt{6}\alpha_4\beta_1+\sqrt{2}\alpha_3\beta_3-\alpha_2\beta_4-\sqrt{3}\alpha_1\beta_5 \\
\end{pmatrix} \\ [0.1in]
\mathbf{5_2} \sim \begin{pmatrix}
\sqrt{6}\alpha_4\beta_3+\sqrt{6}\alpha_1\beta_6 \\
2\alpha_1\beta_1+\sqrt{2}\alpha_4\beta_4+\sqrt{6}\alpha_3\beta_5 \\
\sqrt{3}\alpha_2\beta_2-\sqrt{3}\alpha_2\beta_1-\alpha_1\beta_3-\sqrt{2}\alpha_4\beta_5-\sqrt{3}\alpha_3\beta_6 \\
\sqrt{3}\alpha_3\beta_1+\sqrt{2}\alpha_1\beta_4+\alpha_4\beta_6+\sqrt{3}\alpha_3\beta_2+\sqrt{3}\alpha_2\beta_3 \\
\sqrt{2}\alpha_1\beta_5-2\alpha_4\beta_2-\sqrt{6}\alpha_2\beta_4 \\
\end{pmatrix}
\end{array}
\right.  \\
\left.\begin{array}{c}
\mathbf{5} \otimes \mathbf{5} = \mathbf{1_s} \oplus \mathbf{3_a} \oplus \mathbf{3'_a}  \\
\qquad\quad  \oplus \mathbf{4_s}  \oplus \mathbf{4_a} \oplus \mathbf{5_{1,s}} \\
\qquad \oplus \mathbf{5_{2,s}}
\end{array}\right. &\text{with}&  \left\{
\begin{array}{l}
\mathbf{1_s} \sim \alpha_1\beta_1+\alpha_5\beta_2+\alpha_2\beta_5+\alpha_4\beta_3+\alpha_3\beta_4\\ [0.1in]
\mathbf{3_a}\sim \begin{pmatrix}
 2\alpha_3\beta_4-2\alpha_4\beta_3+\alpha_2\beta_5-\alpha_5\beta_2 \\
\sqrt{3}\alpha_2\beta_1-\sqrt{3}\alpha_1\beta_2+\sqrt{2}\alpha_3\beta_5-\sqrt{2}\alpha_5\beta_3 \\
\sqrt{2}\alpha_2\beta_4-\sqrt{2}\alpha_4\beta_2+\sqrt{3}\alpha_1\beta_5-\sqrt{3}\alpha_5\beta_1 \\
\end{pmatrix}\\ [0.1in]
\mathbf{3'_a} \sim  \begin{pmatrix}
\alpha_4\beta_3-\alpha_3\beta_4+2\alpha_2\beta_5-2\alpha_5\beta_2 \\
\sqrt{3}\alpha_1\beta_3-\sqrt{3}\alpha_3\beta_1+\sqrt{2}\alpha_4\beta_5-\sqrt{2}\alpha_5\beta_4 \\
\sqrt{3}\alpha_4\beta_1-\sqrt{3}\alpha_1\beta_4+\sqrt{2}\alpha_2\beta_3-\sqrt{2}\alpha_3\beta_2 \\
\end{pmatrix}\\ [0.1in]
\mathbf{4_s}\sim \begin{pmatrix}
 4\alpha_4\beta_4+\sqrt{6}\alpha_2\beta_1+\sqrt{6}\alpha_1\beta_2-\alpha_5\beta_3-\alpha_3\beta_5 \\
 4\alpha_2\beta_2+\sqrt{6}\alpha_3\beta_1+\sqrt{6}\alpha_1\beta_3-\alpha_5\beta_4-\alpha_4\beta_5 \\
 4\alpha_5\beta_5+\sqrt{6}\alpha_4\beta_1+\sqrt{6}\alpha_1\beta_4-\alpha_3\beta_2-\alpha_2\beta_3 \\
 4\alpha_3\beta_3+\sqrt{6}\alpha_5\beta_1+\sqrt{6}\alpha_1\beta_5-\alpha_4\beta_2-\alpha_2\beta_4 \\
\end{pmatrix} \\ [0.1in]
\mathbf{4_a}\sim  \begin{pmatrix}
\sqrt{2}\alpha_1\beta_2-\sqrt{2}\alpha_2\beta_1+\sqrt{3}\alpha_3\beta_5-\sqrt{3}\alpha_5\beta_3 \\
\sqrt{2}\alpha_3\beta_1-\sqrt{2}\alpha_1\beta_3+\sqrt{3}\alpha_4\beta_5-\sqrt{3}\alpha_5\beta_4 \\
\sqrt{2}\alpha_4\beta_1-\sqrt{2}\alpha_1\beta_4+\sqrt{3}\alpha_3\beta_2-\sqrt{3}\alpha_2\beta_3 \\
\sqrt{3}\alpha_4\beta_2-\sqrt{3}\alpha_2\beta_4+\sqrt{2}\alpha_1\beta_5-\sqrt{2}\alpha_5\beta_1 \\
\end{pmatrix}\\ [0.1in]
\mathbf{5_{1,s}} \sim \begin{pmatrix}
2\alpha_1\beta_1+\alpha_5\beta_2+\alpha_2\beta_5-2\alpha_4\beta_3-2\alpha_3\beta_4 \\
\alpha_2\beta_1+\alpha_1\beta_2+\sqrt{6}\alpha_5\beta_3+\sqrt{6}\alpha_3\beta_5 \\
\sqrt{6}\alpha_2\beta_2-2\alpha_3\beta_1-2\alpha_1\beta_3 \\
\sqrt{6}\alpha_5\beta_5-2\alpha_4\beta_1-2\alpha_1\beta_4 \\
\alpha_5\beta_1+\alpha_1\beta_5+\sqrt{6}\alpha_4\beta_2+\sqrt{6}\alpha_2\beta_4 \\
\end{pmatrix} \\ [0.1in]
\mathbf{5_{2,s}} \sim \begin{pmatrix}
 2\alpha_1\beta_1+\alpha_4\beta_3+\alpha_3\beta_4-2\alpha_5\beta_2-2\alpha_2\beta_5 \\
\sqrt{6}\alpha_4\beta_4-2\alpha_2\beta_1-2\alpha_1\beta_2 \\
\alpha_3\beta_1+\alpha_1\beta_3+\sqrt{6}\alpha_5\beta_4+\sqrt{6}\alpha_4\beta_5 \\
\alpha_4\beta_1+\alpha_1\beta_4+\sqrt{6}\alpha_3\beta_2+\sqrt{6}\alpha_2\beta_3 \\
\sqrt{6}\alpha_3\beta_3-2\alpha_5\beta_1-2\alpha_1\beta_5 \\
\end{pmatrix}
\end{array}
\right.  \\
\left.\begin{array}{c}
\mathbf{5} \otimes \mathbf{\widehat{6}} = \mathbf{\widehat{2}} \oplus \mathbf{\widehat{2}'} \oplus \mathbf{\widehat{4}'_1} \\
\qquad\quad \oplus \mathbf{\widehat{4}'_2} \oplus \mathbf{\widehat{6}_1} \oplus \mathbf{\widehat{6}_2} \\
\qquad \oplus \mathbf{\widehat{6}_3}
\end{array}\right. &\text{with}& \left\{
\begin{array}{l}
\mathbf{\widehat{2}} \sim \begin{pmatrix}
\alpha_3\beta_1+\sqrt{3}\alpha_1\beta_4+\alpha_4\beta_6+2\alpha_3\beta_2-2\alpha_2\beta_3-\sqrt{2}\alpha_5\beta_5 \\
2\alpha_4\beta_1+\alpha_3\beta_3+2\alpha_5\beta_6-\sqrt{3}\alpha_1\beta_5-\alpha_4\beta_2-\sqrt{2}\alpha_2\beta_4 \\
\end{pmatrix} \\ [0.1in]
\mathbf{\widehat{2}'} \sim \begin{pmatrix}
3\alpha_2\beta_1+\alpha_2\beta_2+\sqrt{6}\alpha_1\beta_3+\sqrt{2}\alpha_5\beta_4+2\sqrt{2}\alpha_4\beta_5-2\alpha_3\beta_6 \\
\sqrt{6}\alpha_1\beta_6+3\alpha_5\beta_2+\sqrt{2}\alpha_2\beta_5+2\alpha_4\beta_3-2\sqrt{2}\alpha_3\beta_4-\alpha_5\beta_1 \\
\end{pmatrix}\\ [0.1in]
\mathbf{\widehat{4}'_1} \sim \begin{pmatrix}
\sqrt{6}\alpha_1\beta_3+\alpha_3\beta_6+\sqrt{2}\alpha_5\beta_4-\sqrt{2}\alpha_4\beta_5-2\alpha_2\beta_2 \\
\sqrt{3}\alpha_3\beta_1+\sqrt{3}\alpha_3\beta_2+\sqrt{6}\alpha_5\beta_5-\sqrt{3}\alpha_4\beta_6 \\
\sqrt{3}\alpha_4\beta_1+\sqrt{6}\alpha_2\beta_4-\sqrt{3}\alpha_4\beta_2-\sqrt{3}\alpha_3\beta_3 \\
\alpha_4\beta_3-\sqrt{2}\alpha_3\beta_4-2\alpha_5\beta_1-\sqrt{2}\alpha_2\beta_5-\sqrt{6}\alpha_1\beta_6 \\
\end{pmatrix} \\ [0.1in]
\mathbf{\widehat{4}'_2} \sim \begin{pmatrix}
\sqrt{3}\alpha_5\beta_4-\sqrt{6}\alpha_2\beta_1-\sqrt{6}\alpha_3\beta_6 \\
\sqrt{2}\alpha_3\beta_2-\sqrt{2}\alpha_2\beta_3-\sqrt{2}\alpha_3\beta_1-\sqrt{6}\alpha_1\beta_4-\alpha_5\beta_5-\sqrt{2}\alpha_4\beta_6 \\
\sqrt{2}\alpha_4\beta_1+\sqrt{6}\alpha_1\beta_5+\sqrt{2}\alpha_5\beta_6+\sqrt{2}\alpha_4\beta_2-\alpha_2\beta_4-\sqrt{2}\alpha_3\beta_3 \\
\sqrt{6}\alpha_5\beta_2-\sqrt{3}\alpha_2\beta_5-\sqrt{6}\alpha_4\beta_3 \\
\end{pmatrix} \\ [0.1in]
\mathbf{\widehat{6}_1} \sim \begin{pmatrix}
\alpha_1\beta_1+\sqrt{3}\alpha_4\beta_4-\sqrt{6}\alpha_2\beta_6 \\
\alpha_1\beta_2+\sqrt{6}\alpha_5\beta_3+\sqrt{3}\alpha_3\beta_5 \\
\sqrt{6}\alpha_2\beta_2+\alpha_1\beta_3-\sqrt{3}\alpha_4\beta_5 \\
\sqrt{3}\alpha_3\beta_1+\sqrt{3}\alpha_4\beta_6-2\alpha_1\beta_4 \\
\sqrt{3}\alpha_4\beta_2-\sqrt{3}\alpha_3\beta_3-2\alpha_1\beta_5 \\
\sqrt{3}\alpha_3\beta_4+\alpha_1\beta_6-\sqrt{6}\alpha_5\beta_1 \\
\end{pmatrix} \\ [0.1in]
\mathbf{\widehat{6}_2} \sim \begin{pmatrix}
\sqrt{3}\alpha_1\beta_1+\sqrt{2}\alpha_5\beta_3-2\alpha_3\beta_5-\alpha_4\beta_4 \\
\sqrt{3}\alpha_1\beta_2+2\alpha_4\beta_4+\sqrt{2}\alpha_2\beta_6-\alpha_3\beta_5 \\
\sqrt{2}\alpha_2\beta_1+2\alpha_5\beta_4-\alpha_4\beta_5-\sqrt{3}\alpha_1\beta_3 \\
\alpha_4\beta_6+2\alpha_3\beta_2+2\alpha_2\beta_3-\alpha_3\beta_1 \\
2\alpha_5\beta_6-2\alpha_4\beta_1-\alpha_4\beta_2-\alpha_3\beta_3 \\
\alpha_3\beta_4+\sqrt{2}\alpha_5\beta_2+2\alpha_2\beta_5-\sqrt{3}\alpha_1\beta_6 \\
\end{pmatrix} \\ [0.1in]
\mathbf{\widehat{6}_3}\sim \begin{pmatrix}
\sqrt{2}\alpha_1\beta_1+\sqrt{3}\alpha_2\beta_6+\sqrt{3}\alpha_5\beta_3-\sqrt{6}\alpha_3\beta_5-\sqrt{6}\alpha_4\beta_4 \\
\sqrt{2}\alpha_1\beta_2+\sqrt{6}\alpha_4\beta_4+\sqrt{3}\alpha_2\beta_6-\sqrt{3}\alpha_5\beta_3-\sqrt{6}\alpha_3\beta_5 \\
\sqrt{3}\alpha_2\beta_1+\sqrt{6}\alpha_5\beta_4-\sqrt{3}\alpha_2\beta_2-2\sqrt{2}\alpha_1\beta_3 \\
\sqrt{2}\alpha_1\beta_4+\sqrt{6}\alpha_3\beta_2+\sqrt{6}\alpha_2\beta_3-\sqrt{6}\alpha_3\beta_1 \\
\sqrt{2}\alpha_1\beta_5+\sqrt{6}\alpha_5\beta_6-\sqrt{6}\alpha_4\beta_1-\sqrt{6}\alpha_4\beta_2 \\
\sqrt{3}\alpha_5\beta_1+\sqrt{3}\alpha_5\beta_2+\sqrt{6}\alpha_2\beta_5-2\sqrt{2}\alpha_1\beta_6 \\
\end{pmatrix}
\end{array}
\right.  \\
\left.\begin{array}{c}
\mathbf{\widehat{6}} \otimes \mathbf{\widehat{6}} = \mathbf{1_a} \oplus \mathbf{3_{1,s}} \oplus \mathbf{3_{2,s}}  \\
\qquad\quad \oplus \mathbf{3'_{1,s}} \oplus \mathbf{3'_{2,s}} \oplus \mathbf{4_s} \\
\qquad\quad \oplus \mathbf{4_a} \oplus \mathbf{5_{1,s}} \oplus \mathbf{5_{2,a}} \\
\qquad \quad \oplus \mathbf{5_{3,a}}
\end{array}\right. &\text{with}& \begin{small}\left\{
\begin{array}{l}
\mathbf{1_a}\sim\alpha_2\beta_1-\alpha_1\beta_2+\alpha_6\beta_3-\alpha_3\beta_6+\alpha_5\beta_4-\alpha_4\beta_5 \\ [0.1in]
\mathbf{3_{1,s}} \sim\begin{pmatrix}
\alpha_6\beta_3+\alpha_3\beta_6-\alpha_2\beta_1-\alpha_1\beta_2-\alpha_5\beta_4-\alpha_4\beta_5 \\
\sqrt{2}\alpha_5\beta_5-\sqrt{2}\alpha_3\beta_2-\sqrt{2}\alpha_2\beta_3 \\
-\sqrt{2}\alpha_4\beta_4-\sqrt{2}\alpha_6\beta_1-\sqrt{2}\alpha_1\beta_6 \\
\end{pmatrix}\\ [0.1in]
\mathbf{3_{2,s}} \sim \begin{pmatrix}
2\alpha_2\beta_2+2\alpha_5\beta_4+2\alpha_4\beta_5-2\alpha_1\beta_1 \\
\sqrt{2}\alpha_3\beta_1+\sqrt{2}\alpha_1\beta_3+\sqrt{2}\alpha_3\beta_2+\sqrt{2}\alpha_2\beta_3-2\alpha_6\beta_4-2\alpha_4\beta_6 \\
\sqrt{2}\alpha_6\beta_1+\sqrt{2}\alpha_1\beta_6-\sqrt{2}\alpha_6\beta_2-\sqrt{2}\alpha_2\beta_6-2\alpha_5\beta_3-2\alpha_3\beta_5 \\
\end{pmatrix} \\ [0.1in]
\mathbf{3'_{1,s}} \sim \begin{pmatrix}
\alpha_2\beta_1+\alpha_1\beta_2+\alpha_6\beta_3+\alpha_3\beta_6 \\
\alpha_6\beta_5+\alpha_5\beta_6-\alpha_4\beta_2-\alpha_2\beta_4 \\
-\alpha_5\beta_1-\alpha_1\beta_5-\alpha_4\beta_3-\alpha_3\beta_4 \\
\end{pmatrix} \\ [0.1in]
\mathbf{3'_{2,s}} \sim \begin{pmatrix}
\alpha_2\beta_2-\alpha_1\beta_1-\alpha_6\beta_3-\alpha_3\beta_6-\alpha_5\beta_4-\alpha_4\beta_5 \\
\alpha_4\beta_2+\alpha_2\beta_4-\sqrt{2}\alpha_3\beta_3-\alpha_4\beta_1-\alpha_1\beta_4 \\
\sqrt{2}\alpha_6\beta_6+\alpha_5\beta_1+\alpha_1\beta_5+\alpha_5\beta_2+\alpha_2\beta_5 \\
\end{pmatrix} \\ [0.1in]
\mathbf{4_s} \sim \begin{pmatrix}
2\sqrt{2}\alpha_3\beta_1+2\sqrt{2}\alpha_1\beta_3+\alpha_6\beta_4+\alpha_4\beta_6-2\sqrt{2}\alpha_5\beta_5-\sqrt{2}\alpha_3\beta_2-\sqrt{2}\alpha_2\beta_3 \\
2\sqrt{2}\alpha_3\beta_3-3\alpha_4\beta_1-3\alpha_1\beta_4-\alpha_4\beta_2-\alpha_2\beta_4-\alpha_6\beta_5-\alpha_5\beta_6 \\
2\sqrt{2}\alpha_6\beta_6+\alpha_5\beta_1+\alpha_1\beta_5-3\alpha_5\beta_2-3\alpha_2\beta_5-\alpha_4\beta_3-\alpha_3\beta_4 \\
\sqrt{2}\alpha_6\beta_1+\sqrt{2}\alpha_1\beta_6+2\sqrt{2}\alpha_6\beta_2+2\sqrt{2}\alpha_2\beta_6-2\sqrt{2}\alpha_4\beta_4-\alpha_5\beta_3-\alpha_3\beta_5 \\
\end{pmatrix} \\ [0.1in]
\mathbf{4_a} \sim \begin{pmatrix}
\sqrt{2}\alpha_2\beta_3-\sqrt{2}\alpha_3\beta_2+\alpha_4\beta_6-\alpha_6\beta_4 \\
\alpha_4\beta_1-\alpha_1\beta_4+\alpha_2\beta_4-\alpha_4\beta_2+\alpha_6\beta_5-\alpha_5\beta_6 \\
\alpha_5\beta_1-\alpha_1\beta_5+\alpha_5\beta_2-\alpha_2\beta_5+\alpha_3\beta_4-\alpha_4\beta_3 \\
\sqrt{2}\alpha_6\beta_1-\sqrt{2}\alpha_1\beta_6+\alpha_3\beta_5-\alpha_5\beta_3 \\
\end{pmatrix} \\ [0.1in]
\mathbf{5_{1,s}} \sim \begin{pmatrix}
-\sqrt{6}\alpha_1\beta_1-\sqrt{6}\alpha_2\beta_2 \\
2\alpha_5\beta_5+\alpha_3\beta_1+\alpha_1\beta_3+\alpha_3\beta_2+\alpha_2\beta_3+\sqrt{2}\alpha_6\beta_4+\sqrt{2}\alpha_4\beta_6 \\
2\alpha_3\beta_3+\sqrt{2}\alpha_4\beta_2+\sqrt{2}\alpha_2\beta_4+\sqrt{2}\alpha_6\beta_5+\sqrt{2}\alpha_5\beta_6 \\
2\alpha_6\beta_6+\sqrt{2}\alpha_4\beta_3+\sqrt{2}\alpha_3\beta_4-\sqrt{2}\alpha_5\beta_1-\sqrt{2}\alpha_1\beta_5 \\
2\alpha_4\beta_4+\alpha_6\beta_2+\alpha_2\beta_6-\alpha_6\beta_1-\alpha_1\beta_6-\sqrt{2}\alpha_5\beta_3-\sqrt{2}\alpha_3\beta_5 \\
\end{pmatrix} \\ [0.1in]
\mathbf{5_{2,a}} \sim \begin{pmatrix}
\alpha_1\beta_2-\alpha_2\beta_1+2\alpha_5\beta_4-2\alpha_4\beta_5+\alpha_3\beta_6-\alpha_6\beta_3 \\
\sqrt{6}\alpha_1\beta_3-\sqrt{6}\alpha_3\beta_1 \\
\sqrt{3}\alpha_4\beta_2-\sqrt{3}\alpha_2\beta_4+\sqrt{3}\alpha_6\beta_5-\sqrt{3}\alpha_5\beta_6 \\
\sqrt{3}\alpha_3\beta_4-\sqrt{3}\alpha_4\beta_3+\sqrt{3}\alpha_1\beta_5-\sqrt{3}\alpha_5\beta_1 \\
\sqrt{6}\alpha_2\beta_6-\sqrt{6}\alpha_6\beta_2 \\
\end{pmatrix} \\ [0.1in]
\mathbf{5_{3,a}}\sim \begin{pmatrix}
\sqrt{2}\alpha_2\beta_1-\sqrt{2}\alpha_1\beta_2+\sqrt{2}\alpha_5\beta_4-\sqrt{2}\alpha_4\beta_5+2\sqrt{2}\alpha_3\beta_6-2\sqrt{2}\alpha_6\beta_3 \\
\sqrt{3}\alpha_1\beta_3-\sqrt{3}\alpha_3\beta_1+\sqrt{3}\alpha_2\beta_3-\sqrt{3}\alpha_3\beta_2+\sqrt{6}\alpha_6\beta_4-\sqrt{6}\alpha_4\beta_6 \\
\sqrt{6}\alpha_4\beta_1-\sqrt{6}\alpha_1\beta_4+\sqrt{6}\alpha_4\beta_2-\sqrt{6}\alpha_2\beta_4 \\
\sqrt{6}\alpha_5\beta_2-\sqrt{6}\alpha_2\beta_5+\sqrt{6}\alpha_1\beta_5-\sqrt{6}\alpha_5\beta_1 \\
\sqrt{3}\alpha_6\beta_1-\sqrt{3}\alpha_1\beta_6+\sqrt{6}\alpha_5\beta_3-\sqrt{6}\alpha_3\beta_5+\sqrt{3}\alpha_2\beta_6-\sqrt{3}\alpha_6\beta_2 \\
\end{pmatrix}
\end{array}
\right.
\end{small}
\end{eqnarray}

\subsubsection{\label{sec:MF-level5}Integer weight modular forms of level 5}

The linear space of modular forms of positive integer weight $k$ and level 5 has dimension $5k+1$. A general vector $\mathcal{M}_{k}(\Gamma(5))$ in the linear space can be explicitly constructed through the Dedekind eta-function and Klein form as follow~\cite{schultz2015notes,Yao:2020zml}:
\begin{equation}
\label{eq:MF-Gamma5}\mathcal{M}_{k}(\Gamma(5))=\sum_{m+n=5k,\,m,n\ge0} c_{mn} \frac{\eta^{15k}(5\tau)}{\eta^{3k}(\tau)} \mathfrak{k}^m_{\frac{1}{5},0}(5\tau)\mathfrak{k}^n_{\frac{2}{5},0}(5\tau)\,,
\end{equation}
where the Dedekind eta-function $\eta(\tau)$ is defined in Eq.~\eqref{eq:eta-func} and Klein form $\mathfrak{k}_{(r_1,r_2)}(\tau)$ is given by~\cite{K_lang1981,lang1987elliptic,lang2012introduction,eum2011modularity}:
\begin{equation}
\label{eq:KleinForm}
\mathfrak{k}_{(r_1,r_2)}(\tau)=q^{(r_1-1)/2}_z(1-q_z)\prod_{n=1}^\infty(1-q^nq_z)(1-q^nq_z^{-1})(1-q^n)^{-2}\,,
\end{equation}
with $q_z=e^{2\pi iz}$ and $z=r_1\tau+r_2$. From Eq.~\eqref{eq:MF-Gamma5}, we known that the modular space of level 5 can be generated by $F_1(\tau)$ and $F_2(\tau)$,
\begin{equation}
F_1(\tau)= \dfrac{\eta^3(5\tau) \mathfrak{k}_{\frac{2}{5}, 0}(5\tau) }{\eta(\tau)^{3/5}},~~~F_2(\tau)= \dfrac{\eta^3(5\tau) \mathfrak{k}_{\frac{1}{5}, 0}(5\tau)}{\eta(\tau)^{3/5}}\,,
\end{equation}
which turn out to be weight $1/5$ modular forms of level 5~\cite{ibukiyama2000modular,Yao:2020zml}. Each modular form of integer weight $k$ and level 5 can be written as a polynomial of degree $5k$ in $F_1(\tau)$ and $F_{2}(\tau)$:
\begin{equation}
\sum_{n=0}^{5k} c_n F^n_{1}(\tau)F^{5k-n}_2(\tau)\,.
\end{equation}
Because $F_1(\tau)$ and $F_{2}(\tau)$ are algebraically independent, all terms in above polynomial are linearly independent, and obviously the number of independent terms matches with the correct dimension $5k+1$. Without loss of generality, we can choose a set of basis vectors of the weight 1 modular space as
\begin{equation}
F^5_1(\tau), ~~F^4_1(\tau) F_2(\tau), ~~F^3_1(\tau) F^2_2(\tau), ~~F^2_1(\tau) F^3_2(\tau), ~~F_1(\tau) F^4_2(\tau),~~F^5_2(\tau)\,.
\end{equation}
In the representation basis of Eq.~\eqref{eq:irr-A5p}, the above six modular forms can be organized into a six dimensional representation $\mathbf{\widehat{6}}$ of $\Gamma'_5=A'_5$:
\begin{equation}
Y^{(1)}_{\mathbf{\widehat{6}}}(\tau)=
\begin{pmatrix}
F_1^5 + 2 F_2^5  \\
2 F_1^5 - F_2^5 \\
5 F_1^4 F_2 \\
5\sqrt{2}F_1^3 F_2^2 \\
-5\sqrt{2} F_1^2 F_2^3 \\
5F_1 F_2^4
\end{pmatrix}\equiv
\begin{pmatrix}
Y_1(\tau) \\
Y_2(\tau) \\
Y_3(\tau) \\
Y_4(\tau) \\
Y_5(\tau) \\
Y_6(\tau)
\end{pmatrix}\,,
\end{equation}
which fulfills the identities in Eqs.~(\ref{eq:MF-decomp},\ref{eq:decom-ST}). The $q$-expansion of the modular forms $Y_{1,2,3, 4,5,6}$ reads as
\begin{eqnarray}
\nonumber&&Y_1(\tau)=1+5q+10q^3-5q^4+5q^5+10q^6+5q^9+\ldots\,, \\
\nonumber&&Y_2(\tau)=2+5 q+10 q^2+5 q^4+5 q^5+10 q^6+10q^7-5q^9+\ldots\,, \\
\nonumber&&Y_3(\tau)=5q^{1/5}\left(1+2q+2q^2+q^3+2q^4+2q^5+2q^6+q^7+2q^8+2q^9+\ldots\right)\,,\\
\nonumber&&Y_4(\tau)=5\sqrt{2}q^{2/5}\left(1+q+q^2+q^3+2q^4+q^6+q^7+2 q^8+q^9+\ldots\right)\,,\\
\nonumber&&Y_5(\tau)=-5\sqrt{2}q^{3/5}\left(1+q^2+q^3+q^4-q^5+2 q^6+q^8+q^9+\ldots\right)\,,\\
\label{eq:q_series_wt1}&&Y_6(\tau)=5q^{4/5}\left(1-q+2q^2+2q^6-2q^7+2q^8+q^9+\ldots\right)\,.
\end{eqnarray}
The higher weight modular forms can be constructed from tensor products of $Y^{(1)}_{\mathbf{\widehat{6}}}(\tau)$. For example, there are 11 independent weight 2 modular forms of level 5, which can be decomposed into two triplets and a quintet transforming in the $\mathbf{3}$, $\mathbf{3'}$ and $\mathbf{5}$ irreducible representations of $A'_5$. Without loss of generality we can choose the weight 2 and level 5 modular forms as
\begin{equation}
  \begin{split}
Y_{\mathbf{3}}^{(2)}&=(Y_{\mathbf{\widehat{6}}}^{(1)}Y_{\mathbf{\widehat{6}}}^{(1)})_{\mathbf{3}_{1, s}}=
\begin{pmatrix}
 -2 \left(Y_1 Y_2+Y_4 Y_5-Y_3 Y_6\right) \\
\sqrt{2} \left(Y_5^2-2 Y_2 Y_3\right) \\
 -\sqrt{2} \left(Y_4^2+2 Y_1 Y_6\right) \\
\end{pmatrix}\,,\\
Y_{\mathbf{3'}}^{(2)}&=(Y_{\mathbf{\widehat{6}}}^{(1)}Y_{\mathbf{\widehat{6}}}^{(1)})_{\mathbf{3'}_{1, s}}=
\begin{pmatrix}
 2 \left(Y_1 Y_2+Y_3 Y_6\right) \\
 2 Y_5 Y_6-2 Y_2 Y_4 \\
 -2 \left(Y_3 Y_4+Y_1 Y_5\right) \\
\end{pmatrix}\,,\\
Y_{\mathbf{5}}^{(2)}&=(Y_{\mathbf{\widehat{6}}}^{(1)}Y_{\mathbf{\widehat{6}}}^{(1)})_{\mathbf{5}_{1, s}}=
\begin{pmatrix}
 -\sqrt{6} \left(Y_1^2+Y_2^2\right) \\
 2 \left(Y_5^2+Y_1 Y_3+Y_2 Y_3+\sqrt{2} Y_4 Y_6\right) \\
 2 \left(Y_3^2+\sqrt{2} \left(Y_2 Y_4+Y_5 Y_6\right)\right) \\
 2 \left(Y_6^2+\sqrt{2} Y_3 Y_4-\sqrt{2} Y_1 Y_5\right) \\
 2 \left(Y_4^2-\sqrt{2} Y_3 Y_5+\left(Y_2-Y_1\right) Y_6\right) \\
\end{pmatrix}\,.
\end{split}
\end{equation}
The linear space of modular forms of weight $k=3$ and level 5 has dimension $5k+1=16$, and they can be decomposed into a quartet and two sextets transforming as $\mathbf{\widehat{4}'}$ and $\mathbf{\widehat{6}}$ under $A'_5$,
\begin{equation}
  \begin{split}
Y_{\mathbf{\widehat{4}'}}^{(3)}&=(Y_{\mathbf{\widehat{6}}}^{(1)}Y_{\mathbf{3'}}^{(2)})_{\mathbf{\widehat{4}'}}=
\begin{pmatrix}
 -\sqrt{6} Y_3 Y_{\mathbf{3}',1}^{(2)}-\sqrt{3} Y_6 Y_{\mathbf{3}',2}^{(2)}+\sqrt{6} Y_5 Y_{\mathbf{3}',3}^{(2)} \\
 -2 Y_4 Y_{\mathbf{3}',1}^{(2)}+Y_1 Y_{\mathbf{3}',2}^{(2)}-3 Y_2 Y_{\mathbf{3}',2}^{(2)}+Y_6 Y_{\mathbf{3}',3}^{(2)} \\
 -2 Y_5 Y_{\mathbf{3}',1}^{(2)}-Y_3 Y_{\mathbf{3}',2}^{(2)}+\left(3 Y_1+Y_2\right) Y_{\mathbf{3}',3}^{(2)} \\
 -\sqrt{6} Y_6 Y_{\mathbf{3}',1}^{(2)}-\sqrt{6} Y_4 Y_{\mathbf{3}',2}^{(2)}+\sqrt{3} Y_3 Y_{\mathbf{3}',3}^{(2)} \\
\end{pmatrix}\,, \\
Y_{\mathbf{\widehat{6}}I}^{(3)}&=(Y_{\mathbf{\widehat{6}}}^{(1)}Y_{\mathbf{3}}^{(2)})_{\mathbf{\widehat{6}}_{1}}=
\begin{pmatrix}
 -Y_1 Y_{\mathbf{3},1}^{(2)}-\sqrt{2} Y_3 Y_{\mathbf{3},3}^{(2)} \\
 Y_2 Y_{\mathbf{3},1}^{(2)}+\sqrt{2} Y_6 Y_{\mathbf{3},2}^{(2)} \\
 Y_3 Y_{\mathbf{3},1}^{(2)}-\sqrt{2} Y_1 Y_{\mathbf{3},2}^{(2)} \\
\sqrt{2} Y_5 Y_{\mathbf{3},3}^{(2)}-Y_4 Y_{\mathbf{3},1}^{(2)} \\
 Y_5 Y_{\mathbf{3},1}^{(2)}+\sqrt{2} Y_4 Y_{\mathbf{3},2}^{(2)} \\
\sqrt{2} Y_2 Y_{\mathbf{3},3}^{(2)}-Y_6 Y_{\mathbf{3},1}^{(2)} \\
\end{pmatrix} \,, \\
Y_{\mathbf{\widehat{6}}II}^{(3)}&=(Y_{\mathbf{\widehat{6}}}^{(1)}Y_{\mathbf{5}}^{(2)})_{\mathbf{\widehat{6}}_{3}}=
\begin{pmatrix}
\sqrt{2} Y_1 Y_{\mathbf{5},1}^{(2)}+\sqrt{3} \left(Y_6 Y_{\mathbf{5},2}^{(2)}-\sqrt{2} Y_5 Y_{\mathbf{5},3}^{(2)}-\sqrt{2} Y_4 Y_{\mathbf{5},4}^{(2)}+Y_3 Y_{\mathbf{5},5}^{(2)}\right) \\
\sqrt{2} Y_2 Y_{\mathbf{5},1}^{(2)}+\sqrt{3} \left(Y_6 Y_{\mathbf{5},2}^{(2)}-\sqrt{2} Y_5 Y_{\mathbf{5},3}^{(2)}+\sqrt{2} Y_4 Y_{\mathbf{5},4}^{(2)}-Y_3 Y_{\mathbf{5},5}^{(2)}\right) \\
\sqrt{3} \left(Y_1 Y_{\mathbf{5},2}^{(2)}-Y_2 Y_{\mathbf{5},2}^{(2)}+\sqrt{2} Y_4 Y_{\mathbf{5},5}^{(2)}\right)-2\sqrt{2} Y_3 Y_{\mathbf{5},1}^{(2)} \\
\sqrt{2} Y_4 Y_{\mathbf{5},1}^{(2)}+\sqrt{6} \left(Y_3 Y_{\mathbf{5},2}^{(2)}+\left(Y_2-Y_1\right) Y_{\mathbf{5},3}^{(2)}\right) \\
\sqrt{2} Y_5 Y_{\mathbf{5},1}^{(2)}-\sqrt{6} \left(Y_1 Y_{\mathbf{5},4}^{(2)}+Y_2 Y_{\mathbf{5},4}^{(2)}-Y_6 Y_{\mathbf{5},5}^{(2)}\right) \\
\sqrt{3} \left(\sqrt{2} Y_5 Y_{\mathbf{5},2}^{(2)}+\left(Y_1+Y_2\right) Y_{\mathbf{5},5}^{(2)}\right)-2\sqrt{2} Y_6 Y_{\mathbf{5},1}^{(2)} \\
\end{pmatrix}\,,
  \end{split}
\end{equation}
where we denote the notation $Y_{\mathbf{3}}^{(2)}\equiv (Y_{\mathbf{3}, 1}^{(2)}, Y_{\mathbf{3}, 2}^{(2)}, Y_{\mathbf{3}, 3}^{(2)})^{T}$, $Y_{\mathbf{3}'}^{(2)}\equiv (Y_{\mathbf{3}', 1}^{(2)}, Y_{\mathbf{3}', 2}^{(2)}, Y_{\mathbf{3}', 3}^{(2)})^{T}$ and $Y_{\mathbf{5}}^{(2)}\equiv (Y_{\mathbf{5}, 1}^{(2)}, Y_{\mathbf{5}, 2}^{(2)}, Y_{\mathbf{5}, 3}^{(2)}, Y_{\mathbf{5}, 4}^{(2)}, Y_{\mathbf{5}, 5}^{(2)})^{T}$, and similar notations are adopted for the modular forms in the following.

The weight 4 modular multiplets can be generated from the tensor product of $Y^{(1)}_{\mathbf{6}}$ and the modular forms of weight 3. We find there are 21 linearly independent modular forms which can be chosen to be
\begin{equation}
\begin{split}
Y_{\mathbf{1}}^{(4)}&=(Y_{\mathbf{\widehat{6}}}^{(1)}Y_{\mathbf{\widehat{6}}I}^{(3)})_{\mathbf{1}_{a}}=Y_2 Y_{\mathbf{6}II,1}^{(3)}-Y_1 Y_{\mathbf{6}II,2}^{(3)}+Y_6 Y_{\mathbf{6}II,3}^{(3)}-Y_3 Y_{\mathbf{6}II,6}^{(3)}+Y_5 Y_{\mathbf{6}II,4}^{(3)}-Y_4 Y_{\mathbf{6}II,5}^{(3)}\,, \\
Y_{\mathbf{3}}^{(4)}&=(Y_{\mathbf{\widehat{6}}}^{(1)}Y_{\mathbf{\widehat{6}}II}^{(3)})_{\mathbf{3}_{1,s}}=
\begin{pmatrix}
-Y_1Y_{\mathbf{6}II,2}^{(3)}-Y_2Y_{\mathbf{6}II,1}^{(3)}+Y_3Y_{\mathbf{6}II,6}^{(3)}-Y_4Y_{\mathbf{6}II,5}^{(3)}-Y_5Y_{\mathbf{6}II,4}^{(3)}+Y_6Y_{\mathbf{6}II,3}^{(3)}\\
-\sqrt{2}Y_2Y_{\mathbf{6}II,3}^{(3)}-\sqrt{2}Y_3Y_{\mathbf{6}II,2}^{(3)}+\sqrt{2}Y_5Y_{\mathbf{6}II,5}^{(3)}\\
-\sqrt{2}Y_1Y_{\mathbf{6}II,6}^{(3)}-\sqrt{2}Y_4Y_{\mathbf{6}II,4}^{(3)}-\sqrt{2}Y_6Y_{\mathbf{6}II,1}^{(3)}
\end{pmatrix}\,,\\
Y_{\mathbf{3'}}^{(4)}&=(Y_{\mathbf{\widehat{6}}}^{(1)}Y_{\mathbf{\widehat{6}}I}^{(3)})_{\mathbf{3'}_{2, s}}=
\begin{pmatrix}
 -Y_1 Y_{\mathbf{6}I,1}^{(3)}+Y_2 Y_{\mathbf{6}I,2}^{(3)}-Y_6 Y_{\mathbf{6}I,3}^{(3)}-Y_5 Y_{\mathbf{6}I,4}^{(3)}-Y_4 Y_{\mathbf{6}I,5}^{(3)}-Y_3 Y_{\mathbf{6}I,6}^{(3)} \\
 Y_4 \left(Y_{\mathbf{6}I,2}^{(3)}-Y_{\mathbf{6}I,1}^{(3)}\right)-\sqrt{2} Y_3 Y_{\mathbf{6}I,3}^{(3)}+\left(Y_2-Y_1\right) Y_{\mathbf{6}I,4}^{(3)} \\
 Y_5 \left(Y_{\mathbf{6}I,1}^{(3)}+Y_{\mathbf{6}I,2}^{(3)}\right)+(Y_1+Y_2) Y_{\mathbf{6}I,5}^{(3)}+\sqrt{2} Y_6 Y_{\mathbf{6}I,6}^{(3)} \\
\end{pmatrix} \,, \\
Y_{\mathbf{4}}^{(4)}&=(Y_{\mathbf{\widehat{6}}}^{(1)}Y_{\mathbf{\widehat{6}}I}^{(3)})_{\mathbf{4}_{a}}=
\begin{pmatrix}
 -\sqrt{2} Y_3 Y_{\mathbf{6}I,2}^{(3)}+\sqrt{2} Y_2 Y_{\mathbf{6}I,3}^{(3)}-Y_6 Y_{\mathbf{6}I,4}^{(3)}+Y_4 Y_{\mathbf{6}I,6}^{(3)} \\
 Y_4 \left(Y_{\mathbf{6}I,1}^{(3)}-Y_{\mathbf{6}I,2}^{(3)}\right)-Y_1 Y_{\mathbf{6}I,4}^{(3)}+Y_2 Y_{\mathbf{6}I,4}^{(3)}+Y_6 Y_{\mathbf{6}I,5}^{(3)}-Y_5 Y_{\mathbf{6}I,6}^{(3)} \\
 Y_5 \left(Y_{\mathbf{6}I,1}^{(3)}+Y_{\mathbf{6}I,2}^{(3)}\right)-Y_4 Y_{\mathbf{6}I,3}^{(3)}+Y_3 Y_{\mathbf{6}I,4}^{(3)}-Y_1 Y_{\mathbf{6}I,5}^{(3)}-Y_2 Y_{\mathbf{6}I,5}^{(3)} \\
\sqrt{2} Y_6 Y_{\mathbf{6}I,1}^{(3)}-Y_5 Y_{\mathbf{6}I,3}^{(3)}+Y_3 Y_{\mathbf{6}I,5}^{(3)}-\sqrt{2} Y_1 Y_{\mathbf{6}I,6}^{(3)} \\
\end{pmatrix}  \,, \\
Y_{\mathbf{5}I}^{(4)}&=(Y_{\mathbf{\widehat{6}}}^{(1)}Y_{\mathbf{\widehat{6}}I}^{(3)})_{\mathbf{5}_{1, s}}=
\begin{pmatrix}
 -\sqrt{6} \left(Y_1 Y_{\mathbf{6}I,1}^{(3)}+Y_2 Y_{\mathbf{6}I,2}^{(3)}\right) \\
 Y_3 \left(Y_{\mathbf{6}I,1}^{(3)}+Y_{\mathbf{6}I,2}^{(3)}\right)+Y_1 Y_{\mathbf{6}I,3}^{(3)}+Y_2 Y_{\mathbf{6}I,3}^{(3)}+\sqrt{2} Y_6 Y_{\mathbf{6}I,4}^{(3)}+2 Y_5 Y_{\mathbf{6}I,5}^{(3)}+\sqrt{2} Y_4 Y_{\mathbf{6}I,6}^{(3)} \\
\sqrt{2} Y_4 Y_{\mathbf{6}I,2}^{(3)}+2 Y_3 Y_{\mathbf{6}I,3}^{(3)}+\sqrt{2} \left(Y_2 Y_{\mathbf{6}I,4}^{(3)}+Y_6 Y_{\mathbf{6}I,5}^{(3)}+Y_5 Y_{\mathbf{6}I,6}^{(3)}\right) \\
 -\sqrt{2} Y_5 Y_{\mathbf{6}I,1}^{(3)}+\sqrt{2} Y_4 Y_{\mathbf{6}I,3}^{(3)}+\sqrt{2} Y_3 Y_{\mathbf{6}I,4}^{(3)}-\sqrt{2} Y_1 Y_{\mathbf{6}I,5}^{(3)}+2 Y_6 Y_{\mathbf{6}I,6}^{(3)} \\
 Y_6 \left(Y_{\mathbf{6}I,2}^{(3)}-Y_{\mathbf{6}I,1}^{(3)}\right)-\sqrt{2} Y_5 Y_{\mathbf{6}I,3}^{(3)}+2 Y_4 Y_{\mathbf{6}I,4}^{(3)}-\sqrt{2} Y_3 Y_{\mathbf{6}I,5}^{(3)}-Y_1 Y_{\mathbf{6}I,6}^{(3)}+Y_2 Y_{\mathbf{6}I,6}^{(3)} \\
\end{pmatrix} \,, \\
Y_{\mathbf{5}II}^{(4)}&=(Y_{\mathbf{\widehat{6}}}^{(1)}Y_{\mathbf{\widehat{6}}I}^{(3)})_{\mathbf{5}_{2, a}}=
\begin{pmatrix}
 -Y_2 Y_{\mathbf{6}I,1}^{(3)}+Y_1 Y_{\mathbf{6}I,2}^{(3)}-Y_6 Y_{\mathbf{6}I,3}^{(3)}+2 Y_5 Y_{\mathbf{6}I,4}^{(3)}-2 Y_4 Y_{\mathbf{6}I,5}^{(3)}+Y_3 Y_{\mathbf{6}I,6}^{(3)} \\
\sqrt{6} \left(Y_1 Y_{\mathbf{6}I,3}^{(3)}-Y_3 Y_{\mathbf{6}I,1}^{(3)}\right) \\
\sqrt{3} \left(Y_4 Y_{\mathbf{6}I,2}^{(3)}-Y_2 Y_{\mathbf{6}I,4}^{(3)}+Y_6 Y_{\mathbf{6}I,5}^{(3)}-Y_5 Y_{\mathbf{6}I,6}^{(3)}\right) \\
\sqrt{3} \left(-Y_5 Y_{\mathbf{6}I,1}^{(3)}-Y_4 Y_{\mathbf{6}I,3}^{(3)}+Y_3 Y_{\mathbf{6}I,4}^{(3)}+Y_1 Y_{\mathbf{6}I,5}^{(3)}\right) \\
\sqrt{6} \left(Y_2 Y_{\mathbf{6}I,6}^{(3)}-Y_6 Y_{\mathbf{6}I,2}^{(3)}\right) \\
\end{pmatrix}\,.
\end{split}\nonumber
\end{equation}
Notice that each entry of these weight four modular forms can be expressed as quartic polynomial of $Y_{1,2,3,4,5,6}$. In a same manner, we can obtain the weight 5 modular multiplets as follow,
\begin{equation}
\begin{split}
Y_{\mathbf{\widehat{2}}}^{(5)}&=(Y_{\mathbf{\widehat{6}}}^{(1)}Y_{\mathbf{5}II}^{(4)})_{\mathbf{2}}=
\begin{pmatrix}
\sqrt{3} Y_4 Y_{\mathbf{5}II,1}^{(4)}-2 Y_3 Y_{\mathbf{5}II,2}^{(4)}+Y_1 Y_{\mathbf{5}II,3}^{(4)}+2 Y_2 Y_{\mathbf{5}II,3}^{(4)}+Y_6 Y_{\mathbf{5}II,4}^{(4)}-\sqrt{2} Y_5 Y_{\mathbf{5}II,5}^{(4)} \\
 -\sqrt{3} Y_5 Y_{\mathbf{5}II,1}^{(4)}-\sqrt{2} Y_4 Y_{\mathbf{5}II,2}^{(4)}+Y_3 Y_{\mathbf{5}II,3}^{(4)}+2 Y_1 Y_{\mathbf{5}II,4}^{(4)}-Y_2 Y_{\mathbf{5}II,4}^{(4)}+2 Y_6 Y_{\mathbf{5}II,5}^{(4)} \\
\end{pmatrix} \,,\\
Y_{\mathbf{\widehat{2}'}}^{(5)}&=(Y_{\mathbf{\widehat{6}}}^{(1)}Y_{\mathbf{5}II}^{(4)})_{\mathbf{2'}}=
\begin{pmatrix}
\sqrt{6} Y_3 Y_{\mathbf{5}II,1}^{(4)}+3 Y_1 Y_{\mathbf{5}II,2}^{(4)}+Y_2 Y_{\mathbf{5}II,2}^{(4)}-2 Y_6 Y_{\mathbf{5}II,3}^{(4)}+2\sqrt{2} Y_5 Y_{\mathbf{5}II,4}^{(4)}+\sqrt{2} Y_4 Y_{\mathbf{5}II,5}^{(4)} \\
\sqrt{6} Y_6 Y_{\mathbf{5}II,1}^{(4)}+\sqrt{2} Y_5 Y_{\mathbf{5}II,2}^{(4)}-2\sqrt{2} Y_4 Y_{\mathbf{5}II,3}^{(4)}+2 Y_3 Y_{\mathbf{5}II,4}^{(4)}-Y_1 Y_{\mathbf{5}II,5}^{(4)}+3 Y_2 Y_{\mathbf{5}II,5}^{(4)} \\
\end{pmatrix}\,, \\
Y_{\mathbf{\widehat{4}'}}^{(5)}&=(Y_{\mathbf{\widehat{6}}}^{(1)}Y_{\mathbf{5}II}^{(4)})_{\mathbf{4'}_{1}}=
\begin{pmatrix}
\sqrt{6} Y_3 Y_{\mathbf{5}II,1}^{(4)}-2 Y_2 Y_{\mathbf{5}II,2}^{(4)}+Y_6 Y_{\mathbf{5}II,3}^{(4)}-\sqrt{2} Y_5 Y_{\mathbf{5}II,4}^{(4)}+\sqrt{2} Y_4 Y_{\mathbf{5}II,5}^{(4)} \\
\sqrt{3} \left(Y_1 Y_{\mathbf{5}II,3}^{(4)}+Y_2 Y_{\mathbf{5}II,3}^{(4)}-Y_6 Y_{\mathbf{5}II,4}^{(4)}+\sqrt{2} Y_5 Y_{\mathbf{5}II,5}^{(4)}\right) \\
\sqrt{3} \left(\sqrt{2} Y_4 Y_{\mathbf{5}II,2}^{(4)}-Y_3 Y_{\mathbf{5}II,3}^{(4)}+\left(Y_1-Y_2\right) Y_{\mathbf{5}II,4}^{(4)}\right) \\
 -\sqrt{6} Y_6 Y_{\mathbf{5}II,1}^{(4)}-\sqrt{2} Y_5 Y_{\mathbf{5}II,2}^{(4)}-\sqrt{2} Y_4 Y_{\mathbf{5}II,3}^{(4)}+Y_3 Y_{\mathbf{5}II,4}^{(4)}-2 Y_1 Y_{\mathbf{5}II,5}^{(4)} \\
\end{pmatrix}\,, \\
Y_{\mathbf{\widehat{6}}I}^{(5)}&=(Y_{\mathbf{\widehat{6}}}^{(1)}Y_{\mathbf{1}}^{(4)})_{\mathbf{6}}=
\begin{pmatrix}
 Y_1 Y_{\mathbf{1},1}^{(4)} \\
 Y_2 Y_{\mathbf{1},1}^{(4)} \\
 Y_3 Y_{\mathbf{1},1}^{(4)} \\
 Y_4 Y_{\mathbf{1},1}^{(4)} \\
 Y_5 Y_{\mathbf{1},1}^{(4)} \\
 Y_6 Y_{\mathbf{1},1}^{(4)} \\
\end{pmatrix}\,, \quad\quad
Y_{\mathbf{\widehat{6}}II}^{(5)}=(Y_{\mathbf{\widehat{6}}}^{(1)}Y_{\mathbf{3'}}^{(4)})_{\mathbf{6}_{1}}=
\begin{pmatrix}
 Y_1 Y_{\mathbf{3}',1}^{(4)}-Y_4 Y_{\mathbf{3}',3}^{(4)} \\
 Y_5 Y_{\mathbf{3}',2}^{(4)}-Y_2 Y_{\mathbf{3}',1}^{(4)} \\
 Y_3 Y_{\mathbf{3}',1}^{(4)}+Y_5 Y_{\mathbf{3}',3}^{(4)} \\
 Y_6 Y_{\mathbf{3}',3}^{(4)}-Y_1 Y_{\mathbf{3}',2}^{(4)} \\
 Y_3 Y_{\mathbf{3}',2}^{(4)}+Y_2 Y_{\mathbf{3}',3}^{(4)} \\
 Y_4 Y_{\mathbf{3}',2}^{(4)}-Y_6 Y_{\mathbf{3}',1}^{(4)} \\
\end{pmatrix} \,, \\
Y_{\mathbf{\widehat{6}}III}^{(5)}&=(Y_{\mathbf{\widehat{6}}}^{(1)}Y_{\mathbf{5}II}^{(4)})_{\mathbf{6}_{2}}=
\begin{pmatrix}
\sqrt{3} Y_1 Y_{\mathbf{5}II,1}^{(4)}-2 Y_5 Y_{\mathbf{5}II,3}^{(4)}-Y_4 Y_{\mathbf{5}II,4}^{(4)}+\sqrt{2} Y_3 Y_{\mathbf{5}II,5}^{(4)} \\
\sqrt{3} Y_2 Y_{\mathbf{5}II,1}^{(4)}+\sqrt{2} Y_6 Y_{\mathbf{5}II,2}^{(4)}-Y_5 Y_{\mathbf{5}II,3}^{(4)}+2 Y_4 Y_{\mathbf{5}II,4}^{(4)} \\
 -\sqrt{3} Y_3 Y_{\mathbf{5}II,1}^{(4)}+\sqrt{2} Y_1 Y_{\mathbf{5}II,2}^{(4)}-Y_5 Y_{\mathbf{5}II,4}^{(4)}+2 Y_4 Y_{\mathbf{5}II,5}^{(4)} \\
 2 Y_3 Y_{\mathbf{5}II,2}^{(4)}-Y_1 Y_{\mathbf{5}II,3}^{(4)}+2 Y_2 Y_{\mathbf{5}II,3}^{(4)}+Y_6 Y_{\mathbf{5}II,4}^{(4)} \\
 -Y_3 Y_{\mathbf{5}II,3}^{(4)}-2 Y_1 Y_{\mathbf{5}II,4}^{(4)}-Y_2 Y_{\mathbf{5}II,4}^{(4)}+2 Y_6 Y_{\mathbf{5}II,5}^{(4)} \\
 -\sqrt{3} Y_6 Y_{\mathbf{5}II,1}^{(4)}+2 Y_5 Y_{\mathbf{5}II,2}^{(4)}+Y_4 Y_{\mathbf{5}II,3}^{(4)}+\sqrt{2} Y_2 Y_{\mathbf{5}II,5}^{(4)}
\end{pmatrix}\,.
  \end{split}\nonumber
\end{equation}
In the end, we give the linearly independent weight 6 modular multiplets of level 5,
\begin{align*}
Y_{\mathbf{1}}^{(6)}&=(Y_{\mathbf{\widehat{6}}}^{(1)}Y_{\mathbf{\widehat{6}}II}^{(5)})_{\mathbf{1}_{a}}=
 Y_2 Y_{\mathbf{6}II,1}^{(5)}-Y_1 Y_{\mathbf{6}II,2}^{(5)}+Y_6 Y_{\mathbf{6}II,3}^{(5)}+Y_5 Y_{\mathbf{6}II,4}^{(5)}-Y_4 Y_{\mathbf{6}II,5}^{(5)}-Y_3 Y_{\mathbf{6}II,6}^{(5)}\,, \\
Y_{\mathbf{3}I}^{(6)}&=(Y_{\mathbf{\widehat{6}}}^{(1)}Y_{\mathbf{\widehat{6}}I}^{(5)})_{\mathbf{3}_{1, s}}=
\begin{pmatrix}
 -Y_2 Y_{\mathbf{6}I,1}^{(5)}-Y_1 Y_{\mathbf{6}I,2}^{(5)}+Y_6 Y_{\mathbf{6}I,3}^{(5)}-Y_5 Y_{\mathbf{6}I,4}^{(5)}-Y_4 Y_{\mathbf{6}I,5}^{(5)}+Y_3 Y_{\mathbf{6}I,6}^{(5)} \\
 -\sqrt{2} \left(Y_3 Y_{\mathbf{6}I,2}^{(5)}+Y_2 Y_{\mathbf{6}I,3}^{(5)}-Y_5 Y_{\mathbf{6}I,5}^{(5)}\right) \\
 -\sqrt{2} \left(Y_6 Y_{\mathbf{6}I,1}^{(5)}+Y_4 Y_{\mathbf{6}I,4}^{(5)}+Y_1 Y_{\mathbf{6}I,6}^{(5)}\right) \\
\end{pmatrix}\,,\\
Y_{\mathbf{3}II}^{(6)}&=(Y_{\mathbf{\widehat{6}}}^{(1)}Y_{\mathbf{\widehat{6}}III}^{(5)})_{\mathbf{3}_{1, s}}=
\begin{pmatrix}
 -Y_2 Y_{\mathbf{6}III,1}^{(5)}-Y_1 Y_{\mathbf{6}III,2}^{(5)}+Y_6 Y_{\mathbf{6}III,3}^{(5)}-Y_5 Y_{\mathbf{6}III,4}^{(5)}-Y_4 Y_{\mathbf{6}III,5}^{(5)}+Y_3 Y_{\mathbf{6}III,6}^{(5)} \\
 -\sqrt{2} \left(Y_3 Y_{\mathbf{6}III,2}^{(5)}+Y_2 Y_{\mathbf{6}III,3}^{(5)}-Y_5 Y_{\mathbf{6}III,5}^{(5)}\right) \\
 -\sqrt{2} \left(Y_6 Y_{\mathbf{6}III,1}^{(5)}+Y_4 Y_{\mathbf{6}III,4}^{(5)}+Y_1 Y_{\mathbf{6}III,6}^{(5)}\right) \\
\end{pmatrix}\,,\\
Y_{\mathbf{3'}I}^{(6)}&=(Y_{\mathbf{\widehat{6}}}^{(1)}Y_{\mathbf{\widehat{6}}I}^{(5)})_{\mathbf{3'}_{1, s}}=
\begin{pmatrix}
 Y_2 Y_{\mathbf{6}I,1}^{(5)}+Y_1 Y_{\mathbf{6}I,2}^{(5)}+Y_6 Y_{\mathbf{6}I,3}^{(5)}+Y_3 Y_{\mathbf{6}I,6}^{(5)} \\
 -Y_4 Y_{\mathbf{6}I,2}^{(5)}-Y_2 Y_{\mathbf{6}I,4}^{(5)}+Y_6 Y_{\mathbf{6}I,5}^{(5)}+Y_5 Y_{\mathbf{6}I,6}^{(5)} \\
 -Y_5 Y_{\mathbf{6}I,1}^{(5)}-Y_4 Y_{\mathbf{6}I,3}^{(5)}-Y_3 Y_{\mathbf{6}I,4}^{(5)}-Y_1 Y_{\mathbf{6}I,5}^{(5)} \\
\end{pmatrix}\,,\\
Y_{\mathbf{3'}II}^{(6)}&=(Y_{\mathbf{\widehat{6}}}^{(1)}Y_{\mathbf{\widehat{6}}III}^{(5)})_{\mathbf{3'}_{1, s}}=
\begin{pmatrix}
 Y_2 Y_{\mathbf{6}III,1}^{(5)}+Y_1 Y_{\mathbf{6}III,2}^{(5)}+Y_6 Y_{\mathbf{6}III,3}^{(5)}+Y_3 Y_{\mathbf{6}III,6}^{(5)} \\
 -Y_4 Y_{\mathbf{6}III,2}^{(5)}-Y_2 Y_{\mathbf{6}III,4}^{(5)}+Y_6 Y_{\mathbf{6}III,5}^{(5)}+Y_5 Y_{\mathbf{6}III,6}^{(5)} \\
 -Y_5 Y_{\mathbf{6}III,1}^{(5)}-Y_4 Y_{\mathbf{6}III,3}^{(5)}-Y_3 Y_{\mathbf{6}III,4}^{(5)}-Y_1 Y_{\mathbf{6}III,5}^{(5)} \\
\end{pmatrix}\,, \\
Y_{\mathbf{4}I}^{(6)}&=(Y_{\mathbf{\widehat{6}}}^{(1)}Y_{\mathbf{\widehat{2}'}}^{(5)})_{\mathbf{4}}=
\begin{pmatrix}
 Y_4 Y_{\mathbf{2}',2}^{(5)}-\sqrt{2} Y_1 Y_{\mathbf{2}',1}^{(5)} \\
\sqrt{2} Y_3 Y_{\mathbf{2}',1}^{(5)}+Y_5 Y_{\mathbf{2}',2}^{(5)} \\
 Y_4 Y_{\mathbf{2}',1}^{(5)}+\sqrt{2} Y_6 Y_{\mathbf{2}',2}^{(5)} \\
 -Y_5 Y_{\mathbf{2}',1}^{(5)}-\sqrt{2} Y_2 Y_{\mathbf{2}',2}^{(5)} \\
\end{pmatrix}\,, \\
Y_{\mathbf{4}II}^{(6)}&=(Y_{\mathbf{\widehat{6}}}^{(1)}Y_{\mathbf{\widehat{4}'}}^{(5)})_{\mathbf{4}_{1}}=
\begin{pmatrix}
 Y_1 Y_{\mathbf{4}',1}^{(5)}-\sqrt{3} Y_6 Y_{\mathbf{4}',2}^{(5)}-\sqrt{2} Y_4 Y_{\mathbf{4}',4}^{(5)} \\
 -Y_3 Y_{\mathbf{4}',1}^{(5)}-\sqrt{3} Y_2 Y_{\mathbf{4}',2}^{(5)}-\sqrt{2} Y_5 Y_{\mathbf{4}',4}^{(5)} \\
\sqrt{2} Y_4 Y_{\mathbf{4}',1}^{(5)}-\sqrt{3} Y_1 Y_{\mathbf{4}',3}^{(5)}+Y_6 Y_{\mathbf{4}',4}^{(5)} \\
 -\sqrt{2} Y_5 Y_{\mathbf{4}',1}^{(5)}-\sqrt{3} Y_3 Y_{\mathbf{4}',3}^{(5)}-Y_2 Y_{\mathbf{4}',4}^{(5)} \\
\end{pmatrix}\,, \\
Y_{\mathbf{5}I}^{(6)}&=(Y_{\mathbf{\widehat{6}}}^{(1)}Y_{\mathbf{\widehat{4}'}}^{(5)})_{\mathbf{5}_{1}}=
\begin{pmatrix}
\sqrt{6} \left(Y_5 Y_{\mathbf{4}',2}^{(5)}+Y_4 Y_{\mathbf{4}',3}^{(5)}\right) \\
\sqrt{6} Y_2 Y_{\mathbf{4}',1}^{(5)}+\sqrt{2} Y_6 Y_{\mathbf{4}',2}^{(5)}+Y_5 Y_{\mathbf{4}',3}^{(5)}-\sqrt{3} Y_4 Y_{\mathbf{4}',4}^{(5)} \\
\sqrt{2} \left(-\sqrt{3} Y_3 Y_{\mathbf{4}',1}^{(5)}+Y_1 Y_{\mathbf{4}',2}^{(5)}+Y_2 Y_{\mathbf{4}',2}^{(5)}+Y_6 Y_{\mathbf{4}',3}^{(5)}\right) \\
\sqrt{2} \left(-Y_3 Y_{\mathbf{4}',2}^{(5)}+Y_1 Y_{\mathbf{4}',3}^{(5)}-Y_2 Y_{\mathbf{4}',3}^{(5)}+\sqrt{3} Y_6 Y_{\mathbf{4}',4}^{(5)}\right) \\
 -\sqrt{3} Y_5 Y_{\mathbf{4}',1}^{(5)}-Y_4 Y_{\mathbf{4}',2}^{(5)}+\sqrt{2} Y_3 Y_{\mathbf{4}',3}^{(5)}+\sqrt{6} Y_1 Y_{\mathbf{4}',4}^{(5)} \\
\end{pmatrix}\,, \\
Y_{\mathbf{5}II}^{(6)}&=(Y_{\mathbf{\widehat{6}}}^{(1)}Y_{\mathbf{\widehat{6}}II}^{(5)})_{\mathbf{5}_{2, a}}=
\begin{pmatrix}
 -Y_2 Y_{\mathbf{6}II,1}^{(5)}+Y_1 Y_{\mathbf{6}II,2}^{(5)}-Y_6 Y_{\mathbf{6}II,3}^{(5)}+2 Y_5 Y_{\mathbf{6}II,4}^{(5)}-2 Y_4 Y_{\mathbf{6}II,5}^{(5)}+Y_3 Y_{\mathbf{6}II,6}^{(5)} \\
\sqrt{6} \left(Y_1 Y_{\mathbf{6}II,3}^{(5)}-Y_3 Y_{\mathbf{6}II,1}^{(5)}\right) \\
\sqrt{3} \left(Y_4 Y_{\mathbf{6}II,2}^{(5)}-Y_2 Y_{\mathbf{6}II,4}^{(5)}+Y_6 Y_{\mathbf{6}II,5}^{(5)}-Y_5 Y_{\mathbf{6}II,6}^{(5)}\right) \\
\sqrt{3} \left(-Y_5 Y_{\mathbf{6}II,1}^{(5)}-Y_4 Y_{\mathbf{6}II,3}^{(5)}+Y_3 Y_{\mathbf{6}II,4}^{(5)}+Y_1 Y_{\mathbf{6}II,5}^{(5)}\right) \\
\sqrt{6} \left(Y_2 Y_{\mathbf{6}II,6}^{(5)}-Y_6 Y_{\mathbf{6}II,2}^{(5)}\right)
\end{pmatrix}\,.
\end{align*}
We see that the linear space of modular forms of level 5 and weight 6 has dimension 31 which is in agreement with the dimension formula $5k+1=5\times6+1=31$. The above modular multiplets and their representations under $A'_5$ are listed in table~\ref{tab:MF-L5-W6}.

For higher level $N=6$ and $N=7$, the interested readers can refer to Refs.~\cite{Li:2021buv,Ding:2020msi} for the details of relevant group theory and modular forms.

\end{appendix}

\clearpage


\providecommand{\href}[2]{#2}\begingroup\raggedright\endgroup

\end{document}